\documentclass[smallextended]{svjour3}
\usepackage[utf8]{inputenc}
\usepackage{aas_macros}
\usepackage{color}
\usepackage{amsmath}
\usepackage{amssymb}
\usepackage{graphicx}
\usepackage[]{natbib}
\usepackage{hyperref}

\newcommand\ull[1]{\bar{\bar{#1}}}
\newcommand{\bx}{{\bf x}}
\newcommand{\bv}{{\bf v}}
\newcommand{\bp}{{\bf p}}
\newcommand{\bF}{{\bf F}}
\newcommand{\bE}{{\bf E}}
\newcommand{\bB}{{\bf B}}
\newcommand{\bJ}{{\bf J}}

\begin{document}
\title{Multi-scale simulations of particle acceleration in astrophysical systems}
\titlerunning{Multi-scale particle acceleration studies}
\author{A.~Marcowith$^1$ \and G.~Ferrand$^2$ \and M.~Grech$^3$ \and Z.~Meliani$^4$ \and I.~Plotnikov$^5$ \and R.~Walder$^6$}
\institute {1: Laboratoire Univers et Particules de Montpellier (LUPM) Universit\'e Montpellier, CNRS/IN2P3, CC72, place Eug\`ene Bataillon, 34095, Montpellier Cedex 5, France 
\and 2: Astrophysical Big Bang Laboratory (ABBL), RIKEN Cluster for Pioneering Research, and Interdisciplinary Theoretical and Mathematical Sciences Program (iTHEMS), RIKEN, Wak\={o}, Saitama, 351-0198 Japan
\and 3: Laboratoire d'Utilisation des Lasers Intenses (LULI), CNRS, Ecole Polytechnique, Sorbonne Universit\'e, CEA, F-91128 Palaiseau cedex, France 
\and 4: LUTH, Observatoire de Paris, PSL Research University, CNRS UMR 8102, Universit\'e Paris Diderot, Meudon, France 
\and 5: IRAP, Universit\'e de Toulouse, UPS-OMP, 9 av. Colonel Roche, BP 44346, F-31028 Toulouse Cedex 4, France 
\and 6: Ecole Normale Sup\'erieur de Lyon, Universite de Lyon, CNRS, Centre de Recherche d'Astrophysique de Lyon (CRAL), UMR 5574 }

\maketitle

\begin{abstract}
This review aims at providing an up-to-date status and a general introduction to the subject of the numerical study of energetic particle acceleration and transport in turbulent astrophysical flows. The subject is also complemented by a short overview of recent progresses obtained in the domain of laser plasma experiments. We review the main physical processes at the heart of the production of a non-thermal distribution in both Newtonian and relativistic astrophysical flows, namely the first and second order Fermi acceleration processes. We also discuss shock drift and surfing acceleration, two processes important in the context of particle injection in shock acceleration. We analyze with some details the particle-in-cell (PIC) approach used to describe particle kinetics. We review the main results obtained with PIC simulations in the recent years concerning particle acceleration at shocks and in reconnection events. The review discusses the solution of Fokker-Planck problems with application to the study of particle acceleration at shocks but also in hot coronal plasmas surrounding compact objects. We continue by considering large scale physics. We describe recent developments in magnetohydrodynamic (MHD) simulations. We give a special emphasize on the way energetic particle dynamics can be coupled to MHD solutions either using a multi-fluid calculation or directly coupling kinetic and fluid calculations. This aspect is mandatory to investigate the acceleration of particles in the deep relativistic regimes to explain the highest Cosmic Ray energies.  
\end{abstract}

\newpage
\section{Introduction}\label{S:INT}
Particle acceleration is a widespread process in astrophysical, space and laser plasmas. Acceleration results from the effect of electric fields, but supra-thermal particles gain energy because their residence time in the acceleration zone increases due to magnetic confinement. Hence, particle acceleration is an electromagnetic process. Particle acceleration can be classified into three main sub-types \citep{1994ApJS...90..515B, 1994plas.conf..225K, 1996Ap&SS.242..209M}: acceleration at flow discontinuities among which shock waves, stochastic acceleration, acceleration by direct electric fields. This last mechanism occurs in the environment of fast rotating magnetized objects like pulsars or planetary magnetosphere. It will not be discussed in this review, interested readers can refer to \citet{2017SSRv..207..111C} and references therein for what concerns pulsar magnetospheric physics. Some recent discussions concerning particle acceleration in Jupiter, the fastest rotator among solar system planets and other giant planet magnetospheres can be found in \citet{2017GeoRL..44.4410M, 2015SSRv..187...51D}.

As stated above acceleration processes can be classified into different categories. Let us give here a short overview of the main mechanisms.
\begin{enumerate}

\item \underline{Stochastic Fermi acceleration} (SFA). This is historically the first discovered acceleration process. SFA is at the heart of Fermi's work on the origin of cosmic rays (CRs) \citep{1949PhRv...75.1169F, 1954ApJ...119....1F}. In its original description particles with speed v gain energy through a stochastic interaction with the convective electric field carried by magnetic clouds moving randomly at a speed $U \ll v$ in the interstellar space (see Sect.~\ref{S:DSA-Fermi}). This process has some well-known caveats: it is inefficient as the mean relative energy gain $\langle \Delta E/E \rangle \propto (U/v)^2$ \citep{1958PhRv..109.1328P}, it produces non-universal power-law solutions (see Eq.~\ref{eq:powerlaw}) which can not explain the power-law distribution of CRs observed at Earth. A modern description of SFA includes randomly moving electromagnetic waves \citep{1967PhFl...10.2620H}, usually in the magnetohydrodynamic (MHD) limit if we want to consider the issue of CR acceleration \citep{1955PhRv...99..241P, 1969ApJ...156..445K}. For SFA by MHD waves the relevant speed for the scattering centers usually are proportional to the local Alfv\'en speed $U_{\rm A}$ \footnote{$U_{\rm A}= B/\sqrt{4\pi \rho_{\rm i}}$,  where B, $\rho_{\rm i}$ are respectively the local magnetic field strength, the ion mass density.}. SFA is more efficient if the speed of the scattering center is close to the speed of light \citep{1997A&A...323..271M, 2004A&A...425..395G} or for particles with speeds close to $U_{\rm A}$,  or other plasma characteristic wave phase speeds as it is likely the case for low-energy CRs which propagate in the interstellar medium (ISM), e.g. \citet{2006ApJ...642..902P}, in the solar corona, e.g. \citep{1996ApJ...461..445M, 1998ApJ...495..377P} or in the heliosphere \citep{2013SSRv..176..133Z}. SFA is a multi-scale process if the scattering waves are distributed over large wave-number bands as it is the case in turbulent flows. SFA is an important process at the origin of particle acceleration and gas heating in hot corona which develop around compact objects (see Sect.~\ref{S:KIN}).

To finish, we also mention the betatron-magnetic pumping process \citep{1958PhRv..109.1328P}. In this process particles propagate along a magnetic field slowly varying with time. As the field strength increases, the particle momentum increases due to the conservation of the adiabatic invariant\footnote{$p_\perp = p \sin\alpha$, where p and $\alpha$ are the particle momentum and pitch-angle (the angle between the particle velocity and magnetic field) respectively.} $p_\perp^2/B$. If the particle suffers some scattering of its pitch-angle or some elastic collisions, a decrease of the magnetic field strength while keeping particle isotropization by pitch-angle scattering produces a net gain in energy during a magnetic field variation cycle. 

\item \underline{Shock or shear-flow acceleration}. One way to cure the inefficiency of the SFA is to allow scattering centers to have a mean direction of motion \citep{1958PhRv..109.1328P, 1963ApJ...137..135W}. Then the mean relative energy gain scales as $\langle \Delta E/E \rangle \propto (U/v)$. This case occurs for a shock because of the advection of the scattering centers towards the shock front but also in the configuration of a shearing, for instance in jets. We do not explicitly for now discuss the case of shear-flow acceleration, the interested reader can refer to \citet{2006ApJ...652.1044R}, and we consider hereafter the case of shock acceleration. Shear-flow acceleration will be reviewed in a forthcoming version of the text. The acceleration process is more efficient if particles can reside for a sufficient amount of time around the shock front \citep{1994plas.conf..225K}. In fact, particle acceleration at shock waves covers three basic different processes [see \citet{1994plas.conf..225K, 2008arXiv0806.4046T, 2016RPPh...79d6901M}]: diffusive shock acceleration (DSA), shock drift acceleration (SDA), shock surfing acceleration (SSA), all described below. 

Before, let us introduce some elements of vocabulary associated with different descriptions of the shock front. First, particle acceleration requires the shock to be collisionless, i.e. to be mediated by electromagnetic processes rather than collisions otherwise collisions being the fastest process force the shocked particle distribution to be Maxwellian. This condition is usually fulfilled in astrophysical and space plasmas. At a macroscopic level a shock wave is characterized by a discontinuity in the thermodynamical variables of the flow. A shock occurs when the flow is supersonic, with a sonic Mach number ${\cal M}_{\rm s}=U_{\rm sh}/c_{\rm s} > 1$, where $U_{\rm sh}$ is the shock speed in the upstream medium restframe. Rankine-Hugoniot conditions give the jump in the flow density, velocity, pressure, temperature and entropy at the shock front \citep{2008arXiv0805.2162T}. At a microscopic level a shock front is a complex, dynamical structure where multi-scale instabilities can develop \citep{2016RPPh...79d6901M}. The so-called supercritical magnetized shock front is composed of three sub-structures: the foot, a bump in gas and magnetic field pressures due to the accumulation of ions reflected at the ramp, the ramp which marks the fast rise of the electromagnetic field potential and gas density and finally the overshoot-undershoot produced by the gyromotion of reflected ions moving in the post-shock gas. Shocks in astrophysics also include precursors in the upstream gas which can have very different origins: radiation, mixture of ions and neutrals or CRs. The size of these precursors make them usually impossible to explore numerically with the shock front structure as a single complex dynamical system. 

As stated above particle acceleration in shocks proceeds through three different mechanisms. In DSA particles repeatedly gain energy by crossing the shock front back and forth \citep{1983RPPh...46..973D}. SDA results from the effect of the convective electric field $\vec{E} = -\vec{U}/c \times \vec{B}$ upstream the shock front due to the motion of the flow at a speed $\vec{U}$ \citep{1994plas.conf..225K, 1985JGR....90...47D}. The particle guiding-center drifts due to the effect of the electric field and to the gradient of the magnetic field in the ramp. SSA results from the trapping of the particle at the shock front because of the combined effect of shock potential raise at the ramp and the convective upstream electric field \citep{1966RvPP....4...23S}. We will come back with more details on these mechanisms in Sects.~\ref{S:DSA} and \ref{S:SDASSA}.

\item \underline{Magnetic reconnection} (REC). 
Magnetic reconnection is the process which transfers magnetic field energy into kinetic energy in an explosive event by re-arranging the magnetic field topology. The most simple 2D picture is sketched in the left panel of Fig.~\ref{fig:ReconnectionTopology_Sweet-Parker}. Separatrices (green, dashed lines) divide the 2D plane into 4 different regions: in the left region, the magnetic field connects points A and B. In the right region, the field connects points A' and B' and is oriented in opposite direction to the field in the left region. No field is present in the upper and lower region between the separatrices. In 2D, due to Amp\`ere's law, a current pointing normal to the plane is necessarily present between the oppositely oriented fields. The REC process then leads to a re-arrangement of the field lines, lowering the magnetic energy. The field now connects the points A and A' (B and B' respectively). These field lines are highly bent and will relax, accelerating the plasma upwards and downwards. The term magnetic reconnection was first coined in~\citet{Dungey1958} and was later adopted by the community.  More details about the REC process are provided in Sect.~\ref{S:REC}. 

REC induces a transfer of magnetic energy into: heat, plasma and particle acceleration and hence radiation \citep{2016ASSL..427.....G, 1994plas.conf....1P}. Particle acceleration in reconnection sites can either occur by a direct acceleration in electric fields in the current sheet, or because of Fermi first order acceleration in the plasma converging towards the reconnection zone or if particles are trapped in a contracting magnetic islands \citep{2015ASSL..407..373D}. The physics of particle acceleration in kinetic reconnection is discussed in Sect.~\ref{Sec:Kinetic_Reconnection}.
\end{enumerate}
\begin{figure}
\begin{centering}
\centerline{
\includegraphics[width=1\linewidth]{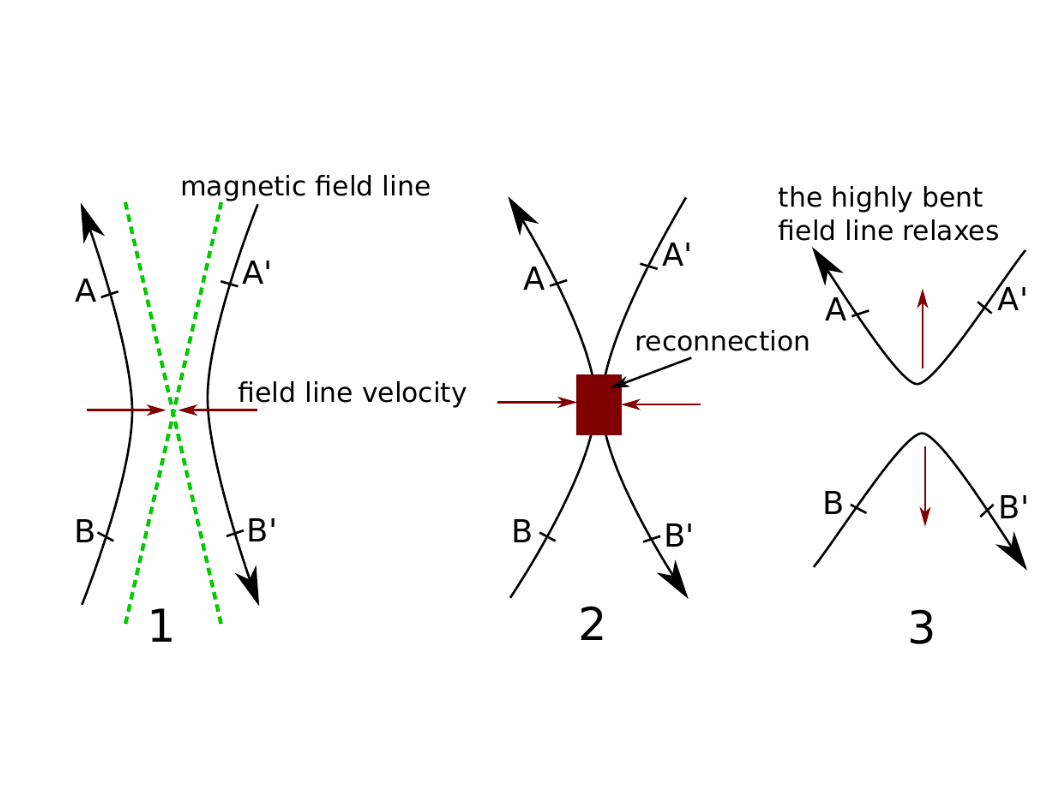}
}
\par\end{centering}
\caption{The re-arrangement of the field topology in magnetic
reconnection in a 2D model. Before the event (Image 1), points A and B
(A' and B' respectively) are located on the same field line. After the
event (3), field lines now are connecting points A and A' (B and B'
respectively). Strongly accelerated outflow is driven in the
directions where the highly bent magnetic field lines are
relaxing. The dashed green lines are called separatrices, lines which
separate the regions of field which are topologically not
connected. (Adapted
from \citet{melzani:tel-01126912}.)}
\label{fig:ReconnectionTopology_Sweet-Parker}
\end{figure}

\begin{figure}
\begin{centering}
\includegraphics[width=1\linewidth]{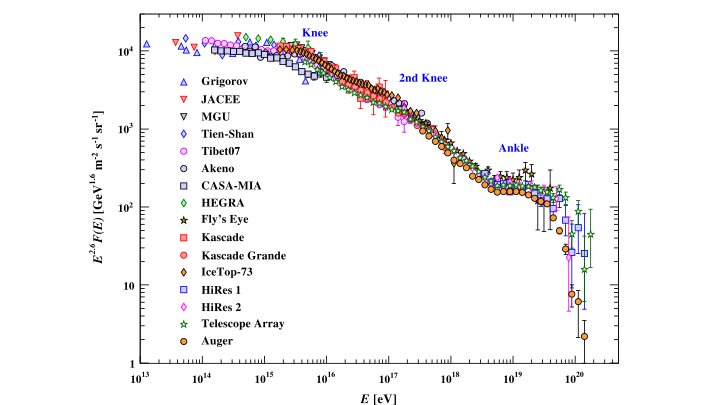}
\par\end{centering}
\caption{The Cosmic Ray spectrum observed at the Earth multiplied by $E^{2.6}$, from \citet{2016ChPhC..40j0001P}  \label{fig:crspectrum}}
\end{figure}

Acceleration mechanisms, as we have seen from the above rapid descriptions, intrinsically involve multi-scale processes which bring particles from the thermal to supra-thermal speeds. In astrophysics these processes have to explain the CR spectrum observed at Earth which extends at least over 15 orders of magnitude in energy (from MeV to ZeV) and more than 30 orders of magnitude in flux (see figure \ref{fig:crspectrum}). Note that in space plasmas, maximum energies reached by the energetic particles are more modest but still supra-thermal, and the particle distributions cover about 5 orders of magnitude (from keV to GeV) (see for instance \citet{2011SSRv..159..357Z} in the context of solar flares). The investigation of particle acceleration then requires different numerical approaches to probe the different inter-connected scales involved in the process of acceleration. Multiple techniques are also required as actually it is not possible to account for such large dynamical spatial, time and energy scales even with modern computers. It is the main object of this review to address these different techniques.

\subsection{Layout}
This review is organized as follows. Section~\ref{S:PAC} addresses the scientific context. It describes the main acceleration processes at work in astrophysical plasma systems. It also a short review on the on-going experimental efforts to reproduce collisionless shocks, magnetic reconnection and particle acceleration in laser-plasma-based experiences. The next sections treat the different numerical approaches to investigate particle acceleration from microscopic scales to macroscopic scales. Section~\ref{S:SMA} describes the different numerical methods adapted to the description of plasma kinetics. Section~\ref{S:MIC-MES} discusses particle acceleration and transport at micro- and meso- plasma scales. Section~\ref{S:MAC} describe numerical techniques developed to follow macroscale dynamics and detail recent results on particle acceleration and transport in astrophysical flows. We conclude in Sect.~\ref{S:CON}.

\paragraph{How to read this review:} The scientific questions and the numerical experiments developed to investigate them are entangled. We have decided to describe this complex modelling in two steps. The first step presents a general description of the main acceleration processes in astrophysical plasmas. This presentation is the main purpose of Sect.~\ref{S:PAC}. Notice that we complement it by a dedicated section addressing some recent studies of particle acceleration at collisionless shocks and magnetic reconnection in the context of laser plasmas in Sect.~\ref{S:LASACC}. The second step describes technical numerical aspects. They are presented in Sects.~\ref{S:VlasovCodes} to \ref{S:HYB} and in all sub-sections of Sect.~\ref{S:MAC}. Sections \ref{S:SMA}, \ref{S:MIC-MES}, \ref{S:MAC} then include discussions which connect the numerical work and scientific questions exposed in Sect.~\ref{S:PAC}.

\subsection{List of acronyms and notations}

All quantities are in cgs Gaussian units.

\begin{table*}[ht]\label{T:ACR}
\begin{tabular}{|c | c|}
\hline 
\rm{Acronym name} & \rm{Definition} \\
\hline
AGN & active galactic nucleus \\
AMR & adaptive mesh refinement \\
CFL & Courant--Friedrichs--Lewy \\
CR & Cosmic Ray \\
DCE & diffusion-convection equation \\
DSA & diffusive shock acceleration \\
EP & energetic particle \\
FDM & finite difference method \\
FVM & finite volume method \\
FP & Fokker--Planck \\
GRB & Gamma-ray burst \\
HD & hydrodynamics \\
ISM & interstellar medium \\
MFA & magnetic field amplification \\
MHD & magneto-hydrodynamics \\
NLDSA & non-linear diffusive shock acceleration \\
PDE & partial differential equation \\
PIC & particle-in-cell \\
PWN & pulsar wind nebula \\
REC & magnetic reconnection \\
SDA & shock drift acceleration \\
SFA & stochastic Fermi acceleration \\
SNR & supernova remnant \\
SSA & shock surfing acceleration \\
\hline
\end{tabular}
\end{table*}

We recall here the definition of the different quantities used to construct the main parameters involved in shock acceleration processes: B is the magnetic field strength, $\rho$ is the gas mass density, $\rho_{\rm i}$ is the ion mass density, $\gamma_{\rm ad}$ is the gas adiabatic index, v and p are the charged particle speed and momentum and Ze is the charge.

\begin{table*}[h]\label{T:NOT}
\begin{tabular}{|c | c|}
\hline 
\rm{Notation} & \rm{Definition} \\
\hline
$\theta_{\rm B}$ & shock magnetic field obliquity \\&(angle between field lines and shock normal) \\
\hline
$U_{\rm sh}$ & shock velocity in the upstream (observer)frame \\
\hline
$U_{\rm A}=B/\sqrt{4\pi \rho_{\rm i}}$ & local Alfv\'en speed \\
\hline
$M_{\rm A} = U_{\rm sh}/U_{\rm A}$ & Alfv\'enic Mach number \\
\hline
$c_{\rm s}=\sqrt{\gamma_{\rm ad} P/\rho}$ & local sound speed \\
\hline
$M_{\rm s} = U_{\rm sh}/c_{\rm s}$ & sonic Mach number \\
\hline
$\sigma =B^2/4\pi \rho c^2$ & local magnetization \\& (ratio of the upstream magnetic pressure to the upstream gas kinetic energy)(shocks)\\
\hline
$\sigma = \omega_{\rm ci}^2/\omega_{\rm pi}^2$ & local magnetization \\& (ratio of the square of cyclotron to plasma frequencies) (reconnection)\\
\hline
$\beta_{\rm p}$ & plasma parameter \\
\hline
$r_{\rm L}= p v/ Ze B$ & gyro-radius (Larmor radius) \\
\hline
\end{tabular}
\end{table*}

Notice of caution: There is not a unique way to define the magnetization parameter $\sigma$. In shock studies $\sigma$ is the ratio of the magnetic pressure to the kinetic energy of the ambient gas, all quantities being measured in the upstream rest-frame. In relativistic shock studies the magnetization parameter is sometimes defined as $\sigma= B^2/4\pi \Gamma U c^2$, where $\Gamma= \sqrt{1+U^2}$ is the Lorentz factor of the shock and U is the four velocity of the flow \citep{2016RPPh...79d6901M}. In reconnection studies $\sigma$ is the ratio of the square of cyclotron to plasma frequencies. It is related to the the ratio $r_{\rm A}= U_{\rm A}/c$ by $\sigma= r_{\rm a}^2/(1-r_{\rm A}^2)$ (eg. \citet{2019arXiv190808138S})

\clearpage
\newpage
\section{Astrophysical and physical contexts}\label{S:PAC}
Following the introductory remarks in Sect.~\ref{S:INT}, we present below an overview of the astrophysical and physical contexts where the numerical tools discussed in this review are actively developed. We aim here at a short description of the basic concepts necessary to describe particle acceleration. In particular, we show that particle acceleration involves a large range in scale/time/energy which justifies the use of very different numerical techniques detailed in the next sections.

First, in Sect.~\ref{S:SAC} we briefly overview the mechanism of stochastic acceleration (SA). Then the three next sections cover different aspects of the physics of particle acceleration at collisionless shocks. In Sect.~\ref{S:DSA} we provide a general and rather detailed presentation of the physics of diffusive shock acceleration (DSA) which is one of the main frameworks to study particle acceleration in astrophysical systems. Beyond a standard description of the process itself we discuss specific issues connected with the acceleration of cosmic rays (CRs) at fast astrophysical shock waves: the injection problem and non-linear back-reaction of CRs over the flow solution. These two difficulties require the development of specific scale-dependent numerical techniques described in the next sections. Sect.~\ref{S:SDASSA} is a short presentation of the other two shock acceleration processes, namely the shock drift acceleration (SDA) and the shock surfing acceleration (SSA) which are especially relevant for particle injection in the DSA process. Sect.~\ref{S:REL} discusses the specific case of Fermi acceleration at relativistic shocks and the development of micro turbulence at these shock fronts. Magnetic reconnection (REC) is discussed in some detail in Sect.~\ref{S:REC}, where we present the most relevant vocabulary necessary to understand particle acceleration in reconnection structures. Section~\ref{S:LASACC} reviews the most important undergoing or planned laser experiments to study particle acceleration. This rapidly growing field of research starts now to investigate astrophysically relevant conditions for particle acceleration at collisionless shocks and magnetic reconnection. Notice that we decided to not include any review of acceleration processes in space plasmas, this will deserve a special section in a forthcoming version.

It should be stressed that \emph{by no means this section is intended to be exhaustive}. It has to be understood as a short introduction to the scientific cases where the different simulation techniques discussed hereafter are developed. For each type of acceleration/transport mechanism we refer the interested reader to more complete dedicated reviews.

\subsection{Stochastic acceleration}\label{S:SAC}
As discussed in Sect.~\ref{S:DSA-Fermi} stochastic acceleration occurs because, on average, energetic particles at a speed v interact with scattering centers moving at a speed U more often through head-on collisions than through rear-on collisions if $v \gg U$. This results in a broadening of the particle distribution and an increase of the mean particle energy \citep{1980panp.book.....M}. In astrophysical plasmas the scattering centers often\footnote{Note however that, similarly to Fermi's original ideas, some models invoke finite amplitude waves or shocks as the origin of stochastic energy exchanges with energetic particles, see in different astrophysical contexts e.g. \cite{1987Ap&SS.138..341B, 1990A&A...231..251A, 1998ApJ...502..598P, 2004A&A...425..395G}.} can be described as plasma waves, and when we deal with high-energy CRs these waves can be described using the MHD approximation \citep{1955PhRv...99..241P, 1966PhRv..141..186S, 1971Ap&SS..12..302K}. But it is necessary to go beyond MHD if we want to consider the acceleration of non-relativistic or mildly relativistic particles \citep{1997A&A...323..271M, 1998ApJ...495..377P}.

As explained in Sect.~\ref{S:INT}, well-known caveats prevent the interpretation of the CR spectrum observed at the Earth as resulting from stochastic acceleration by MHD waves: 1)~the non-universality of the distribution of accelerated particles, 2)~a weak relative energy gain at each wave-particle interaction scaling as $(U/v)^2$. The second issue can be partly overcome if we consider the case of low energy (sub-GeV) CR propagation in the ISM, as in that case the ratio v/U drops. Still, an important problem results in the prohibitive amount of ISM turbulence necessary to re-accelerate the low energy end of CR spectrum \citep{2006ApJ...642..902P, 2014MNRAS.442.3010T, 2017A&A...597A.117D}. 
Nevertheless, SA has been invoked to be an important source of turbulence damping and particle acceleration in solar flares \citep{2012SSRv..173..535P}, in active corona above the accretion discs around compact objects \citep{1996ApJ...456..106D, 2004ApJ...611L.101L, 2008A&A...491..617B, 2009ApJ...698..293V}, in SNRs or their associated superbubbles \citep{1992MNRAS.255..269B, 1996A&A...314.1010K, 2010A&A...515A..90M, 2010A&A...510A.101F}, in galaxy clusters \citep{2007MNRAS.378..245B}, or in the case the wave phase (Alfv\'en) speed gets close to the speed of light as can be the case in AGNs \citep{1999APh....11..347H}, in GRBs \citep{2000A&A...360..789S}, or in pulsar winds \citep{2012MNRAS.421L..67B}.

\subsection{Diffusive Shock Acceleration}\label{S:DSA}
DSA is probably the favored production mechanism of CRs. It is thought to be a natural outcome of collisionless shocks, and so is believed to be at work in astrophysical shocks at all scales, active in the bow shocks in the solar system, in the blast waves of supernova remnants (SNRs) or in the internal shocks in the jets of gamma-ray bursts (GRBs) or active galactic nuclei (AGNs). DSA is rooted in the early ideas of Fermi (\citeyear{1949PhRv...75.1169F, 1954ApJ...119....1F}); it was developed independently in the late 1970s by \cite{1977DoSSR.234.1306K, 1977ICRC...11..132A, 1978MNRAS.182..147B, 1978MNRAS.182..443B, 1978ApJ...221L..29B}, see \cite{1983RPPh...46..973D,1991SSRv...58..259J,2001RPPh...64..429M} for comprehensive reviews. A key feature of DSA is that it produces power-law distributions as a function of energy (although this spectrum can be altered by non-linear effects), which is similar to the CR spectrum as observed from the Earth, modulated by CR transport and escape from the Milky Way. DSA requires two ingredients to accelerate particles: a converging flow (the shock wave), and scattering centers (perturbations of the magnetic field). In this mechanism individual microscopic particles can be accelerated up to very high energies because they are interacting a large number of times with the macroscopic shock discontinuity before escaping the system. Again, one major difficulty in simulating this process becomes apparent: DSA is intrinsically a multi-scale problem. 

\subsubsection{Fermi processes and building power-laws}\label{S:DSA-Fermi}
In his original model, Fermi considered the interaction of charged particles with moving magnetized clouds. In the cloud frame, the particle (of velocity $\mathbf{v}$) is elastically deflected around the $\mathbf{B}$ field. In the Galactic frame (with respect to which the cloud is moving at velocity $\mathbf{U}$), the energy~$E$ of the particle changes according to
\begin{equation}
\frac{\Delta E}{E}=-2\frac{\mathbf{v}\cdot\mathbf{U}}{c^{2}}\,.\label{eq:deltaE}
\end{equation}
The effect depends on the geometry of the encounter: for head-on collisions ($\mathbf{v\cdot U}<0$) the particle gains energy ($\Delta E>0$), whereas for overtaking collisions ($\mathbf{v\cdot U}>0$) it loses energy ($\Delta E<0$). The exchange is mediated by the magnetic field, even though $\mathbf{B}$ does not appear in the formula ($\Delta E$ is nothing but the work of the Lorentz force exerted on the particle by the electric field~$\mathbf{E}$ induced by the moving~$\mathbf{B}$). For a random distribution of moving clouds, after many interactions the particle experiences a net energy gain, because head-on collisions are more likely. This is only an average gain, hence the name \emph{stochastic} acceleration, and it scales as~$\beta^{2}$ where $\beta=v/c$, hence the name \emph{second-order Fermi acceleration} (or simply \emph{Fermi II}). Fermi himself realized that this process was probably not efficient enough to produce the bulk of Galactic CRs. Now if somehow only face-on collisions occur, then the energy gain is systematic, hence the name \emph{regular acceleration}, and it scales as~$\beta$, hence the name \emph{first-order Fermi acceleration} (or simply \emph{Fermi I}). 

A~shock wave (the S in DSA) provides such a configuration: both the upstream and the downstream medium see the opposite side arriving at the same speed $\Delta U=\frac{r-1}{r}V_{\rm sh}\:\left(=\frac{3}{4}\,U_{\rm sh}\:\mathrm{if}\:r=4\right)$ where $r$ is the compression ratio and $U_{\rm sh}$ is the speed of the shock (with respect to the unperturbed upstream medium). Let us further assume that magnetic turbulence in the vicinity of the shock efficiently scatters off the particles (leading to the D in DSA), so that they are effectively isotropized on each side of the shock, meaning that their mean velocity follows the local flow velocity\footnote{More precisely, the velocity that matters for particles is that of the magnetized waves present in the flow, the difference with respect to the fluid flow is called the Alfv\'en drift, in the case the main scattering waves are Alfv\'en waves, see sections \ref{S:MHDM} and \ref{S:SAN-Blasi}.}. Then, the particles experience a regular Fermi acceleration, the clouds being replaced by a reflecting wall moving at velocity $\Delta U$. Averaging Eq.~(\ref{eq:deltaE}) over all angles, one gets a mean energy gain
\begin{equation}
\langle \frac{\Delta E}{E} \rangle
= \frac{4}{3} \frac{\Delta U}{c}
= \frac{4}{3} \frac{r-1}{r} \frac{U_{\rm sh}}{c}
\:\left(=\beta_{s}\:\mathrm{if}\:r=4\right)\,.\label{eq:deltaE_DSA}
\end{equation}
By considering the duration of a complete reflection from the opposite medium, and the probability of escape from the acceleration region, one can derive the final distribution in energy of particles. To do so, two approaches are possible: a microscopic approach, where one considers the fate of individual particles, and a kinetic approach, where one reasons with their distribution function as a function of energy or rather momentum (see details of the calculations for both approaches in \cite{1983RPPh...46..973D} and references therein). This basic choice will also apply to the numerical methods presented in this review. 
From a general point of view an acceleration mechanism can be characterized by its acceleration time $\tau_{\mathrm{acc}}$ (defined so that particles are accelerated at a rate $\partial E/\partial t=E/\tau_{\mathrm{acc}}$) and its escape time $\tau_{\mathrm{esc}}$ (defined so that particles escape the accelerator at a rate $\partial N/\partial t=N/\tau_{\mathrm{esc}}$). If particles are injected at an energy $E_0$ with a rate $Q(E_0)$, after a time t a number density $N(E) dE = Q(E_0) \exp(-t(E)/\tau_{\rm esc}) dt$ of particles will have escaped. Now energy and time are linked by $t(E)=\tau_{\rm acc} \ln(E/E_0)$, inserting this time in the previous relation leads to the steady-state solution.
\begin{equation}
N(E)\propto E^{-s}\quad\rm{with}\quad s=1+\frac{\tau_{\rm{acc}}}{\tau_{\rm{esc}}}\,.\label{eq:powerlaw}
\end{equation}
In the limit where escape never occurs ($\tau_{\mathrm{esc}}=\infty$), the hardest spectrum one can obtain is $N(E)\propto E^{-1}$. In DSA the ratio $\tau_{\mathrm{acc}}/\tau_{\mathrm{esc}}$ turns out to be independent of~$E$, and so one gets a power-law distribution, of index
\begin{equation}
s=\frac{r+2}{r-1}\:\left(=2\:\mathrm{if}\:r=4\right).\label{eq:slope_DSA}
\end{equation}
The spectral index is controlled by the compression ratio of the shock, and so is a universal value for strong shocks.

\subsubsection{The transport equation and the diffusion coefficient}\label{S:DSA-D}
Assuming the particle distribution is isotropic (to first order) in momentum~$\vec{p}$, and considering here for simplicity a plane-parallel shock along direction~$x$, in the kinetic description we may work with the quantity $f=f\left(x,p,t\right)$, which is defined so that the number density is $n\left(x,t\right)=\int f\left(x,p,t\right)\;4\pi p^{2}\;\mathrm{dp}$, and which obeys the convection-diffusion equation (CDE) \footnote{This equation is obtained by averaging the full Fokker--Plank equation (FPE) over particle pitch-angle and azimuthal gyromotion angle (see section~\ref{S:KIN}).}:
\begin{equation}
\frac{\partial f\ensuremath{}}{\partial t}+\frac{\partial (Uf)}{\partial x}=\frac{\partial}{\partial x}\left(D\frac{\partial f}{\partial x}\right)+\frac{1}{3p^{2}}\frac{\partial p^{3}f}{\partial p}\frac{\partial U}{\partial x}\,.\label{eq:Fokker-Plank}
\end{equation}
On the right-hand side of the equation, the second term represents advection in momentum, powered by the fluid velocity divergence \textbf{$\partial U/\partial x$}, while the first term models the spatial diffusion of the particles, resulting from their scattering off magnetic waves, and described by a diffusion coefficient $D(x,p)$. This coefficient, together with the shock speed $U_{\rm sh}$, sets the space- and time-scales of DSA. Upstream of the shock front, particles can counter-stream the flow up to a distance 
\begin{equation}
\ell_{\rm prec}(p)=\frac{D(p)}{U_{{\rm sh}}}\label{eq:x_upst(p)}
\end{equation}
which sets the scale of what is called the CR precursor region. The acceleration timescale, defined so that $dp/dt=p/t_{{\rm acc}}$, goes as
\begin{equation}
t_{{\rm acc}}(p)\propto\frac{D(p)}{U_{{\rm sh}}^{2}}\label{eq:t_acc(p)}
\end{equation}
where the proportionality factor is of order 8-20 depending on the shock obliquity and the Rankine--Hugoniot conditions linking up- and downstream magnetic field strengths, see \citet{1998ApJ...493..375R}. For DSA to work requires $\ell_{\rm prec}$ to be less than the accelerator's size, and $t_{{\rm acc}}$ to be less than the accelerator's age, which puts limits on the maximum momentum $p_{\max}$ that the particles can reach (for particles that radiate efficiently like electrons, $p_{\max}$ may also be limited by losses). The diffusion law is thus a critical ingredient in DSA. This aspect will however be treated in very different ways in different kinds of numerical simulations. In kinetic approaches, where one solves one equation of the kind of Eq.~(\ref{eq:Fokker-Plank}), the diffusion coefficient $D(x,p)$ or some equivalent quantity must be specified. In microscopic approaches, where one directly integrates the equation of motion of individual particles, the diffusion coefficient may actually be measured from the observed paths of an ensemble of particles. 

Computing the value of the diffusion coefficient from theory is a difficult problem (see \citealt{2009ASSL..362.....S} for a review).
Very generally, the diffusion coefficient can be expressed as $D=\frac{1}{3}\ell.v$ where again $v$ is the particle velocity and $\ell$ its mean free path. When charged particles are deflected by Alfv\'en waves, $\ell$~is inversely proportional to the energy density $\delta B^{2}$ in waves present with the resonant wavelength $\lambda\simeq r_{\rm g}$, where $r_{\rm g}=\frac{pc}{qB}$ is the particle gyroradius. A special case of interest is the so-called ``Bohm limit'' (see e.g. \citealt{1991MNRAS.249..439K, 1997APh.....7..183B, 2013APh....43...56B}) reached when $\ell\simeq r_{\rm g}$, that is when the particles are scattered within one gyroperiod, meaning that the turbulence is random ($\delta B\sim B$) on the scale~$r_{\rm g}$. This constitutes a lower limit on the value of the (parallel) diffusion coefficient, and so on the acceleration time-scales. In that case, $D\propto pv$ so that
\begin{equation}
D_{B}(p)=D_{0}\:\frac{p^{2}}{\sqrt{(1+p^{2})}}\label{eq:D_Bohm(p)}
\end{equation}
where one can evaluate $D_{0} \simeq 3\times10^{22}/B$~cm$^{2}$.s$^{-1}$
with $B$ in $\mu G$, and $p$ is expressed in $m_{p}c$ units. 

Historically the Bohm limit has been widely favored in the literature, in its true form in Eq.~(\ref{eq:D_Bohm(p)}) or with a free normalization and only keeping the ``Bohm scaling'' in~$p$, and using the exact dependence on~$p$, or only the relativistic scaling $D(p)\propto p$, or a parametrized scaling of the form $D(p)=D_{0}\:p^{\alpha}$ with free index~$\alpha$ (commonly informally called ``Bohm-like'' coefficients).
This choice of the Bohm limit may be in part due to the fact that it is the most favorable case, and also may just stem from habit given the lack of clear theoretical alternatives. Analytically it has only been derived under the assumption of strong turbulence \cite{2009APh....31..237S}, and numerical studies have found that it does not generally hold \citep{2002PhRvD..65b3002C, 2004JCAP...10..007C, 2004NuPhS.136..169P}. The validity of this assumption in the context of supernova remnant studies has been regularly questioned \citep{2001JPhG...27.1589K, 2006A&A...453..387P}, and in the context of interplanetary shocks other models have been used according to the turbulence properties and shock obliquity \citep{2010AdSpR..46.1208D, 2012AdSpR..49.1067L}. In any case, when it comes to simulations, a key aspect is how strongly the diffusion coefficient depends on the particle's energy, given relations~(\ref{eq:x_upst(p)}) and~(\ref{eq:t_acc(p)}).

\subsubsection{The injection problem}\label{S:DSA-inj}

DSA is a bottom-up acceleration mechanism, whereby (a fraction of)
the particles from a plasma get boosted to very large energies. The
particles are accelerated from a non-thermal distribution that extends
beyond the thermal distribution of the plasma (often assumed to be
a Maxwellian). The way these two populations are connected is a delicate
problem. The discussion of DSA above assumes that particles are sufficiently
energetic that they can leap over the shock wave and perceive it as
a discontinuity, meaning that their mean free path in the magnetic
turbulence is already larger than the physical width of the shock
wave, which is typically of the order of a few Larmor radii of
thermal ions. The way by which particles from the background
plasma enter the acceleration process is known as the ``injection''
mechanism. The general idea, referred to as ``thermal leakage'' is
that particles heated at the shock may be able to re-cross the shock
to start the DSA cycle (\citealt{1995A&A...300..605M, 1998PhRvE..58.4911M}).
The efficiency of the injection determines the fraction of the available
shock energy that is channeled into energetic particles. It is widely
expected to vary as a function of parameters such as the shock obliquity,
although there is no firm agreement yet on what are the most favorable
configurations.

Injection is treated in much different ways according to the level
of the numerical modeling. In kinetic approaches that decouple the
non-thermal population (obeying Eq.~\ref{eq:Fokker-Plank})
and the thermal population (obeying classical conservation laws), the injection process is parametrized. The
simplest way to do this is to postulate that some fraction $\eta$
of the particles crossing the shock enter the acceleration process, at some
momentum $p_{\mathrm{inj}}$ above the typical thermal momentum. The
requirement that the particles power-law matches the plasma Maxwellian
at $p_{\mathrm{inj}}$ actually implies that these quantities are
related, as shown in \cite{2005MNRAS.361..907B}. A~more advanced
approach is to use a ``transparency function'' to inject particles
at the shock, as done by \cite{2000A&A...364..911G}. In contrast,
in the Monte-Carlo simulations of the kind of \cite{1984ApJ...286..691E}
no formal distinction is made between the thermal and non-thermal
populations, allowing for a more consistent treatment of injection, for a given
scattering law. Only PIC simulations are able to directly address
the formation of the collisionless shock concomitantly with the energization
of particles, and it has been only very recently that the computational
power has become sufficient to see the DSA power-law naturally emerge
(e.g., \citealt{2014ApJ...783...91C}) -- although still on very small space-
and time-scales compared to any astrophysical object of interest. The results associated to these simulations are described with more details in Sect.~\ref{S:MIC-MES}.
This again illustrates the need for a model at different scales and
their entanglement, a given numerical approach often relying on results
obtained from other approaches for the aspects it cannot describe.

\subsubsection{Back-reaction and non-linear effects}\label{S:DSA-NL}

\begin{figure}[htb]
\begin{centering}
\includegraphics[width=1\linewidth]{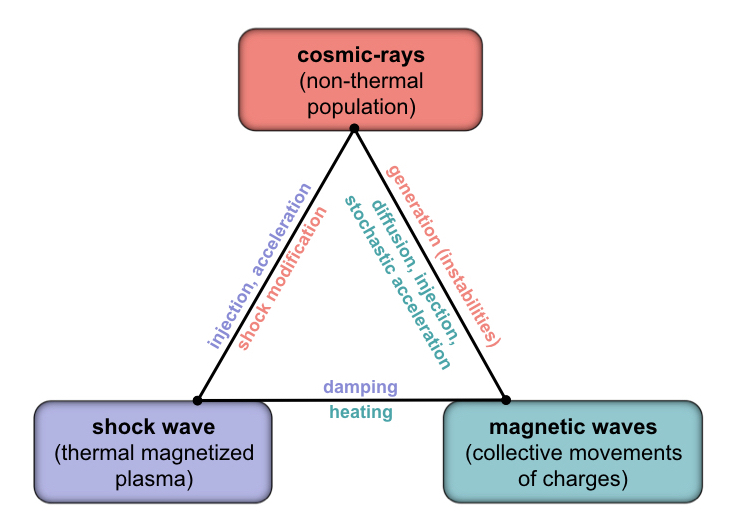}
\par\end{centering}
\caption{Sketch of the physical components and their couplings in the diffusive
shock acceleration mechanism.\label{fig:DSA-coupled-system}}
\end{figure}

DSA at astrophysical shocks involves three kinds of actors: energetic particles, a plasma
flow, and magnetic waves (see Fig.~\ref{fig:DSA-coupled-system}).
Charged particles are being injected from the plasma and accelerated
at the shock, thanks to their confinement by magnetic turbulence.
In our discussion so far we have assumed a prescribed background plasma
flow and magnetic turbulence, that is, we were implicitly discussing
the ``test particle'' regime. But if the acceleration process is
efficient (meaning that a substantial fraction of the available energy
ends up into particles), then the particles will play a role in the
dynamics of the plasma and in the evolution of the magnetic field.
This in turn will affect the way they are being accelerated, so that
the DSA process becomes non-linear (NLDSA). The time-dependent problem
is intractable analytically in the general case, which is the reason
why studies of (efficient) particle acceleration rely on numerical
techniques, as described in this review. To end this section on DSA, 
we summarize the main aspects of the two back-reaction loops: of the particles
on the plasma flow, and of the particles on the magnetic turbulence.

Even before DSA theory was established, \cite{1958ApJ...128..677P}
noted that CRs modify the medium in which they propagate: being relativistic
they lower the overall adiabatic index of the flow. We can make this
more precise by considering the CR pressure, defined as
\begin{equation}
P_{\mathrm{cr}} 
= \int_{p}\frac{p\,v}{3}\,f(p)\,4\pi p^{2}dp
= \frac{4\pi}{3}m_{p}c^2\,\int_{p}\frac{p^{4}}{\sqrt{1+p^{2}}}\,f(p)\,dp
\label{eq:CR-pressure}
\end{equation}
(where in the right expression momenta are expressed in $m_{p}c$
units). This quantity can grow without limit in the linear regime.
But as CRs diffuse upstream of the shock in an energy dependent way,
a gradient of CR pressure forms in the precursor, which produces a
force acting on the plasma. This CR-induced force pre-accelerates
the plasma ahead of the shock front, leading to the formation of a
smooth, spatially extended velocity ramp upstream of the shock front.
The shock itself is thus progressively reduced to a so-called ``subshock'',
whose compression ratio is $r_{\mathrm{sub}}<4$, while the overall
compression ratio $r_{\mathrm{tot}}$ (measured from far upstream
to far downstream) becomes $>4$ from mass conservation -- the plasma
is more compressible when particles become relativistic, and even
more so when the particles escape the system \citep{1999ApJ...526..385B}.
As particle of different energies can explore regions of different
extent ahead of the shock, they will feel different velocity jumps,
so that the spectral slope defined by Eq.~\ref{eq:slope_DSA}
becomes energy-dependent. The spectrum is thus no longer the canonical
power-law, but gets concave. Particles of low energy ($p\ll m_{p}c$)
only sample the sub-shock, feeling a compression $r_{\mathrm{sub}}$
that produces a slope larger than the canonical value of~2 (that
is, a steeper spectrum), whereas particles at the highest energies
sample the whole shock structure, feeling a compression $r_{\mathrm{tot}}$
that produces a slope that is smaller (that is, a flatter spectrum).
So one of the historically most attractive features of DSA -- its
ability to naturally produce power-law spectra -- cannot strictly
hold when it is efficient.

Turning to magnetic turbulence, it was also observed early on (e.g., \citealt{1975MNRAS.173..245S})
that the CRs can generate themselves the waves that will scatter them
off. They can indeed trigger various instabilities by streaming upstream
of the shock, which generates magnetic turbulence, which is then advected
to the shock front and the downstream region. Denoting by $W(\vec{k},x,t)$
the power spectrum of the magnetic waves where $\vec{k}$ is the wave vector, 
its evolution obeys a transport equation of the general form 
(here written for simplicity along one dimension). Assuming for simplicity that U is constant we have:
\begin{equation}\label{eq:W}
\frac{\partial W}{\partial t}+U\frac{\partial W}{\partial x}=\Gamma_{g}-\Gamma_{d} \, ,
\end{equation}
where $\Gamma_{g}$ is the growth rate of the waves, which is dictated
by the particles, and $\Gamma_{d}$ is their damping rate in the plasma.
For a CR-induced streaming instability the growth rate $\Gamma_{g}$
scales as the gradient in the CR pressure. Using hybrid
MHD+particle simulations, \cite{2000MNRAS.314...65L,2004MNRAS.353..550B}
showed that the seed magnetic field can be amplified by up to two
orders of magnitude, which is important because in turn the magnetic
field controls the diffusion and confinement of the particles and
thus the maximum energy that they can reach \citep{2001MNRAS.321..433B}.
This discovery prompted a slew of work on this complex topic (see
\citealt{2012SSRv..173..491S} for a review). Two regimes of streaming instabilities
can operate, a resonant instability at work when the Larmor radius
of a particle matches the wavelength of a perturbation, and a non-resonant
instability driven by the current of particles (see also \citealt{1991SSRv...56..373G}). 
The amplified field saturates at an energy density 
$\delta \vec{B}^{2}$ that scales as~$U_{\rm sh}^{2}$,
or possibly even~$U_{\rm sh}^{3}$ \citep{2012A&ARv..20...49V}. 
Other longer wavelength instabilities can be triggered too. 
Combined with observational evidence for high
magnetic fields at the shocks of young supernova remnants \citep{2012SSRv..166..231R},
this has lead to the current view that magnetic field amplification
(MFA) is a critical ingredient of DSA. This effect has been integrated
in numerical simulations in different ways depending on the level
of description of the particles and waves. For instance, in the kinetic
description of DSA, one can in principle compute the diffusion coefficient
$D(x,p,t)$ self-consistently from the cosmic-rays distribution $f(x,p,t)$.
Obtaining a fully consistent description of the dynamical evolution of
the particles, the plasma flow, and the magnetic turbulence, is still
a work in progress.

\subsection{Shock drift and shock surfing acceleration}\label{S:SDASSA}
\begin{figure}
\begin{centering}
\includegraphics[width=1.\linewidth]{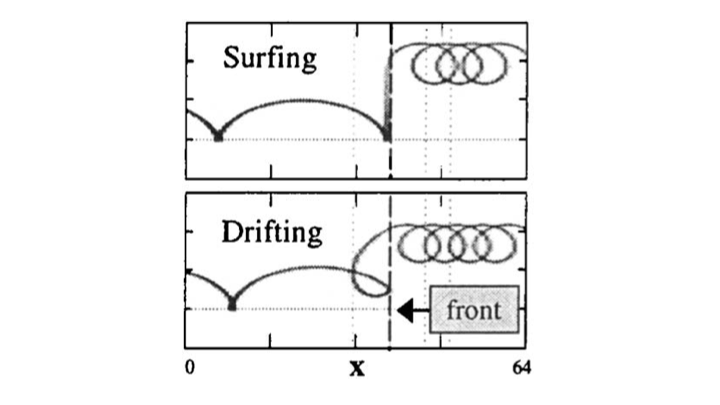}
\par\end{centering}
\caption{Particle trajectories along the shock front due to SDA or SSA mechanisms (from \citet{2003P&SS...51..665S}). The flow is directed along the $x$-direction, and particles are injected at the left of the box. The magnetic field is perpendicular to the plane of the plot, as revealed by the gyro-motion of the particles. In the SDA process the particle is forced to cross the shock several times. In the SSA process the particle moves along the shock front.
\label{F:SDASSA}}
\end{figure}

These two acceleration processes rely on the effect of the convective electric field $\vec{E} = -\vec{U}/c \times \vec{B}$ induced by magnetized fluid motions towards the shock. The difference between the two processes results from the way the particles are either confined at the shock front in the case of shock surfing or move up- and downstream in the case of shock drift \citep{1965MNRAS.131...23H}. 
Figure~\ref{F:SDASSA} shows different trajectories adopted by particles due to these two different processes. 

\subsubsection{Shock Drift Acceleration: SDA}
Acceleration associated to the drift of the particle's guiding center depends strongly on whether the shock is super- or subluminal. The super- or subluminal character of a shock depends on the speed of the intersection point of the upstream magnetic field with the shock front: in superluminal shocks this speed is larger than the speed of light, in subluminal shocks it is smaller. 

In subluminal shocks it is always possible to find a frame where the convective electric field vanishes (so where the fluid velocity lies along the magnetic field line direction), this is the so-called de~Hoffmann-Teller (HT) frame \citep{1994plas.conf..225K}. In the HT frame particle energy is conserved. As an upstream field line intersects the shock the particle guiding center drifts along the shock and can either be transmitted or reflected at the shock front because the magnetic field is compressed there. The energy gain is the highest for particles reflected at the shock front \citep{1988SSRv...48..195D}. This effect is similar to a reflection at the edge of a magnetic bottle. A calculation assuming that adiabatic theory applies uses a Lorentz transformation between a frame at which the shock is stationary to the HT frame to derive the energy gained by a particle reflected at the shock. Averaging over initial particle pitch-angles gives a ratio of the particle energy after the reflection to the initial energy of  
\begin{equation}
\left \langle \frac{E_{\rm f}}{E_{\rm i}} \right \rangle = \frac{1+\sqrt{1-b}}{1-\sqrt{1-b}} \ ,
\end{equation}
where $b=B_{\rm u}/B_{\rm d}$ is the ratio of the upstream to downstream magnetic field strengths.  
For $b=0.25$ (for a compression of the magnetic field by a factor 4) we find a maximum ratio $\langle {E_{\rm f}/ E_{\rm i}} \rangle_{\max} \simeq 15.8$ and only half of this for a transmitted particle.

In superluminal shocks, a configuration obtained in perpendicular shocks, the adiabatic invariant $p_\perp^2/B$ is conserved even while the particle crosses the shock, as long as its speed is much larger than the shock speed \citep{1983ApJ...270..537W, 1986JGR....91.4149W}. The maximum  value of the ratio  $E_{\rm f}/E_{\rm i}$ is $1/\sqrt{b} = 2$. 
Superluminal shocks are the rule in relativistic flows \footnote{the shock is subluminal only if the obliquity angle between the upstream magnetic field and the shock normal is $\theta_{\rm u} \le 1/\Gamma_{\rm sh}$, where $\Gamma_{\rm sh}$ is the shock Lorentz factor}. \citet{1990ApJ...353...66B} investigate the condition under which SDA process operates at relativistic perpendicular shocks associated with the synchrotron emission of radio galaxy hot spots. As the flow speed gets closer to the speed of light the condition for the adiabatic theory breaks down. \citet{1990ApJ...353...66B} propose an alternative method by following individual particle orbits. Due to shock dynamics, particles can cross the shock front at maximum three times before being advected downstream. 

\subsubsection{Shock Surfing Acceleration: SSA}
Shock surfing is produced when a particle is trapped between the shock electrostatic potential $e\phi$ ($+e$ is the particle charge in case of a proton) which appears at the shock ramp and the upstream Lorentz force along the shock normal which carries the particle back to the front. Original ideas about this mechanism can be found in \citet{1966RvPP....4...23S, 1973JETPL..17..279S}. The particle is accelerated under the action of the convective electric field until its kinetic energy along the shock normal exceeds $e\phi$. The trapped particle accelerates essentially along the shock front like a surfer's motion along a wave; hence \citet{1983PhRvL..51..392K} name this process \emph{shock surfing} acceleration. The SSA mechanism is often invoked at quasi-perpendicular shocks as a pre-acceleration process. It allows to inject particles beyond the energy threshold for DSA to operate \citep{1996JGR...101..457Z, 1996JGR...101.4777L}. Particles gain energy as long as they stay trapped at the shock front. The acceleration ceases if they escape either upstream if the magnetic field has some obliquity or downstream if the Lorentz force exceeds the electrostatic force at the shock layer. Particle acceleration can operate also in the relativistic regime where particles with small initial speeds are trapped the longest. \citet{2003P&SS...51..665S} for instance find that particles can have 10 bounces and reach Lorentz factors $\sim 2$ through this mechanism.


\subsection{Fermi acceleration process at relativistic shocks}\label{S:REL}
We now discuss how particle acceleration is affected when the flow itself is relativistic.

\subsubsection{General statements}
If we consider a relativistic shock front moving with a Lorentz factor $\Gamma_{\rm sh}=1/\sqrt{(1-(U_{\rm sh}/c)^2)}$, the relative energy gain as the particle is doing a shock crossing cycle (e.g., up-down-upstream) can be obtained from relativistic kinematics by imposing a double Lorentz transformation between the upstream and downstream rest frames. The relative energy variation between the final $E_{\rm f}$ and the initial $E_{\rm i}$ particle energies is \citep{1999MNRAS.305L...6G}
\begin{equation}
\frac{\Delta E}{E}=\frac{(E_{\rm f}-E_{\rm i})}{E_{\rm i}}=\Gamma_{\rm r}^2(1-\beta_{\rm r} \mu_{u\rightarrow d})(1+\beta_{\rm r} \mu'_{d \rightarrow u}) - 1 \ ,
\end{equation}
where unprimed and primed quantities mark upstream and downstream rest frame quantities respectively. The relative Lorentz factor between upstream and downstream is $\Gamma_{\rm r} = \Gamma_{\rm sh}/\sqrt{2}$. The mean energy gain is obtained by averaging over the cosine $\mu_{\rm u\rightarrow d}$ and $\mu'_{\rm d \rightarrow u}$ of the penetration angles of the particle from upstream to downstream and downstream to upstream with respect to the direction of the boost. In the most optimistic case a high relative energy gain $\Delta E/E \sim \Gamma_{\rm sh}^2$ can be achieved \citep{1995ApJ...453..883V}. However, this gain is restricted to the first cycle if the initial particle distribution is isotropic. The particle distribution after one cycle becomes highly anisotropic, beamed in a cone of size $1/\Gamma_{\rm sh}$ and due to particle kinematics, the average relative gain drops to $\sim 2$ for the next crossings \citep{1999MNRAS.305L...6G}. Particle deflection in the cone can either proceed through \emph{its motion in a uniform magnetic field} in the absence of scattering waves or by scattering with resonant waves with $kr_{\rm g} \sim 1$. Resonant scattering occurs only if the amplitude of the magnetic perturbations is small enough \citep{2001MNRAS.328..393A}. The shock particle distribution in the test-particle limit shows a universal energy spectrum $N(E) \propto E^{-2.2}$ whatever the deflection upstream if scattering is effective downstream. This result is consistent with the index of the relativistic electron distribution producing synchrotron radiation in GRB afterglow \citep{1997ApJ...485L...5W}. Note that this result has been obtained for an isotropic turbulence downstream. A more general formulation in terms of shock speed gives an energy index \citep{2005PhRvL..94k1102K} of
\begin{equation}
s = \frac{\beta_{\rm u}-2\beta_{\rm u}\beta_{\rm d}^2 + \beta_{\rm d}^3 +2\beta_{\rm d} }{ (\beta_{\rm u}-\beta_{\rm d})} \ ,
\end{equation}
where $\beta_{\rm u/d}$ is the upstream/downstream fluid velocity normalized to c; $s = 2 + 2/9 \simeq 2.2$ is recovered in the ultra-relativistic limit ($\beta_{\rm u} \rightarrow 1$ and $\beta_{\rm d} \rightarrow 1/3$).
The value of the ultra-relativistic index has also been assessed by numerical simulations using a Monte-Carlo method \citep{1998PhRvL..80.3911B, 2001MNRAS.328..393A} or a semi-analytical method based on the derivation of eigenfunctions in the particle pitch-angle cosine of the solution of the diffusion-convection equation \citep{2000ApJ...542..235K}. However, this spectrum is not properly universal in the sense that the index depends on the geometry of the turbulence \citep{2006MNRAS.366..635L}.

\begin{figure}
\begin{centering}
\includegraphics[width=1\linewidth]{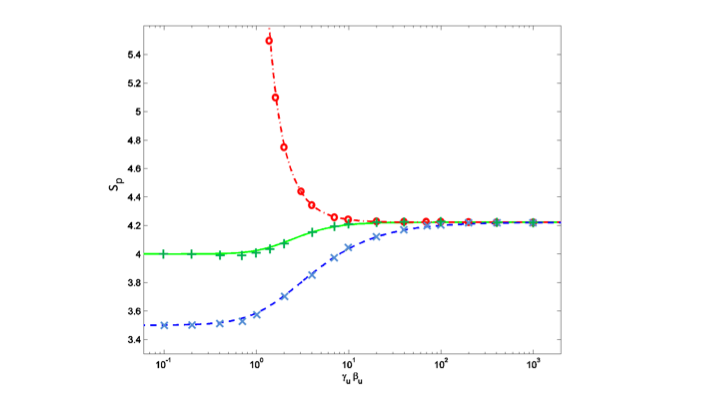}
\par\end{centering}
\caption{Spectral index of the shock particle distribution as function of the shock Lorentz factor (from \citealt{2015SSRv..191..519S}). Solid lines are the solutions obtained in \citet{2005PhRvL..94k1102K} compared with the solutions of \citet{2000ApJ...542..235K} plotted with symbols. A strong shock solutions with the J\"uttner/Synge equation of state is shown in solid line and crosses, a strong shock with fixed adiabatic index $\gamma_{\rm ad}= 4/3$ is shown in dashed line and x-marks, and a shock with a relativistic gas where $\beta_{\rm u}\beta_{\rm d} = 1/3$ is shown in dash-dotted line and circles.\label{F:RELSH}}
\end{figure}

Figure~\ref{F:RELSH} shows the index of the shock downstream particle distribution as a function of the shock Lorentz factor from mildly relativistic to ultra-relativistic regimes for different equation of state of the relativistic gas (the quantity plotted is $s+2$ with our notation). 

It appears that unless some particular turbulence develops around the shock front, Fermi acceleration associated with repeated shock crossings does not operate at relativistic shocks because of  particle kinematic condition to cross the shock front \citep{1990ApJ...353...66B, 2006ApJ...645L.129L, 2009MNRAS.393..587P}.
This fact results from that relativistic shocks are generically perpendicular unless the magnetic field upstream is oriented within an angle $1/\Gamma_{\rm sh}$ along the shock normal. If the background turbulence around the shock is absent or if its coherence length is larger than the particle's gyroradius it can be shown that while returning to the shock from downstream to upstream the particle is unable to do more than one cycle and a half before being advected downstream \citep{2006ApJ...645L.129L}. There are two necessary conditions for an efficient scattering and particle acceleration: 1) the turbulence which develops around the shock has to be strong, with a perturbed magnetic field such that $\delta B/B_0 > 1$, where $B_0$ is the background upstream magnetic field strength, 2) the perturbations have to be at scales smaller than the particle  gyroradius (with respect to the total magnetic field) \citep{2009MNRAS.393..587P}. One drawback of these conditions is that, as the micro-turbulence develops with a coherence scale smaller than $r_{\rm g}$, the spatial diffusion coefficient scales as $D \propto r_{\rm g}^2$, and as the acceleration timescale scales as $t_{\rm acc} \propto D/U_{\rm sh}^2 \propto E^2$ even when $U_{\rm sh} \rightarrow$ c, the time required to reach extremely high energies can become very large. This is the main reason which explains why ultra-relativistic GRB shocks $\Gamma_{\rm sh} \gg 1$ cannot be at the origin of CRs with energies $\sim 100$ EeV \citep{2010MNRAS.402..321L, 2013MNRAS.430.1280P, 2014MNRAS.439.2050R}.

\paragraph{One note on the origin of the micro-turbulence.}
We have seen that the onset of micro-turbulence is necessary for the Fermi process to operate at relativistic shocks. What type of micro-turbulence develops depending on shock velocity regimes and upstream medium properties? 

The characteristic scale on which the micro-turbulence can develop is given by the CR precursor scale $\ell_{\rm prec}$. It is either set by the regular gyration of particles reflected by the shock front, in which case $\ell_{\rm prec}=r_{\rm g}/\Gamma_{\rm sh}^2$, or by diffusion, in which case $\ell_{\rm prec}=r_{\rm g}^2/\ell_{\rm c}$, where $\ell_{\rm c}$ is the turbulence coherence scale. As the energy spectrum is softer than $E^{-2}$, low energy particles carry the bulk of free energy and generate most of magnetic perturbations. Hence, the scale over which micro-instabilities can develop in relativistic shocks is small. This restricts the number of instabilities to a few \citep{2010MNRAS.402..321L}. The nature of the dominant instability depends on two main parameters: the shock Lorentz factor $\Gamma_{\rm sh}$ (or momentum  $\Gamma_{\rm sh}\beta_{\rm sh}$ for mildly relativistic shocks), and the upstream magnetization $\sigma_{\rm u}= B_0^2/4\pi (\Gamma_{\rm sh}(\Gamma_{\rm sh}-1) \rm{n^\star mc}^2$, where $\rm{n}^\star$ is the ambient (upstream) proper gas density composed of particles of mass $m$ \citep{2015SSRv..191..519S,2016RPPh...79d6901M}.
In weakly magnetized shocks ($\sigma_{\rm u} \le 10^{-3}$) the dominant instability is the electromagnetic filamentation or Weibel instability \citep{1959PhRvL...2...83W}. Filamentation/Weibel instabilities grow due to the presence of two counter-streaming population of particles and produce modes in the direction perpendicular to the streaming direction \citep{1959PhFl....2..337F, 2009ApJ...699..990B}. \citet{2013MNRAS.430.1280P} discuss also the case of the oblique two-stream instability which can have a competitive growth rate with respect to the filamentation instability. Finally, the Buneman instability has been discussed as a source of electron heating in relativistic shock precursors \citep{2011MNRAS.417.1148L}. At higher magnetization ($0.1 > \sigma_{\rm u} > 10^{-3}$) a current driven instability either in subluminal \citep{2004MNRAS.353..550B, 2005MNRAS.358..181B, 2006PPCF...48.1741R, 2006ApJ...651..979M} or superluminal shocks \citep{2009MNRAS.393..587P, 2013MNRAS.433..940C, 2014MNRAS.440.1365L} can develop in the precursor. If $\sigma_{\rm u} > 0.1$ then the gyration of particles in the background magnetic field gains in coherence and the shock is mediated by the synchrotron maser instability. This instability produces a train of semi-coherent large amplitude electromagnetic waves that escapes into the upstream medium \citep{1992ApJ...391...73G, 1992ApJ...390..454H, 2019MNRAS.485.3816P}. The interaction of this wave with the background plasma is a source of an efficient electron pre-heating up to equipartition with protons \citep{2006ApJ...652.1297L, 2011ApJ...726...75S}. The question of whether this wave can lead to a significant particle acceleration is debated \citep[e.g., ][]{2008ApJ...672..940H, 2017ApJ...840...52I, 2018MNRAS.474.1135L}.

\subsubsection{Progress with fully kinetic simulations}
Until late 2000s, most progress on the understanding particle acceleration at relativistic shocks were supported by Monte-Carlo simulations \citep[e.g., ][]{1998PhRvL..80.3911B, 2003ApJ...589L..73L, 2004APh....22..323E, 2006MNRAS.366..635L} or semi-analytic approaches \citep{2000ApJ...542..235K, 2001MNRAS.328..393A, 2005PhRvL..94k1102K}. As pointed out by \citet{2011A&ARv..19...42B}, recent advances were possible by employing fully kinetic PIC simulations. Self-consistent build-up of Fermi process was observed and a survey in which parameter space it operates was done \citep{2008ApJ...682L...5S, 2009ApJ...695L.189M, 2009ApJ...698.1523S, 2011ApJ...726...75S, 2013ApJ...771...54S, 2018MNRAS.477.5238P}. For instance, these simulations demonstrate that the Weibel-filamentation instability dominates in controlling the shock structure in weakly magnetized shocks, as predicted by \citet{1999ApJ...526..697M} and \citet{1999ApJ...511..852G}. Particle acceleration is correlated to the efficiency of triggering this instability. Typically, the non-thermal particles contain about 1\% of total particle number and about 10\% of total energy. The tail develops into a power-law with spectral slope $s \simeq 2.4$ that is close to the semi-analytic prediction of 2.2. The maximum energy of particles evolves in time as  $E_{\max} \propto \sqrt{t}$ \citep{2013ApJ...771...54S} due to the small-scale nature of the magnetic turbulence (see above). In the small-angle scattering regime, the spatial diffusion coefficient of particles is $D \propto E^2$, unless the external magnetic field imposes a saturation that sets the maximal particle energy to be $E_{\max}>e \delta B^2/B_0 \ell_{\rm c}$ \citep{2011A&A...532A..68P, 2016RPPh...79d6901M}. 
For typical parameters in ultra-relativistic shocks with $\Gamma_{\rm sh} \sim 100$ propagating in the ISM with $B_0 \sim 3~\mu$G, this energy does not exceed $10^{16}$~eV. Section~\ref{S:MIC-MES} presents more detailed discussions of these studies.

\subsubsection{Long term evolution}
The fate of the micro turbulence and, more generally, the long-term evolution of the weakly magnetized shocks remains the major unanswered question in relativistic (but also in non-relativistic) shock physics. As this micro turbulence is composed of initially short wavelength perturbations, these are expected to be rapidly damped by Landau damping downstream \citep{2001ApJ...563L..15G, 2008ApJ...674..378C, 2015JPlPh..81a4501L}. One possibility to overcome this effect would be to have some amounts of inverse cascade to generate large wavelengths or to rely one the perturbations generated upstream by high-energy particles and transmitted downstream. These possibilities, and others are detailed in Sect.~3.2 of \citet{2015SSRv..191..519S}.

\subsection{Reconnection in astrophysical flows}\label{S:REC}
Magnetic reconnection can occur in a collisional or in a collisionless plasma. The bulk part of the particles are accelerated to order of the Alv\'en speed and heated in the same process. A fraction of the particles can be accelerated to much higher velocities and form a power-law up to very high Lorentz factors. The power-law slopes can be harder than power-laws produced by a Fermi process, e.g. in collisionless shocks. In astrophysics, reconnection (REC) is a highly important process to accelerate particles.

In this introductory section, we briefly summarize some important concepts of magnetic reconnection. For any further details, we refer to recent reviews on the subject by \citet{2009ARA&A..47..291Z}, \citet{2011SSRv..159..357Z} (solar flares), \citet{melzani:tel-01126912}, \citet{2016ASSL..427.....G}, \citet{2018arXiv180501347J}. A more deep insight into subjects directly related to this review, the acceleration of particles, will be given in Sect.~\ref{Sec:Kinetic_Reconnection}.

\paragraph{Astrophysical objects where reconnection takes place:}
\label{S:REC_Objects}

\begin{figure}[htb]
\begin{centering}
\includegraphics[width=1\linewidth]{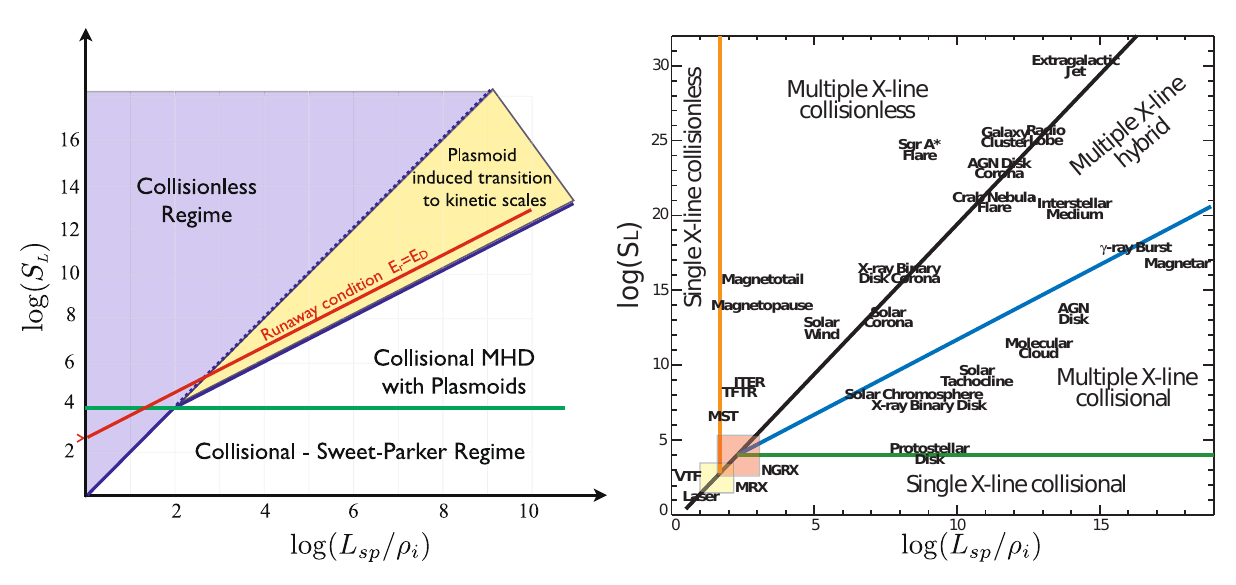}
\par\end{centering}
\caption{Different REC regimes, as derived by \citet{Daughton2012} (left panel) and \citet{Ji2011} (right panel). $S_{\rm L}$ is the Lundquist number defined by Eq.~(\ref{eq:Lundquist}), $L_{\rm sp}$ the macroscopic system size, $\rho_{\rm i}$ the ion Larmor radius in the asymptotic magnetic field (including the guide field), not to be confused with the ion mass density. The red curve is computed for $\beta_{\rm p} = 0.2$ and a REC rate of $R = 0.05$. \label{fig:ReconnectionParameterSpace}}
\end{figure}

Well known REC sites in the solar systems are the upper chromosphere and the corona of the Sun as well as the magnetotail and the magnetopause of planets. There, predominately electrons are accelerated to very high speeds. REC may be partly responsible for heating the solar corona and thus for the existence of the solar wind. But highly accelerated particles are also a severe threat for spacecrafts and astronauts and even aircraft passengers. They are at the source of geo-magnetic storms which severely endanger communication and power grids on Earth. The demand for a better understanding of space-weather is one reason why in recent years the effort to understand REC has intensified and brought decisive new insights.

In outer space, REC was found to play a crucial role in the understanding of high-energy objects such as pulsars and their winds and nebulae, as well as magnetars, (micro-)quasars, and GRBs. In most of these objects, REC is partly a driver of their dynamics. Besides shocks and wave-turbulence, REC can accelerate particles to highest energies under such conditions. These particles and their interaction with the environment also inevitably contribute to the emission spectrum of such objects. In addition, they may be at the source of the production of high-energy neutrinos observed on Earth. REC in such objects is mostly relativistic in that the energy stored in the associated magnetic fields exceeds the rest mass energies of electrons and protons.

Figure~\ref{fig:ReconnectionParameterSpace} shows the parameter space of magnetic REC present in astrophysics and relates it to concrete systems. As can be taken from the figure, collisional and non-collisional REC equally contribute to the overall picture of magnetic REC in space. Indicated are as well different other regimes which will be discussed below. Fusion devices like ITER and TFTR and experimental setups like MRX, NGRX, MST, VTF are also shown.

\subsubsection{Collisional reconnection models}
\paragraph{Sweet--Parker model:} 
The first theory of magnetic reconnection was presented by \citep{1958IAUS....6..123S} for a \emph{collisional} plasma with resistivity $\eta; \mathbf{J} = \eta \mathbf{E}$. \citet{1957JGR....62..509P} worked out the scaling relations presented below. It was soon clear that it is not always applicable, as this model predicts too slow events as compared to observations. The question how to make REC fast is still today in the center of the discussion and not generally solved.

Assume a collisional plasma with a certain resistivity. Then the induction equation and Ohm's law become, with $\mathbf{U}$ the flow velocity,
\begin{equation}
\frac{\partial \mathbf{B}}{\partial t} = \nabla \wedge (\mathbf{U} \wedge \mathbf{B})
    + \eta \nabla^2 \mathbf{B} ;
\hspace{1cm} \mathbf{E} + {\mathbf{U} \over c} \wedge \mathbf{B} =  {\mathbf{J} \over \sigma}.
\end{equation}
The non-ideal terms are the resistive diffusivity $\eta= c^2/4\pi \sigma$ in the induction equation and the resistive current in Ohm's law expressed in terms of the electrical conductivity $\sigma$. The non-ideal induction equation makes clear that there is a competition between the diffusion of the magnetic field (governed by the resistive time-scale $\tau_{\rm R}$) and the ideal evolution (governed by the Alfv\'enic time scale $\tau_{\rm A}$). This balance is expressed by the magnetic Reynold's number, $R_{\rm m} \equiv \ell \cdot \tilde{U} / \eta$ with $\ell$ a characteristic length scale and $\tilde{U}$ a typical velocity of the system. In ideal MHD $R_{\rm m} >> 1$ and REC is suppressed; whenever $R_{\rm m} << 1$ field diffusivity wins and REC becomes possible though not mandatory.

\begin{figure}[htb]
\begin{centering}
\includegraphics[width=\linewidth]{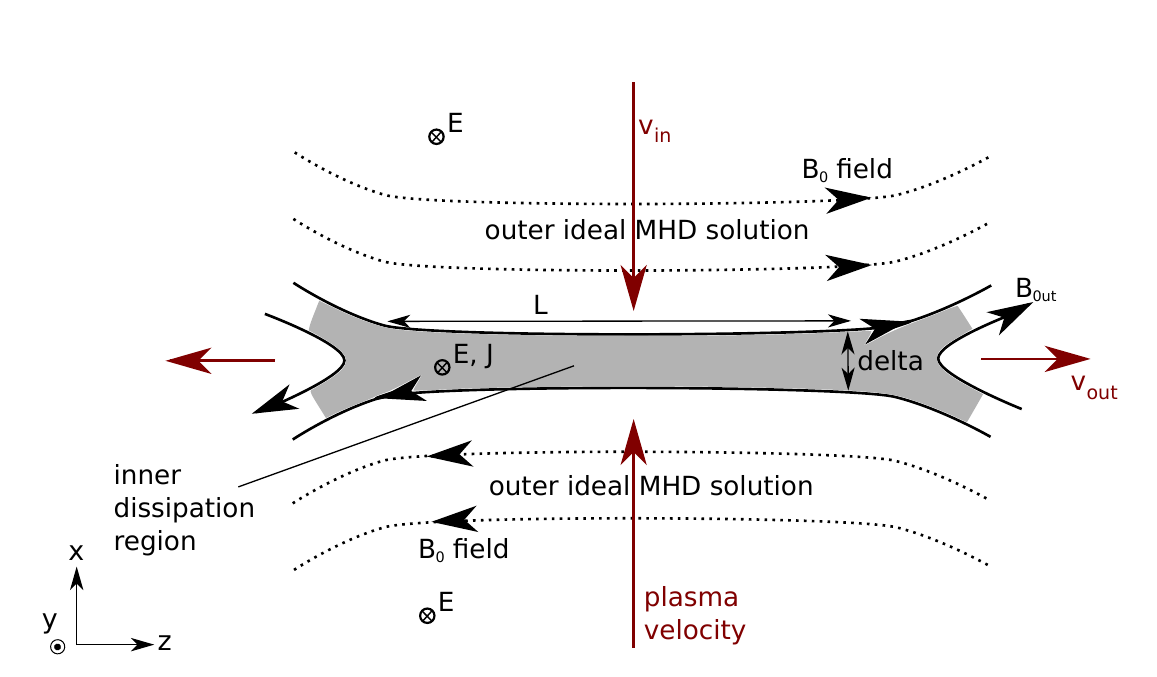}
\par\end{centering}
\caption{Sweet--Parker model of magnetic reconnection. See details in the text. (Adapted from \citet{melzani:tel-01126912}.)  \label{fig:Reconnection_Sweet-Parker}}
\end{figure}

Figure~\ref{fig:Reconnection_Sweet-Parker} describes a 2D steady situation. Plasma from an outer ideal region flows in parallel to the x-direction towards a dissipation region, which has a length-scale, $L$, and a thickness, $\delta$. The inflow velocity is just given by the $E \times B$-drift in the plasma. In the outer ideal region ($R_{\rm m} >> 1$), the plasma is frozen to the magnetic flux. This is no longer true in the diffusion region where the resistivity is dominant: the plasma decouples from the magnetic field. This opens the possibility that the field reconnects and plasma is expelled in the z-direction. These outflows are called exhausts.

Applying mass and energy conservation, non-compressibility, and that the field energy is dominant at the inflow and the kinetic energy of the particles at the outflow, two important relations follow:
\begin{equation}
 U_\mathrm{\rm out} = \sqrt{2}\, \dfrac{B_0}{\sqrt{4\pi m\,n_\mathrm{in}}} = \sqrt{2}\, U_\mathrm{A,in},
 \label{Eq:Reconnection_Speed_Exhaust}
\end{equation}
\begin{equation}
 \frac{\delta}{L} = \frac{U_\mathrm{in}}{U_\mathrm{A,in}} = M_{\mathrm{a, in}},
\end{equation}
where $U_\mathrm{A,in}$ is the Alf\'en speed and $M_{\mathrm{a, in}}$ the Alf\'enic Mach-number of the inflow. Thus, the outflow speed in the exhausts is of order of the Alf\'enic speed of the inflow. Assuming non-forced REC, the inflow is just $E \times B$-drift, $U_\mathrm{in}= cE_{\rm y} /B_0 = \eta/\delta$ and thus generically very sub-Alf\'enic. The Lundquist-number, $S_{\rm L}$, and the REC rate $R$ are defined as
\begin{equation}
S_L \equiv \frac{LU_\mathrm{A,in}}{\eta} \sim \left(\frac{L}{\delta} \right)^2
\sim \left(\frac{U_\mathrm{A,in}}{U_\mathrm{in}}\right)^2 \sim \left(M_{\mathrm{a, in}}\right)^2,
\label{eq:Lundquist}
\end{equation}
\begin{equation}
R \equiv \frac{U_\mathrm{in}}{U_\mathrm{out}} \sim \frac{\delta}{L} \sim 1/S_{L}^{1/2}.
\label{eq:REC_rate}
\end{equation}
The Lundquist-number is equal to the magnetic Reynolds-number, $R_{\rm m}$, for the case where the typical velocity is equal the Alfv\'enic speed of the inflow. Highly conducting plasmas as found in astrophysics have high Lundquist numbers: laboratory plasma experiments typically have Lundquist numbers between $10^2$ and $10^8$. In astrophysics, they are higher, up to $10^{20}$ (Fig.~\ref{fig:ReconnectionParameterSpace}, right panel) and thus Sweet--Parker reconnection rates very low -- in contrast with what is observed.

The ratio between the incoming and outgoing energy flux in Sweet--Parker reconnection is $\propto 1/S_L$. So, energy is indeed transferred from the magnetic field (incoming flux) to particles (outgoing flux). In the diffusive region with de-magnetized particles, the electric field can accelerate them. But there are other, most probably even more important acceleration mechanisms as will be discussed in Sect.~\ref{Sec:Kinetic_Reconnection}.

REC rates of such high Lundquist numbers are much too low as compared to what is observed. This result can be translated to time-scales. The magnetic Reynolds-number can also be expressed as $R_{\rm m} = \tau_{\rm R} / \tau_{\rm A}$, where $\tau_{\rm R}$ is the resistive diffusion time-scale and $\tau_{\rm A}$ Alfv\'en time-scale. In astrophysical environments with high Lundquist numbers it is thus found that $\tau_{\rm R} << \tau_{\rm A}$. The typical Sweet--Parker REC time-scale is $\sim \tau_{\rm A}^{1/2}\,\tau_{\rm R}^{1/2}$. This indicates that Sweet--Parker REC is indeed faster than resistive diffusion of the magnetic field (scaling with $\tau_{\rm R}$). However, it is much too slow when compared to REC observed in astrophysical plasmas which scales as $10-100\;\tau_{\rm A}$. For instance, in flares of the solar corona, $S_{\rm L}\sim 10^8$ (see Fig.~\ref{fig:ReconnectionParameterSpace}), $U_{\rm A}\sim 100\,{\rm km}\,{\rm s}^{-1}$, and $L\sim 10^4\,{\rm km}$. Thus, the Sweet--Parker-timescale is a few tens of days. Observed is a magnetic energy release within a few minutes to an hour. This major discrepancy is known as the fastness problem of Sweet--Parker REC.

On the other hand, numerical simulations based on resistive MHD as well as experiments such as the MRX, the Magnetic Reconnection Experiment \citep{1998PhRvL..80.3256J} are in good agreement with the Sweet--Parker model. Clearly, the Sweet--Parker model has its deficits, in that it neglects dimensionality and any time-dependence, as well as viscosity, compressibility, downstream pressure, and, in particular, turbulence and is strictly valid only for a collisional plasma. There have been numerous papers addressing the fastness problem. Only in recent years, significant progress in this question has been made, see below.

\paragraph{Petschek model:}

\begin{figure}[htb]
\begin{centering}
        \includegraphics[width=\linewidth]{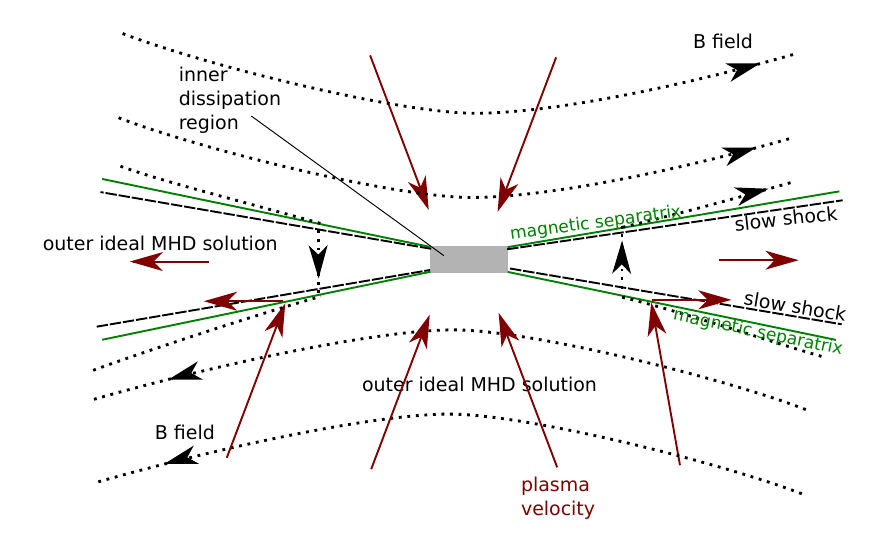}
\par\end{centering}
\caption{Illustration of the fast REC Petschek model. (Adapted
from \citet{melzani:tel-01126912}.)  \label{fig:Reconnection_Petschek}}
\end{figure}

\citet{1964NASSP..50..425P} proposed a REC model in which the reconnection rate is nearly independent of the Lundquist number, $v_{\mathrm{in}} / v_{\mathrm{A,in}} \approx \pi / 8 \ln S_L$ and thus REC is fast. The trick is to add, in close neighborhood to the separatrices slow shocks (left panel of Fig.~\ref{fig:Reconnection_Petschek}) into the configuration. In this way, particles can be accelerated without having to pass through the inner dissipation region with resistive dissipation. Instead, magnetic energy can be conversed to kinetic particle energy in the shocks.

However, all resistive MHD simulation are in agreement with the Sweet--Parker-model unless a localized anomalously large resistivity is used, mimicking that the mean free particle path becomes larger than the reconnection layer. Otherwise, shocks are not observed in MHD simulations. Therefore, the Petschek-model is likely not a model of resistive MHD -- though this is still a controversial question. However, recent PIC simulations in a collisionless plasma show an X-point and separatrix-structure in reconnection, which resembles somehow the Petschek-model \citep{2012JGRA..117.1220H,
2012PhPl...19b2110L, 2015JPlPh..81a3209L} -- at least at scales larger than kinetic scales.  There is also observational evidence that in the reconnection region of Earth's magnetotail, slow shocks are present \citep{2004JGRA..10910212E}. This discussion will be resumed in Sect.~\ref{Sec:Kinetic_Reconnection}.

\begin{figure}[htb]
\begin{centering}
        \includegraphics[width=\linewidth]{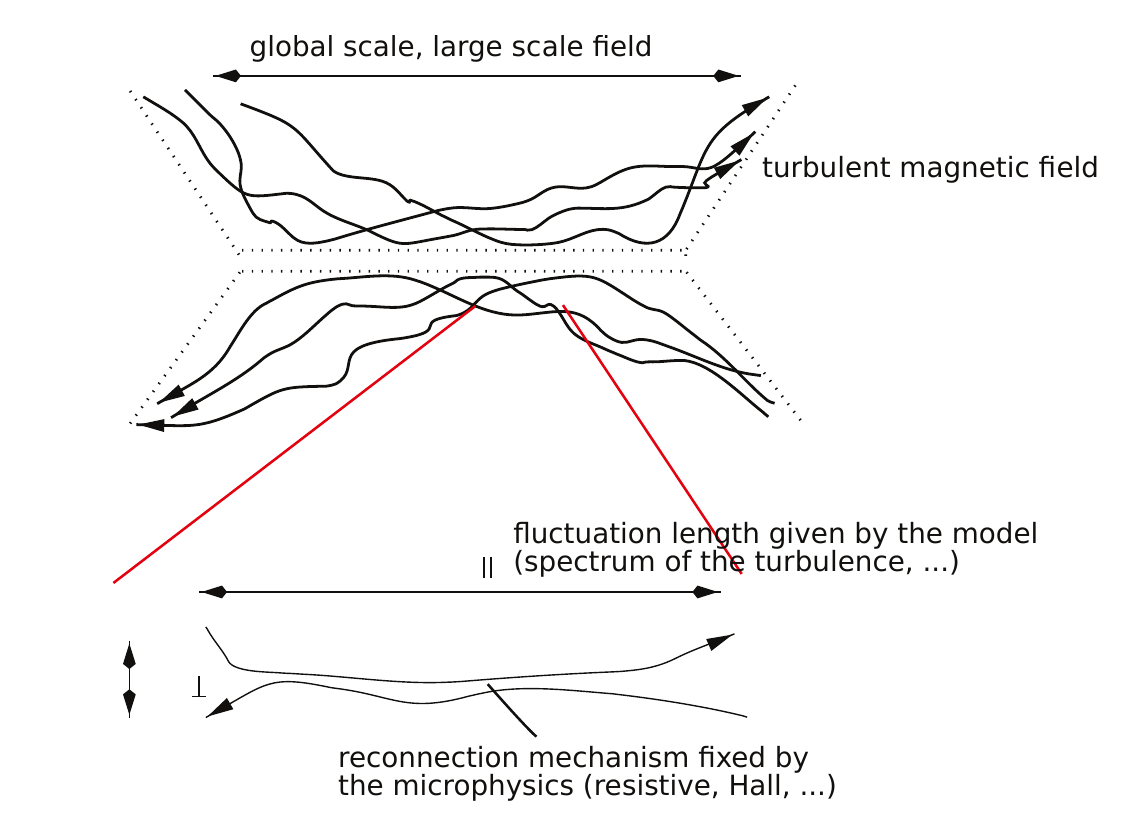}
\par\end{centering}
\caption{Illustration of the fast REC stochastic model
by \citet{1999ApJ...517..700L}. (Adapted
from \citet{melzani:tel-01126912}.)  \label{fig:Reconnection_LazarianVishniac}}
\end{figure}

\paragraph{Turbulence:} -- external or self-generated in the REC process -- seems to be the key process which allows resistive MHD REC to be fast. As indicated in Fig.~\ref{fig:Reconnection_LazarianVishniac} turbulent fluctuations allow to form many, much smaller scaled, reconnection spots along the global length, L, of the sheet. As shown in \citet{1999ApJ...517..700L}, the REC becomes thus much faster and is independent of the exact REC mechanism at each of these spots (Sweet--Parker, collisionless, ...). The exact result depends, however, on the nature of the turbulence and its fluctuation. Numerical simulations show good agreement with the analytic result \citep{2009ApJ...700...63K,
2012NPGeo..19..297K}.

Simulations show that a Sweet--Parker like current sheet generates islands above a critical Lundquist number $S_c \sim 10^4$ \citep{Daughton2012}. In relativistic flows, this critical number may be higher, $S_c \sim 10^8$ \citep{2011MNRAS.418.1004Z}. This is confirmed by newer investigations and linked to an extremely fast growing tearing instability of the current sheet \citep{2016MNRAS.460.3753D, 2018arXiv180110534P}. This limit is indicated as the green line in the left panel of
Fig.~\ref{fig:ReconnectionParameterSpace}.

\citet{2008PhRvL.100w5001L} showed that a Sweet--Parker sheet setup in a Harris or force-free equilibrium sheet develops slow REC. On a much longer time-scale, tearing modes start to fragment the sheet and several X-points form. The exhausts of these X-points generate turbulence leading to multiple short lived REC regions, popping up randomly, frequently and at multiple locations simultaneously. Consequently, fast REC sets in. Similar findings for 3D resistive reconnection are presented by \citet{2015ApJ...806L..12O}. By linking such self-generated turbulence with external turbulence, \citet{2012NPGeo..19..251L} formulate a united approach. So one may, with still some care, conclude that also collisional, resistive REC is fast, at least under certain conditions.

\subsubsection{Collisionless reconnection}

\begin{figure}
\begin{centering}
\includegraphics[width=1\linewidth]{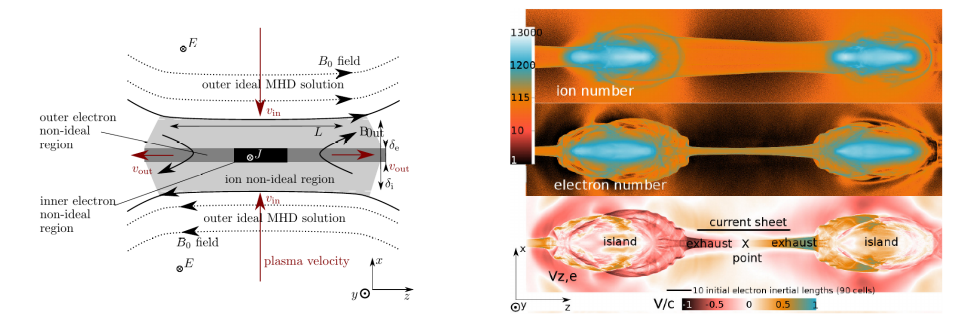}
\par\end{centering}
\caption{Collisionless REC in a electron-ion plasma. Left panel: sketch of REC on a scale smaller than the ion inertial length. The diffusion regions for ions are much larger than that for electrons. The current sheet is more like an X-point than a double y-point. Ion trajectories normally do not pass through the electron non-ideal region (adapted from \citet{melzani:tel-01126912}). Right panel: X-point, exhausts and islands from a collisionless electron-ion PIC-simulation of REC using a mass-ration $m_{\rm i}/m_{\rm e} = 25$. Visibly, the ion diffusion regions is about a factor of 5 ($\delta_{\rm k} \sim \sqrt{m_{\rm k}}, k=e,i)$ larger than the electron diffusion region  (adapted  from \citet{2014A&A...570A.111M}).  \label{fig:Collisionless_Reconnection}}
\end{figure}

On length scales shorter than the ion inertial length
$c/\omega _{\rm {p,i}}$ where $ \omega _{\rm {pi}}\equiv {\sqrt {{ {4\pi n_{\rm i}Z^{2}e^{2}}/{m_{\rm i}}}}}$ is the ion plasma frequency, ions decouple from electrons and the magnetic field becomes frozen into the electron fluid rather than the bulk plasma. Consequently, other terms than just resistivity start to contribute to the Ohm's-law. For instance, based on a two-fluid non-relativistic plasma model, \citet{melzani:tel-01126912} derives a more complex Ohm's-law for electrons:
\begin{equation}\label{Equ:Reconnection_Ohm2}
\begin{aligned}
 \underbrace{\mathbf{E}+{\mathbf{v}_{\rm i} \over c}\wedge \mathbf{B}}_{\mathrm{E-field\,in\,the\,ion\,plasma\,frame}} &= 
       \underbrace{\dfrac{1}{n_{\rm e} \rm{e}} \mathbf{J} \wedge \mathbf{B}}_\mathrm{Hall\,term} 
     - \underbrace{\dfrac{m_e}{\rm e}\left(\dfrac{\partial \mathbf{v}_{\rm e}}{\partial t} +
     \mathbf{v}_{\rm e}\cdot\nabla\mathbf{v}_{\rm e}\right)}_\mathrm{electron\,bulk\,inertia}
     -\underbrace{\dfrac{1}{n_{\rm e} \rm{e}} \nabla\cdot {P}_{\rm e}}_{\mathrm{e^-\,thermal\,inertia}} \\
 ~~~&+ \underbrace{\dfrac{\chi}{(n_{\rm e} \rm{e})^2} \mathbf{J}}_\mathrm{e-i\,collisions} 
     + \underbrace{\dfrac{\chi_e}{n_{\rm e} \rm{e}} \nabla^2\mathbf{v}_{\rm e}}_\mathrm{e-e\,collisions}.
\end{aligned}
\end{equation}
Here, $n_e$ is the electron number density, $v_i$, $v_e$ the ion and the electron velocity respectively, c the speed of light, $e$ the elementary charge, $\chi$ accounts for the effect of collisions between electrons and ions which, in general can be anisotropic and depend on the magnetic field orientation. $\chi_e \nabla^2\mathbf{v}_{\rm e}$ describes the electron viscosity due to electron-electron collisions. $P_{\rm e}$ is the pressure tensor
\begin{equation}
P_{\rm e} = \int\mathrm{d}^3\!\mathbf{v}\, m_{\rm e}(v_{\rm a}-\bar{v}_{\rm a})(v_{\rm b}-\bar{v}_{\rm b})\;
\end{equation}
with $a,\,b=x,\,y,\,z$, and $\bar{v}$ is the mean velocity.  
Electron inertia, both thermal and bulk, now contribute to the non-ideal
terms. In particular, if the plasma is completely collisionless,
($\chi=\chi_e=0$), these are the only contribution of non-idealness of
the plasma.

The sketch in the left panel of Fig.~\ref{fig:Collisionless_Reconnection} shows that the dissipation region now is subdivided into a larger ion dissipation region with a size of $\delta_{\rm i}$ and a smaller electron dissipation region, sized to $\delta_{\rm e}$. Here, $\delta_{\rm i, e}$ denotes the ion and electron inertial length.

On these scales, the Hall effect becomes important, because now the magnetic field lines are advected with the electrons while the ions no longer follow this motion. The Hall term is not responsible for REC as it appears when the magnetic flux is still frozen to the motion of electrons. However, there is a debate whether it may contribute to the fastness of REC as it allows to accelerate electrons to higher speeds, increasing the bulk inertia. As can be taken from the right panel of Fig.~\ref{fig:Collisionless_Reconnection} this two-layer picture derived from a two-fluid model is quite accurately reproduced by full kinetic simulations though the two-fluid model will not provide the full picture as effects like wave turbulence, Landau-damping and particle acceleration to speeds much higher than the Alfv\'en speed. Fully collisionless REC is found to be always fast. It will be further addressed in Sect.~\ref{Sec:Kinetic_Reconnection}.

\subsubsection{Other effects}
\paragraph{Dimensionality:} REC in 3D shows a variety of new features. In some cases still a separatrix-like reconnection as in 2D can be observed, but there are also many other cases. It is not the place to discuss this here in detail. A good summary can be found in \citet{melzani:tel-01126912}.

\paragraph{Fat tails and high-energy power-laws:} Magnetic reconnection is efficient to accelerate particles, both in the collisional and collisionless regime. The typical speed of accelerated particles is the local Alfv\'en speed. If the flow is highly magnetized, this speed can be close to the speed of light. But kinetic simulations have revealed that the distribution function of accelerated particles have fat tails and power-laws up to very large relativistic Lorentz factors \citep{2013ApJ...770..147C, 2014A&A...570A.112M, 2014ApJ...783L..21S, 2018MNRAS.473.4840W, 2018ApJ...862...80B}. Different acceleration mechanism are here at work which will discussed in Sect.~\ref{Sec:Kinetic_Reconnection}.

\paragraph{Driven reconnection:} reconnection sites are normally embedded in a large scale environment which is dynamic as well: jets, accretion disks, stellar winds, stellar atmospheres and coronae. Some of these environments are turbulent flows. As seen above, this can decisively accelerate the REC process. But also directed large scale flows -- as compared to the diffusion region or X-point where REC actually happens -- can significantly accelerate REC in that they provide significantly higher inflow velocities. Therefore, much more magnetic flux can be carried from larger scales to the reconnection site. The timescale of the forcing also proves to be important \citep{2001PhPl....8.3251P, 2005JGRA..11010213P, 2009PFR.....4...24O,
2010PhPl...17k2904K, 2014JPhCS.561a2021U, 2018PFR....1301025U}.

\paragraph{Multi-scale and Multi-physics problem:} As was seen so far, REC is a multi-scale problem. Large scale MHD flows can have a significant impact on the rate and the energetics of REC. Another scale is the transition to a diffusive regime which `prepares' for REC, e.g., a Sweet--Parker reconnection sheet. Such sheets may break apart, introducing even smaller scales. This cascade in scales likely ends on kinetic scales. There also the physics may change, from a collisonal to a collisionless regime. Another important point is the scale-difference in mass between electrons and ions, which also translates into differently scaled diffusive regions, the ion-diffusion region being about $42.85\,(\equiv\,\sqrt{m_p/m_e})$ times bigger than the electron diffusion region. And the different spatial lengths translate into equally different temporal scales. Magnetization and with it the ratio between an inertial length and the gyroradius yet complicates the situation.

But one has to address also other physical processes which influences the REC process. Outstanding here are radiative processes like synchrotron emission which directly changes the gyroradius. In an environment which is rich of photons, Compton scattering and Bremsstrahlung become important. More and more such processes are being addressed \citep{2003ApJ...591..366K, 2009PhRvL.103g5002J, 2013ApJ...770..147C, 2017ApJ...850..141B, 2016ASSL..427..473U, 2019MNRAS.482L..60W}.

Multi-scale, multi-physics simulations demand for special techniques which are now in the course of being developed \citep{2005JGRA..11012226T, 2014JCoPh.268..236D, 2012JCoPh.231..870T, 2013JCoPh.238..115I, 2014JCoPh.271..415M, 2015JGRA..120.4784A, 2015JCoPh.283..436R, 2016GeoRL..43..515L, 2016JGRA..121.1273T, 2017CoPhC.221...81M, 2017JPlPh..83b7005L, 2018FrP.....6..113G, 2018CoPhC.229..162G, 2018PFR....1301025U}.
We will come back to the issue in Sect.~\ref{Sec:Kinetic_Reconnection}.




\subsection{Laser plasma experiments}\label{S:LASACC}
Over the past four decades, tremendous progress in the development of high-energy and high-power laser systems has brought the scientific community with the possibility to reproduce, in the laboratory, various scenarios relevant to astrophysics, space physics and planetology. 
This opened a new avenue for the development of so-called \emph{Laboratory Astrophysics}, a field of growing activity that federates several communities [among which but not restricted to astrophysicists and (laser-)plasma physicists] and relies on the  joint development of novel experimental and numerical capabilities.

In this section, we briefly review some key experiments focusing
on the study of collisionless shocks and magnetic reconnection
in laser-created plasmas\footnote{Other experimental facilities such as the Z-pinch machines~\citep{remington2006} 
or the Large Plasma Device (LAPD) at UCLA (CA, USA)~\citep{gekelman1991,gekelman2016} 
also offer interesting opportunities for laboratory astrophysics, but are not discussed in this review.}
The reader, interested in other branch of laboratory astrophysics using laser-plasma experiments,
will find interesting material covered in the review articles by \cite{ripin1990,rose1994,takabe1999,remington1999,remington2006,takabe2008}.
These reviews cover topics ranging from warm dense matter, 
to equation of states and their application to planetology, 
opacities relevant to stellar interiors, 
or experiments investigating the hydrodynamics and magnetohydrodynamics of supernovae and (collisional) shocks.

In addition to presenting some of the main experimental results on collisionless shocks and magnetic reconnection, this Section also aims at providing the reader
with the characteristic parameters and conditions that can be created in the laboratory. 
To do so, we first introduce, in Sect.~\ref{S:LASACC_laser}, the two main classes of lasers used for laboratory astrophysics.
Then, in Sect.~\ref{S:LASACC_coll} we discuss the conditions that have to be met to ensure that collisional effects can be neglected.
Finally, Sects.~\ref{S:LASACC_shock} and \ref{S:LASACC_reconn} summarize some of the key experimental results obtained 
on collisionless shock formation and magnetic reconnection.

\subsubsection{Overview of laser facilities \& characteristic parameters} \label{S:LASACC_laser}

Two main classes of laser systems are today used to support laboratory astrophysics research.
First, high-energy density laser facilities delivering long (nanosecond) energetic (from few kJ up to 10s of kJ) light pulses have already allowed to reproduce various astrophysics-relevant scenarios, from warm dense matter studies, to the physics of hydrodynamic (radiative or not) shocks~\citep{remington2006}.
Second, ultra-high intensity laser facilities deliver short (from few tens of femtoseconds to few picoseconds) light pulses that once focused onto a target allow to reach very high intensities.
Even though laboratory astrophysics studies on this second class of laser systems is still in its infancy, recent developments of petawatt (and multi-petawatt) laser systems worldwide open new possibilities.

\begin{figure}
\begin{centering}
\includegraphics[width=0.75\linewidth]{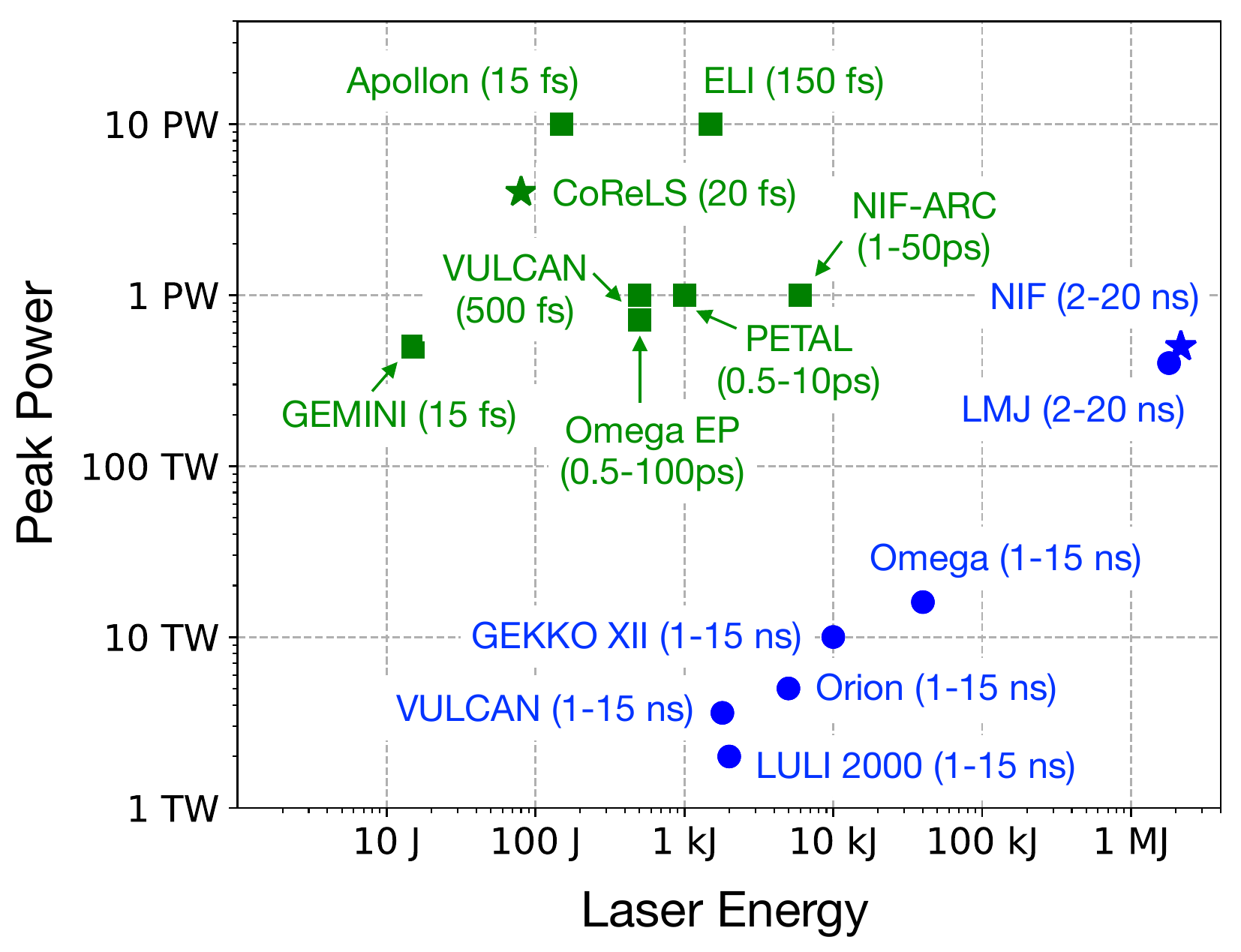}
\par\end{centering}
\caption{
Prominent laser facilities presented as a function of the delivered laser energy and peak power. 
High-energy density (HED) laser facilities are shown in blue, ultra-high-intensity (UHI) laser facilities in green.
The laser characteristics (energy, power and typical pulse duration) are indicative.
\label{fig:lasers}
}
\end{figure}

In this section, we report on some of the prominent laser facilities that are currently operating or will soon operate.
Figure~\ref{fig:lasers} lists these facilities 
as a function of the delivered energy and peak-power (the corresponding pulse durations are also indicated).

\emph{High-energy density lasers --} The development of high-energy density (HED) laser systems delivering energies of few tens of kilo-Joule (kJ) up to the Mega-Joule (MJ) 
over few to tens of nanoseconds (ns) has been strongly pushed forward by inertial confinement fusion programs~\citep{atzeni_meyertervehn}. 
Most of the experimental work that will be discussed in what follows has been performed on such laser systems.
Various HED laser systems are today available, most of which are multi-beam facilities. 
Each beam can deliver ns pulses with few to 10 kJ (i.e., hundreds of beams are used on MJ-class laser systems) that, once focused onto target, 
allow to  reach moderately high intensities of $10^{13}$ to a few $10^{15}~{\rm W/cm^2}$.
HED laser technology is based mainly on Nd:Glass amplifiers, which provide light beams at a (central) wavelength of $\sim 1.05~{\rm \mu m}$, 
but often use frequency doubling of tripling techniques, so that the operating wavelength can be decreased to $\sim 0.53{\rm \mu m}$ (doubling)
or $\sim 0.35{\rm \mu m}$ (tripling).

Among the prominent facilities are -- at the multi-kJ level -- the LULI 2000 laser\footnote{\url{https://portail.polytechnique.edu/luli/fr/installations/luli2000} (in French)} in France, 
the Orion\footnote{\url{https://www.awe.co.uk/what-we-do/science-engineering-technology/orion-laser-facility/}} and VULCAN\footnote{\url{https://www.clf.stfc.ac.uk/Pages/Vulcan-laser.aspx}} facilities in the UK,
the GEKKO XII facility\footnote{\url{http://www.ile.osaka-u.ac.jp/eng/facilities/gxii/index.html}} in Japan,
and the Omega laser\footnote{\url{http://www.lle.rochester.edu/omega_facility/omega/}} in Rochester, US.
Two mega-joule-class lasers are also operating or under construction:
the National Ignition Facility (NIF)\footnote{\url{https://lasers.llnl.gov/}} in Livermore, California, started operating in the early 2010s.
The Laser-MegaJoule (LMJ)\footnote{\url{http://www-lmj.cea.fr/}} is still under construction in the South-West of France.
Note that the MJ-energy level is achieved by combining hundreds of 10~kJ nanosecond laser beams.

\emph{Ultra-high intensity lasers --} High-power ultra-high intensity (UHI) laser facilities provide light pulses of moderate energy (from few tens of Joule to few kilo-Joule) 
but of a very short duration (from tens of femtoseconds to a few picosecondes) that, 
when focused onto a target, allow to reach tremendous intensities (beyond $10^{18}~{\rm W/cm^2}$). 
At such intensities, electrons  rapidly -- in less than an optical cycle -- become relativistic, and such UHI laser can help drive extremely fast, 
potentially relativistic, flows of plasmas. 

Among the UHI lasers that are today considered for laboratory astrophysics studies, many are coupled to HED facilities.
This is the case for instance of the PETAL\footnote{\url{http://www.enseignementsup-recherche.gouv.fr/cid99515/petawatt-aquitaine-laser-petal.html} (in French)} 
and NIF-ARC\footnote{\url{https://lasers.llnl.gov/science/photon-science/arc}} petawatt-class lasers coupled to the LMJ and NIF facilities, respectively,
which deliver petawatt-level light pulses with energy of a few kJ and duration in the picosecond range.
LULI 2000, VULCAN and ORION also have short (picosecond) pulse beamlines that deliver energy up to a few 100s J.

Other UHI facilities are also available that are not coupled to HED laser systems.
This is the case e.g. of the femtosecond laser GEMINI\footnote{\url{https://www.clf.stfc.ac.uk/Pages/Laser-system-Gemini.aspx}} in UK.
Delivering 15 J in 30 fs, GEMINI has been used e.g. to produce dense electron-positron clouds which were used to drive current instabilities in a Helium plasma~\citep{warwick2017}.
Let us further note that the most powerful laser is today the CoReLS\footnote{\url{https://www.ibs.re.kr/eng/sub02_03_05.do}} in Gwanju, South-Korea, 
that has recently delivered 4~PW pulse~\citep{nam2018}.
In addition the Apollon\footnote{\url{http://cilexsaclay.fr/Phocea/Vie_des_labos/Ast/ast_technique.php?id_ast=9} (in French)} laser (in construction on the Plateau de Saclay, 20 km south of Paris, France) and the ELI project\footnote{\url{https://eli-laser.eu/}} aim at reaching the unprecedented 
power of 10~PW within the next few years (in light pulses of a few tens to few 100s fs).
Laboratory astrophysics studies are envisioned on these facilities.

\subsubsection{The collisionless regime} \label{S:LASACC_coll}

As previously stated, this section focuses on collisionless laser-plasma experiments 
and on the physics of collisionless shocks and magnetic reconnection in particular.
The first observations of a collisionless coupling in laser-created plasmas date back to the early 1970s \citep{cheung1973},
and very early the question collisionality effects arose,
see, e.g., the work by \citet{dean1971} and following exchange \citep{wright1972,dean1972}.

The first (theoretical) investigations of collisionless shock experiments actually addressed this issue~\citep{drake2012,park2012,ryutov2012}.
These works proposed the first designs and scaling laws to reproduce electrostatic or Weibel-mediated shocks (see Sect.~\ref{S:MIC-MES} for complementary definitions) in laser-plasma experiments,
and addressed the potential effects of particle collisions (and how to mitigate them) in counter-streaming plasma flows.

Of particular importance are collisions in between counter-streaming ions of the two flows (inter-flow collisions) that can have a dramatic effect on the shock formation.
Indeed, and as stressed by \citet{drake2012}, the mean-free-path for ion-ion collisions measures the length over which an ion (subject to multiple scatterings/collisions)
sees its velocity deflected by $90^{\circ}$. Hence, collisional effects will effectively isotropize the flow over a characteristic given by this mean-free-path
and collisional shocks are known to develop on this spatial scale.
Conducting a collisionless shock experiment thus requires that this (inter-flow) collision \emph{greatly} exceeds the characteristic length of shock formation.
As will be detailed in the following Sect.~\ref{S:LASACC_shock}, it turns out that this condition can be ``easily'' met in electrostatic and to some extent magnetized shock experiments.
In the case of Weibel-mediated shocks, entering the collisionless regime requires extremely fast (several 1000s km/s) flows overlapping for a sufficient time accessible only on the most energetic (MJ-class) laser systems~\citep{park2012}.

In addition to inter-particle collisions, internal collisions between ions of the same flow can also be of importance,
in particular as the ion temperature in the flow is quite low (at least before shock formation).
This issue is briefly addressed by \citet{drake2012} and \citet{ryutov2012}.
Yet their impact on the development of instabilities such as the ion Weibel instability, or shock formation remains unclear.
While it is difficult to claim that these (internal) collisions may not strongly modify the physics of shock formation, this may be checked by careful numerical modeling,
e.g. relying on kinetic (Particle-In-Cell) simulations including collisional effects (see Sec.~\ref{PICCodes:additionalModules}).

Last, \citet{drake2012} discussed the possible impact of electron-ion collision on the dissipation of the magnetic structures that play a central role in the 
formation of Weibel-mediated shocks; and on longer time on particle acceleration. 
The authors showed that such collisions may indeed impact the small scale magnetic structures, but will most
likely no impact the larger scale structures that develop on the scale of the ion skin-depth and thus the formation of Weibel-mediated shocks.

\subsubsection{Collisionless shock experiments}\label{S:LASACC_shock}

As previously mentioned, evidences of collisionless processes in the presence of 
counter-streaming laser-produced plasmas were reported is the early 1970s.
The first reported observation of a collisionless shock in a laser-created plasma\footnote{Collisionless shock waves were obtained in plasma experiments, albeit not using lasers, since the mid-1960s~[see \citet{strokin1985} and 
references therein]. Already these studies where motivated by space-plasmas and astrophysics.}
dates back to~\citet{bell1988}. 
This experiment was conducted on the VULCAN laser~\citep{ross1981} at the Rutherford Laboratory (UK) where two laser
pulses, each delivering 120J~ over 18~ns (FWHM), were focused in a 50~${\rm \mu m}$-diameter spot onto a flat carbon target.
The resulting laser-produced ablation plasma had a density $\sim 10^{18}~{\rm cm^{-3}}$ and velocity of a few 100s of km/s.
It collided with an obstacle (located 250~${\rm \mu m}$ away from the ablated target).
The experiment led to the formation of density structures that were interpreted as collisionless bow shocks. 
In this experiment, all mean-free-paths were larger than the mm, while the width of the observed shock front  
ranged from 0.01 to 0.05 mm.
The nature of the shock -- either electrostatic or weakly magnetized -- was however not fully defined.

\emph{Electrostatic shocks --} 
Following this pioneering work, and since the late 2000s in particular,
collisionless electrostatic shocks have been abundantly produced in laser-plasma experiments.
These later developments were accompanied by both strong developments in diagnostics,
and the use of kinetic (Particle-In-Cell) simulations to support the experimental effort.

\begin{figure}
\begin{centering}
\includegraphics[width=0.95\linewidth]{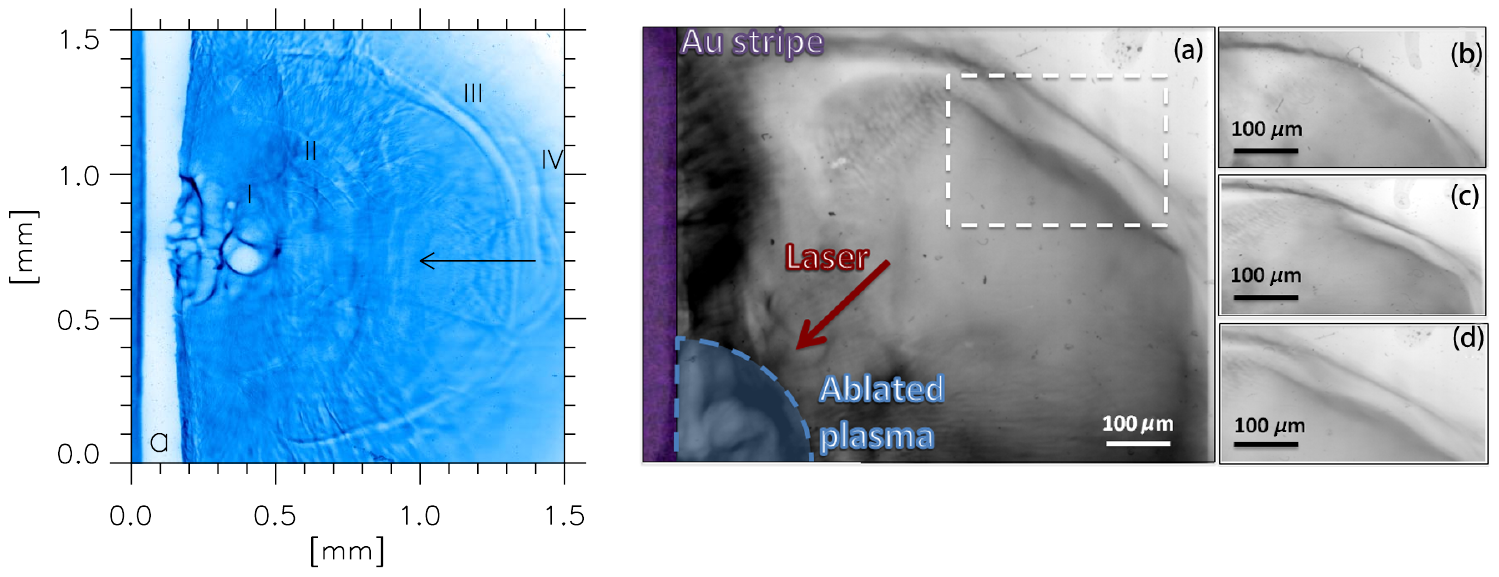}
\par\end{centering}
\caption{
Typical proton radiographies of laser-driven electrostatic shocks. 
[\emph{Left}, taken from  \citet{romagnani2008}] 
Region~I shows strong modulations associated with the ablating plasma, regions II and III show different structures that are interpreted as shock waves propagating ahead of the ablating plasma,
while the modulated pattern in Region~IV is located ahead of the shock front and possibly associated with a reflected ion bunch. 
The arrow indicates the laser beam direction. 
[\emph{Right}, taken from \citet{ahmed2013}] Shock structure 150 (b), 160 (c) and 170 ps (d) after the beginning of the interaction.
The different times are accessible, for a single shot, by selecting protons with different energies [11.5 MeV for panel (b), 10 MeV for panel (c) and 9 MeV for panel (d)] 
as their time of flight from their source to the shock structure is different.}
\label{fig:romagnani2008}
\end{figure}

For instance, \citet{romagnani2008} demonstrated the creation of an electrostatic shock 
following the sudden expansion of a plasma into a rarefied gas.
In this experiment carried out on the LULI 100~TW laser facility, one laser pulse with duration 470~ps and energy of a few tens of J
was focused onto a Tungsten or Aluminium foil. Quickly heated, the ablated foil expanded in the surrounding 
media and drove the formation of a collisionless electrostatic shock wave, about 1 mm away from the target,
that was propagating at a velocity close to the ion acoustic velocity $\sim 200-400$ km/s.
The shock was diagnosed using proton radiography~\citep{borghesi2001}, a technique that is now central to the study of collisionless
shock in laser-plasma experiments. It relies on the deflection (in the electromagnetic fields developed 
at the shock front) of protons created by a second ultra-short ($\sim 300$ fs) ultra-intense ($\gtrsim10^{18}~{\rm W/cm^2}$)
laser pulse. The proton radiography is recorded onto dosimetrically calibrated radiochromic films (RCFs),
as shown in Fig.~\ref{fig:romagnani2008}.

Other experiments~\citep{kuramitsu2011,ahmed2013,morita2013} have similarly reported the formation 
of electrostatic collisionless shock waves using ablating plasmas, either in direct interaction with a 
standing (background) plasma, or in counter-streaming plasma flows.

\begin{figure}
\begin{centering}
\includegraphics[width=0.65\linewidth]{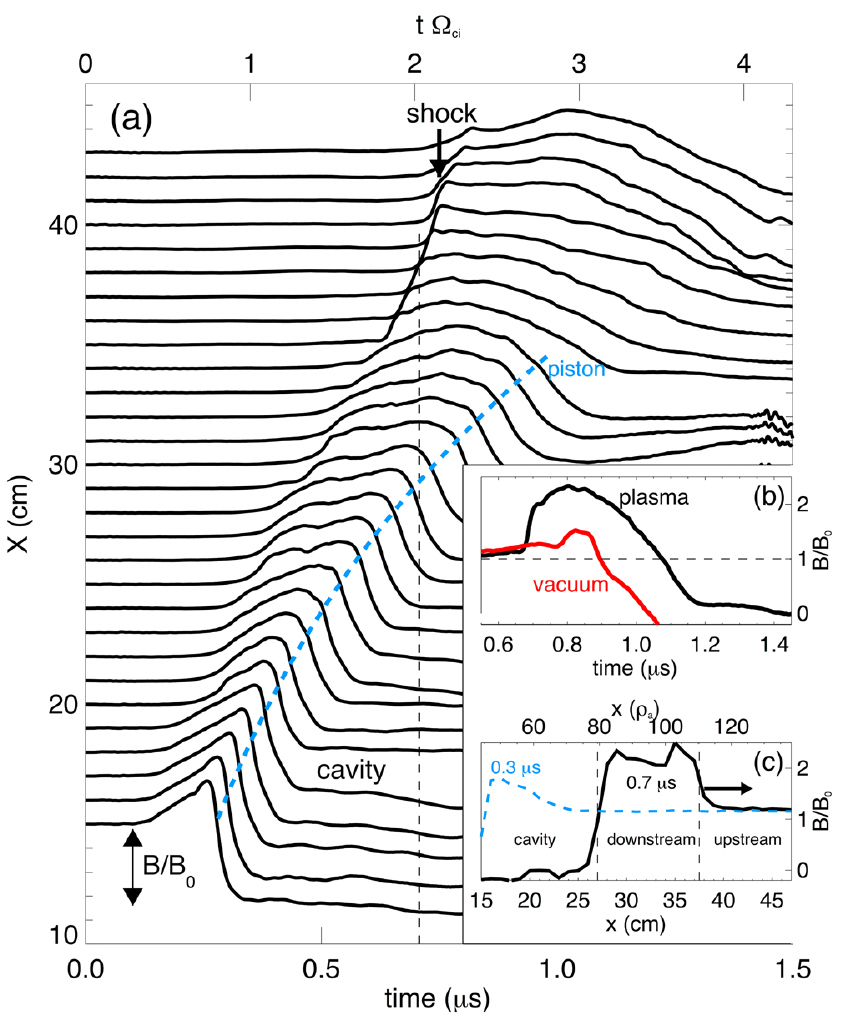}
\par\end{centering}
\caption{
Magnetized shock experiment at the Large Plasma Device (LAPD, University of California Los Angeles).
(a) Magnetic ($B$) field structure as a function of position ($x$ is the direction of the carbon plasma flow, transverse to the direction of the background magnetic field) and time.
(b) Temporal evolution of the magnetic field at $x = 35$ cm with (black) and without (red) the ambient plasma.
(c) Spatial profile of the magnetic at two different times, $0.3~{\rm \mu s}$ (before shock formation, dashed light blue line) and $0.7~{\rm \mu s}$ (after shock formation, solid black line).
Taken from~\citet{niemann2014}.
\label{fig:niemann2014}
}
\end{figure}

\emph{Magnetized shocks --} 
The first observation of a magnetized collisionless shock motivated by astrophysics studies was claimed
by \citet{niemann2014} combining the use of a 25 ns - 200~J laser and the LAPD. 
In this experiment, the LAPD was used to produce a large scale (17m x 0.6m) low density ($10^{12} - 10^{13}~{\rm cm^{-3}}$)
and temperature ($T_{\rm i}=1~{\rm eV}$, $T_{\rm e}=6~{\rm eV}$)
hydrogen plasma embedded in an external magnetic field $B_0=300~{\rm G}$.
The $10^{13} {\rm W/cm^2}$ ns laser pulse was fired at a solid polyethylene target embedded inside the magnetized plasma, 
which launched a denser ($8\times10^{16}~{\rm cm^{-3}}$) slightly warmer ($T_{\rm e} \sim 7.5~{\rm eV}$) carbon ion plasma at a velocity
of $\sim 500$ km/s directed perpendicular to the magnetic field. 
The interaction of this super-Alfv\'enic plasma with the ambient (LAPD) plasma led -- through a collisionless coupling --
to the formation of a magnetic piston and then to the formation of self-sustained magnetosonic shock, supported by the ambient ions and
 propagating away from the piston at a velocity of $\sim 370$ km/s (corresponding to an Alfv\'enic Mach number $M_A \sim 2$).
 The reported measurements (shock velocity and magnetic field compression $B/B_0 \sim 2$) were found to be consistent with
 Rankine-Hugoniot conditions as well as with two-dimensional, collisionless, simulations performed using an electromagnetic
 Darwin code \citep{winske2007}.
Note that, as illustrated in Fig.~\ref{fig:niemann2014}, for this particular experiment, the use of the LAPD allowed to follow the magnetic piston and shock formation
over large spatial (few tens of cm) and temporal (few microseconds) scales, well beyond what is usually accessible using
HED or UHI laser systems.

\begin{figure}
\begin{centering}
\includegraphics[width=0.95\linewidth]{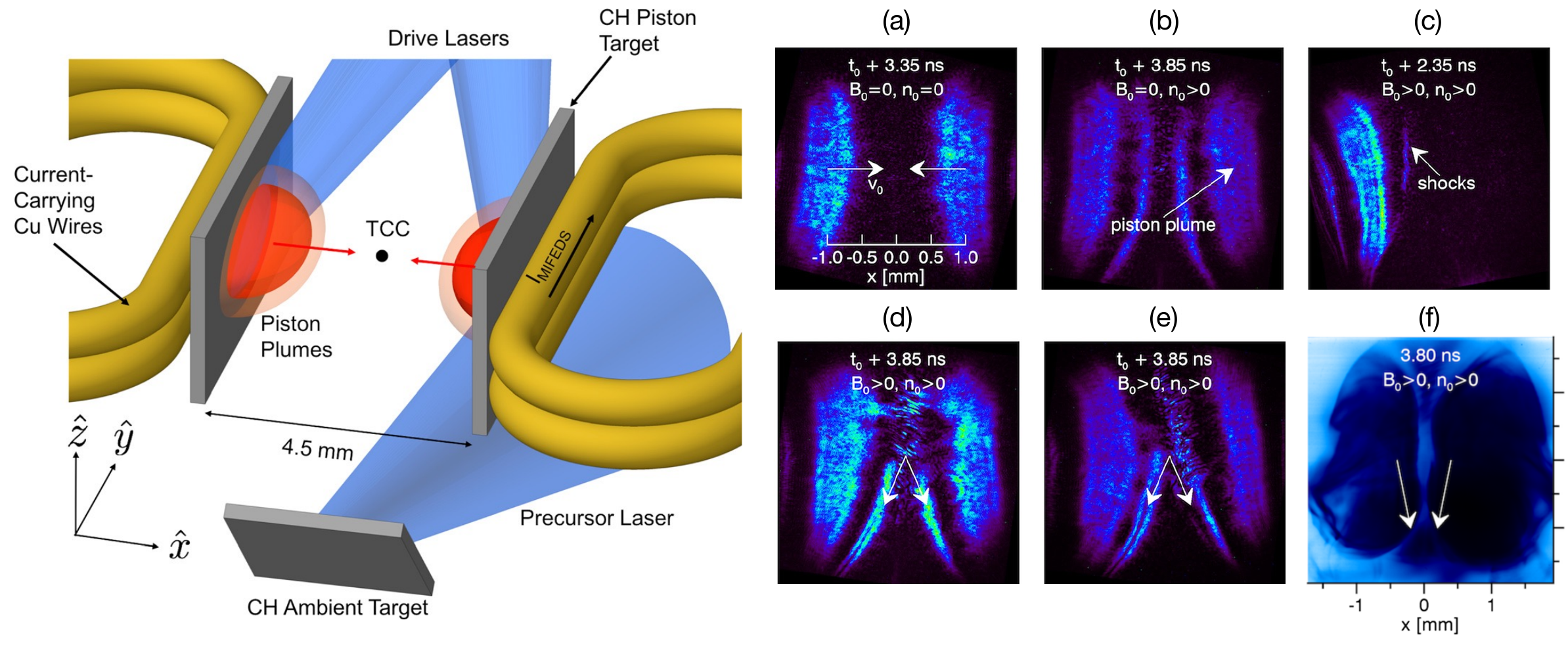}
\par\end{centering}
\caption{Magnetized shock experiment at the Omega EP laser facility [adapted from \citet{schaeffer2017,schaeffer2017b}]. 
(\emph{Left}) Simulation setup.
(\emph{Right}) Panels (a-e) present the angular filter refractometry measurements for different configurations: 
(a) no background plasma, no external magnetic field (no shock is observed);
(b) with both background plasma and external magnetic field but only one piston plume; 
(c) presence of a background plasma but no external magnetic field;
(d,e) in the presence of the two piston plumes and background plasma, with external magnetic fields in parallel and anti-parallel configuration, respectively.
(f) Proton radiography signal revealing the strong magnetic field compression at the location of the shock structure.
\label{fig:schaeffer2017}
}
\end{figure}

The first laboratory observation of a laser-driven high-Mach-number magnetized collisionless shock was reported by \citet{schaeffer2017}.
This experiment was conducted on the Omega EP laser facility at Rochester (US) and built up on the concept of magnetic piston used in, e.g, the previous experiment by \citet{niemann2014}.
However, it relied solely on the use of the HED laser Omega EP and allowed to create super-critical magnetized shocks with (magnetosonic) Mach number 
$M_{\rm ms} \equiv u_{\rm sh}/c_{\rm ms} \sim 12$, with $u_{\rm sh} \sim 700$ km/s the measured shock velocity ($c_{\rm ms}^2 = u_A^2+c_s^2$, $u_A$ and $c_s$ being
the (upstream) Alfv\'en and ion acoustic velocities, respectively).
To do so, various beams of  the Omega EP laser were used.
A first beam, with energy 100 J and duration 1 ns, was focused (at intensity $\sim 10^{12}~{\rm W/cm^2}$) onto a CH target thus
producing a background plasma.
A second and third beam, with energy 1.5 kJ and duration 2 ns, were then focused onto two opposing CH targets, 
leading to the production of two counter-propagating ablation plasmas.
Even though the details of the resulting three plasma flows are not fully documented in either \citep{schaeffer2017} or the companion paper \citep{schaeffer2017b},
the density and temperature of the overlapping plasma were estimated to $\sim 6 \times 10^{18} {\rm cm^{-3}}$ and $T_{\rm e} \sim 15~{\rm eV}$, respectively. 
The whole set-up was embedded into a 80 kG perpendicular magnetic field produced by a pulsed current passing through Copper wires located
behind the 2 opposing CH targets (for the readers convenience, the experimental set-up is reproduced in the left panel of Fig.~\ref{fig:schaeffer2017}).
The resulting magnetic piston and shock structures are evidenced in the right panels of Fig.~\ref{fig:schaeffer2017}.
The importance of the background plasma (created by the first 100 J-beam) is made clear by comparing panel (a) to panels (b-e).
Without background plasma, panel (a), no shock is observed.
In the presence of a background plasma, panels (b-e), a shock-like structure is observed in all cases, strongest in the presence of the external magnetic field [panels (c-e)],
but still present when no external magnetic field is applied [panel (b)].
The authors advance the possibility, supported by PIC simulations, that in this latter case, the Biermann-Battery process was responsible for seeding a large scale magnetic 
field even though no external one is applied.
Note also, that in panel (c), the authors report the creation of a shock when only one of the 1.5 kJ-beam was used, demonstrating that only one piston plume interacting
with the background plasma was needed to produce a magnetized shock.
Last, panel (f) reports the measurement obtained using proton radiography and reveals a strong magnetic field compression at the location of the shock structure.

Note that the overall experimental campaign strongly relied on advanced diagnostics [shadowgraphy, angular filter refractometry~\citep{haberberger2014}, and proton radiography]  
as well as the combined used of hydrodynamic (for the plasma characterization) and PIC (for the shock formation and evolution) simulations.

Since these experiments, the effort in producing and studying magnetized shocks has continued, exploring e.g. the possibility to produce parallel shocks~\citep{weidl2017},
or to use the Biermann-Battery process to magnetize the plasmas~\citep{umeda2019}.

\emph{Weibel-mediated shocks --}  
Weibel-mediated collisionless shocks are certainly the most sought after collisionless shocks in laser-based laboratory astrophysics experiments. 
Even though important progress have been made in the last decade, recreating such shocks in the laboratory has not yet been achieved. 
The main difficulty in recreating such shocks stems from the need to achieve flow densities that are, on the one hand, sufficiently small to ensure
that one operates in the collisionless regime and, on the other hand, large enough for the ion Weibel instability to develop and the resulting magnetic 
turbulence to build up. 
The combined experimental and theoretical effort in this endeavor has been started since the early 2010s and HED NIF-class laser systems,
allowing to produce plasma flows with densities of a few $10^{19}\,{\rm cm^{-3}}$ and velocities of several 1000s km/s, have been identified 
as the most promising path toward collisionless Weibel-mediated shock formation.

\begin{figure}
\begin{centering}
\includegraphics[width=0.95\linewidth]{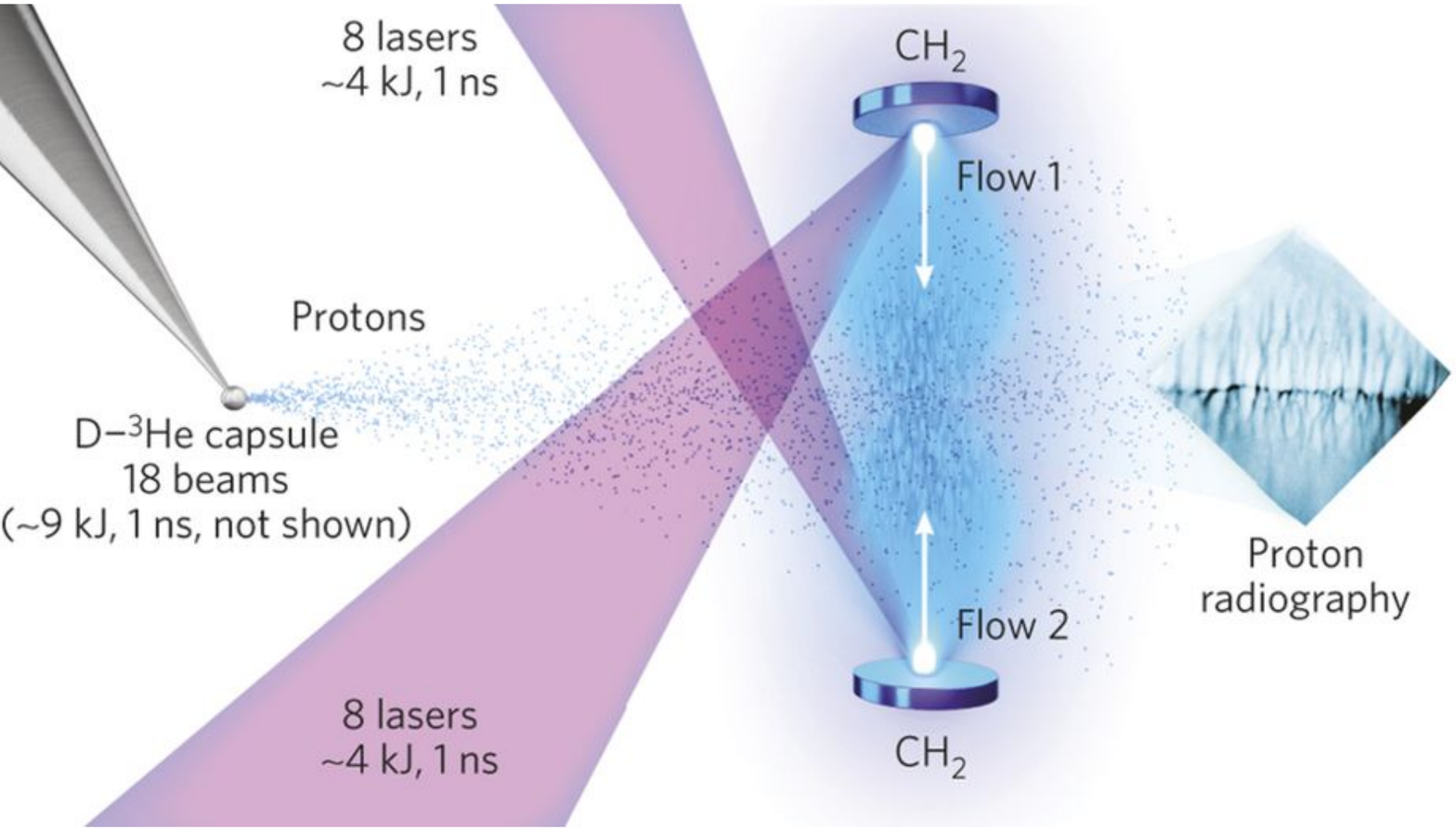}
\par\end{centering}
\caption{Demonstration of the ion Weibel instability in counter-streaming plasmas on the Omega EP facility [taken from \citet{huntington2015}]. 
Simulation setup: two ablation plasmas are formed in counter-streaming configuration by irradiating two opposing plastic targets.
A $^3$He-D target is imploded to drive fusion reactions that allow for the production of 3 MeV and 14.7 MeV protons that are used to radiograph
the electromagnetic fields developed in the region where the two counter-streaming plasma overlap.
The resulting proton radiograph (here taken about 5~ns after the beginning of interaction using the 14.7 MeV protons) shows evidence of the growth
of filamentary structures associated to the Weibel magnetic fields.
\label{fig:huntington2015}
}
\end{figure}

The first step toward creating plasma flows relevant for such studies was taken by \citet{park2012}, demonstrating the possibility
to drive plasma flows with velocities of $\sim 1000$ km/s and densities of $\sim 10^{18}\,{\rm cm^{-3}}$ from plasma ablation 
at the Omega laser facility. The production of large-scale electromagnetic structures~\citep{kugland2012} and later clear demonstration 
of Weibel-type ion filamentation instabilities~\citep{fox2013,huntington2015} in the presence of counter-streaming plasmas were obtained by irradiating a pair 
of opposing plastic (CH) foils with few kJ,  few ns laser pulses on the Omega EP laser system. Figure~\ref{fig:huntington2015} reproduces 
the schematic experimental set-up used by \citet{huntington2015} together with a typical proton radiography measurement of the magnetic field filamentary structures 
following from the development of the ion Weibel instability. The region imaged by the proton radiograph is about 3 mm wide, and 
the filamentary structures have a typical width of $\sim 150 - 300~{\rm \mu m}$, consistent with the ion skin-depth for the reported plasma
density of a few $10^{18}\,{\rm cm^{-3}}$. 

The plasmas flows achievable at Omega are unfortunately too low density, and short life, to allow creating a Weibel-mediated collisionless shock.
A recent theoretical model and 2D PIC simulations by \citet{ruyer2016} indeed predict that the isotropization necessary for shock formation
may be achievable if the two counter-streaming flows overlap over a length of the order of at least: 
\begin{eqnarray}\label{eq:Liso}
L_{\rm iso} \simeq 35\,\left[m_i/(Zm_e)\right]^{0.4}\,\left(c/\omega_{pi}\right) \rightarrow 5~{\rm cm} \times \left(\frac{A}{Z}\right)^{0.9}\, \sqrt{\frac{10^{19}\,{\rm cm^{-3}}}{n_0}}\,,
\end{eqnarray}
where $c/\omega_{\rm pi}$ is the ion skin-depth associated to the plasma flow density $n_0$, $m_{\rm i}$ ($m_{\rm e}$) the ion (electron) mass,
and $Z$ ($A$) the ion charge (mass) number.
Conversely, the two flows shall overlap for a time of the order of $\tau_{\rm iso} = L_{\rm iso}/v_0 \simeq $, with $v_0$ the relative flow velocity.
A necessary, yet not sufficient, condition for maintaining the collisionless regime imposes
that the isotropization length given by Eq.~\eqref{eq:Liso} is much smaller than the characteristic ion-ion collision mean-free-path~\citep{park2012}:
\begin{eqnarray}
\lambda_{\rm mfp} \simeq 5~{\rm cm} \times \frac{A^2}{Z^4}\,\left(\frac{v_0}{1000\,{\rm km/s}}\right)^4\,\left(\frac{10^{19}\,{\rm cm^{-3}}}{n_0}\right)\,,
\end{eqnarray}
for which are considered only collisions between ions of the two counter-streaming flows. 
These estimates allow to infer that Weibel-mediated collisionless shocks may be achieved in the presence of (hydrogen) plasma flows with density of the order 
of $10^{19}\,{\rm cm^{-3}}$,
colliding at a relative velocity of at least a few 1000s km/s, provided these flows overlap over distance of a few cm during a few to tens of ns.
Such conditions can be met only at the most energetic, MJ-class laser systems such as NIF.

The first experiments to recreate Weibel-mediated shocks at NIF have been started in the framework of the \emph{Discovery Science} program.
The first experimental results were reported by \citet{ross2015} focusing on how to tune the experimental conditions to access the collisionless regime.
These experiments, for which no external magnetic field was used, considered two solid targets, made of a mixture of Carbon and either hydrogen or deuterium (CH or CD),
each irradiated by several beams allowing to deliver energies of $\sim 250$ kJ (much larger than accessible e.g. at Omega) per foil during about 5 ns. 
The resulting ablation plasma flows at velocities of $\sim 1000$ km/s  and (ion) densities of a few $10^{19}~{\rm cm^{-3}}$ in the interaction (overlapping) region.
This work demonstrated that if the foils (in opposing configuration) were sufficiently distant from one another, collisional effects could be strongly mitigated due to the reduction
of the plasma flow density in the overlapping region.
Most importantly, this work reported evidence of a collisionless (collective) heating in the flow interaction region,
which the authors associate with the nonlinear stage of the Weibel instability and thus to the early stage of shock formation.
This result suggests that the scientific community is on the verge of producing Weibel-mediated collisionless shocks in the laboratory.
A new experiment was actually conducted at NIF in the last months increasing the driving laser beam energy to $\sim 500$ kJ delivered to each foil.
This experiment is expected to lead to shock formation, and may also gives the first signs of particle acceleration in the shock.
To this date, the results of this last experimental campaign have not been announced.

\emph{Prospective numerical studies --} The possibility to drive collisionless shocks in the laboratory has prompted 
the laser-plasma community to investigate various laboratory configurations to drive collisionless collective processes
and shocks \textit{in silico}. Indeed, various numerical experiments have been performed using Particle-In-Cell (PIC) simulation.
Even if some of these numerical experiments consider laser parameters not yet within our reach (ultra-short ps-level,
energy at the 100s J level) other have addressed conditions that are or will soon be achievable on the forthcoming
extreme light facilities such as Apollon or ELI.

\citet{fiuza2012} put forward the possibility to drive -- through Weibel-like instabilities -- a collisionless shock in a dense target 
using an ultra-intense light pulse. This scenario was revisited by \citet{ruyer2015} that demonstrated the dominant
role of laser-driven hot electrons in the shock formation. More recently, \citet{grassi2017} showed that tuning the laser-plasma
interaction configuration can help mitigate the hot electron production so that shock formation can be driven by the
ion Weibel instability, as expected in astrophysical scenarios. 

In addition, dense electron-positron flows have been produced in laser-plasma experiments~\citep{chen2015,sarri2015}, 
offering the opportunity to study pair-plasma processes in the laboratory, and motivated various numerical experiment. 
Using QED-PIC simulation, \citet{lobet2015} demonstrated the possibility to drive ultra-relativistic, counter-propagating
electron-positron pair plasmas using extreme light pulses (with intensity beyond $10^{23}~{\rm W/cm^2}$, 100s kJ 
and duration of few tens of fs). Ultra-fast isotropization and thermalization (a first step in shock formation) were observed in the 
simulation, and associated to both the Weibel instability and a remarkable contribution from synchrotron emission by the
ultra-relativistic leptons in the strong (Weibel) magnetic fields. 
More recently, the (collisionless) interaction of midly relativistic pair jets with background (electron-ion) plasma was also investigated
in kinetic simulations~\citep{2018PhPl...25f2122D, 2018PhPl...25f4502D}. Remarkably, these studies are not only motivated by
astrophysics~\citep{dieckmann2019} but also by recent experiments that demonstrated the growth of a current-driven instability developing during the interaction a quasi-neutral
pair beam with a background (electron-ion) plasma~\citep{warwick2017}.

\subsubsection{Magnetic reconnection experiments}\label{S:LASACC_reconn} 

\begin{figure}
\begin{centering}
\includegraphics[width=0.95\linewidth]{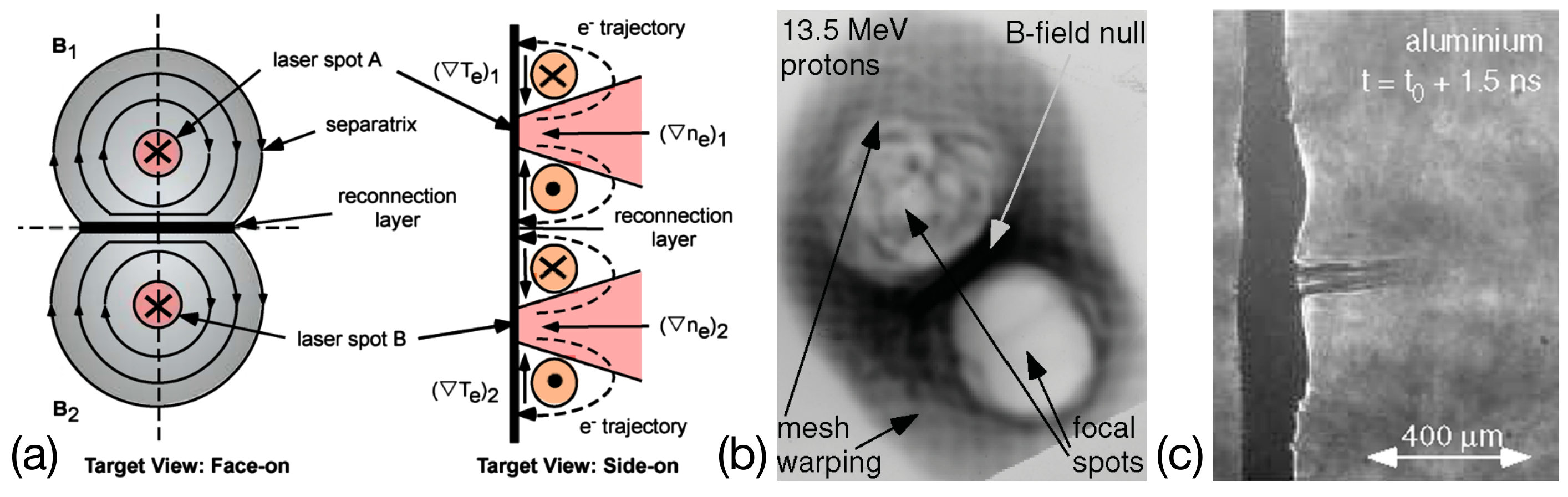}
\par\end{centering}
\caption{Adapted from the work by \citet{nilson2006} presenting the first magnetic reconnection experiment with laser-created plasmas.
(a) Experimental set-up using the two-spot configuration. 
(b) Proton-radiography measurements showing in dark the regions were protons are recorded.
(c) Shadowgraphy measurements indicating the formation of two jets at the reconnection layer.}
\label{fig:Nilson}
\end{figure}

Laser-plasma experiments also provide a test bed for magnetic reconnection studies\footnote{In this review,
we focus once more on laser-plasma experiments. Yet, other experiments have been conducted on various devices:
the Magnetic Reconnection Experiment (MRX) at Princeton \citep{yamada1997},  
the LAPD at UCLA \citep{gekelman2010}, Z-pinch machines \citep{hare2017}.
Magnetic reconnection is also known to affect Tokamak experiments \citep{goetz1991}.
See also \citep{howes2018} for a review of various laboratory experiments for space plasma physics.}.
The first evidences for magnetic reconnection in a laser-plasma experiment were reported by \citet{nilson2006}.
This experiment was performed at the VULCAN laser facility in the UK, and relied on a now popular set-up 
consisting in firing two laser pulses at a solid (here Aluminium or Gold) target [see Fig.~\ref{fig:Nilson}(a)].
The interaction of each pulse leads to plasma ablation and expansion associated with the generation of an azimuthal magnetic field through the Biermann-Battery process.
In between the two pulses, the magnetic fields driven by the two pulses are in an anti-parallel configuration,
and a reconnection layer can form.

This experiment was carried out in the HED regime of interaction, each laser pulse of the VULCAN facility delivering 200J over 1~ns  in a 30--50 ${\rm \mu m}$ focal spot 
(the corresponding laser intensity is moderate $\sim 10^{15}~{\rm W/cm^2}$).
Various complementary diagnostics were used, highlighting features consistent with magnetic reconnection.
(i) Proton radiography \citep{borghesi2001} allowed to probe the generated magnetic fields.
A typical measurement is reproduced in Fig.~\ref{fig:Nilson}(b); and an additional analysis of these measurements is given in \citet{willingale2010}.
Light regions correspond to regions free of protons, i.e., to regions where the strong Biermann-Battery magnetic field 
(estimated to be of the MG-level in this experiment few 100s of ps after the ablated plasma started expanding) 
is present.

The presence of a strong proton signal (dark region) in between the two lighter blobs was identified as the reconnection layer,
where opposite magnetic field lines can reconnect and lead to a null-magnetic field region.
(ii) In addition, the interaction region was also probed by a short (10ps) light pulse allowing to produce a shadowgraphy [as well as an interferogram (not shown here)]
of the interaction region.
Such a shadowgram is reproduced in Fig.~\ref{fig:Nilson}(c) and shows the formation - on the ns-timescale - of two distinct jets 
[see original paper and \citet{nilson2008} for more details] 
with velocities of $\sim 500$ km/s, 
which would not be expected should only hydrodynamic processes govern the plasma evolution.
(iii) Finally, Thomson Scattering measurements (not shown here) showed that, while the electron temperature was of the order of 800 eV (at 1.5 ns) and 700 eV (at 2.5 ns) 
in the ablated plasmas, a much higher electron temperature $\sim 1.7$ keV (at 1.2 ns) was measured in what was identified as the reconnection layer 
(where there is no laser).
Such high electron temperatures were also put forward as a result of magnetic reconnection; and were consistent with supporting hybrid simulations.

\begin{figure}
\begin{centering}
\includegraphics[width=0.4\linewidth]{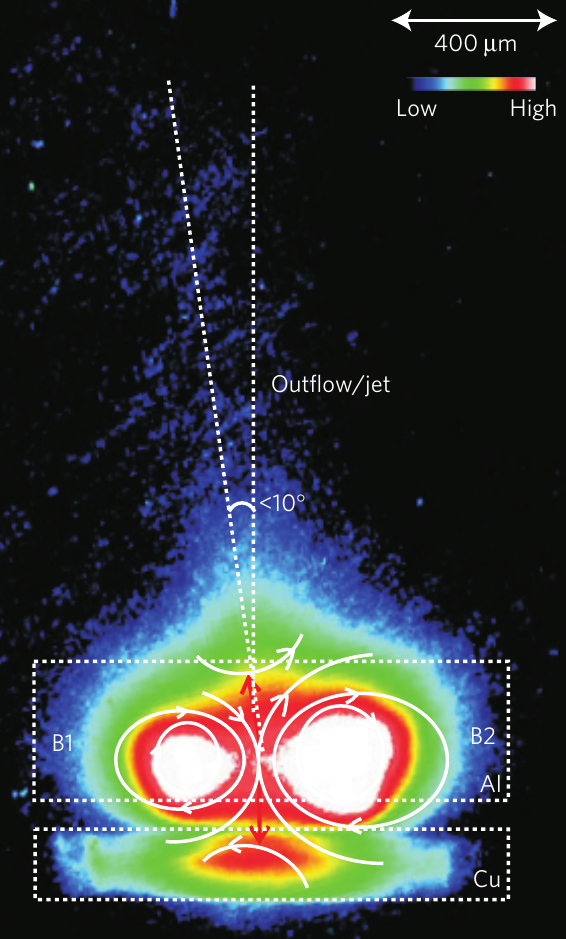}
\par\end{centering}
\caption{Taken from \citet{zhong2010}. x-ray imaging showing two bright spots in the Al target were the expanding plasmas are heated by the laser beams.
Here the asymmetry in the laser intensities (in regions B1 and B2) leads to an inclination of the upward flow.
Bellow the Al-target, a Cu-target is placed. The downward flow impinges on this target and results in a hot x-ray source.
\label{fig:zhong}
}
\end{figure}

Since this first experiment, similar results were obtained on other HED laser facilities.
\citet{li2007} report on an experiment performed at the OMEGA laser facility using a slightly more energetic 1ns laser pulse (500J), 
with a spot diameter of $\sim 800 {\rm \mu m}$ (corresponding to an intensity of $\sim 10^{14} {\rm W/cm^2}$).
This experiment benefited from a high-quality proton radiography (monoenergetic 14.7 MeV protons were produced by fusion reactions from an imploded D$^3$-He target),
which allows the authors to probe the changes in the magnetic field topology as magnetic reconnection proceeds.
See also \citet{rosenberg2015a,rosenberg2015b}.
Similarly, \citet{zhong2010} reported on a similar experiment carried out on the Shenguang II (SG II) laser facility in Shanghai, China.
In this experiment, four laser beams (1ns, few 100s J, $50$--$100 {\rm \mu m}$-wide spots) are used to drive the plasma expansion in the two-spot configuration previously discussed, but shining the lasers on the front and back side of the target simultaneously. This experiment also put the accent on scaling their results with respect to 
reconnection outflows in solar flares. As an exemple, relying on X-ray imaging of the interaction region,
the authors could demonstrate a change in the directionality of the jets due to an asymmetry in the driving laser intensities\footnote{See also the work by \citet{rosenberg2015b} for a study of asymmetric reconnection OMEGA laser facility.},
as shown in Fig.~\ref{fig:zhong}. 

Another experimental set-up was also proposed by \citet{fiksel14} and conducted on the OMEGA EP laser.
In contrast with the previous experiments, this new set-up relies on (i) an head-on configuration with two targets (irradiated by
kJ, ns laser beams), (ii) current-carrying conductors placed behind the two targets to create an external (up to 80 kG) magnetic field imposed 
perpendicular to the expanding plasma flows and designed such that the field a x-type null point in between the two targets;
and (iii) the presence of a background plasma created by a third (100J, 1ns) laser beam.
Proton radiography measurements indicate the formation and collision of magnetic ribbons, pileup of magnetic flux and reconnection
which are found to be in remarkably good agreement with 2D PIC simulations (that include particle collisions).

While the previous experiments were conducted in a collisional regime,
more recent experiments have focused on the collisionless regime.
\citet{dong12} conducted an experiment on the SG II laser, using a similar set-up then previously presented by \citet{zhong2010},
but firing the laser beams (450J on each target) at two Al-targets separated by 150--240 ${\rm \mu m}$,
which, even though the collisionless nature of the reconnection region is not fully addressed, lead the authors to claim 
the study of a structure of collisionless reconnection.
These authors also report on the ejection of a plasmoid that, when it rapidly propagates away, deforms the reconnected magnetic field
and generate a secondary current sheet. This process seems to be well reproduced by PIC simulations, 
and the primary reconnection event is found to be associated with well-collimated plasma outflows containing high-energy (MeV)
electrons.

\begin{figure}
\begin{centering}
\includegraphics[width=0.8\linewidth]{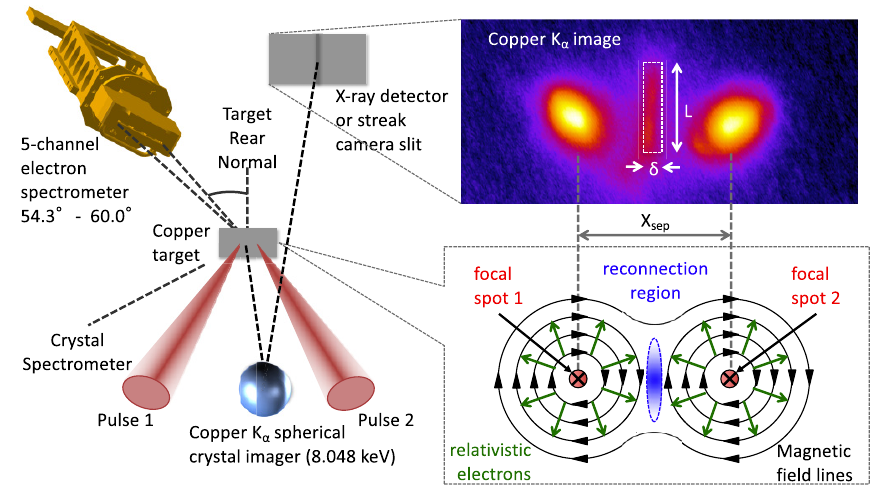}
\par\end{centering}
\caption{Taken from \citet{raymond2018}. Experimental set-up relying on ultra-high-intensity short pulse laser beams allowing to probe relativistic reconnection.
A typical x-ray imaging shows the location of the two hot expanding plasmas and, in between, the reconnection layer.}
\label{fig:raymond}
\end{figure}

In addition, \citet{raymond2018} conducted  experiments, on both the OMEGA EP facility and the HERCULES laser at University of Michigan,
in a regime where magnetic reconnection was not only collisionless but also driven by relativistic electrons.
This was made possible by using short pulse laser beams (20 ps for the OMEGA EP laser, 40 fs for the HERCULES laser)
in a configuration otherwise similar to that (the two-spot experiment) initially proposed by \citet{nilson2006}.
Using short pulses indeed allowed to reach ultra-high intensities ($\sim 10^{18} {\rm W/cm^2}$ on OMEGA EP and $\sim 2\,10^{19} {\rm W/cm^2}$ on HERCULES),
thus allowing to enter the relativistic regime of laser-plasma interaction.
Figures~\ref{fig:raymond} shows the typical experimental set-up as well as a typical X-ray imaging where the two heated and expanding plasmas can be seen, 
together with a reconnection layer in between.
In addition, the authors report the formation of a non-thermal (few MeV) electron population whenever reconnection is expected, 
consistent with supporting 3D PIC simulations. 

\newpage
\section{Solving kinetic problems}\label{S:SMA} 
This section reviews the model equations used to describe particle kinetic physics, i.e., the dynamics of charged particles in the configuration and momentum space under the effect of electro-magnetic forces. The section is divided as follows: in Sect.~\ref{S:VMAX} we describe the Vlasov--Maxwell system of equations, then in Sect.~\ref{S:VlasovCodesNum} we discuss the numerical methods developed to follow the dynamics of such a system. Section~\ref{S:PICCodes} discusses the particle-in-cell (PIC) technique used to study solutions of the Vlasov--Maxwell system. In Sect.~\ref{S:VlasovCodes} we provide a discussion on the comparison between PIC and Vlasov approaches. Section \ref{S:HYB} briefly describes hybrid methods where a fluid approximation is introduced for the electronic component whereas kinetic (PIC) techniques are used to describe ions. In Sect.~\ref{S:KIN} we specifically discuss the Fokker--Planck description of kinetic problems. The Fokker--Planck approach is particularly well-adapted to investigate cosmic ray propagation. Finally, we give a particular focus on Fokker--Planck simulations developed in the context of the study of the radiative transfer in hot plasmas around compact objects. 

\subsection{The Vlasov--Maxwell description of a collisionless plasma}\label{S:VMAX}

\subsubsection{Governing equations}
Let us consider a plasma composed of various species, labeled $s$, corresponding to particles with mass $m_s$ and charge $q_s$.
The kinetic description of this plasma relies on the representation of each species $s$ by its (one-particle) distribution function $f_{\!s}(t,\bx,\bp)$, 
$f_{\!s}(t,\bx,\bp) d^3\!x\,d^3\!p$ measuring, at any time $t$, the (probable) number of particles of species $s$ in a volume element $d^3\!x\,d^3\!p$ 
at a position $(\bx,\bp)$ in phase-space ($\bx$ and $\bp$ standing for the spatial and momentum coordinates, respectively). 
In the absence of collisions, the evolution of the distribution functions $f_{\!s}$ satisfies the Vlasov equation:
\begin{eqnarray}\label{eq:Vlasov}
\frac{\partial}{\partial t}f_{\!s} + (\bv \cdot \nabla ) f_{\!s} + (\bF_s \cdot \nabla_p) f_{\!s} = 0\,,
\end{eqnarray}
where $\bv = \bp/(m_s \gamma)$ is the velocity corresponding to a particle of momentum $\bp$ and Lorentz factor $\gamma = \sqrt{1+\bp^2/(m_s c)^2}$
($c$ is the speed of light in vacuum), 
and $\bF_{\!s}$ is the force acting on the species particles.
As this work focuses on electromagnetic plasmas, this force is the Lorentz force exerted by the collective electric $\bE$ and magnetic $\bB$ fields:
\begin{eqnarray}\label{eq:LorentzForce}
\bF_{\!s} = q_s \left( \bE + \frac{1}{c}\,\bv \times \bB \right)\,.
\end{eqnarray}
It is important to stress that, in the Vlasov equation, the electromagnetic fields are collective fields, 
also referred to as macroscopic fields in the sense that they do not account for the microscopic variations developing at the particle scale. 
Hence collisions are not considered in this description.

The electric \bE(t,\bx) and magnetic \bB(t,\bx) fields are in general functions of both space and time, 
and satisfy Maxwell's equations\footnote{In some cases where only electrostatic fields are important, only Poisson Eq.~\eqref{eq:Poisson} may be used. The Vlasov-Poisson description is of particular importance for the description of cold plasmas in particular, as considered e.g., for plasma propulsion, see Ref.~\citep{garrigues}}:
\begin{subequations}\label{eq:Maxwell}
\begin{align}
\label{eq:Poisson} \nabla \cdot \bE   &=  4\pi \rho \,, \\
\label{eq:divBfree} \nabla \cdot \bB   &=  0 \,, \\
\label{eq:Faraday} \nabla \times \bE &=  -\frac{1}{c}\frac{\partial \bB}{\partial t} \, , \\
\label{eq:Ampere}  \nabla \times \bB &=  \frac{4\pi}{c} \bJ +\frac{1}{c}\frac{\partial \bE}{\partial t} \, .
\end{align}
\end{subequations} 
The electromagnetic fields act onto the plasma through the Lorentz force \eqref{eq:LorentzForce}, 
and are in turn modified by the plasma through the total charge and current densities, $\rho = \sum_s \rho_s$ and $\bJ = \sum_s \bJ_s$, respectively,
where each species charge and current densities are defined as:
\begin{subequations}\label{eq:rhoJ}
\begin{align}
\label{eq:chargeDensity}  \rho_s(t,\bx) &= q_s\,\int \!\!d^3p\, f_{\!s}(t,\bx,\bp) \, ,\\
\label{eq:currentDensity}  \bJ_s(t,\bx)  &= q_s\,\int\!\!d^3p\, \bv f_{\!s}(t,\bx,\bp) \,.
\end{align}
\end{subequations}

The coupled system of Eqs.~\eqref{eq:Vlasov} and \eqref{eq:Maxwell}, together with the Lorentz force~\eqref{eq:LorentzForce} 
and definitions of the charge and current densities~\eqref{eq:rhoJ} form the Vlasov--Maxwell model.
It provides a self-consistent, kinetic description for the evolution of a collisionless plasma and the associated collective electromagnetic fields.

\subsubsection{Initial and boundary conditions}

The Vlasov--Maxwell model relies on a system of partial differential equations and thus requires initial and boundary conditions. The initial condition of the system (at time $t=0$) consists first in defining the initial distribution functions $f_s(t=0,{\bf x},{\bf p})$ for all species $s$ of the system. One usually considers equilibrium\footnote{At least in the sense of hydrodynamic equilibrium.} distribution functions, and Maxwellian or Maxwell--J\"uttner distribution functions (drifting or not) are often considered\footnote{The loading of a species with drifting Maxwell--J\"uttner distribution in Particle-In-Cell codes should be handle with some care, as discussed e.g., in Refs.~\citep{melzani2013,zenitani2015}.}. The initial electromagnetic fields spatial distribution ${\bf E}(t=0,{\bx})$ and ${\bf B}(t=0,{\bx})$ also needs to be prescribed. At $t=0$, these fields have to satisfy Eq.~\eqref{eq:Poisson} and Eq.~\eqref{eq:divBfree}, respectively. Hence, ${\bf B}(t=0,{\bx})$ has to be divergence free, while ${\bf E}(t=0,{\bx})$ can be either divergence free (e.g., if an external electric field is considered) or has to be computed from Poisson's Eq.~\eqref{eq:Poisson} using the initial distribution functions $f_s(t=0,{\bf x},{\bf p})$ to compute the initial charge density.

Various boundary conditions (BCs) can be considered and will strongly depend on the physics at hand. First, BCs on the distribution functions can be used to reflect, thermalize already existing particles or inject new particles at the border of the spatial domain. In addition, when directly solving the Vlasov equation in phase-space (see Sect.~\ref{S:VlasovCodes} on so-called Vlasov codes), BCs on the momentum component have to be considered. Similarly, electromagnetic fields can be reflected, absorbed or injected at the domain border by prescribing the correct BCs for the electric and magnetic fields.

\subsection{Solving the Vlasov--Maxwell system numerically: General considerations}\label{S:VlasovCodesNum}

\begin{figure}
\begin{centering}
\includegraphics[width=0.8\linewidth]{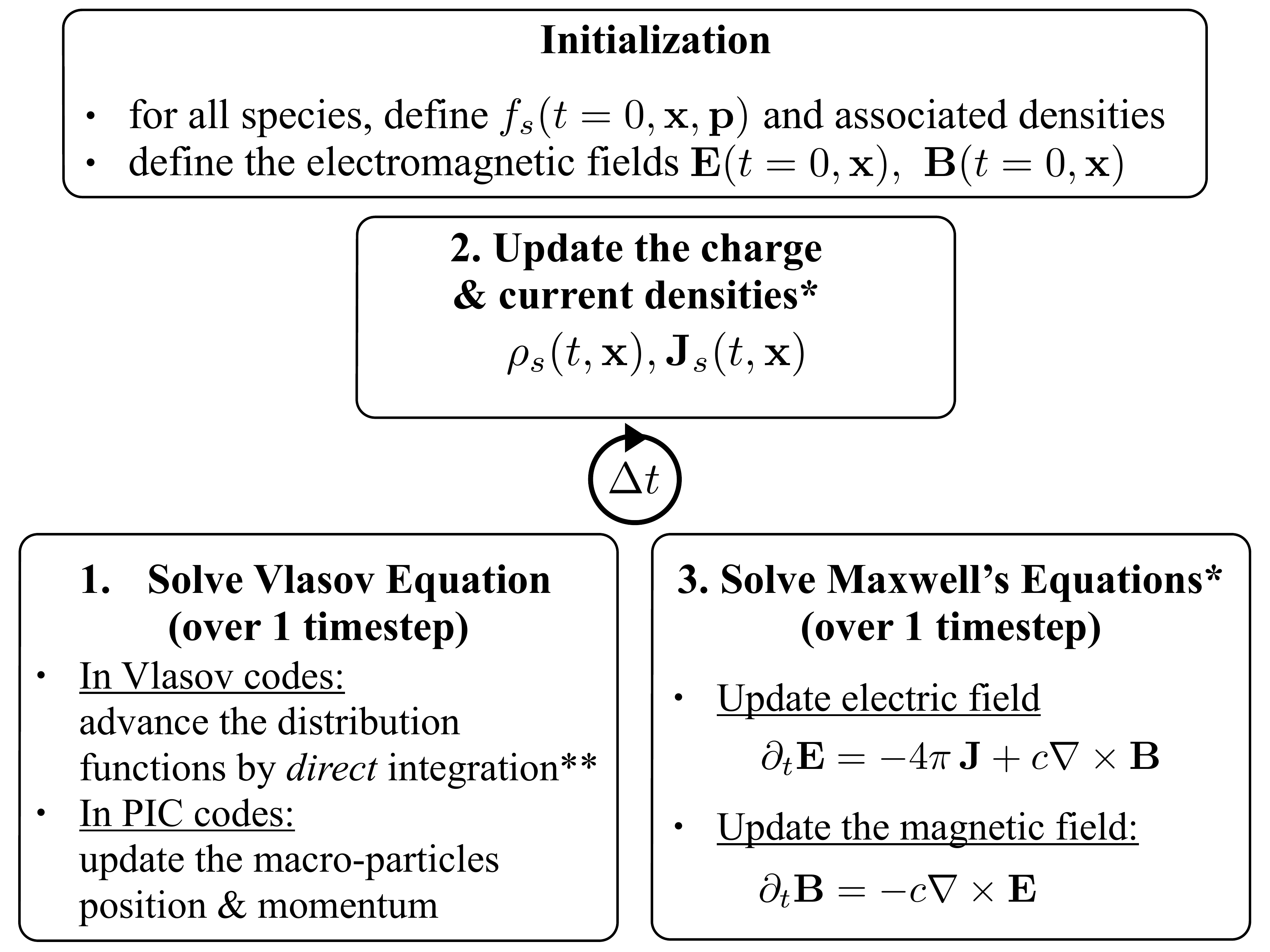}
\par\end{centering}
\caption{Schematic presentation of the numerical procedure used to solve the Maxwell--Vlasov system of equations. *If the computation of the current densities onto the simulation grid is done in such a way that charge is conserved, and considering that the initial electromagnetic fields satisfy Eqs.~\eqref{eq:Poisson} and~\eqref{eq:divBfree}, solving Maxwell--Amp\`ere [Eq.~\eqref{eq:Ampere}] and Maxwell--Faraday [Eq.~\ref{eq:Faraday}] is sufficient to ensure that Eqs.~\eqref{eq:Poisson} and~\eqref{eq:divBfree} remain satisfied (within the machine precision) at all times. **{\it Direct} integration here refers to advancing the distribution function on a grid in phase-space $({\bf x},{\bf p})$, as further discussed in Sect.~\ref{S:VlasovCodes}.}\label{fig:FigNumericalMethodVM}
\end{figure}

Computer simulation is an outstanding tool for solving the Vlasov--Maxwell system of equations together with the prescribed initial and boundary conditions, and most of today's kinetic simulations of plasmas rely on massively parallel tools to do so. 
In what follows, we present two of the main numerical approaches to solve this system. 
The first method is used in so-called Vlasov codes, while the second is used in so-called Particle-In-Cell (PIC) codes. 
The main difference between the two methods lies in the way they solve the Vlasov equation. 
Otherwise, both methods follow the same procedure which rely on discretizing the fields onto a spatial grid (henceforth referred to as the simulation grid), 
advancing the distribution function and then updating the associated charge and current densities onto the simulation grid.
This procedure is here briefly detailed and summarized in Fig.~\ref{fig:FigNumericalMethodVM}. 

\subsubsection{Initialization and time-loop}

First, the initialization step consists in prescribing the distribution functions for all species $s$ at time $t=0$ together with the initial electric and magnetic fields. Again, one should stress that both fields must satisfy Eqs.~\eqref{eq:Poisson} and~\eqref{eq:divBfree}. Thus the initial electric field can be obtained by solving Poisson Eq.~\eqref{eq:Poisson} and adding any divergence-free (e.g., external) electric field. For the magnetic field, one can start either from a zero magnetic field or any non-zero divergence-free magnetic field that will act as an external field applied to the system.

One then enters the time loop of the numerical solver. This time loop consists in advancing the various quantities (defined on the simulation grid) from a timestep $n$ (time $t_n=n\,\Delta t)$ to the next timestep $n+1$ (time $t_{n+1}=t_n+\Delta t$). Various methods are available to do so, some of which rely on defining different quantities at either integer or half-integer timesteps to ensure a centering of the numerical time derivatives. For the sake of simplicity, we will not account for this subtlety here. 

The first step in the time-loop consists in advancing the distribution functions for all species $s$. Knowing the electromagnetic field at timestep $n$, the distribution functions at timestep $n+1$ are computed either by \emph{direct} integration (this is the case in Vlasov codes, see Sect.~\ref{S:VlasovCodes}) or by advancing so-called macro-particles which are in effect discrete element of the distribution function (this is the case in PIC codes, as described in Sect.~\ref{S:PICCodes}). 

The updated distribution functions are then used to compute, onto the simulation grid, the updated charge and current densities (step 2).

These densities are then used, in a third step, to advance the electromagnetic fields from time step $n$ to $n+1$. If the current density deposition onto the grid (step 2) conserves the charge\footnote{In the sense that it satisfies, at the machine precision, the continuity equation ${\partial_t\rho+\nabla\cdot{\bf J}=0}$.}, solving Maxwell--Amp\`ere and Maxwell--Faraday Eqs.~\eqref{eq:Ampere} and~\eqref{eq:Faraday}, respectively, is sufficient to ensure that the electromagnetic fields remain Maxwell-consistent. To solve these equations, different Maxwell solvers are available (and are discussed in Sect.~\ref{S:MaxwellSolvers}).

With the fields updated, the loop gets back to step 1 and is run as long as required to reach the final timestep of the simulation.

\subsubsection{Brief discussion of Maxwell solvers}\label{S:MaxwellSolvers}
Various numerical methods (so-called solvers) can be used to solve Maxwell's equations. 
Here we briefly introduce two methods that are most popular in plasma simulation.

The Finite-Difference Time-Domain (FDTD) method is a time-honoured approach to solve Maxwell's equations~\citep{taflove2005}. 
It relies on a finite-difference discretization of the partial time derivatives and curl operators in Maxwell--Amp\`ere and Maxwell--Faraday's equations.
Most important is that all differential operators are centered.
Centering in space requires the use of a staggered grid, the so-called Yee grid, different components of the electric and magnetic fields being 
defined at different positions in space (onto the grid).
Centering in time requires that the electric and magnetic fields are advanced in a leap-frog way,
e.g., solving first Maxwell--Amp\`ere equation to advance the electric field then using the updated electric field to solve Maxwell--Faraday's equation and advance the magnetic field.
A major advantage of the FDTD method stems from its local nature, which allows for its easy and effective (scalable) implementation in massively parallel environments.
A major drawback of the method is that it is subject to numerical dispersion, the numerical electromagnetic wave propagating with velocities potentially smaller than $c$ \citep{birdsall_langdon,nuter2014}.
This effect is in part responsible for the {\it spurious numerical Cherenkov instability}, see Sec.~\ref{PICCodes:cherenkov}.

Pseudo-Spectral methods, on the other hand, allow to solve Maxwell's equations with an extraordinary level of precision and correctly capture the dispersion relation of electromagnetic waves.
They consist {\rm of} advancing the electromagnetic fields in Fourier space (for the spatial coordinates) 
while relying on an (explicit) finite-difference for the time derivatives \citep{liu1997,vay2013}. 

The increased precision allowed by pseudo-spectral methods however comes with the cost of global communications associated with the use of Fourier transforms over the entire simulation domain.
These global communications have been a major impediment to the adoption of pseudo-spectral methods in massively parallel environments.
Recently, \citet{vay2013} have proposed a domain-decomposition method that allows for the efficient parallelization of pseudo-spectral solvers.
This method takes advantage of the finiteness of the speed of light and relies only on local (over subdomains much smaller than the entire simulation domain) 
fast Fourier transforms and communications between neighbouring subdomains.
\citet{vincenti2018} have demonstrated that this method may allow for unprecedented scalability of pseudo-spectral solvers over tens to hundreds of thousands of computing elements (cores).

\subsection{Particle-In-Cell codes}\label{S:PICCodes}

The Particle-In-Cell (PIC) method was introduced in the mid-1950s by Harlow and collaborators to solve fluid dynamics problems (see \citealt{harlow2004} and references therein). 
Following the pioneering works of \citet{buneman1959}, \citet{dawson1962}, \citet{birdsall1969} and \citet{langdon1970} 
(see \citealt{dawson1983} and \citealt{verboncoeur2005} for a history of the development of PIC codes),
PIC codes have become a central tool for plasma simulation \citep{birdsall_langdon}. 
Indeed, the simplicity of the PIC method together with the possibility to implement it efficiently in a massively parallel environment have established PIC codes as the most popular tool for the kinetic simulation of plasmas. 

\subsubsection{Method}\label{S:PICCodes:method}

The PIC method differs from the Vlasov-code approach in the way it solves the Vlasov equation, 
and by extension, the way it computes the current densities on the simulation grid.
In PIC codes, the distribution function $f_{\!s}$ is approximated as a sum over $N$ macro-particles:
\begin{eqnarray}\label{eq:fs_discretization}
f_{\!s}(t,\bx,\bp) \equiv \sum_{p=1}^N w_p\,S\big(\bx-\bx_p(t)\big)\,\delta\big(\bp-\bp_p(t)\big)\,,
\end{eqnarray}
where $\delta(\bp)$ is the Dirac delta-distribution, $S\big(\bx\big)$ is the so-called particle shape-function, 
and $w_p$, $\bx_p(t)$ and $\bp_p(t)$ are the $p^{th}$ particle numerical weight, position and momentum, respectively.
The macro-particles can be regarded as \emph{walkers} in Monte-Carlo simulations, 
and the PIC method as a Monte-Carlo procedure for solving the Vlasov equation~\citep{lapeyre}.
The initial state of each species of the plasma is obtained from a random sampling of the distribution function $f_{\!s}$ at time $t=0$,
and the Vlasov equation is then solved following the macro-particles/\emph{walkers} motion through the influence of the collective 
electromagnetic fields.

Indeed, introducing the discretized distribution function~\eqref{eq:fs_discretization} in Vlasov equation~\eqref{eq:Vlasov},
one can show [see e.g., \citep{derouillat2018}] that solving Vlasov equation reduces to solving, 
for all macro-particles $p$, their equations of motion:
\begin{subequations}\label{eq:motion}
\begin{align}
\label{eq:momentum}  \frac{d\bp_p}{dt} &= \frac{q_s}{m_s}\,\left(\bE_p + \frac{\bv_p}{c} \times \bB_p\right) \, ,\\
\label{eq:position}        \frac{d\bx_p}{dt} &= \bv_p = \frac{\bp_p}{m_s \gamma_p}\,,
\end{align}
\end{subequations}
with $\gamma_p = \sqrt{1+\bp_p^2/(m_s c)^2}$ the $p^{th}$ macro-particle Lorentz factor 
and where we have introduced the electric and magnetic fields seen by the macro-particle:
\begin{subequations}\label{eq:fieldSeenByParticle}
\begin{align}
\label{eq:Epart}  \bE_p &= \int\!\! d^3x\,S(\bx-\bx_p)\,\bE(\bx)\,, \\
\label{eq:Bpart}  \bB_p &= \int\!\! d^3x\,S(\bx-\bx_p)\,\bB(\bx)\,.
\end{align}
\end{subequations}
Correspondingly, the species charge and current densities on the simulation grid can be obtained by direct deposition onto the grid. Yet, such a direct deposition would not in general satisfy the charge conservation equation, and the electric fields then would required to be corrected to ensure that they verify the Poisson equation~\citep{mardahl1997}. Some current deposition strategies that conserve the charge have, however, been proposed. 
A popular charge conserving deposition scheme has been proposed by ~\cite{esirkepov2000} for PIC code relying on the FDTD Maxwell solver. The macro-particle equations of motion Eq.~\ref{eq:motion} are most commonly solved using Boris pusher \citep{birdsall_langdon}.
 It is a second-order leap-frog integrator where the updated particle momentum is computed knowing the electromagnetic fields at the position of the macro particle as:

\begin{eqnarray}
\bp_p^{(n+\tfrac{1}{2})}=\bp_p^{(n-\tfrac{1}{2})} + \frac{q_s}{m_s} \Delta t \, \left[ \bE_p^{(n)} + \frac{\bv_p^{(n+\tfrac{1}{2})}+\bv_p^{(n-\tfrac{1}{2})}}{2c} \times \bB_p^{(n)}\right],
\end{eqnarray}
and the updated particle position is computed as:
\begin{eqnarray}
\bx_p^{(n+1)}=\bx_p^{(n)} + \Delta t \, \frac{\bp_p^{(n+\tfrac{1}{2})}}{\gamma_p},
\end{eqnarray}

Several alternative solvers where recently developed \citep[e.g.,][]{2008PhPl...15e6701V, 2017PhPl...24e2104H} presenting sometimes better accuracy than the traditional Boris algorithm.

\subsubsection{Additional physics modules}\label{PICCodes:additionalModules}

In their most basic implementation (detailed above), PIC codes describe collisionless plasmas through the self-consistent evolution of the particle distribution functions and collective (macroscopic) electromagnetic fields. 
Additional physics modules can easily be implemented in PIC codes to account for additional processes. 
Here we provide some references for some of this processes: field ionization \citep{nuter2011}, collisions and collisional ionization~\citep{nanbu1997,nanbu1998,perez2012}, high-energy photon (synchrotron or inverse Compton) emission and its back-reaction~\citep{duclous2011,lobet2016,niel2018}, pair production in a strong electromagnetic [Breit--Wheeler process \citep{duclous2011,lobet2016}] or Coulomb [Trident and Bethe--Heitler processes \cite{martinez2018}].
These later processes are of outmost importance for extreme plasma physics as will soon be encountered on multi-petawatt laser facilities (see e.g., Sect.~\ref{S:LASACC_laser}),
but also at play in the most extreme astrophysical environments around, e.g., neutron stars and black holes \citep{uzdensky2019}.

\subsubsection{Stability issues: relativistic flows}\label{PICCodes:cherenkov}
It is worth mentioning that there is one important numerical issue when one deals with relativistic flows in PIC simulations: the \emph{spurious Cherenkov instability} \citep[e.g., ][]{1974JCoPh..15..504G}. This instability results from the resonance of the light-wave modes with the streaming beam when electromagnetic fields are defined on a discrete Eulerian grid. As the numerical light-wave mode is affected by the finite-difference scheme, especially at high-$k$ \citep{birdsall_langdon}, this resonance is nonphysical. The instability appears as well in spectral codes but has a different signature \citep{2015CoPhC.196..221G}. It is practically very difficult to avoid in long-term simulations of relativistic flows or shocks. Nevertheless various methods have been proposed that allow to mitigate, or at least delay the onset of the instability. Some of these methods rely on digital filtering of electromagnetic fields and/or current densities \citep{2004JCoPh.201..665G, 2011JCoPh.230.5908V}; modifying the numerical stencil of the FDTD solver \citep{lehe2013,grassiPhD2017} or upgrading to a semi-implicit scheme \citep{pukhov2019}; special patching of the most unstable modes \citep[e.g.,][]{2014JCoPh.267....1G, 2017CoPhC.214....6L}; artificially increasing the speed of light in the Maxwell solver \citep{nuter2016} or solving the PIC equations in Galilean coordinates \citep{lehe2016}.
Yet, there is no definitive solution to remove it completely. Even if the most unstable modes are `cleaned' or stabilized, the difficulty arises in long term evolution from coupling of the secondary aliasing modes with low-wavenumber oblique modes of the streaming plasma, that is very hard to remove without touching important physical scales \citep{2006PhPl...13k2110D}. Despite this difficulty, several studies managed to push simulations beyond $10^4\,\omega_{pi}^{-1}$ allowing to extract important results from simulations (see Sect.~\ref{S:MIC-MES}).

\subsubsection{Examples}\label{PICCodes:examples}

Various PIC codes are today available and used for astrophysics or space plasma applications\footnote{We restrict our presentation to electromagnetic PIC codes that have been applied to astrophysics and/or space plasma physics studies.}.
Some of these codes are freely distributed under free-software licenses,
this is the case of \textsc{Epoch}\footnote{\url{https://gitlab.com/arm-hpc/packages/wikis/packages/EPOCH}} \citep{epoch} 
\textsc{Piccante}\footnote{\url{https://github.com/ALaDyn/piccante}} \citep{piccante},
\textsc{Smilei}\footnote{\url{www.maisondelasimulation.fr/smilei}}, \citep{derouillat2018} 
\textsc{Tristan-MP}\footnote{\url{https://github.com/ntoles/tristan-mp-pitp}} \citep{tristan-mp}, 
and \textsc{Zeltron}\footnote{\url{http://ipag-old.osug.fr/~ceruttbe/Zeltron/index.html}} \citep{2013ApJ...770..147C}.
Among other proprietary codes used for astrophysics applications are 
\textsc{A-ParT} \citep{2013A&A...558A.133M}, 
\textsc{Calder} \citep{calder}, 
\textsc{Osiris} \citep{osiris}, and
 \textsc{Photon-Plasma} \citep{photonplasma}.
Finally, other PIC codes rely on more advanced numerical schemes;
 e.g., the implicit code \textsc{iPIC3D} \citep{iPIC3D}
or the \textsc{Slurm} code for modeling magnetized fluids or plasmas \citep{slurm}.

\subsection{Vlasov--Maxwell codes}\label{S:VlasovCodes}

\subsubsection{PIC codes vs Vlasov codes}

Particle-In-Cell codes reduce the problem of solving Vlasov equation  to solving the (ordinary differential) equations of motion of many macro-particles. 
It follows that the main advantages of the PIC method are its conceptual simplicity, its robustness and easy implementation on (massively) parallel super computers. 
The simplicity of the PIC method also allows for PIC codes to be multi-purpose simulation tools: 
a single PIC code can address various problems from basics plasma physics, astrophysics studies to the modelling of laser-plasma experiments.

However, due to the introduction of a finite number of macro-particles, PIC simulation suffers from the highly exaggerated level of noise. 
This well known short-coming of the PIC method makes it less adapted to treating problems 
for which regions of phase-space where the distribution function assumes small values (e.g., in its tail) can impact the physics at play.

In contrast, \emph{Vlasov codes} which directly integrate the (partial differential) Vlasov equation on a grid in phase-space are virtually noise-free,
and are thus an interesting alternative to PIC codes whenever low noise simulation is required\footnote{A theoretical discussion
on the relative efficiency of the PIC and (direct) Vlasov approaches to treat a given problem is presented in \citep{feix2005}}.
The fine description allowed by Vlasov codes however comes with the cost of increased numerical complexity.
As a result, Vlasov codes are in general much less multi-purpose than PIC codes, and are usually developed to tackle a definite class of problems.

\subsubsection{The problem of filamentation in phase-space}

One impediment in the development and adoption of Vlasov codes is their computational cost and memory requirement 
when dealing with all 6 dimensions of phase-space. 
This problem can however be mitigated by reducing the number of dimensions
e.g., by relying on conservation laws and symmetries of the problem \citep[see, e.g.,][]{manfredi1995,feix2005}. 
It also becomes less exacting with the fast development of modern high-performance computing.

A more stringent limitation to the development of Vlasov codes stems from the numerical effort necessary to \emph{directly} solve the Vlasov equation, 
and to the problem of filamentation in phase-space in particular. 
Indeed, the time evolution of the distribution function in the Vlasov equation is associated with its breaking - filamentation - into increasingly small structures in phase-space, 
and thus to strong gradients of the distribution function. 
When discretizing the distribution function onto a grid with finite resolution, handling these gradients becomes numerically inaccurate, 
and can lead to spurious oscillations, numerical instabilities and inaccurate rendering of conserved quantities (e.g., non-positive distribution functions).  
Dealing with this issue greatly contributes to the numerical complexity behind Vlasov codes' development.

\subsubsection{Example of different methods for electromagnetic Vlasov codes}

The numerical complexity of Vlasov codes has -- since the seminal work by \citet{cheng1976} introducing the time-splitting technique\footnote{The time-splitting technique
separates advection in (real) space and velocity-space.} -- led to the development of various techniques to \emph{directly} integrate the
Vlasov equation onto a grid in phase-space. 
It is thus beyond the scope of this brief review to detail these techniques
and we here restrict our presentation to some electromagnetic Vlasov codes and their applications.
Review articles by \citep{filbet2003,buchner2007,ghizzo2009,2018LRCA....4....1P} discuss various techniques,
that range from finite-volume type methods \citep{fijalkow1999,filbet2001}, to semi-Langrangian methods \citep{sonnendrucker1999}, 
and spectral methods \citep{klimas1987}.

Each method has its own advantages and limitations, and Vlasov codes are usually designed to tackle a specific class of physical problems.
\citet{ghizzo1990} for instance developed a relativistic electromagnetic 1D Vlasov code to study stimulated Raman scattering;
while \citet{shoucri2015} considered the problem of stimulated Brillouin scattering.
These codes actually used the conservation of canonical momentum to reduce the number of dimension in velocity/momentum-space\footnote{Other methods have been developed that take advantage of the existence of canonical invariants
to solve the Vlasov equation; see e.g., \citet{liseikina2004,inglebert2011}.}.
A somewhat similar approach was used to study the breaking of a relativistic Langmuir wave \citep{grassi2014}
as well as laser-driven (electrostatic) shock acceleration of ions and ion turbulence \citep{grassi2016}.

A (non-relativistic) Eulerian Vlasov--Maxwell solver was developed by \citet{mangeney2002}.
It was applied to various studies ranging from wave propagation in magnetized plasmas \citep{califano2003},
to the study of the nonlinear kinetic regime of the Weibel instability \citep{califano2002}.
An off-spring of this solver is the hybrid (kinetic ions, fluid electrons) Vlasov code \citep{valentini2007}
used in particular to tackle turbulence studies in either 2D3V \citep[see, e.g.,][]{cerri2017} and 3D3V \citep[see, e.g.,][]{cerri2018} geometries.

Another Eulerian Vlasov--Maxwell model was developed by \citet{umeda2009}
and applied to the study of various instabilities, such as the Kelvin--Helmholtz instability \citep{umeda2014} 
or the collisionless Rayleigh--Taylor instability \citep{umeda2016}.

The semi-Lagrangian method introduced by \citet{sonnendrucker1999} \citep[see also][]{crouseilles2010} has also led to a new kind of Vlasov codes.
As an example, a relativistic semi-Lagrangian Vlasov--Maxwell solver (VLEM) was recently developed by \citet{sarrat2017}.
It was used to tackle problems related to streaming instabilities in plasmas, such as the current Weibel-filamentation and two-stream instabilities;
and operates in 1D3V, 2D2V and 2D3V geometries.

\subsection{Hybrid methods}\label{S:HYB}
In this approach thermal electrons are taken to be a massless, neutralizing and are treated as a magnetized fluid. Ions (thermal or even non-thermal) are treated using a PIC approach. The advantage of this method is to eliminate Debye-scale physics while still catching microscopic phenomena.

In hybrid codes, ion positions are advanced using the Boris pusher as in PIC codes (see Sect.~\ref{S:PICCodes:method}). Electron dynamics is the one of a massless fluid then 
\begin{equation}
    m_{\rm e} n_{\rm e} \frac{d\vec{v}_{\rm e}}{dt} = 0 = -en_{\rm e} \left(\vec{E} + \frac{\vec{v}_{\rm e}}{c} \times \vec{B} - \vec{\nabla}.{\bar{\bar P}}_{\rm e} \right) \ .
\end{equation}
This combined with the Amp\`ere law for a non-relativistic flow, hence neglecting the displacement current leads to an equation for the electric field
\begin{equation}
    \vec{E} \simeq -\frac{\vec{v}_{\rm i}}{c} \times \vec{B} - \frac{1}{e n_{\rm e}} \vec{\nabla}.{\bar{\bar P}}_{\rm e} - \frac{1}{4\pi q_{\rm i} n_{\rm i}} \left(\vec{\nabla} \times \vec{B}\right) \times \vec{B} \ ,
\end{equation}
where ${\bar{\bar P}}_{\rm e}$ is the electron pressure rank 2 tensor.  This method will not be reviewed here, the interested reader is invited to read recent references on the subject : \citet{2002hmst.book.....L, Kunz14}.

We note here the case of the \textsc{dHybrid} code \citep{2007CoPhC.176..419G}. This code is explicit fully parallelized code and it uses MPI. \textsc{dHybrid} solves the dynamics of non-thermal particles based on a PIC approach. The code has been used in the context of particle acceleration and transport at collisionless shocks, some of its results are presented in Sect.~\ref{S:PICres}. 

\subsection{Solving Fokker--Planck problems}\label{S:KIN}
The Fokker--Planck equation (FPE) is one of the most important equation in kinetic physics. It describes the evolution in the phase space of the particle distribution function $f(\vec{r}, \vec{p},t)$ under the effect of a diffusion process with small increments in which initial conditions are lost (a.k.a. a Markov process). In this review we are interested in collisionless plasmas, in that case, particle diffusion results from the process of scattering off plasma waves. However, note that FPEs are also well studied in the context of collisional plasmas. We refer the interested reader to \cite{2008JCoPh.227.4308W} for the description of numerical treatments of the collision operator. As is concerning high-energy particles, the FPE describes processes which develop over scales explored by these particles, it is also adapted to the study of macroscopic processes in astrophysical plasmas detailed in Sect.~\ref{S:MAC}. 
The interested reader can advantageously consult \citet{1989fpem.book.....R} for an overview of the properties of the FPE.

For a system of energetic particles in a magnetic field oriented along the z~axis, we can write the FPE as \citep{2002cra..book.....S}:
\begin{eqnarray}
\partial_t f + v\mu \partial_z f - \epsilon \Omega_{\rm s} \partial_{\phi} f &=& \frac{1}{p^2} \partial_{\rm x_\alpha} \left[p^2 \left(D_{\rm x_\alpha x_\delta} \partial_{\rm x_\delta}f + a f\right)\right] + q(\vec{r}, \vec{p},t)
\end{eqnarray}
where the diffusion process runs over the variables: x, y, z, p, $\mu$, $\phi$, hence we have 25 diffusion coefficients $D_{\rm x_\alpha x_\delta}$,\footnote{The Cartesian coordinates x,y,z mark here the position of the particle's guiding center.} and $\mu$ and $\phi$ are the particle pitch-angle cosine and azimuthal gyration angle respectively.
Here particles of charge q and mass m are relativistic (with speeds $v \simeq c$) and gyrate around a magnetic field of strength B with a gyrofrequency $\Omega_{\rm s} \simeq c/r_{\rm L}$. We note $\epsilon = q/\rm{sgn}(q)$. The term $a(p,\vec{r},t)$ describes the momentum change of the particle either due to loss or acceleration and $q(\vec{r},\vec{p},t)$ represents particle injection and/or escape. Although it should be kept in mind that the FPE is deduced from the more general Vlasov equation, we focus below on numerical solutions of this equation. Often, in the context of CR physics, the FPE is not directly solved but rather the convection-diffusion equation (CDE). The CDE results from the former by an averaging procedure over $\phi$ and $\mu$ in the case fast scattering processes build a gyrotropic and an isotropic distribution. 

\subsubsection{The Fokker--Planck equation}\label{S:FPE}

We start by studying 1D diffusion problems as is the case for stochastic acceleration. In that case the FPE can be simplified as
\begin{equation}\label{Eq:FPSFA}
\partial_t F(p,t) = \frac{1}{p^2} \partial_{\rm p} \left[p^2 \left(D_{\rm p p} \partial_{p}F + a(p) F\right)\right] - \frac{F}{\tau_{\rm esc}} + Q(p,t)\, .
\end{equation}
Here the particle distribution $F(p,t) = \int f(\vec{r}, \vec{p},t) \mathrm{d}^3\vec{r} \mathrm{d}\mu \mathrm{d}\phi$ is averaged over the space volume and is assumed to be isotropic (it fulfills the diffusion-convection limit) and diffusive escape is treated by the means of an escape timescale $\tau_{\rm esc}(p)$, the loss/gain $a(p, t)$ term is also averaged. This equation can be solved using finite difference schemes \citep{1996ApJS..103..255P}.

\begin{figure}
\begin{centering}
\includegraphics[width=1\linewidth]{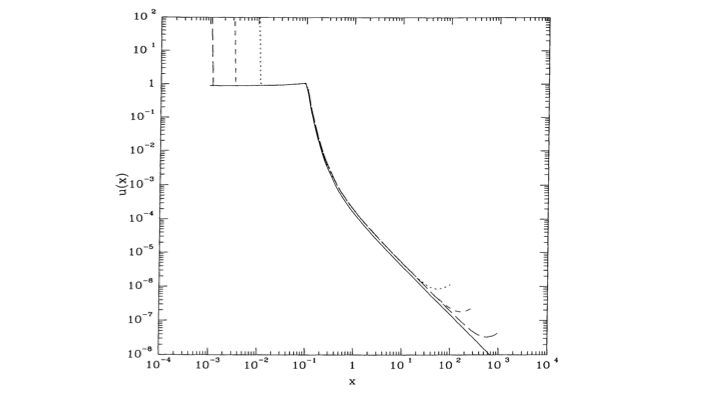}
\par\end{centering}
\caption{Time-dependent solutions of the FP problem $\partial_t F= \partial_p (p^3 \partial_p F -p^2 F) - F+ \delta(p-p_0)\delta(t)$ over a logarithmically-spaced mesh in momentum. The injection momentum is $p_0$. Three different numerical boundaries are compared: $p_1= 10^{-2}$ and $p_2=10^2$ (short-dashed lines);  
$p_1= 10^{-2.5}$ and $p_2=10^{2.5}$ (medium-dashed lines); $p_1= 10^{-3}$ and $p_2=10^{3}$ (long-dashed lines)
The steady state numerical solutions were obtained at t = 10 in normalized units (from \citet{1996ApJS..103..255P}). \label{F:ChangCooper}}
\end{figure}

\paragraph{Boundary conditions:}
As stated by \citet{1996ApJS..103..255P} any boundary condition which is a linear combination of $F$ and $F'(p)=\partial_{\rm p} F$ is viable for Eq.~(\ref{Eq:FPSFA}) if the points $p_1$ and $p_2$ at which they are taken fulfill $0 < p_1 < p < p_2 < \infty$. So we end up with two types of boundary conditions either with no particle $F(p_1)=F(p_2) = 0$ or with no flux $\phi(p_1)= \phi(p_2)= 0$ at the boundaries, where $\phi(p) =-(p^2 D_{\rm pp} F + a(p) F)$. The choice of 
one condition with respect to the other depends on the specific problem under investigation. A drawback of the no-particle condition is that it does not respect the particle number conservation. \\
\paragraph{Numerical schemes}
A simple way to solve Eq.~(\ref{Eq:FPSFA}) is to use an explicit finite difference method (FDM) with fluxes evaluated 
at grid midpoints, namely
\begin{equation}\label{Eq:EFPE}
\frac{F^{n+1}_{j+1} -F^{n}_{j}}{\Delta t} = -\frac{1}{p_j^2} \left(\frac{\phi^{n}_{j+1/2}-\phi^{n}_{j-1/2}}{\Delta p}\right) - \frac{F^n_j}{\tau_{\rm esc}(p_j)} + Q^n(p_j) \, .
\end{equation}
Time is discretized as $\Delta t = t_{n+1}-t_n$ and momentum is discretized following a constant logarithmic mesh where $\Delta p_{\rm j}/p_{\rm j}$= constant. We write $\Delta p_{\rm j} = (p_{\rm j+1}- p_{\rm j-1})/2$. The fluxes are calculated at midpoints defined as $p_{j+1/2} = (p_{j+1} + p_j)/2$. The coefficients entering in the flux calculation are evaluated as, e.g., $a_{j+1/2}=(a(p_{j+1})+ a(p_j))/2$ instead of a direct evaluation at $p_{j+1/2}$. For an explicit scheme the CFL condition (see Sect.~\ref{S:CFL}) $\Delta t/\Delta p_j^2 < p_j^2/D_{\rm pp,j}$ usually produces prohibitively small time steps. Semi-implicit and implicit methods can be used to circumvent this problem; they are obtained by changing $n$ to $n+1/2$ and $n+1$ in the RHS of Eq.~\eqref{Eq:EFPE} respectively. These methods lead to the derivation of a tridiagonal system of equations that can easily be solved once the boundary conditions are selected. For a given class of method, schemes then differ by the way the flux is calculated. One efficient implicit method is due to \citet{1970JCoPh...6....1C} and a well-known semi-implicit calculation is the Crank--Nicholson method (see \citealt{2002nrca.book.....P}). These methods are second order in time and second order in momentum for a uniform grid and first order in momentum for a non-uniform grid. \citet{1996ApJS..103..255P} show that the no-flux condition and the implicit Chang--Cooper scheme ensure positive solutions of 1D FP problems contrary to the Crank--Nicholson method (see an example of a solution of a FP problem in Fig.~\ref{F:ChangCooper}). While accounting for losses in $\phi(p)$ it is useful to adapt the time step to the dominant loss timescale. \citet{2014MNRAS.443.3564D} use a time step $\Delta t = 1/2\;\rm{min}(t_{\rm loss}(p_j))$, where $t_{\rm loss}(p) = a(p)/p$.

\paragraph{A note on particle transport and stochastic acceleration in hot plasmas:} FDM are widely used in CR physics but they are also used to solve radiative transfer problems in hot plasmas which develop in corona or in jets associated with compact objects. The radiative transfer in accretion disk corona can be treated in 1D assuming some particular geometry (usually slab-type or spherical) for the source of high-energy particles which allows to derive an escape probability and hence a simple expression for $\tau_{\rm esc}$ in Eq.~(\ref{Eq:FPSFA}). In this approach the coupled system of electron-positron plasma and its associated photon field can be described by a set of FPEs. However, a major difficulty to simulate such plasma systems is that they involve non-local processes in momentum\footnote{Some processes are also in principle non-local in space, this is the case for instance of the inhomogeneous synchro-Compton effect in jets, i.e., the Inverse Compton scattering of low energy photons generated by synchrotron radiation by a population of relativistic electrons (see \citet{1985A&A...146..204G}).}. This is for instance the case for the Compton scattering process \citep{1998ApJS..114..269N}. In that case, the fluxes are expressed in terms of integrals over lepton and photon populations \citep{2008A&A...491..617B, 2009ApJ...698..293V, 2013arXiv1303.0712M}\footnote{A version of such radiative transfer codes adapted to GRBs can be found in \citet{2011ApJ...738...77V}.}. Aside from Compton scattering, integrals also result from the calculation of other processes: pair production and annihilation, Coulomb losses, synchrotron losses. These codes solve the diffusion problem usually using Chang-Cooper-type methods. However, as noticed by \citet{2008A&A...491..617B}, the radiative transfer problem requires a high accuracy in momentum to preserve a high level of particle number and energy conservation. The Chang-Cooper method which is only first order accurate on non-uniform grid needs to be modified using both grid center and faces and calculating the momentum derivatives as $\partial_{\rm p} F = (F_{\rm{j}+1/2}-F_{\rm{j}-1/2})/\Delta x_{\rm j}$ and $\partial_{\rm p}^2 F = (F_{\rm{j}+1}-F_{\rm{j}})/\Delta p_{\rm j+1/2}-(F_{\rm{j}}-F_{\rm{j}-1})/\Delta p_{\rm j-1/2}$. The integral parts have to be calculated using specific treatments.
First, the different elements of the cross sections are stored and then interpolated during the course of the runs. Then, boundary conditions are different for the FP and the transfer parts. The transfer part has a wall-type boundary condition which includes a modification of the differential cross section (see \citet{2008A&A...491..617B} for details). Finally, integrals used to calculate the Compton process can be treated differently depending on the photon energy with respect to the electron energy from a continuous process at low energy leading to a derivative term and to a full integral calculation in the Klein--Nishina limit. Compton scattering of photons involves the same kind of treatment and is applied depending on the electron energy (see \citet{2009ApJ...698..293V} for details).  

\paragraph{Multi-variable FPE:}
Astrophysical or space plasmas usually involve multi-dimensional diffusion-advection processes. The study of CR propagation in the Milky Way requires to account for several complex effects: CR spallation reactions, radioactive decay, anisotropic diffusion with respect to the background magnetic field direction, etc. Specific numerical tools have been developed to handle this complexity.\footnote{We point towards the corresponding code websites: \textsc{Dragon}: \url{https://github.com/cosmicrays}, \textsc{Galprop}: \url{https://galprop.stanford.edu/code.php}, \textsc{Picard}: \url{http://astro-staff.uibk.ac.at/~kissmrbu/Picard.html} Let us also mention the semi-analytical tool USINE, https://dmaurin.gitlab.io/USINE/, where the FPE is solved using a path integral method.} Most of CR transport codes use multi-dimensional finite difference methods, this is the case for \textsc{GALPROP} and \textsc{DRAGON} which adopt a Crank-Nicholson method. In multi-dimensional problems this method leads to a non-tridiagonal system of equations to solve. It is solved usually adopting an iterative procedure like the Gauss--Seidel relaxation method \citep{2002nrca.book.....P}. The integration is adapted to the specific diffusion problem by starting from a large time step and reducing it as the stationary solution is reached \citep{1998ApJ...509..212S}. Both \textsc{GALPROP} and DRAGON codes also use an operator splitting technique to handle multi-dimensional diffusion problems \citep{2002nrca.book.....P}. The technique of operator splitting consists in splitting the time integration in Eq.~(\ref{Eq:FPSFA}) or its multi-dimensional generalization into a succession of simpler operations involving N different operators $L_{\rm i}$, such that 
\begin{equation}\label{Eq:FPSFA2}
\partial_t F(p,t) = \sum_{i=1}^N L_{\rm i} F(p,t) \, .
\end{equation}

Each operator contributes to move the solution from $F^{n}$ to $F^{n+1}$ as $F^{n+1} = \prod_{i=1}^N {\cal L}_{\rm i} F^n$  where each finite difference operator solves a part of the numerical problem. One difficulty with this method is that operator actions do not commute, hence one usually has to proceed with trials with a guess of the correct solution to select the correct operator ordering. The \textsc{DRAGON} code designed in cylindrical coordinates uses a series of operators associated to each relevant transport process. For instance the operator associated with diffusion along galactic vertical height~$z$ is $L_{\rm z} = D_{\rm zz} \partial_{\rm z}^2 F(z,r,p,t) + \partial_{\rm z} D_{\rm zz} \partial_{\rm z} F(z,r,p,t)$. Similarly other operators are derived for diffusion along other space variable~$r$ (the galactic radius), momentum loss or advection \citep{2017JCAP...02..015E}. Then each derivative is treated using a Crank--Nicholson scheme. The \textsc{Picard} code uses a different numerical approach as it first solves a stationary problem and also as the momentum evolution is treated using an integration instead of a FDM \citep{2014APh....55...37K}.

Multi-dimensional numerical solutions of FP problems is an active research field with a rich variety of solvers based on three main approaches: finite difference methods as we just discussed (FDM), finite element methods (FEM) or path integrals techniques. For most of them they still wait to be applied in the context of astrophysical or space plasma research.

\subsubsection{Stochastic differential equations}\label{S:SDE}
Stochastic differential equations or SDEs are a very efficient way to solve complex multi-dimensional Fokker--Planck problems with simple numerical schemes, although SDE schemes can become themselves rather complex. We invite the interested reader to consult some monographs cited in \citet{2017SSRv..212..151S}. These authors provide an overview of the use of SDE in the fields of DSA, CR transport in the ISM and space plasmas. The interested reader can consult this complete review to what concerns space plasmas problems. Below we bring a complementary discussion on the use of SDE in the context of shock acceleration. The intrinsic idea behind SDE is to derive a set of equations of motion which reproduce the random walk in each of the stochastic variables which describe the phase space evolution of a particle. \citet{1994A&A...286..314K} demonstrate the equivalence between a FPE and a set of SDEs. The simplest SDE scheme is the Ito first order explicit scheme. As an example let us write the SDE for a 1D random walk in a space x direction of a particle represented by a diffusion coefficient D(x,t). Let us assume also that the particle is advected with a speed u(x,t). The first order forward explicit Ito scheme then writes the increment of the position of the particle within a time step $\Delta t$ as
\begin{equation}\label{Eq:SDE1D}
\Delta x =  \left(u(x,t) + \frac{\partial D(x,t)}{\partial x}\right) \Delta t + \xi_{\rm x} \sqrt{2 D(x,t) \Delta t} =\Delta x_{\rm adv} + \xi_{\rm x} \Delta x_{\rm diff} \ .
\end{equation}
We note $V(x,t)=u(x,t) + {\partial D(x,t)}/{\partial x}$.
This equation shows that the particle path has two terms, the first term is deterministic and reproduces a forward Euler increment due to advection. The second term is stochastic and describes a diffusion as $\Delta x \propto (\Delta t)^{1/2}$. The term $\xi_{\rm x}$ is a random variable usually sampled over a Gaussian distribution with 0 mean and variance 1. The particle distribution can then be reconstructed using a large number of particles. The method is simple; however, it can suffer from noise in parts of the phase space sampled by only a few particles. The latter issue can be partly handled using a particle-splitting scheme \citep{2015JHEAp...5....1Y} where the weight attributed to a particle is split over several particles when reaching a region of the phase space with low statistics, as can be the case in the energy space if the particle distribution has an exponential cut-off. 

More generally, the diffusive term in Eq.~(\ref{Eq:SDE1D}) is a Wiener processes $W(t,x)$ which models the Brownian motion of a particle in an homogeneous medium, we write $dW(x,t)/dt = \xi_{\rm x}$. More complex SDE schemes can be interesting to use if necessary, like schemes backward in time in order to start from a known distribution and reconstruct the particle injection at sources. This way has the advantage to improve statistics if we want to have information at a particular location, corresponding for instance to a satellite. Higher order schemes in space and time are possible by doing a Taylor expansion of both advection and stochastic parts of Eq.~(\ref{Eq:SDE1D}). Schemes stable in time can be obtained by searching advection and diffusive terms at a time $t' = t+ \theta \Delta t$, where $\theta$ is to be taken between 0 and 1 \citep{1989PhRvA..39.3511S}. 

\paragraph{DSA with SDEs} The study of shock acceleration using SDEs requires some care in fixing the time step $\Delta t$ \citep{1994A&A...286..314K}. In shock acceleration studies the shock front is usually obtained from a fluid code, so has a finite width $\Delta x_{\rm sh}$ traced by a few grid cells. The condition over the time step to describe the DSA process properly is then $\Delta X_{\rm adv} < \Delta x_{\rm sh} < \Delta X_{\rm diff}$. The first inequality allows particles to stay around the shock to get accelerated whereas the second inequality allows the particle to sample the up- and downstream media correctly. However, if the diffusion coefficient is an increasing function of the particle energy, i.e., $\partial D(x,E,t)/\partial E > 0$ it is possible to find a threshold energy $E^*$ for which the condition $\Delta x_{\rm sh} = \Delta X_{\rm diff}(E^*)$ is fulfilled \citep{2005APh....23...31C, 2010MNRAS.406.2633S}. Below $E^*$ the shock acceleration process can not be properly treated. One possibility to address this problem is to sharpen artificially the shock \citep{2003A&A...404..405C}. This method can be easily handled in 1D \citep{2010A&A...515A..90M} but is difficult to construct in 2 or 3D as the shock front starts to corrugate. Another difficulty is that at an non-parallel shock, the MHD Rankine--Hugoniot conditions induce a discontinuous diffusion coefficient up- and downstream. Quite generally the diffusion transition at the shock can be decomposed into a continuous component $D_{\rm c}$ and a jump at the shock front $\Delta D= D_{\rm u}-D_{\rm d}$ expressed in terms of the up- and downstream diffusion coefficients. The diffusion coefficient can then be written as
\begin{equation}
D(x)=D_{\rm c}(x) + \Delta D \delta(x-x_{\rm sh})  \, , 
\end{equation}
where $x_{\rm sh}$ is the shock position \citep{2010A&A...515A..90M}. \citet{2000ApJ...541..428Z} proposes to account for the discontinuous part using a skewed Brownian motion which introduces an asymmetric shock crossing probability. To proceed we introduce a new variable $\tilde{x} = x \zeta(x)$ where 
\begin{equation}
\zeta(x) = \left\{
           \begin{array}{ll}
                  \epsilon&\rm{x < x_{\rm sh}}\\
                  1/2&\rm{x=x_{\rm sh}}\\
                  (1-\epsilon)&\rm{x > x_{\rm sh}} \, ,
           \end{array}
           \right.
\end{equation}
with $\epsilon = D_{\rm u}(x_{\rm sh})/(D_{\rm d}(x_{\rm sh})+ D_{\rm u}(x_{\rm sh})).$
\citet{2011MNRAS.411.2628A} propose a more general scheme adapted to shock configuration with strong gradients in the diffusion coefficient. This situation occurs especially upstream, in the shock precursor, in case of strong magnetic field amplification. The scheme involves a second-order accuracy predictor-corrector method [see section~4 in \citet{2011MNRAS.411.2628A} for details]. The scheme is however much slower than the simple Ito scheme and it is necessary to switch from one scheme to the other in order to save simulation resources.  

\begin{figure}[htb]
\centerline{\includegraphics[width=1\linewidth]{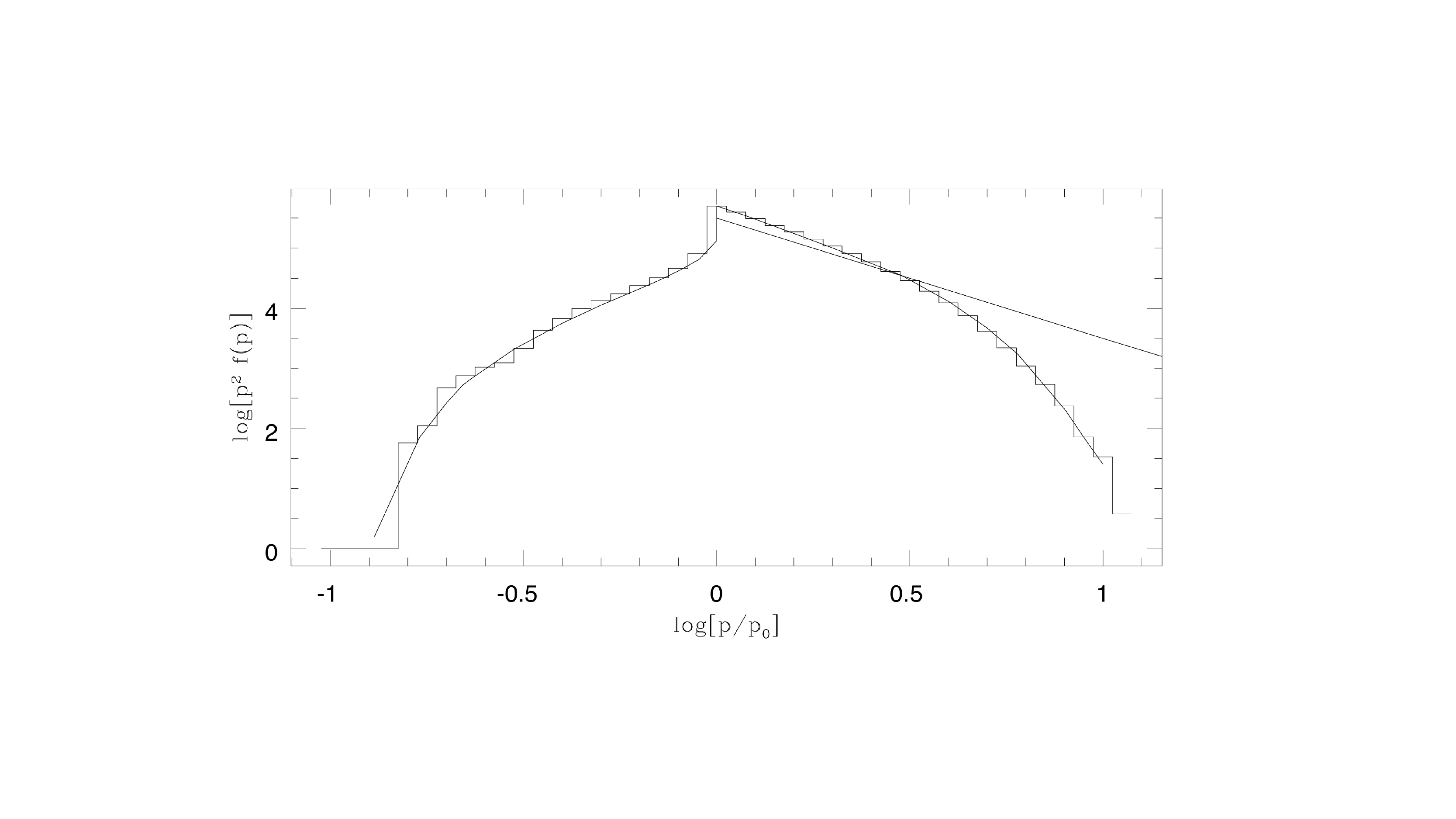}}
\caption{Shock electron distribution including synchrotron losses \citep{1999A&A...347..391M}. Particles are injected at momentum $p_0$ for $a_{\rm s,u} = 10$, $a_{\rm s,d}$ = 1 and a compression ratio r = 4 compared to the analytical solution of \citet{1984A&A...137..185W}. The solid line shows a solution $f(p) \propto p^{-4}$.  \label{fig:shockloss}}
\end{figure}

One also has to account for the particle increment in energy or momentum at each shock crossing. It is also possible to use an explicit Ito scheme for CR energy or momentum similarly to Eq.~(\ref{Eq:SDE1D}). If stochastic acceleration can be neglected, \citet{1999A&A...347..391M} introduce an implicit scheme:
\begin{equation}\label{Eq:SDELOSS}
\Delta \ln(p) = - \left(a_{\rm loss} p + \frac{1}{3} \frac{du}{dx}\right) \Delta t \, ,
\end{equation}
where $a_{\rm loss}$ is a loss rate and the second term accounts for the increase in particle momentum from shock acceleration. The implicit scheme rewrites the particle position with time as a linear interpolation $x = (\Delta x/\Delta t) t$. Eq.~(\ref{Eq:SDELOSS}) has the solution
\begin{equation}\label{Eq:SDELOSSINT}
    \ln(p(t')/p(t)) = -\ln(F_I+L_{\rm s}) \, ,
\end{equation}
where $F_I=\exp((\Delta V/3) \Delta t/\Delta x)$ gives the momentum increment by DSA and
\begin{equation}
L_{\rm s} =a_{\rm s} \frac{\Delta t}{\Delta x} p \int_{x(t)}^{x(t')} \exp\left(\frac{\Delta V}{3} \frac{\Delta t}{\Delta x}\right) dx' \, ,
\end{equation}
where $L_{\rm s}$ accounts for the effect of losses. The increment $\Delta x$ is calculated from the SDE in x, which is evaluated at $t'=t+\Delta t$. The new momentum is obtained from Eq.~(\ref{Eq:SDELOSSINT}) calculated using $x(t')=x(t+\Delta t)$. Figure~\ref{fig:shockloss} gives the shock solution for electrons including synchrotron losses.

The SDE method has proven to be very efficient in calculating particle acceleration by DSA at non-relativistic shocks in 1D and even in 2.5D in the context of jets \citep{2003A&A...404..405C, 2005APh....23...31C}. The method can in principle be coupled to MHD solutions by sub-cycling the MHD timestep (see Sect.~\ref{S:PICMHD}). To our knowledge no scheme has yet included CR back-reaction over the thermal plasma, but solutions proposed by other Monte-Carlo models (see Sect.~\ref{S:MCSHOCK}) should be applicable to this particular technique.

\paragraph{Relativistic shocks and SDEs} In the context of relativistic shock two difficulties emerge if one wants to apply the SDE method to the problem of particle acceleration because the shock is moving almost as fast as the particles. First, a simple scheme as in Eq.~(\ref{Eq:SDE1D}) may lead to a violation of the causality principle, because over the diffusive step the particle speed $\Delta x/\Delta t$ can exceed the speed of light. Second, DSA is based on the diffusive approximation which requires the ratio of the particle speed to the shock speed to be small. \citet{2001MNRAS.328..393A} evaluate particle acceleration in the shock rest frame but simulate the spatial diffusion process from the pitch-angle scattering process and hence reconstruct particle trajectories [see also \citet{1998PhRvL..80.3911B}]. These works retrieve the shock particle distribution produced by scattering by an isotropic turbulence $f(p) \propto p^{-4.2}$ (see Sect.~\ref{S:REL}).

\subsubsection{Simulating shock acceleration using a Monte-Carlo method}\label{S:MCSHOCK}

\citet{1984ApJ...286..691E, 1991SSRv...58..259J} proposed a Monte-Carlo method to simulate particle pitch-angle scattering around non-relativistic shocks. The particle mean free path is assumed to scale as a function of the particle rigidity as $\lambda \propto R^a/\rho$, where $R= pc/q$ is the particle rigidity and $\rho$ is the fluid mass density. The momentum vector follows a random walk which produces a variation of the particle pitch-angle $\delta \alpha$. The scattering is assumed to be elastic and isotropic in the fluid rest frame. The particle are injected upstream from the thermal plasma. The CR pressure is reconstructed at different distances from the shock front; particle-splitting technique is used in order to improve statistics at high energies. The CR pressure term is then included in the Rankine--Hugoniot conditions to account for CR backreaction over the shock solutions. Finally, a far escape boundary is adopted to calculate the escaping energy flux carried by the particles. 
An example of results can be seen in Fig.~\ref{fig:NLDSAcomp}, 
compared with two other methods discussed elsewhere in this review.

Recently the technique has been used to study the effect of magnetic field amplification in NLDSA at non-relativistic SNR shocks \citep{2006ApJ...652.1246V}, as well as acceleration at relativistic shocks in GRB afterglows \citep{2015MNRAS.452..431W}.

\section{Small and meso-scale numerical particle acceleration studies}\label{S:MIC-MES}

The understanding of the initial stages of particle acceleration in astrophysical plasma relies on the non-linear interplay between the particle distribution function and electromagnetic fields. The inherent non-linearity of the process prevents the development of robust analytic models, unless numerical simulations provide basic guidelines of the behavior of the system. This is especially true when particles produce the turbulence responsible for their self-confinement and acceleration around the shock front. The improvement in the computation power during last decades allowed for significant progresses in this field of research using computationally expensive, yet considered as \emph{ab-initio}, PIC simulations. Despite the fact that the astrophysical sources space and time scales are out of reach of PIC simulations, a large number of fundamental questions have been addressed and, sometimes, answers provided using this technique. In this section, we review a number of works that investigate the question of particle acceleration efficiency at shocks (Sect.~\ref{S:PICres}) and in magnetic reconnection (Sect.~\ref{Sec:Kinetic_Reconnection}) processes, using PIC simulations. Before we further proceed, we first introduce some vocabulary concerning the different category of shocks investigated with the help of PIC simulations. 

\paragraph{Collisionless shocks classes} Collisionless shocks are mediated by collective plasma effects \citep{1966RvPP....4...23S}. In this sense, more refined classification is required then for hydrodynamical/MHD shocks (weak, strong, radiative, fast, slow, parallel, oblique, perpendicular) where the shock is expected to be mediated by binary collisions between particles. Seminal \citep{1966RvPP....4...23S} as well as recent studies \citep[e.g., ][]{2014PhRvL.113j5002S, 2014NatSR...4E3934S, 2014PhPl...21g2301B, ruyer2016} considered and demonstrated the dominant role of small scale plasma instabilities in forming and mediating collisionless shocks. Different types of instabilities are dominant depending of plasma beta, composition, shock Mach number and upstream magnetic field orientation with respect to the shock propagation direction. This translates into a bestiary of different plasma instabilities at play when describing the shock structure. Commonly, the separation into electrostatic, Weibel-mediated and magnetised shocks is done.  Non-relativistic, weakly magnetised and low-Mach number shocks are believed to be mediated by electrostatic effects (two-stream, Buneman instabilities). With increasing Mach number ($M_a \gg 1$) or going into relativistic regime, weakly magnetised shocks are mediated by Weibel-filamentation \citep{2014PhPl...21g2301B, huntington2015}. Strongly magnetised shocks are typically mediated by coherent magnetic reflection of particles on the shock barrier. In this case the shock width is of the order of ion gyroradius.

\subsection{Shock acceleration numerical experiments: PIC simulations}\label{S:PICres}

The short chronological version is the following. In the pioneering studies 1D3V geometry was adopted because of numerical cost \citep[e.g.,][]{1972PhRvL..28..410B}. Several physical processes were evidenced in this way, such as shock front self-reformation \citep[e.g.,][]{1987PhFl...30.1767L, 1992PhFlB...4.3533L} or positron acceleration through resonant absorption of ion-cyclotron waves \citep{1992ApJ...390..454H}. However, this configuration was found to be too restrictive to trigger the Fermi process in most of cases, especially for quasi-perpendicular shocks. Next, multidimensional simulations were long enough to form the shock, but no evidence of first-order Fermi acceleration was found \citep{2004ApJ...608L..13F, 2005ApJ...623L..89H, 2007ApJ...668..974K, 2008ApJ...675..586D}.  This raised the question whether shocks can accelerate particles self-consistently or some external source of turbulence was necessary. More recent studies were able to form the shock and follow its propagation long enough to allow several Fermi cycles and produce extended power-law particle distribution \emph{self-consistently} \citep[e.g.,][]{2008ApJ...682L...5S,2009ApJ...695L.189M, 2011ApJ...726...75S, 2018MNRAS.477.5238P, 2019MNRAS.485.5105C, lemoine_PRE_2019}. Yet, even in the longest simulations the power-law spans no more than two orders of magnitude in particle energy, reflecting the challenging nature for fully kinetic simulations to reach astrophysical space and timescales.

\subsubsection{Ultra-relativistic shocks}
The ultra-relativistic regime is particularly interesting for several reasons \footnote{see also the recent review article by \citet{vanthieghem_arXiv_2020} on the physics of weakly magnetized shocks}. (i) Energy gain per cycle is large, $\Delta E/E \simeq 2$ instead of $\Delta E/ E =\beta_{\rm sh} \ll 1$ in the non-relativistic regime and (ii) Scattering time must be short, otherwise particles get advected within the downstream flow as the shock front recedes rapidly (the shock front moves away with velocity equal $c/3$ in the frame where the downstream plasma is at rest). These two reasons mean that the build-up of the non-thermal power law is faster in the ultra-relativistic regime than in the non-relativistic case. Hence, one can diagnose whether relativistic shock accelerate particles efficiently or not for a given parameter regime on a timescale of several $10^3\,\omega_{\rm pe}^{-1}$ (given upstream flow magnetization $\sigma$, magnetic field inclination with respect to the shock propagation direction and plasma composition). On the other hand, fast and efficient particle acceleration prompted early Monte-Carlo and semi-analytical studies to suggest that relativistic shocks are viable candidates for the acceleration of Ultra High Energy CRs (UHECRs).

For the reasons outlined just before, the demonstration of the first order Fermi operability in PIC simulations was firstly done in the regime of relativistic shocks by \citet{2008ApJ...682L...5S}. In the non-relativistic case it was done several years later, as it requires much longer simulation time \citep[e.g., ][]{2015ApJ...802..115K, 2015PhRvL.114h5003P}.

We now turn to the discussion of different studies that used PIC simulations to understand the physics of ultra-relativistic shocks. The studies go from 1D to 3D, deal with different plasma compositions, different magnetic field geometries and magnetization. Table~\ref{T:SIM_relat} provides a (non-exhaustive) list of such studies and presents some relevant numerical parameters that they used. Below we discuss different type of micro-instabilities which are reviewed in \citet{2016RPPh...79d6901M, 2009ApJ...699..990B}.

\begin{table*}
 \begin{footnotesize}
\begin{center}
\begin{tabular}{lccccc} 
\hline
\hline
Authors & \rm{Compos.} & \rm{Dim.} & $m_{\rm i}/m_{\rm e}$ & $\sigma$ & $\theta_{\rm B}^\circ$  \\

\hline
Langdon et al. 1988 & $e^--e^+$ & 1D& 1 & $[0.1; 13.3]$ & 90\\
Gallant et al. 1992  & $e^--e^+$& 1D& 1 & $[3\cdot 10^{-5};5]$ & 90\\
Hoshino et al. 1992  & $e^--e^+ - i$& 1D& 20 & $[5\cdot 10^{-3};0.5]$ & 90\\
Nishikawa et al. 2003 &$e^--i$  & 3D & 20 & 0 & -\\
Frederiksen et al. 2004 &$e^--i$  & 3D & 16 & 0 & -\\
\citet{2004ApJ...617L.107H} &$e^--i$  & 3D & 16 & 0 & -\\
Spitkovsky 2005 & $e^--e^+$ & 3D & 1 & $[0; 0.1]$ & 90\\
Lyubarsky 2006 & $e^--i$ & 1D & 50 & $3\cdot 10^{-3}$& 90\\
Kato 2007 & $e^--e^+$ & 2D & 1 & 0 & -\\
Hoshino 2008 & $e^--i$ & 1D & 50 & $2\cdot 10^{-3}$ & 90\\
Dieckmann et al. 2008 & $e^--i$ & 1D,2D & 400 & $2.5\cdot 10^{-3}$ & $\simeq 10$\\
Chang et al 2008 & $e^--e^+$ & 2D & 1 & 0 & -\\
Spitkovsky 2008a & $e^--e^+$ & 2D & 1 & 0 & -\\
Spitkovsky 2008b & $e^--i$ & 2D & $\leq 10^3$ & 0 & -\\
Keshet et al 2009 & $e^--e^+$ & 2D & 1 & 0 & -\\
Martins et al 2009  & $e^--i$ & 2D & 32 & 0 & -\\
Sironi \& Spit. 2009 & $e^--e^+$ & 2D & 1 & 0.1 & $[0; 90]$\\
Sironi \& Spit. 2011 & $e^--i$ & 2D,3D & \textbf{16} & $[10^{-5}; 0.1]$ & $[0;90]$\\
Haugb{\o}lle 2011 & $e^--i$ & 2D,3D & 16 & 0 & -\\
Sironi et al 2013 & $e^--e^+/i$ & 2D,3D & 1\& \textbf{25} & $[0 ; 0.1]$ & 90\\
\citet{2013JPlPh..79..367B} & $e^--e^+$ & 2D & 1 & 0 & -\\
Ardaneh et al 2015 & $e^--i$ & 3D & 16 & 0 & -\\
Plotnikov et al 2018 & $e^--e^+$ & 2D & 1 & $[0 ; 5]$ & 90\\
Crumley et al 2019 & $e^--i$ & 2D & \textbf{64} & 0.007 & 10 \& 55 \\
\hline
\hline
\end{tabular}
\end{center}
\caption{Table of different PIC studies of relativistic shocks, referenced by the authors names and publication dates. Mass ratio values indicated with bold font indicate that larger values were also explored.}
\end{footnotesize}
\label{T:SIM_relat}
\end{table*}

\paragraph{1D studies} The first exploration of relativistic shocks was motivated by the study of the termination shock physics in Pulsar Wind Nebulae. \citet{1988PhRvL..61..779L} performed 1D PIC simulations of perpendicular magnetized shocks in pair plasma with upstream Lorentz factors $\gamma_0=20$ and $40$ and Alfv\'enic Mach numbers going from $24$ to $154$.  The size of the simulation box was about 10 Larmor radii for $\sigma=0.1$ and about $100$ Larmor radii for $\sigma=13.3$. The shock front formed by magnetic reflection between the incoming and wall-reflected plasma. Well-formed shocks exhibited a soliton-like structure where most of dissipation occurs through maser synchrotron instability. Strong electromagnetic precursor emission was observed in these shocks but no particle acceleration. More systematic study was performed by \citet{1992ApJ...391...73G} where electromagnetic precursor energy was systematically derived. The most efficient precursor emission was observed for $\sigma \sim 0.1$ where it carries about 10\% of the incoming kinetic energy. Based on these results, the authors concluded that magnetized perpendicular relativistic shocks in pair plasma are not efficient particle accelerators. An interesting particle acceleration mechanism was observed by \citet{1992ApJ...390..454H} in the case where a small fraction of the plasma are ions (electron-positron-ion composition). In this case the shock structure is modified by ions, even if their number fraction is small compared to positrons. Magnetosonic waves emitted by ions are resonantly absorbed by upstream positrons that produces a non-thermal tail in positron distribution function. Electrons and ions distributions are still Maxwellian. 

Concerning the studies of shocks in electron-ion plasma using 1D simulations, \citet{2006ApJ...652.1297L} and \citet{2008ApJ...672..940H} explored mildly magnetized regimes. Lyubarsky explored the effect of the electromagnetic precursor on the incoming electrons and found that the relativistic oscillation of electrons in the field of the wave results in temperature equipartition between electrons and ions, once they reach downstream medium. The theoretical model was confirmed by a 1D PIC simulation with the mass fraction $m_i/m_e=200$, upstream Lorentz factor $\gamma_0=50$, upstream magnetization $\sigma=0.003$ and $0.6$.  No supra-thermal tail in electron distribution was found. This finding was contrasted by the study of \citet{2008ApJ...672..940H} where it was found that the precursor wave produces a non-thermal tail in ion and electron distribution functions. The mechanism for particle acceleration is expected to be of wakefield nature.

\paragraph{Early multi-dimensional studies}
The opening of additional degrees of freedom in directions transverse to the shock propagation is essential for relativistic shock physics. For instance, the dominant instability in low-$\sigma$ regime, Weibel-filamentation, is artificially suppressed in 1D because it is only triggered by a non-zero $k_{\perp}$, where $k_{\perp}$ is the transverse wavenumber to the beam propagation direction. This regime is particularly relevant for astrophysics (for instance in the case of the GRB or AGN studies of the propagation of the forward shock) because the magnetization in the ISM is $\sigma_{\rm ISM} \sim 5~10^{-11} B_{\mu \rm G}^2/n_{\rm cm^{-3}}$, where the magnetic field strength is in units of $\mu$ Gauss and the ambient density in units of $\rm{cm}^{-3}$. Consequently, the observation of filamentation and concomitant trigger of the Fermi process relies on multi-dimensional configuration of the simulation. This restriction is less severe in non-relativistic case for quasi-parallel shocks, where non-resonant streaming or Bell instability can be triggered in 1D and sustains particle scattering (or mirroring) on both sides of the shock front. 

Additional step from 2D to 3D is important to correctly deal with particle scattering properties as the topology of the turbulent magnetic field is different.

The early multi-dimensional exploration of unmagnetized relativistic shocks was done with full 3D3V but short simulations. \citet{2003ApJ...595..555N}  simulated a relativistic jet with $\Gamma_{\rm jet}=5$ propagating into an unmagnetized electron-ion plasma at rest. These simulations were done using \textsc{TRISTAN} code \citep{buneman93}. The simulation box was small $[15 \times 8 \times 8 \times 15] (c/\omega_{\rm pe})^3$ but still large enough to capture the electron-scale Weibel-filamentation instability and the simulation time was $T_{\rm sim}=23.4 \,\omega_{\rm pe}^{-1}=5.23 \,\omega_{\rm pi}^{-1}$. Capturing only the initial stage (shock not formed) the authors still demonstrated the importance of the Weibel-filamentation instability. \citet{2004ApJ...608L..13F} performed 3D simulation of unmagnetized colliding plasma clouds with density ratio $n_{inj}/n_0=3$, relative Lorentz factor $\Gamma_{\rm jet}=3$ and ion to electron mass ratio $m_i/m_e=16$. The simulation box was larger $L_x \times L_y \times L_z = [200 \times 200 \times 800] (c/\omega_{\rm pe})^3 = [10 \times 10 \times 40] (c/\omega_{\rm pi})^3$ and the simulation time longer $T_{\rm sim}=480 \,\omega_{\rm pe}^{-1}=120 \,\omega_{\rm pi}^{-1}$. The main result of this work is that the initial filamentation grows from electron to ion scales. However, the simulation was just long enough to reach the ionic scale, the saturation was just reached and the shock was not completely formed. \citet{2005ApJ...623L..89H} continue in the same direction by producing 3D simulation with similar parameters but longer time $T_{\rm sim}=360 \,\omega_{\rm pi}^{-1}$. The authors observed an interesting electron acceleration mechanism when electrons cross the ionic current channels. This produced a non-thermal tail in the electron distribution function $d N/d E \propto E^{-2.7}$. However, here again the shock was not fully formed because downstream ion distribution was still far from isotropy. The same electron energization mechanism was later found by \citet{2015ApJ...811...57A}, who studied the jet-ambient medium interaction by means of 3D simulations. These authors suggested that electrons can also be pre-accelerated by SSA mechanism during the shock formation.

\citet{2005AIPC..801..345S} considered a pure $e^--e^+$ plasma where there is no scale separation. In this way the typical shock formation and evolution time is much shorter than for electron-ion plasma. Different magnetizations were explored $\sigma \in [0, 0.1]$ with the relative Lorentz factor $\gamma_{0}=15$, in 3D3V configuration. The shock was triggered by reflection of incoming flow on a conducting wall. In this way, the interaction of wall-reflected and incoming flows produces the shock [simulation frame = downstream rest frame]. The box size was similar to previous studies $ [200 \times 40 \times 40] (c/\omega_{\rm pe})^3$. The shock was formed as the downstream plasma reached the expected Rankine--Hugoniot jump conditions and the distribution function isotropized in the overlap region. As previously, for the unmagnetized case $\sigma=0$ the shock is mediated by Weibel-filamentation but strongly magnetized shock $\sigma=0.1$ has a very different structure shaped by the perturbation of the upstream magnetic field: the incoming flow is coherently reflected on the magnetic barrier at the shock front position. The shock is then mediated by magnetic reflection. In all cases the author did not find evidence of non-thermal part in particle distribution that suggested that the acceleration is either slow to setup or not present at all. Very similar conclusions were found by \citet{2007ApJ...668..974K} where 2D simulations of a shock with $\gamma_{0}=2.24$ in pair plasma were performed. This study demonstrated that the small scale magnetic field fluctuations, self consistently produced by Weibel-filamentation, is able to mediate unmagnetized collisionless shocks. The ratio of magnetic energy to the incoming kinetic energy of the upstream flow, $\xi_{\rm B} = \delta B^2 / (4\pi \gamma_{\rm sh}^2 \rho c^2)$, peaks at the shock front ($\xi_{\rm B} = 0.14$) where the incoming mono-directional flow is isotropized and rapidly decreases downstream by phase-mixing.

\paragraph{Recent multi-dimensional studies}
The main difference with the early studies is (i)~full formation of pair \emph{and} electron-ion shocks and (ii)~the realization of efficient particle acceleration through first-order Fermi process in a self-consistent way by following the evolution of shocks on longer timescale.

\citet{2008ApJ...674..378C} addressed the question of the fate of the magnetic turbulence downstream of relativistic unmagnetized shock in pair plasma. The authors used the same code as in \citet{2005AIPC..801..345S} (\textsc{TRISTAN-MP}) with relatively long simulation time $T_{\rm sim}=3500\,\omega_{\rm p}^{-1}$, where $\omega_{\rm p}$ is the total plasma frequency.
As expected, the unmagnetized shock is mediated by Weibel-filamentation in the precursor. The filamentary magnetic field in the precursor becomes almost isotropic in the downstream medium once the filaments break at the shock front. The coherence scale of the field just behind the front is small $\ell_c \sim 10\,c/\omega_{\rm p}$ but grows with increasing distance from the shock front downstream. At the same time magnetic field intensity is found to decrease rapidly. The authors compared the simulations with an analytic model where magnetic field decreases by linear response of the plasma. The intensity is predicted to decrease as  $\xi_B \propto \delta B^2/8\pi \propto (x_{\rm front} - x)^{-q}$, with $q=2/3$. The simulations suggested however that $q=1$ close to the front located at $x_{\rm front}$ and becomes closer to $2/3$ far from the shock front. Due to numerical noise in PIC simulations at finite time of the simulations it is still not clear whether the field strength drops to 0 far from the shock, on macroscopic scales.

The work of \citet{2008ApJ...682L...5S} presented the first self-consistent demonstration of first-order Fermi process in shocks. The same code as in \citet{2005AIPC..801..345S} was used, but in 2D configuration for an unmagnetized pair plasma with upstream Lorentz factor of $\gamma_{\rm 0}=15$. The simulation time extended up to $T_{\max}=10^4\,\omega_{\rm p}^{-1}$ and the box size was  $[10^4 \times 400] (c/\omega_{\rm p})^2$. In the intermediate times the shock structure is identical to \citet{2007ApJ...668..974K} and \citet{2008ApJ...674..378C}. At late time, a supplementary population of particles builds up as a small fraction of  particles in the bulk downstream plasma is able to scatter back into upstream and participate in a standard DSA (in the relativistic regime). The energy gain per cycle is consistent with the analytic prediction $\Delta E / E \simeq 1 $. The non-thermal population carries typically 1\% by number and 10\% energy fraction of the total incoming plasma (i.e., the ratio of non-thermal electron energy to the total is $\varepsilon_{\rm e} \sim 0.1$). 

\citet{2008ApJ...673L..39S} presented the first study of electron-ion relativistic shocks where the shock is fully formed. Several mass ratios were explored $m_i/m_e = [16, 30, 100, 500, 1000]$ and the upstream plasma was unmagnetized. The most important result of this study is that electrons are brought to sub-equipartition with ions during their crossing of the precursor where they are substantially heated inside the ionic filamentary structures. In the downstream medium one gets $T_e \simeq T_i = (\gamma_0/3) m_p c^2$. This result implies an empiric similarity between shocks in pair plasma and in electron-ion plasma because the relativistic mass of particles in the downstream medium is equal. The simulations where still too short to observe the formation of a non-thermal tail in particle distributions.

\citet{2009ApJ...693L.127K} addressed the long term evolution of $\sigma=0$ shocks in pair plasma by performing the longest possible simulations allowed by numerical stability. The simulation box was $[63000 \times 1024] (c/\omega_{\rm pe})^2$ and the simulation time was $T_{\rm sim}= 12600\,\omega_{\rm pe}^{-1}$. They demonstrated that, as particles accelerate to larger energies with time, the precursor size increases and the width of the zone filled with magnetic turbulence increases both upstream and downstream. No convergence was reached, which leaves the question of long-term evolution open.

\citet{2009ApJ...695L.189M}, by means of 2D PIC simulations with the \textsc{OSIRIS} code, demonstrated that DSA works in electron-ion unmagnetized plasma. The mechanism is very similar to the pair plasma case since electrons are at sub-equipartition with ions in the downstream medium.

The influence of magnetic field orientation with respect to the shock normal in strongly magnetized relativistic shocks ($\sigma=0.1$ and $\gamma_0=15$) was studied by \citet{2009ApJ...698.1523S} in pair plasma and by \citet{2011ApJ...726...75S} in electron-ion plasma, using 2D and 3D PIC simulations. These works provide a first survey of parameter space and show in which conditions relativistic shocks are efficient accelerators or not. Even if relativistic shocks are known to be generically quasi-perpendicular, the magnetic field inclination parameter, $\theta_{B}$, is important for the shock physics. The very special case of quasi-parallel (or subluminal) shocks, even if very rare, is interesting as it shows very different behavior. In all cases, simulations were carried out for long enough time to form the shock and see whether particle acceleration is present or not. For strongly magnetized shocks ($\sigma=0.1$) the authors demonstrate an important difference between sub-luminal and super-luminal shocks in terms of structure and particle acceleration efficiency. In parallel shocks, the relativistic version of Bell instability is triggered and sustains an efficient DSA process. In oblique, but still sub-luminal, shocks an important contribution from the SDA mechanism was observed in competition with standard DSA. This contribution comes from the fact that the upstream plasma carries a motional electric field that can energize particles when they are reflected on the shock front. Consequently, the power-law slope of the non-thermal particle distribution function, where ${\rm d}N / {\rm d} E \propto E^{-\alpha}$, is not equal to the standard prediction but varies between $2.2$ and $2.8$.  In superluminal configuration these authors did not find any particle acceleration. In this case shocks are mediated by the emission of semi-coherent electromagnetic wave from the shock front. 


\begin{figure}
\begin{centering}
\includegraphics[width=1\linewidth]{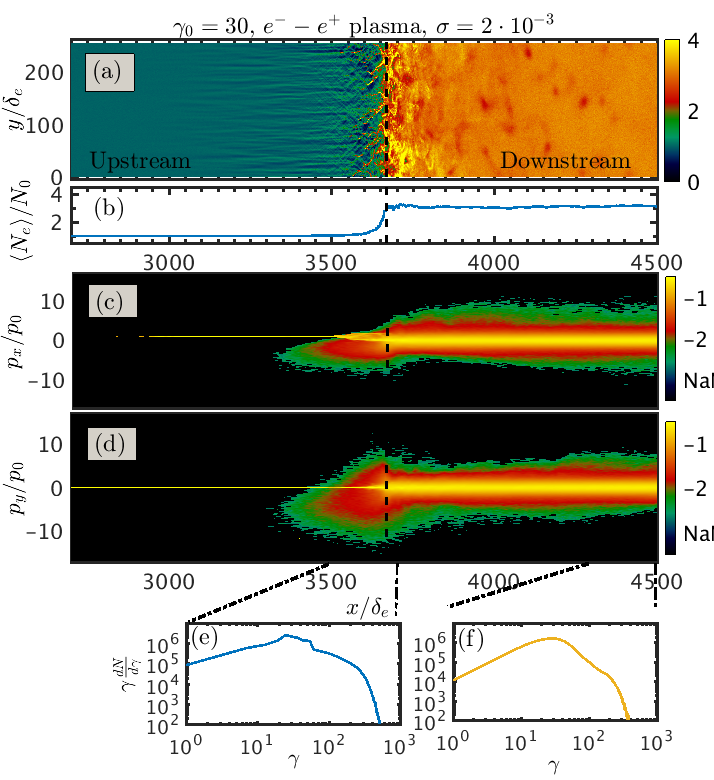}
\par\end{centering}
\caption{Structure of a relativistic perpendicular shock in a pair plasma, at the simulation time $t\omega_{\rm p}=500$, obtained using 2D3V PIC code \textsc{SMILEI}. The Lorentz factor of the upstream incoming flow is $\gamma_0=30$ and  the upstream plasma magnetization is $\sigma=2 \times 10^{-3}$. Panel (a) shows the electron number density in the simulation plane. Panel (b) shows the transversely averaged electron density normalized to the upstream value. Panels (c) and (d) show the longitudinal phase space $x-u_x$ and transverse phase space $x-u_y$, respectively. Panels (e) and (f) present the particle distribution function in energy around the shock front and far downstream. } 
\label{fig:relat_gam30_shock_structure}
\end{figure}

In order to illustrate the output from PIC simulations of shocks, in Fig.~\ref{fig:relat_gam30_shock_structure} we present the structure of perpendicular ($\theta_{\rm B}=90^\circ$) relativistic shock in pair plasma for mildly magnetized case $\sigma=2\times 10^{-3}$, obtained with the PIC code \textsc{SMILEI}. This structure is similar to the one found by \citet{2009ApJ...698.1523S} for superluminal shocks or, more closely, to the mildly magnetized case in \citet{2013ApJ...771...54S}. The magnetization is chosen so that particle acceleration is efficient but maximal energy is limited by the precursor size being of the order of the Larmor radius of incoming particles $R_{L,0} = \gamma_0 m_e c^2 / (e B_0)$: $\gamma_{\max} \sim 20 \gamma_0$. Panels (a) and (b) present the electron density in the simulation plane and the transversely averaged profile, respectively. The shock front position is delimited by the vertical dashed line and the front propagates from the right to the left side with a velocity $\upsilon_{\rm sh | d} \simeq 0.5 c$ as measured in the downstream (simulation) frame. Ahead of the shock front oblique filamentary density structures emerge as a result of the interaction between the incoming flow and the cloud of accelerated particles. This region defines the shock precursor. Panels (c) and (d) show the longitudinal phase space $x-u_x$ and transverse phase space $x-u_y$, respectively. The transition at the shock front is clearly seen at the position where the flow becomes isotropic and hot. The cloud of energetic particles ahead of the shock front corresponds to the accelerated population. Finally, panels (e) and (f) present the particle distribution function in energy around the shock front and far downstream, respectively. Particle acceleration operates mainly around the shock front, while far downstream distribution exhibits a Maxwellian part and the start of non-thermal tail at the highest energies.


\citet{2011ApJ...739L..42H} explored the differences in the structure of unmagnetized electron-ion shocks between 2D and 3D simulations. While very similar, some quantitative differences emerged in 3D simulations: the cross shock electrostatic field is slightly larger than in 2D, magnetic energy density in the shock transition region is smaller and the index of the power-law tail is closer to $2.2$, instead of $2.4$ in 2D. The latter is more consistent with analytical expectation \citep[e.g., ][]{2001MNRAS.328..393A}.

Maximal energy of accelerated particles in perpendicular shocks was investigated by \citet{2013ApJ...771...54S} by means of 2D and 3D long-term simulations. Both pair plasma and electron-ion plasma were explored for a range of magnetizations from unmagnetized case $\sigma=0$ to strongly magnetized $\sigma=0.1$ and for different Lorentz factors of the upstream flow ($\gamma_{0}=[3,240]$). The simulation box transverse size was $100~c/\omega_{\rm pi}$ in electron-positron case and  $25~c/\omega_{\rm pi}$ in electron-ion case, allowing to capture at least several filaments when Weibel-filamentation mediates the shock. It was found that the maximum particle energy increases in time as $E_{\max} \propto t^{1/2}$ for both electron-positron and electron-ion shocks. This result emerges from small-angle scattering regime of the accelerated particles in the self-excited micro-turbulence, where one expects the spatial diffusion coefficient to scale as $D \propto E^2$. The other important result of \citet{2013ApJ...771...54S} study is evidencing the critical magnetization above which relativistic perpendicular shocks are not accelerating particles. For electron-positron composition the critical magnetization value is $\sigma_{\rm crit} \approx 3\times 10^{-3}$ and for electron-ion composition it is $\sigma_{\rm crit} \approx 3\times 10^{-5}$. Weakly magnetized shocks with $\sigma < \sigma_{\rm crit}$ were found to be mediated be Weibel-filamentation that generates strong small-scale magnetic field in the vicinity of the front. In this regime DSA is efficient, with the maximum particle energy scaling as $E_{\max} \propto \sigma^{-1/4}$, and a fraction of energy transmitted to the supra-thermal particles $\xi_{\rm \tiny CR} \sim 10\%$. On the other side, for $\sigma > \sigma_{\rm crit}$ DSA is inhibited as the shock structure is no longer dominated by the filamentation instability.


\begin{figure}
\begin{centering}
\includegraphics[width=1\linewidth]{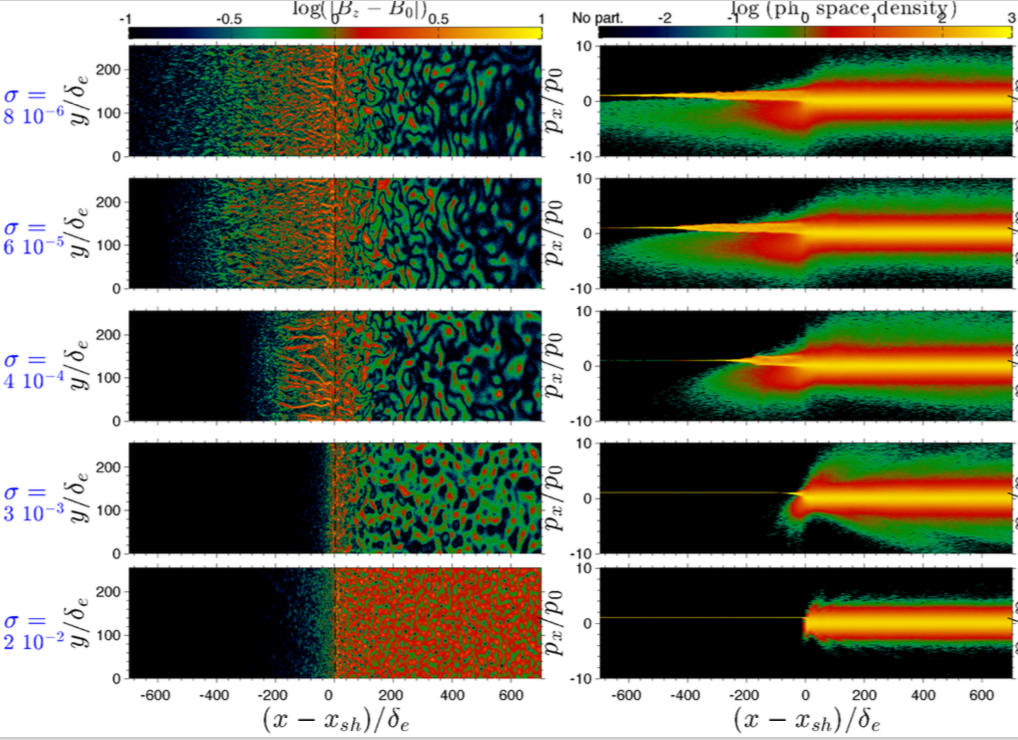}
\par\end{centering}
\caption{Dependence of the structure of relativistic perpendicular shocks in a pair plasma on the flow magnetization $\sigma$, at the simulation time $t\omega_{\rm p}=900$. Five representative cases are shown, from top to bottom, $\sigma=8\times 10^{-6},6\times 10^{-5}, 4\times 10^{-4}, 3\times 10^{-3}$ and $2\times 10^{-3}$. The left column shows the absolute value of the transverse magnetic field increment and the right column shows the longitudinal $x-p_x$ phase space distribution in the shock transition region. Images adapted from \citet{2018MNRAS.477.5238P}.}
\label{fig:relat_magnetization}
\end{figure}


\citet{2014PhPl...21g2301B} studied the shock formation mechanism in the unmagnetized $\sigma=0$ case for pair plasma. An analytical model was developed, based on the growth and saturation time of the Weibel-filamentation instability. The formation time is estimated as a multiple of the instability $e$-folding time, $3\tau_{\rm sat}$. At saturation, the density in the overlap region is 2, then the phase of density accumulation up to Rankine--Hugoniot conditions is expected to be linear in time as the incoming plasma supplies the downstream region. The analytic model is then compared with 2D PIC simulation obtaining a reasonably good agreement.

Using 2D simulations, \citet{2018MNRAS.477.5238P} provided a more systematic investigation over $\sigma$ than previously done, from unmagnetized to strongly magnetized shocks. Two different PIC codes were used, Finite Difference Time Domain (FDTD) and pseudo-spectral. Shock formation time, jump conditions, shock structure transition from low-$\sigma$ to high-$\sigma$ were investigated. The shock structure evolution for five different values of $\sigma$ is presented in Fig.~\ref{fig:relat_magnetization}. It shows the gradual transition from filamentation mediated shocks ($\sigma < 10^{-3}$), where particle acceleration is efficient, to magnetic reflection-shaped shocks ($\sigma > 10^{-2}$) where particle acceleration is inhibited, confirming the findings of \citet{2013ApJ...771...54S}. The shock formation time was found to be significantly longer than predicted by \citet{2014PhPl...21g2301B}. This points out the importance of other physical process than only saturation of Weibel-filamentation instability. For example, the studies of \citet{2018PhPl...25g2115V, 2018PhRvL.120x5002R} demonstrate a dominant role of the drift-kink instability in the non-linear phase during which the shock front really forms. The particular focus of the study of \citet{2018MNRAS.477.5238P} was on particle scattering properties, directly extracted by following self-consistent particle dynamics. The results demonstrated that the particle diffusion coefficient scales as $D=\langle \Delta x^2 \rangle /2 \Delta t \propto E^2$ in weakly magnetized shocks, which justifies the increase in particle maximum  energy of accelerated particles as $\gamma_{\max} \propto \sqrt{t}$, evidenced by \citet{2013ApJ...771...54S}. In moderately magnetized shocks, the diffusion coefficient is modified by the presence of the ordered component that imposes a saturation of the maximum particle energy once particles get advected downstream under the effect of regular gyration.



\subsubsection{Mildly relativistic shocks}

The current consensus is that ultra-relativistic shocks are not very efficient particle accelerators for particle energies above PeV energies, mainly because of the quadratic dependence of the spatial diffusion coefficient on the particle energy, $D \propto E^2$. In the non-relativistic case, supernova remnants are considered to accelerate protons up to several hundreds of TeV or a few PeV at best and iron nuclei at energies 26 times higher. The question is then: how do the particles get accelerated to $10^{20}$~eV, maximal energy of CRs as measured at Earth?
One of the promising scenarios considers mildly relativistic shocks as viable candidates (for example, trans-relativistic phases of supernova explosions or internal shocks in GRBs and jets of the AGNs). The reason is that the energy gain approaches the relativistic limit ($\Delta E /E \simeq 1$ per cycle), while a number of intrinsic limitations of the ultra-relativistic regime are alleviated, such as the generic superluminal configuration imposed by strong contraction of the pre-shock magnetic field by the shock front. Also, recent non-linear Monte-Carlo simulations demonstrated the efficiency of trans-relativistic shocks to accelerate particles to very high energies \citep{2013ApJ...776...46E}.

The mildly relativistic regime is still poorly explored with kinetic simulations as of now. In the review by \citet{2016RPPh...79d6901M}, section 4.3 was devoted to the discussion of mildly relativistic shocks. The studies discussed there concerned mainly the plasma physics of shock formation, but not the long term evolution. Here we provide a short update in light of the most recent studies.

\paragraph{Electron-positron plasma}
Using the PIC code \textsc{Epoch} \citet{2017JPlPh..83a9004D, 2018MNRAS.473..198D} investigate the generation of instabilities in a 2D configuration in the case of two interpenetrating pair plasma clouds. One of the beams is produced by the reflection at a wall of the incoming beam. The simulations focus on the generation of micro-instabilities and the shock formation process, but are not long enough to investigate non-thermal particle production.  \citet{2018MNRAS.473..198D} consider a pair plasma moving at a speed of c/2 and perform three simulations, one in 1D with a resolution of $67500c/\omega_{\rm pe}$ and $3.4 \times 10^7$ macro particles, and two in 2D with the best resolution at [$67500\times1500$]$(c/\omega_{\rm pe})^2$ and using 1 billion macro particles. The two-stream and Weibel instabilities are found to rule the wave growth at the shock transition layer in this regime and to take over the filamentation instability, which nevertheless may develop upstream. We note that the Debye length has to be resolved in order to accurately capture the two-stream instability, which limits the spatial and temporal extension of the simulations. 

Another interesting configuration can be found in the study of the expansion of a mildly relativistic pair plasma in a background electron-proton plasma. This setup approaches the scientific case studied in the so-called two-flow model developed to investigate gamma-ray emission in blazar jets \citep{1989MNRAS.237..411S}. Two setups have been considered either in an unmagnetized plasma \citep{2018PhPl...25f2122D, 2018PhPl...25f4502D} or with a guiding magnetic field oriented along the pair plasma drift direction \citep{2019A&A...621A.142D}. While in the former work the pair were hot with a mildly relativistic temperature of 1 MeV, the two latter works have a similar setup: the simulations are 1D with a cold pair plasma with a temperature of 400 keV moving at 0.9c, the background plasma has a realistic proton/electron mass ratio of 1836. The study follows the formation of the pair jet and the interaction between the two plasmas. It results from the free expansion of the pair beam the production of an electromagnetic piston that expels and compresses ambient electrons. The excess of negative current decelerates further the electrons but accelerates the positrons than can drift ahead the jet's head, and reach kinetic energies of $\sim$ MeV. In the meantime in both configurations (unmagnetized and magnetized) the pair beam and the background plasma interact through the filamentation instability which builds up a turbulent electro-magnetic field and contributes to accelerate the ambient protons also to MeV energies.

\paragraph{Electron-ion plasma}
Early studies of mildly relativistic shocks in electron-ion plasma \citep[e.g., ][]{2008ApJ...675..586D} presented important insights on the shock formation process but unfortunately their simulations were not long enough to follow-up on particle acceleration efficiency. The longest (and largest in transverse dimension) 2D PIC simulations to date were performed by \citet{2019MNRAS.485.5105C} allowing the full formation and mid-term evolution of the shock. These authors studied the regime where the shock front velocity (in the pre-shock frame) is $\beta_{\rm sh} \approx 0.83 c$, Lorentz factor $\gamma_{\rm sh} \approx 1.8$, and the Aflv\'enic Mach number of $M_{\rm A} = 15$. Two different magnetic field inclinations to the shock-normal where investigated: $\theta_{\rm Bn}=15^\circ$ (sub-luminal) and $\theta_{\rm Bn}=55^\circ$ (super-luminal). The main finding is that sub-luminal (quasi-parallel) shocks are efficient particle accelerators (for both electrons and ions) but not super-luminal shocks \footnote{We note that some caution has to be taken when assimilating super-luminal shocks with quasi-perpendicular shocks. In non-relativistic cases, quasi-perpendicular -- but still sub-luminal -- shocks can be efficient electron accelerators.}. When particle acceleration is efficient, the energy fraction transferred from the shock to supra-thermal ions was found to be $\varepsilon_{\rm p} \simeq 0.1$ (same as in non-relativistic and ultra-relativistic cases) and the energy fraction in accelerated electrons $\varepsilon_{\rm e} \simeq 5\times 10^{-4}$ was found to be higher than in the non-relativistic shocks ($\varepsilon_{\rm e} \sim 10^{-4}$; see next subsection) but still much smaller than in ultra-relativistic shocks where electrons are in equipartition with ions, hence $\varepsilon_{\rm e}\sim 0.1$. 

Some details of plasma physics underlying the particle acceleration efficiency were also addressed by \citet{2019MNRAS.485.5105C}. The presence of whistler waves was found in the simulation of the quasi-parallel shock, confirming the finding of \citet{2008ApJ...675..586D}, but their role in electron acceleration or injection was found to be sub-dominant, i.e., with increasing mass ratio $m_{\rm i}/m_{\rm e}$ from 64 to 160 whistler dynamics were expected to play more important role in electron acceleration efficiency but this effect was not observed. The maximal energy of the accelerated particles was found to increase linearly in time similarly to non-relativistic shocks, implying that the diffusion coefficient scales as $D \propto E$, resulting from efficient excitation of Bell instability in the shock precursor. On numerical side, these authors also evidenced the importance of large transverse size of the simulation box, showing that too narrow box suppresses the electron acceleration efficiency. The reason is that too narrow box suppresses Bell modes, which dominate the non-linear physics of the shock precursor. 



\subsubsection{Non-relativistic shocks}
\label{S:NR-PIC-simus}

If compared to ultra-relativistic (UR) or mildly-relativistic (MR) shocks, the difficulty of capturing the full development of non-thermal tail in non-relativistic (NR) shocks comes from the fact that the energy gain per Fermi cycle (upstream $\to$ downstream $\to$ upstream) is much smaller than in UR and MR cases, as $\Delta E/E \simeq u_{\rm sh}/c$. As a consequence, the duration of simulations must be long enough to capture at least a dozen of cycles in order to get a well-developed power-law tail while in UR shocks only a couple of cycles provides a distinguishable tail.

Despite significant efforts, only the initiation and very early stages of particle acceleration process were conveniently addressed using PIC and hybrid-PIC simulations. DSA is the most accepted model for particle acceleration at shocks. As already discussed in Sect.~\ref{S:DSA-inj} one major difficulty for DSA to operate is the process requires particle to have a Larmor radius larger than the shock width, typically of the order of a few thermal ion Larmor radii. This concern is particularly stringent for electrons which at sub-relativistic energies have very small Larmor radii. We here discuss recent PIC simulations which address the problem of injection of electrons and ions, while the acceleration performances on dynamical timescale of the shock will be discussed in the following sections.

Table \ref{T:SIM} presents a (non-exhaustive) list of PIC numerical experiments applied to NR shock studies. All these works are discussed in the text below. 

\begin{footnotesize}
\begin{table*}[h]
\begin{tabular}{ccccccc}
\hline
\hline

\rm{Authors} & \rm{Dim.} & $m_{\rm i}/m_{\rm e}$ & $\theta_{\rm B}^o$ & $M_{\rm A}$ & $M_{\rm s}$ &$u_{\rm sh}/c$ \\
\hline
\rm{Shimada \& Hosh. (2000)} & 1D & 20 & 90 & 3.4-10.5 & 1.3-4 & - \\
\rm{Hoshino \& Shim. (2002)} & 1D & 20 & 90 & 32 & 3.3 & 0.375 \\
\rm{Amano \& Hosh. (2007)} & 1D & 100 & 80 & 15 & 188 & 0.075 \\
\rm{Amano \& Hosh. (2009)} & 2D & 25 & 90 & 14 & 28 & 0.28 \\
\rm{Lemb\`ege et al (2009)} & 2D & 400 & 90 & 5 & - & 0.42 \\
\rm{Riquelme \& Spit. (2011)} & \rm{2D,3D} & 40-1600 & 45-90 & 3.5-14 & - &0.042-0.14 \\
\rm{Mastumoto et al (2012)} & \rm{2D} & 100 & 90 & 14-30 & 28-60 & 0.3\\
\rm{Guo et al (2014a)} & 2D,3D & 100,400 & 63 & $\leq 5$ & 3 & 0.225 \\
\rm{Guo et al (2014b)} & 2D,3V & 100 & 13-80 & $\leq 5$ & 3 & 0.225 \\
\rm{Park et al (2015)} & 1D & 100 & 30 & 20 & 40 & 0.1 \\
\rm{Kato (2015)} & 1D & 30 & 30 & 23& - & 0.37 \\
\rm{Wieland et al (2016)} & 2D & 50 & 45 & 27.6 & 755 & 0.39 \\
\rm{Bohdan et al (2017)} & 2D & 100 & 90 & 31-35& 50, 1550 & 0.26-0.3 \\
\rm{Mastumoto et al (2017)} & \rm{3D} & 64 & 75 & 21 & 23 & - \\
\hline
\hline
\end{tabular}
\caption{The different PIC experiments discussed in the text referenced by author's names and publication dates. The different columns display: simulation dimensionality in space (in velocity space all simulations are `3V', tracking all three components), the ion to electron mass ratio, the magnetic field obliquity angle, the Alfv\'enic and sonic Mach numbers and/or the ratio of the shock velocity to the speed of light when they are available.}
\label{T:SIM}
\end{table*}
\end{footnotesize}


\paragraph{Electron injection at non-relativistic shocks.} 
Non-thermal electrons are responsible for the radio synchrotron emission from SNRs \citep{2012A&ARv..20...49V} and are observed at interplanetary shocks \citep{2016ApJ...826...48M}. Several PIC simulations have explored the injection of electrons for different shock regimes and magnetic field obliquity.

At quasi-perpendicular shocks electrons can be accelerated by SSA if large amplitude electrostatic waves can develop at the shock front. These waves can be produced in the non-linear regime of the Buneman instability triggered by the relative streaming of reflected ions and incoming electrons \citep{1984SSRv...37...63W, 2000ApJ...543L..67S}. \citet{2002ApJ...572..880H} and \citet{2007ApJ...661..190A} using 1D PIC simulations investigate the acceleration of electrons in a perpendicular fast shock. Electrons are heated by the interplay of the Buneman instability and trapped in these electrostatic waves and reflected back to the shock front by the motional upstream electric field. The maximum energy is expected to be at best $E_{\max} = m_{\rm i}c^2 (u_{\rm sh}/c)$. These simulations have been generalized to multi-dimensions (2D in configuration space 3D in velocity space) by \citet{2009ApJ...690..244A, 2012ApJ...755..109M, 2012PPCF...54h5015D, 2016ApJ...820...62W, 2017ApJ...847...71B}. These works first set the criteria for SSA to occur\footnote{\citet{2017ApJ...847...71B} provide an update of these conditions for the orientation of the magnetic field with respect to the simulation plane.}: 1)~the thermal speed of electrons has to be smaller than the drift speed between ions and electrons, hence the shock Mach number must satisfy $M_{\rm s} \ge (1+\alpha)/\sqrt{2} \sqrt{m_{\rm i} T_{\rm e}/m_{\rm e} T_{\rm i}}$ where $\alpha$ is the density ratio of reflected to incoming ions and $T_{\rm i,e}$ are the background ion and electron temperatures, and 2)~Buneman modes have to be destabilized, this requires the shock Alfv\'enic Mach number to satisfy $M_{\rm A} \ge (1+\alpha) (m_{\rm i}/m_{\rm e})^{2/3}$. These conditions depend on the ion to electron mass ratio adopted in the PIC simulations. The development of the Buneman instability and the intensity of electrostatic waves vary considerably with the number of reflected ions in the shock reformation process. Acceleration of electrons to non-thermal energies is confirmed \citep[but see the discussion in][]{2012PPCF...54h5015D}, but the efficiency of the process depends on the background magnetic orientation with respect to the simulation plane (recall that simulations are 2D). Acceleration is the most efficient when Buneman instability-generated waves have the highest intensities, which happens when the magnetic field is out of the plane of the simulation \citep{2017ApJ...847...71B}. Electron acceleration also depends on shock non-stationarity associated to its reformation \citep{2009JGRA..114.3217L}. \citet{2017arXiv170903673M} perform 3D perpendicular shock PIC simulations of an oblique shock ($\theta_{\rm B}\simeq 75^o$) with a high Alfv\'en Mach number ($M_{\rm A} \simeq 21$). They find a two-step electron acceleration: first electrons gain energy via SSA in the electrostatic waves driven by the Buneman instability as above, but then they further gain energy by interacting with turbulent fields produced by the Weibel ion-ion instability triggered by the interaction of reflected and background ions. The downstream electron distribution shows the formation of a power-law energy spectrum with an index $\sim -3.5$.

\citet{2011ApJ...733...63R} perform an extensive survey of shock conditions to investigate electron acceleration. They study the effect of variations of the shock speed, ambient medium magnetization, electron to ion mass ratio and magnetic field obliquity over non-thermal electron injection at shocks. However their simulations are restricted to rather modest Alfv\'enic Mach numbers $M_{\rm A} < 14$. One important issue raised by the authors is that a small ion to electron mass ratio suppresses the propagation of oblique whistler waves \citep{2004AnGeo..22.2345S}\footnote{whistler waves are likely excited due to the cross-field drift of background electrons with respect to either reflected or background ions, this is the so-called modified two-stream instability or MTSI.}, whereas these waves can become over-dominant to heat/energize electrons in the foot. A criterion for whistler wave to grow is $M_{\rm A}/(m_{\rm i}/m_{\rm e})^{1/2} < 1$ \citep{2003JGRA..108.1459M}. The acceleration mechanism relies on the property of oblique whistler waves to have an electric field component parallel to the magnetic field. Particles are then first accelerated by this electric field before the complementary action of the convective electric field. Electrons are preferentially accelerated at high obliquity $\theta_{\rm B} \sim 70^o$ (at $M_{\rm A} = 7$), where the downstream energy index is $\sim 3.6$. At smaller obliquities particles are not sufficiently confined at the shock front whereas for quasi-perpendicular shocks particles can not propagate in the foot. Electron acceleration efficiency depends mostly on the Alfv\'en Mach number first through the condition on whistler wave production recalled above. The electron distribution is the hardest for Alfv\'enic Mach numbers in the range 3-7. The energy index changes form 2.6 to 4 as $M_{\rm A}$ changes from 3.5 to 14. 
A complementary study of electron acceleration in low Mach number ($M_a \leq 5$) shocks was performed by \citet{2014ApJ...794..153G, 2014ApJ...797...47G} using 2D PIC simulations in order to get better understanding of electron acceleration in galaxy cluster shocks. These authors found that a measurable fraction of incoming upstream electron (up to 15\%) bounces back upstream and formes a non-thermal tail in the distribution function with power-law index in energy $p 
\simeq 2.4$. These particles scatter back to the shock front on self-generated waves via firehose instability and participate in the SDA process. This acceleration process was found to be efficient if upstream plasma is high beta ($\beta \geq 20$) for nearly any magnetic field obliquity.

High Alfv\'enic Mach number, quasi-parallel shocks could also allow electron injection. These shocks are likely good proton injectors (see below). In turn protons (ions) can trigger magnetic perturbations, as the magnetic field grows in the pre-shock medium then lowering $M_{\rm A}$ and its transverse component can be compressed at the front. The conditions then resemble the case of highly-oblique moderate Alfv\'en Mach number shocks discussed by \citet{2011ApJ...733...63R} [see also \citet{2014ApJ...794...46C}].
\citet{2015PhRvL.114h5003P} perform long term 1D PIC simulations of high $M_{\rm A}$ quasi-parallel shocks \citep[see also ][]{2015ApJ...802..115K}. Protons destabilize non-resonant streaming (Bell) modes and electrons are accelerated by a combination of SDA and Fermi processes as they are scattered by the non-resonant streaming modes. Interestingly, non-thermal electrons entering in the relativistic regime show a $E^{-2}$ energy spectrum and a non-thermal electron to proton ratio $\sim 10^{-3}$ roughly proportional to $u_{\rm sh}/c$.

\paragraph{Ion injection at non-relativistic shocks.}

The major drawback of full-PIC simulations to address the ion injection into DSA is the need to resolve both electron- and ion-scale physics, that bakes typical simulation not longer than a $\sim 10 \omega_{ci}^{-1}$. This difficulty is partly bypassed using hybrid-PIC simulations where ions are still treated kinetically but electrons are treated as massless fluid  \citep[see, e.g.,][]{2002hmst.book.....L, 2007CoPhC.176..419G, Kunz14}. In this approach all ion-scale kinetic physics are preserved while the global numerical cost is about two orders of magnitude lower than in full-PIC simulations.

The injection of thermal ions into the acceleration process was investigated by \citet{2013ApJ...773..158G} and  \citet{2015ApJ...798L..28C} using multi-dimensional hybrid-PIC simulations. The first main finding of these studies is that protons are not injected by `thermal leakage' of downstream thermalized distribution into pre-shock medium but by specular reflection on time-varying shock barrier.  For quasi-parallel shocks ($\theta_{\rm Bn}\leq 45^\circ$) and high $M_a>5$,  the injection efficiency is larger than 10\%. As evidenced by \citet{2015ApJ...798L..28C}, protons gain energy through SDA in consecutive reflections on the shock front and inject into DSA when their energy is large enough to escape upstream. They also propose a quantitative model that accounts for the drop in injection efficiency of quasi-perpendicular shocks with  $\theta_{\rm Bn} \geq 45^\circ$, as more than 4 SDA cycles are required for injection into DSA, while at each SDA cycle a large fraction of ions ($\sim 75\%$) is lost downstream. This effect explains the rapid drop in injection efficiency of $\theta_{\rm Bn} \geq 45^\circ$ shocks \citep[see, however, ][]{2016ApJ...827...36O}.

Other studies addressed the thermalization of heavy ions in post-shock medium and chemical enhancement in shock accelerated particles. It was found that each species acquires downstream temperature proportional to its mass, $T_d \propto A_i$, where $A_i$ is the atomic number \citep{2016JTePh..61..517K, 2017PhRvL.119q1101C}. The efficiency of ion injection into DSA increases with $A/Z$ ratio, where $Z$ is the charge. \citet{2017PhRvL.119q1101C} show that there is preferential acceleration of ions with large $A/Z$ in quasi-parallel shocks. For $M_a > 10$ these authors find that the fraction of DSA-accelerated ions scale as $(A/Z)^2$, in quantitative agreement with abundance ratios in Galactic Cosmic Rays. The injection mechanism of heavy ions is different from proton injection, since they do not have any dynamical impact on the shock structure. Instead of reflecting specularly on the shock front, heavy ions directly thermalize in the post-shock medium. If the downstream isotropization time is shorter than advection time, a small fraction can back stream into pre-shock medium and participate in DSA. We note that \citet{2019ApJ...872..108H}, using 2D hybrid-PIC simulations as well, confirm the preferential injection of heavy ions up to $A/Z \sim 10$ but find that there is saturation for higher $A/Z$ values, in contrast with \citet{2017PhRvL.119q1101C} findings. For quasi-perpendicular shocks with $\theta_{\rm Bn}>50^\circ$, similarly to pure electron-proton shocks \citep{2014ApJ...783...91C}, there is no injection into DSA of any ion species, since advection time becomes shorter than isotropization time.

\subsubsection{Discussion: }
\label{S:Crit_discuss_Shock-PIC-simus}
Let us summarise and briefly discuss the micro-physical studies of particle acceleration at collisionless shocks. Several key points emerge in recent studies: 

\begin{itemize}
\item The self-consistent shock structure produces non-thermal particle distributions for a broad range of parameter space. The main requirement for efficient particle acceleration is the ability of particle impinging the front to escape back into the pre-shock medium and trigger wave growth through kinetic instabilities. This condition is generally met in high-Mach number sub-luminal shocks (or weakly magnetized shocks).
\item Unmagnetized relativistic  shocks are efficient particle accelerators in both electron-positron and electron-ion plasma. Here, it was demonstrated that the Weibel-filamentation instability mediates the shock transition region. It leads to significant magnetic field amplification and concomitant particle acceleration. Typically, about 1\% by particle number fraction and 10\% by the shock kinetic energy fraction is channelled to non-thermal particles. However, the acceleration rate is slow as it scales quadratically with particle energy: $t_{\rm acc} \propto E^2$. Thus, one infers the maximum energy of protons achievable in ultra-relativistic shocks of GRBs as $E_{\max} \sim 10^{16}$~eV. 
\item When the relativistic shock front propagates into highly magnetised medium ($\sigma > 0.01$) one has to distinguish between sub-luminal and superluminal configurations. In the former case, particle acceleration was found to be efficient, sometimes even more than in unmagnetised case. Yet, the sub-luminal configuration is statistically disfavoured in relativistic case. When the configuration is superluminal, shock-processed particles are advected downstream and are unable to undergo Fermi process. 
\item At intermediate magnetization, $\sigma_{\rm crit} < \sigma < 0.01$, limited particle acceleration occurs and the maximum energy scales as $E_{\max} \propto \sigma^{-1/4}$.
\item The interesting case of mildly-relativistic shocks, e.g., $\gamma_{\rm sh} \geq 1$, where energy gain per Fermi cycle is large and the shock can easily be subluminal is poorly studied. Recent study by \citet{2019MNRAS.485.5105C} found that shock physics in quasi-parallel case is similar to non-relativistic shocks.
\item Non-relativistic shocks are the most common and studied in literature. The most representative case is the external shock of the Supernova Remnants. In this regime, when particles are efficiently reflected on the front, several instabilities can be in competition in the precursor region (e.g., Buneman, firehose, whistler, Weibel, gyroresonant, Bell), depending on the Mach number, magnetic field strength and obliquity. Hence, the phenomenology is more complex then in ultra-relativistic shocks. For example, quasi-parallel shocks are common, contrary to the ultra-relativistic regime, and lead to efficient magnetic field amplification through resonant and Bell instability in the shock precursor.
\end{itemize}

There are several open questions under active investigation or to be addressed in the near future. 

\begin{enumerate}
    \item Are ultra-relativistic shocks always locked in the slow acceleration rate, i.e., $t_{\rm acc} \propto E^2$, or an additional source of magnetic turbulence can produce faster acceleration? How is the long term evolution of unmagnetized shocks where the shock transition is governed by self-excited microturbulence ? No steady state was reached with current simulations. For recent progress in this field, see \citet{PhysRevLett.123.035101}.
    \item The question of which configuration, quasi-parallel or quasi-perpendicular, in non-relativistic shocks is more efficient for ion/electron acceleration? The regime of quasi-perpendicular but still sub-luminal shocks is a particular case that requires clarification.
    \item How promising is the mildly relativistic regime?
    \item On numerical side, important efforts are undertaken to push PIC simulations to the largest scales and longest time benefiting from modern computational resources. While largely needed, gaining one order of magnitude in system size and simulation time becomes rapidly prohibitive even with the largest available supercomputers. In this respect, hybrid approaches such as MHD-PIC are promising when one is mainly interested in dynamics of supra-thermal particles (see Sect.~\ref{S:PICMHD}). Yet, this requires to robustly prescribe how some part of thermal particles (simulated using the fluid approach) are promoted to non-thermal status (simulated using PIC or Vlasov approach). For example, in the shock problem one prescribes a fixed fraction of shock-processed particles to be injected into the non-thermal pool \citep{2015ApJ...809...55B, 2018MNRAS.473.3394V} but more accurate parametrisation is required when the shock structure becomes modified by non-thermal particles.
\end{enumerate}

In conclusion, PIC simulations provide detailed non-linear solutions of the shock problem. Therefore, they are an efficient tool to probe the efficiency of particle acceleration for a given parameter set. They are, however, of limited duration (a couple of ion gyro-periods for full-PIC and $\sim 10^3$ ion gyro-periods for hybrid-PIC) and box sizes are typically less than thousands of ion skin depths. Even with the most powerful current computational facilities, simulating the global astrophysical systems is unachievable with this approach. The goal and common approach is to provide robust scalings which can be included in fluid simulations as sub-grid prescriptions. 

\subsection{Kinetics of magnetic reconnection}
\label{Sec:Kinetic_Reconnection}

There is a vast literature on magnetic reconnection, both for the
collisional case based on resistive MHD and the non-collisonal case
based on the Maxwell-Vlasov equations. Nevertheless, many questions
are still far from understood, including {\it `what triggers
  reconnection events in real astrophysical objects'}, {\it `what are
  the physical processes which accelerate particles to super-high
  Lorentz-factors'}, {\it `do associated energy spectra always show
  power-law slope and what is the spectral index'}, {\it `can ions be
  accelerated to equal energies as electrons'}, {\it `is there an
  upper limit for the Lorentz-factors that can be achieved and which
  process sets this upper limit'}\,?  A deeper understanding of these
questions will definitely help to answer relevant astrophysical
questions: 1) To what degree is magnetic reconnection important for
the dynamics of large scale flows like the launching of jets from
compact objects or driven shock waves? 2) To what degree is
REC responsible for the production of thermal and non-thermal
high-energy photons observed from the Sun to AGNs? 3) To what degree
can REC accelerate ions to relativistic speeds and can thus
contribute to the cosmic ray flux and the hadronic channel of
emissivity of photons and energetic neutrinos?

As the literature is vast there is no chance to refer to all
papers. In a hopefully not too biased view, basic ideas are thus
presented on the basis of selected papers in 4 subsections. 1)
  What kinetic simulations can achieve and why we decisively need
  them, 2) Some key results based on kinetic simulations, 3) The most prominently discussed physical processes able to
  accelerate particles, 4) A critical discussion and outlook.
For further important points, which are not discussed due to lack of
space, we refer to the reviews given at the beginning of
Sect.~\ref{S:REC}.

\subsubsection{What kinetic simulations can achieve and why we decisively need them}
 
Microphysical studies have the great advantage that they rely only on
fundamental physics. Difficult questions -- like which equations of
the MHD family best model reconnection events and which values of the
transport coefficients (resistivity, viscosity, Hall-parameters, or
even higher moments) are most appropriate -- can be omitted. All this
comes self-consistently from the kinetic physics solved, described by
Vlasov--Maxwell equations given in Sect.~\ref{S:VMAX}. The numerical
method most widely used in astrophysics is the PIC method described in
Sect.~\ref{S:PICCodes}, though solvers for the Vlasov
equations in the 6+1 dimensional phase space
start to appear.

The price to pay is that the computational costs to solve kinetic
equations are much higher than those to solve any MHD model -- even
complex ones like the popular 10-moment closure model, see for instance \citet{2015PhPl...22a2108W} or \citet{2018FrP.....6..113G}. With kinetic
simulations, even huge ones, only local aspects can be addressed, on
spatial scales which measure at most some thousand electron inertial
lengths and thus up to about hundred proton inertial lengths. This is
sufficient to study a single X-point or even a small current sheet
breaking up into a plasmoid chain. It is, however, not sufficient to
answer questions like how these reconnection sites are embedded in the
large scale environment and how they have been formed. One way out,
which is included in the discussion below, is to apply some splitting
between MHD models and localized particle models or to post-process
MHD solutions by propagation of test-particles. In the future, more
numerical codes will probably be available which dynamically couple
MHD and particle dynamics. Some references to such codes are given at
the end of Sect.~\ref{S:REC}.

Another limitation of kinetic simulations is that they do
not include photons and particle physics. Such aspects are ultimately
relevant for high-energy plasma processes. Power law slopes and
cutoffs may change when considering energy losses by the emission of
synchrotron radiation or inverse Compton scattering of electrons on
colder photons originating from the reconnection site itself or from
other, external processes, possibly far away from the reconnection
site. In relativistic REC, where the energy involved exceeds
the rest masses of electrons (and maybe even of protons), the building
of electron-positron-pairs is very likely to take place, which
significantly back-react on the reconnection dynamics. The same may be
true if accelerated protons can create pions and higher hadronic
resonances and neutrinos. Attempts to account for radiative losses
have recently been made, but are only at the very beginning.

Despite these limitations, kinetic simulations have brought, in about
the last 10 to 15~years, an immense progress in our understanding of
both, magnetic reconnection and shock waves, as well as of associated
particle acceleration. This progress could not have been achieved on
the basis of pure MHD simulations.

\begin{figure}[htb]
\centerline{\includegraphics[width=1.\linewidth]{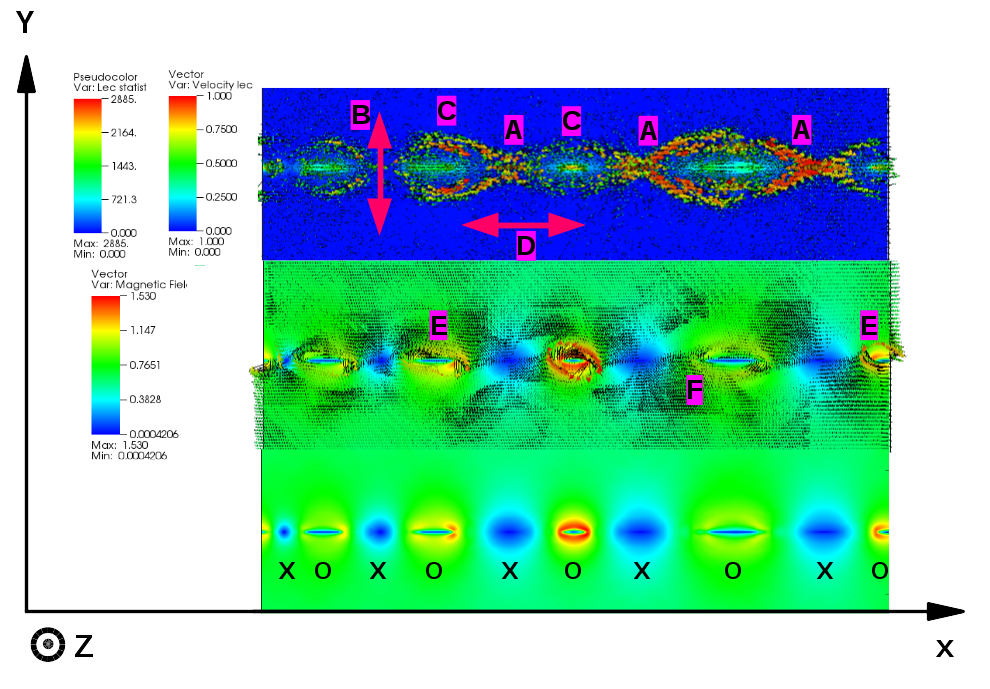}}
\caption{Acceleration sites in a Harris-sheet broken into a chain of
  plasmoids and X-points (from a un-published PIC simulation by
  D.~Folini and R.~Walder). Shown are accelerated electrons on top of
  the electron number density (top panel), magnetic field orientation
  (middle panel), and magnetic field strength (bottom panel). Possible
  acceleration sites are indicated by different labels: A)
  Reconnecting electric field. B) First order Fermi-mechanism between
  the converging inflows to the reconnection site. C) Contracting
  plasmoids. D) Merging plasmoids. E) Drifts in inhomogeneous magnetic
  fields. F) Turbulence.
  \label{Fig:Reconnection_Acceleration_Mechanism}}
\end{figure}

\paragraph{Simulation setup:}

Most kinetic simulations of magnetic reconnection have been performed
in two spatial dimensions or in simple prolongations into the third
dimension (but see the paragraph on dimensionality) of the setup
sketched below. In two space-dimensions, there are known two analytical,
stationary, though unstable configuration of current sheets for
the Vlasov--Maxwell equations. These solutions are typically used to
setup a kinetic simulation to study magnetic reconnection.

1) {\it Harris-sheet:} the reconnecting magnetic field $B_0$ is
oppositely oriented along one axis (say the x-axis in
Fig.~\ref{Fig:Reconnection_Acceleration_Mechanism}). Normal to it
(along the y-axis), the field strength varies as $\mathbf{B}_{\rm x,
  rec}(y) = \hat{x} B_0 \tanh{(y/\delta)}$. The thickness of the
sheet, $\delta$, typically is a few electron inertial lengths and
thus, for realistic mass ratios between electrons and ions, less than
a proton inertial length. The plasma density varies as $n(y) = n(0)/
\cosh^2(y/\delta)$ along the y-axis. Together with an appropriate
temperature, the thermal pressure within the sheet balances the
magnetic pressure from outside the sheet. The induced current then
points into the normal direction and writes $J(x) = c B_0 / 4 \pi
\delta\, \mbox{sech}^2(x/ \delta) \hat{z}$.

2) {\it Force free equilibrium:} a basically similar equilibrium can
be obtained using the force-free equations (see
e.g., \citet{2015ApJ...806..167G, doi:10.1063/1.4997703} for details).

The parameters of both configurations are symmetric with respect to
the center of the sheet, except that the direction of the magnetic
field is reversed. Some authors use for their setup many such current
sheets aligned in parallel, (see \citet{2000mrp..book.....B,
  2010ApJ...709..963D, 2011ApJ...735..102K, 2013ApJ...774...41K,
  2018MNRAS.473.4840W}). This can be quite natural, for instance
within the frame of striped wind models of the solar corona
\citep{2014ApJ...793....7L} or at the termination shock of the solar
wind at the heliopause \citep{2010ApJ...709..963D} or in neutron star
nebulae \citep{2007A&A...473..683P, 2009ASSL..357..421K,
  2012CS&D....5a4014S, 2014ApJ...780....3U, 2016MNRAS.457.2401C,
  2017A&A...607A.134C}.

An extension, and a step towards more realistic models, is to use
asymmetric initial configurations \citep{2007PhPl...14j2114C,
  2012PhPl...19b2108B, 2013PhPl...20f1210H, 2013PhPl...20b2902A,
  2013PhPl...20f1204P, 2013PPCF...55l4001E}. Within this review we do
not further elaborate on this situation. We concentrate on the 2D
Harris-sheet instead, what can be learned from it and where the
limitations of this toy setup lie.

If the thickness of such current sheets is less than a typical field
diffusion length, they are linearly unstable to tearing
modes. Eventually, X-points will develop where REC will
start. In simulations, one often uses overpressured inflows to
accelerate the development of the instability. As described below,
each X-point will develop its exhausts. Exhausts of different X-points
may collide to form magnetic islands (also called plasmoids or
O-points), see
Fig.~\ref{Fig:Reconnection_Acceleration_Mechanism}. Waves are
generated and with them turbulence develops within the sheet and
outside of it, on both its sides in the diffusion regions. These
ingredients turn out to be important drivers of non-thermal particle
acceleration.

Some authors trigger one single X-point in their simulations, from
which huge exhausts develop. Within the exhausts, secondary X-points
and islands can develop. Many features described below are valid for
both, the triggered and the un-triggered approach. Both approaches
develop a power-law spectrum of accelerated non-thermal particles
(Figs.~\ref{Fig:Reconnection_Results_melzani},
\ref{Eq:Reconnection_Results_WERNER},
\ref{Fig:Reconnection_Results_Ball}). Differences between the two
approaches and whether one approach is more close to a reconnection
event in astrophysics are not yet worked out in detail (but see some
points discussed below).

Different boundaries can be used and these will be discussed when presenting
simulations and associated results below.

\paragraph{The parameters governing the physics:}
It turns out that the evolution of the sheet and the acceleration of
particles depend on the strength of the reconnecting magnetic field,
the particle density, and the field temperatures of electrons and
ions. Often, the magnetization is given as the ratio $\sigma_{\rm s}$ of
magnetic energy density to enthalpy density, where $s$ stands for the
particle species, either electrons or ions, $s=e, i$. To account for
temperature effects, one may define $\beta_{\rm s}$ as the ratio of particle
thermal pressure and magnetic pressure. Other, equivalent,
characterizations used are $\sigma_{\rm hot, s}$, the ratio of the energy
flux in the reconnecting magnetic field to that in the particles for
each species (thermal and bulk, and particle rest mass) and
$\sigma_{\rm cold, s}$, which does not consider temperature and bulk flow.

A particular parameter is the background field. The term refers
to the field component which is not going to reconnect, i.e., a
component which encloses a certain angle, $\theta_{\rm B}$, with the
inverted field components that drive the reconnection process. If a
background field is present, the reconnection physics is going to
alter drastically.

As discussed in Sect.~\ref{S:REC}, the physics of collisionless
REC is different for pair plasmas on the one hand and for an
electron-ion plasma on the other hand. The much larger masses of ions
as compared to those of electrons results in their much earlier
de-magnetization when the plasma flows into the reconnection region
(consult~Fig.\ref{fig:Collisionless_Reconnection}), with a series of
consequences \citep{2014A&A...570A.112M}. Relativistic settings also
result in peculiarities \citep{2014A&A...570A.111M,
  2014A&A...570A.112M}. In the ultrarelativistic limit (i.e., when for
both, electrons and ions, the energy largely exceeds the rest-mass),
plasma time- and lengths-scales (cyclotron frequencies, skin depths,
Larmor radii) become independent of the particle rest mass and depend
only on the particle energy. \citet{2016ApJ...818L...9G} showed that,
in this ultra-relativistic limit, a pair and ion/electron plasma
behave essentially similarly.

Regarding PIC simulations themselves, care must be taken that the
number of super-particles per discretization cell is sufficiently high
to ensure that collisionless kinetic processes remain faster than
collisional effects, e.g., for thermalization. For this, the PIC-plasma
parameter, $\Lambda^{PIC}$, the number of superparticles per Debye
sphere, has to be sufficiently large, see \citet{2013A&A...558A.133M}
for a more thorough discussion. In addition, when using an explicit
PIC scheme on a Yee grid (in particular of low order), care must be
taken that energy conservation is sufficiently well guaranteed and
that artificial Cherenkov radiation \citep{2004JCoPh.201..665G} does
not dominate the scene. The problem appears in particular for
particles traveling close to the speed of light, where the
high-frequency waves from artificial Cherenkov radiation actively
interact with the particles and thus may influence the acceleration
process in a non-physical way (see section \ref{PICCodes:cherenkov}).

\subsubsection{Some key results based on kinetic simulations}
\label{Sec:Kinetic_Reconnection_Results}

In this section, selected results are presented regarding the energy
spectra of accelerated particles, the energy partitioning between ions
and electrons, and the overall efficiency of the energy transfer from
the reconnecting magnetic field to the accelerated particles. These
results set the stage for a more detailed view on the physical
mechanisms behind particle acceleration, to be presented in the next
section.

We concentrate on the case of a relativistic electron-ion plasma in an
idealized 2D Harris-sheet set up. The non-relativistic case is
important to understand processes on the Sun, space-weather, but also
technical devices, from plasma thrusters to tokomaks. The case of
relativistic pairs is interesting in pulsar winds and perhaps in
certain regions of a black hole corona -- in every environment where
pair-cascades could develop. However, these two cases will not be
discussed due to lack of space and because many results have been
presented in other reviews (references given at beginning of
Sect.~\ref{S:REC}).

We call REC semi-relativistic when the energy of the magnetic
field exceeds the electron rest mass energy but is smaller than the
ion rest mass energy, otherwise we call it relativistic. Ultra-relativistic
REC terms the situation where the magnetic energy largely
exceeds the rest mass energy of both species, electrons and ions.

\newcommand{\lec}{\mathrm{e}}

\begin{table*}[tbp]
  \label{Tab:Parameters_RelElIonRecon}
\caption{Order of magnitude for physical parameters in astrophysical
  environments. Adapted from~\citet{2014A&A...570A.112M}. }
\begin{tabular}{l|c|c|c|c|c}
 Objects with ion-electron plasmas (with also pairs) & $B$ (G) & $n_\lec$ ($\mathrm{cm^{-3}}$) & $T_\lec$ (K) & $\sigma^\mathrm{cold}_{\lec}$ & $U^\mathrm{R}_\mathrm{A,in}/c$ \\
\hline
 Microquasar coronae, X-ray emitting region $^a$                           & $10^5$-$10^7$       & $10^{13}$-$10^{16}$  & $10^9$         & $10^{-1}$-$10^5$& 0.003-1 \\
 AGN coronae, X-ray emitting region$^b$                        &                     &                      & $10^9$         & 1.7-180         & 0.03-$0.3$ \\
 Giant radio galaxy lobes$^c$                                 & $10^{-6}$-$10^{-5}$ & $3\times10^{-6}$     & $10^6$         & 0.8-80          & 0.02-$0.2$ \\
 Extragalactic jet, $\gamma$-ray emitting region ($<0.05$\,pc)$^d$ & 12             & 80                   &                & $2\times10^5$   & $\sim 1$ \\
 Extragalactic jet, radio emitting region (kpc scales)$^e$     & 1-$3\times10^{-5}$  & 0.8-$5\times10^{-8}$ &                & 500-2500        & $\sim 1$ \\
 GRB jet, at radius of fast reconnection$^f$                   & $7\times10^8$       & $10^{10}$            & $10^8$         & $5\times10^{12}$& $0.9$ \\
 & & & & & \\
 Objects with pair plasmas & $B$ (G) & $n_\lec$ ($\mathrm{cm^{-3}}$) & $T_\lec$ (K) & $\sigma^\mathrm{cold}_{\lec}$ & $V^\mathrm{R}_\mathrm{A,in}/c$ \\
\hline
 At the termination shock of pulsar winds$^g$                  & $10^4$              & 0.1-10               &                & $10^{13}$      & $\sim 1$ \\
 In a pulsar wind nebulae$^h$                                  & $5\times10^{-3}$    & 5 to $10^3$          & $\gamma\sim 10$-$10^9$& $<0.5$    & $0.6$ \\
 \hline
\end{tabular}
$^{a}$\,Analytical disk and corona models, \citet{2005A&A...441..845D, 1998MNRAS.299L..15D, 2001MNRAS.321..549M,2013ApJ...769L...7R};
matching observed spectra with radiation models \citet{2013MNRAS.430..209D,2014A&A...562L...7R}. \\
\noindent
$^{b}$\,Analytical disk and corona models, \citet{1998MNRAS.299L..15D, 2001MNRAS.321..549M,2013ApJ...769L...7R}. \\
\noindent
$^{c}$\,Observations \citet{2004ApJ...604L..77K} \\
\noindent
$^{d}$\,Analytical model assuming $\sigma_{ion}^\mathrm{cold}=100$, \citet{2009MNRAS.395L..29G}, See also \citet{2004ApJ...600..127G} for magnetic field measurements (0.2\,G, but on larger scales). \\
\noindent
$^{e}$\,Observations, \citet{2006ApJ...640..592S}. See also \citet{1992A&A...262...26R}. \\
\noindent
$^{f}$\,Analytical model, \citet{2012MNRAS.419..573M}. Pairs are also present, with $n_\mathrm{pair}\sim10n_\lec$. \\
\noindent
$^{g}$\,Analytical model and observations, \citet{2011MNRAS.410..381B,2011ApJ...726...75S}. \\ 
\noindent
$^{h}$\,Analytical model and observations,
\citet{1996A&AS..120C.453A,2010A&A...523A...2M,2011ApJ...737L..40U,2013ApJ...770..147C}. The plasma
distribution function is a broken power-law with Lorentz factors
$\gamma$ in the indicated range. We note that \citet{2013ApJ...770..147C}
considers only the high-energy electron population, and hence has
larger magnetizations.
\end{table*}
There are only a few papers that have addressed the relativistic
electron-ion regime though it is decisive for our understanding
astrophysical high-energy objects: the dynamics and emission of
accretion disks around and jets from either black holes or neutron
stars (see Table~\ref{Tab:Parameters_RelElIonRecon}).

\begin{figure}
\begin{centering}
\centerline{
  \includegraphics[width=1.05\linewidth]{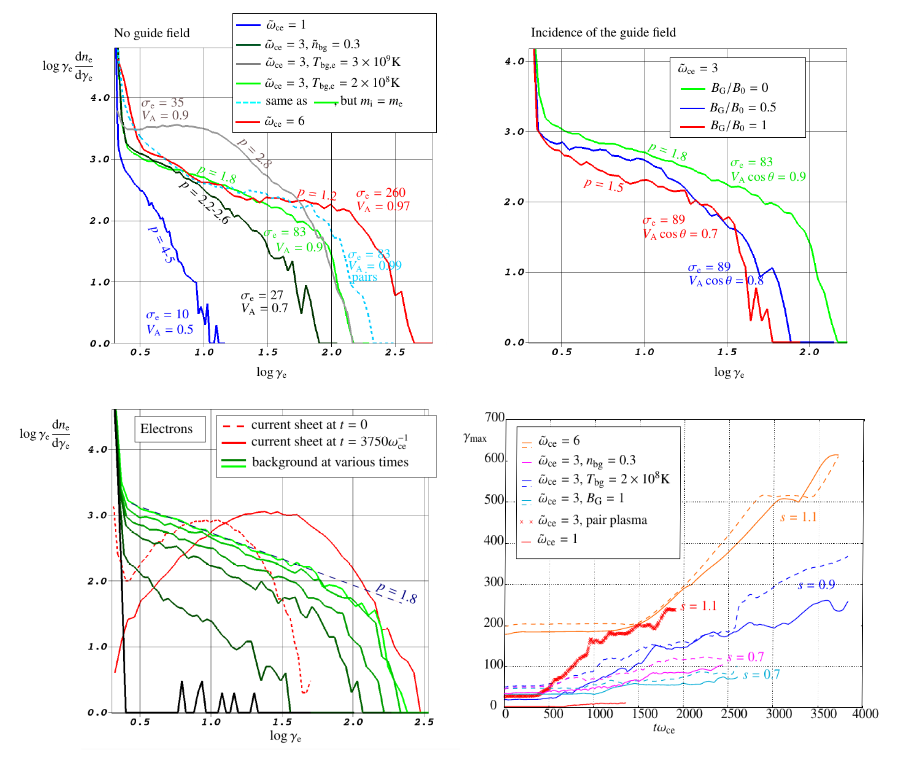}
  }
\par\end{centering}
\caption{Adapted from~\citet{2014A&A...570A.112M}. Upper row: final
  electron spectrum for various simulations, without (left panel) and
  with a guide field (right panel). Lower row, left panel: Lorentz
  factor distributions of electrons of a simulation
  by~\citet{2014A&A...570A.112M}, using $\omega_{\rm ce}/\omega_{\rm pe}=3$
  and a magnetization in the background plasma $\sigma_{\rm hot, i,e} =
  3.6, 83$. Red curves indicate particles originally found in the
  setup current sheet, green curves indicate particles from the plasma
  which flows into the sheet during the reconnection process. For the
  green curves, times are ordered as dark to light green, with values
  0, 750, 1500, 2250, 3000, 3750 $\omega_{\rm ce}^{-1}$, i.e., one curve
  every  $750\,\omega_{\rm ce}^{-1} = 250\, \omega_{\rm pe}^{-1} = 50\,
  \omega_{\rm pi}^{-1} = 30\, \omega_{\rm ci}^{-1}$. The blue dashed line
  indicates the final power-law slope of the background accelerated
  particles. Lower right panel: the time-evolution of the maximum
  Lorentz factor of the background particles for various simulations
  with $m_{\rm i} /m_{\rm e} = 25$ or $1$. Solid lines are for electrons, dashed
  lines for ions and represent $m_{\rm i}/m_{\rm e} \times
  \gamma_{\rm i,max}$. $\tilde{\omega}_{\rm c,e} = \omega_{\rm c,e}/
  \omega_{\rm p,e}$. The index $s$ refers to the power law index $t^{\rm s}$.}
  \label{Fig:Reconnection_Results_melzani}
\end{figure}

\paragraph{\citet{2014A&A...570A.111M} and \citet{2014A&A...570A.112M}} present a
first study of collisonless relativistic magnetic reconnection of an
electron-proton-plasma in two space dimensions. They use reduced mass
ratios $m_{\rm p}/m_{\rm e} = 25$ and $50$ respectively, and started from a
Harris-equilibrium. Periodic boundary conditions normal to the current
sheet and reflecting conditions parallel to the sheet are
used. Tearing instability and subsequent REC develop from
numerical noise. The magnetization varies as $10 \le \sigma_{\rm e} \le
260$ and as $0.4 \le \sigma_{\rm i} \le 14$. Except for one simulation $ 2.8
\cdot 10^{-5} \le \beta_{\rm p,  i} = \beta_{\rm p, e} \le 2\cdot 10^{-3}$. They present
cases without a guide field and with a guide field of $B_{\rm G}= 0.5 B_0$
and $B_{\rm G}= B_0$, where $B_0$ is the reconnecting component of the
field, the component (anti-)parallel to the current sheet.
 
\citet{2014A&A...570A.111M} describe the general structure of the
reconnection process for both, the cases with and without guide
field. The bulk inertia is identified as the main non-ideal process,
which de-magnetizes both, electrons and ions. In the energy flux of
the exhausts the thermal component dominates over the bulk component.
Protons are generally hotter than electrons in the exhausts. The
numerical results correspond to analytical estimates given. They
identified a good measure for the relativistic reconnection rate,
\begin{equation}
  E^* = E_{\rm Rec} / B_0 U^{\rm R}_{\rm in},
  \label{Eq:Relativistic_ReconnectionRate}
\end{equation}
with $U^{\rm R}_{\rm in}$ the relativistic Alfv\'en speed, 
$B_0$ the reconnecting field, and $E_{\rm Rec}$ the reconnection electric
field. This rate varies between $0.14$ and $0.25$, showing higher
values with lower background densities. Generally, the rate is higher
than in the non-relativistic case (for which $0.07 \le E^* \le 0.14$),
which is in line with the findings for pair plasmas, e.g., by
\citet{2007ApJ...670..702Z} ($E^* = 0.2$),
by~\citet{2012ApJ...754L..33C} ($E^* = 0.17$), or by
\citet{2012ApJ...750..129B} ($E^* = 0.19$ and $0.36$). It follows
directly that the reconnection electric field is very large,
$E_{\rm Rec}/B_0 \sim 0.2U^{\rm R}_{\rm A,in} \sim 0.2c$, which turns out to be
important for the acceleration process of particles.

\citet{2014A&A...570A.112M} further elaborate on associated results,
notably on the acceleration of non-thermal particles to
very high Lorentz-factors. The results are summarized in
Fig.~\ref{Fig:Reconnection_Results_melzani}.

REC produces a population of non-thermal particles. This
population shows a power-law spectrum, which gets harder for
increasing magnetization (upper left panel). To first order, the slope
of the power law is independent of the presence of a background field
(upper right panel). There is a clear difference in the spectrum of
accelerated particles (lower left panel), depending on whether the
particles have been present in the initial current sheet ({\it
  population CS}) or whether particles have floated into the sheet
after REC has started ({\it population BG}). Population {\it
  BG} forms a power-law, population {\it CS} is substantially heated
but its spectrum is still Maxwellian. This points to a different
acceleration mechanism of the two populations, see next section.

The lower right panel of Fig.~\ref{Fig:Reconnection_Results_melzani}
shows the temporal evolution of the maximum Lorentz-factor
$\gamma_{\max}(t) \sim t^{\rm s}$, which is close to an exponential
growth. Higher magnetization $\sigma$ in a larger coefficient $s$ and
thus a faster growth of the spectrum to high Lorentz-factors. While,
as said above, non-thermal accelerated particles have the same
spectrum, independently whether a background field is present or not,
the growth time of the power-law cutoff $\gamma_{\max}$ is clearly
slower in the presence of a background field. Also, to first order,
$\gamma_{\max}$ for protons grow slower than electrons by a factor
$m_{\rm i}/m_{\rm e}$.

Finally, while in the thermal exhausts there is much more energy in
the ions than in the electrons, the situation is slightly different
for the accelerated particles. Without a background field, there is
more energy in the ions. But the situation reverses when a background
field is present, when more energy is carried by the electrons. We add
a note of caution to this result as the mass-ratio between electrons
and ions used is either $25$ or $50$. Simulations with a realistic
mass ratio should be undertaken and their results be confronted with
the findings of~\citet{2014A&A...570A.112M}.

\citet{2016ApJ...818L...9G}\, perform PIC simulations of relativistic
electron-ion reconnection (magnetically dominated in their terms)
without a guide field, starting from a force-free equilibrium of the
current sheet. They use different mass-ratios between electrons and
ions, between 1 and 1836 and use different domain sizes and different
inflow temperatures. They explore magnetizations reaching from the
relativistic to the ultra-relativistic case, but cover not really the
semi-relativistic case, in contrast to the other work discussed in
this section. This should be kept in mind when looking at the
following results. They find that for low mass ratios, ions gain slightly (1.1 times)
more energy than electrons while for a real mass ratio, the ions gain
1.5 times more energy than the electrons. All power-law slopes are
hard -- when put into the form $f \propto (\gamma - 1)^{-s}$, $s$ is
between $1$ and $2$ and very close to $1$ for high magnetizations. For
their high magnetizations, the electron power-law slope does not
depend on the mass ratio and is the same for a pure pair plasma, for a
mass ratio of 100 and for a realistic mass ratio. In energy space, the slopes for electrons and ions slightly differ,
e.g., 1.35 for electrons and 1.2 for ions for a simulation with
$\sigma_0 = B_0^2 /(4\,\pi (m_{\rm i} + m_{\rm e}) n c^2 = 100$ ($n$ is the
particle number density) and $m_{\rm i}/m_{\rm e}=100$. However, the momentum
distribution shows $s_{\rm p} = 1.35$ for both species. They argue that this
can only be achieved by a Fermi-like acceleration mechanism. They find a slight dependence of the power laws on the size of the
current sheet, not so much for the all over slope, but for secondary
variations of the slope. Smaller domain sizes show more variations,
with a significant change of the slope for different energies
($\gamma_{\rm e} - 1$). Larger domain sizes show a much more smooth, unique
slope over the entire energy range. Finally, these authors emphasize that all their power-laws show an
exponential cutoff.

\begin{figure}
\begin{centering}
\centerline{
  \includegraphics[width=1.\linewidth]{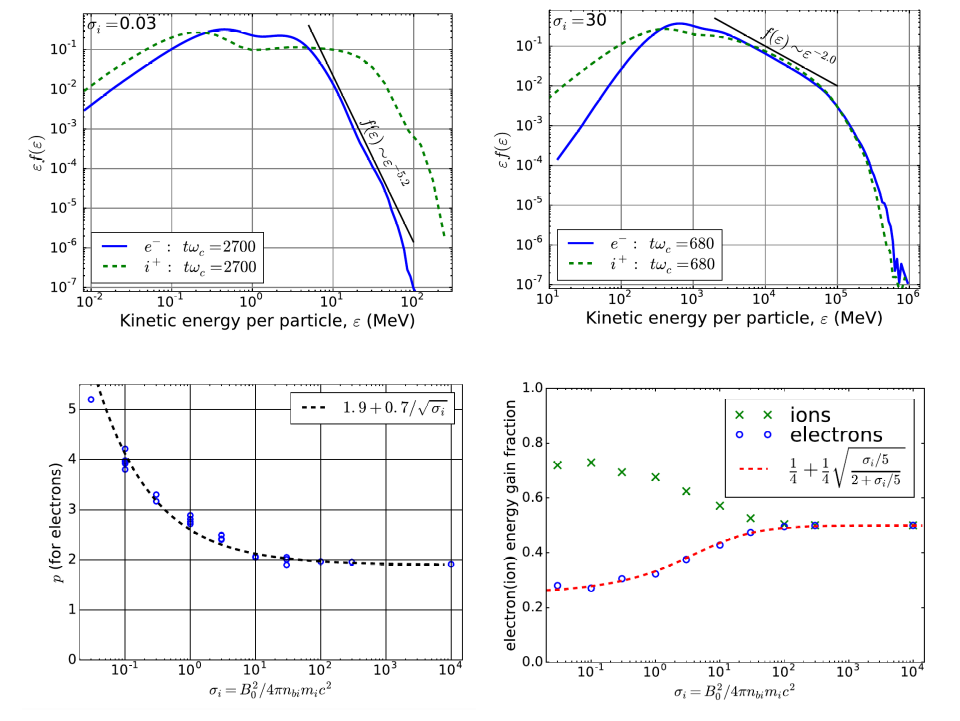}
  }
\par\end{centering}
\caption{Adapted from \citet{2018MNRAS.473.4840W}. Upper row: final
  electron (blue-solid) and proton (green-dotted) spectrum of
  non-thermal particles for a low ($\sigma_{\rm i}=0.03$, left panel) and a
  high magnetization ($\sigma_{\rm i}=30$, right panel). Lower left panel:
  Fit for the electron power law index $p$
  (Eq.~\ref{Eq:Reconnection_WERNER_PowerLawIndex}). Lower right
  panel: final energy partition between background electrons and
  ions at different ionizations $\sigma_{\rm i}$. As expected, there is no
  difference between electrons and ions in the ultra-relativistic
  regime. The fit relates to
  Eq.~(\ref{Eq:Reconnection_WERNER_EnergyRepartionIonElectrons}).}
  \label{Eq:Reconnection_Results_WERNER}
\end{figure}
\citet{2018MNRAS.473.4840W} present an extensive 2D study of
collisionless REC in the semi-relativistic and relativistic
regime, varying the ion magnetization, $\sigma_{\rm i}$, from $10^{-3}$ to
$10^4$. The plasma-beta value of ions was 0.01 for all simulations. A
realistic mass ratio between ions and electrons is used. For their
setup, they use a doubled Harris sheet with two field reversals an
they use period boundary conditions in both directions. The authors analyze the plasma flows in the 'thermal' regime and describe the
Hall signatures due to the different sizes of the electron and ion
diffusion region (see Fig.\ref{fig:Collisionless_Reconnection}). They
find a reconnection rate of about 0.1 of the Alfv\'enic rate (their
slightly different definition of the reconnection rate) across all
regimes, slightly below (0.08) in the semi-relativistic regime and
slightly above (0.12) in the ultra-relativistic regime. In the
ultra-relativistic limit, the release of magnetic energy during
REC is distributed equally between electrons and protons, but
protons gain more in the semi-relativistic regime, up to 75 \% for the
weakest $\sigma_i$ (see Fig.~\ref{Eq:Reconnection_Results_WERNER},
lower right panel).  The authors present a formula for the fractional
energy gain of the electrons, $q_{\rm e}$,
\begin{equation}
  q_{\rm e} = \frac{1}{4} \left( 1 + \sqrt{\frac{\sigma_{\rm i}/5}{1+\sigma_{\rm i}/5}} \right).
  \label{Eq:Reconnection_WERNER_EnergyRepartionIonElectrons}
\end{equation}

Integration is done till $2000$ Larmor time-scales. Particles are
accelerated to a non-thermal regime. Power-laws and energy-cut off
seem to saturate till the end of the simulation. By fitting
their data to a power-law, they find for electrons the
time-saturated relation
\begin{equation}
  f(\epsilon) \sim \epsilon^{-p}; \hspace{1.cm}
  p(\sigma_{\rm i} ) \approx 1.9 + 0.7 \sigma_{\rm i}^{-1/2},
  \label{Eq:Reconnection_WERNER_PowerLawIndex}
\end{equation}
where $\epsilon$ denotes the kinetic energy of the electrons without
their rest-mass (see Fig.~\ref{Eq:Reconnection_Results_WERNER}, lower
left panel). They emphasize that this index can be understood on the
basis of the bouncing of electrons between approaching islands (see
next section). The normalized cutoff energy $\epsilon_{\rm c} /\sigma_{\rm e} m_{\rm e}
c^2$ rises slowly with $\sigma_{\rm i}$ in the semi-relativistic regime,
from around 2.5 to 4 or 4.5 as $\sigma_{\rm i}$ goes from 0.1 to 10. The authors emphasize that it is not yet clear, whether the computed
power-law indices and cut-off energies are truly independent of the
simulation length  $L$ of the current sheet. In the ultra-relativistic regime, $\sigma_i > 10$, the ion spectra
show a power law which closely matches that of the electrons. For
lower $\sigma_i$, the situation is somehow puzzling (see
Fig.~\ref{Eq:Reconnection_Results_WERNER}, upper row). A possible
power law only appears at high ion energies while at lower energies
the spectrum is rather flat and much harder, with a slope of about
1. This part is neither a clear power law nor, as it is much broader,
a Maxwellian. This flat region in the spectrum turns downward
significantly below the electron cut-off energy. However, at these
energies, there are always more ions than electrons. The authors find
no explanation for this behavior but speculate that indeed the power
laws may show a break at energies where protons become
trans-relativistic ($\omega \approx m_{\rm i} c^2 \approx 103$~MeV). We add
that yet another possibility is different dominant acceleration
process for protons depending on energy.

\begin{figure}
\begin{centering}
\centerline{
\includegraphics[width=1.\linewidth]{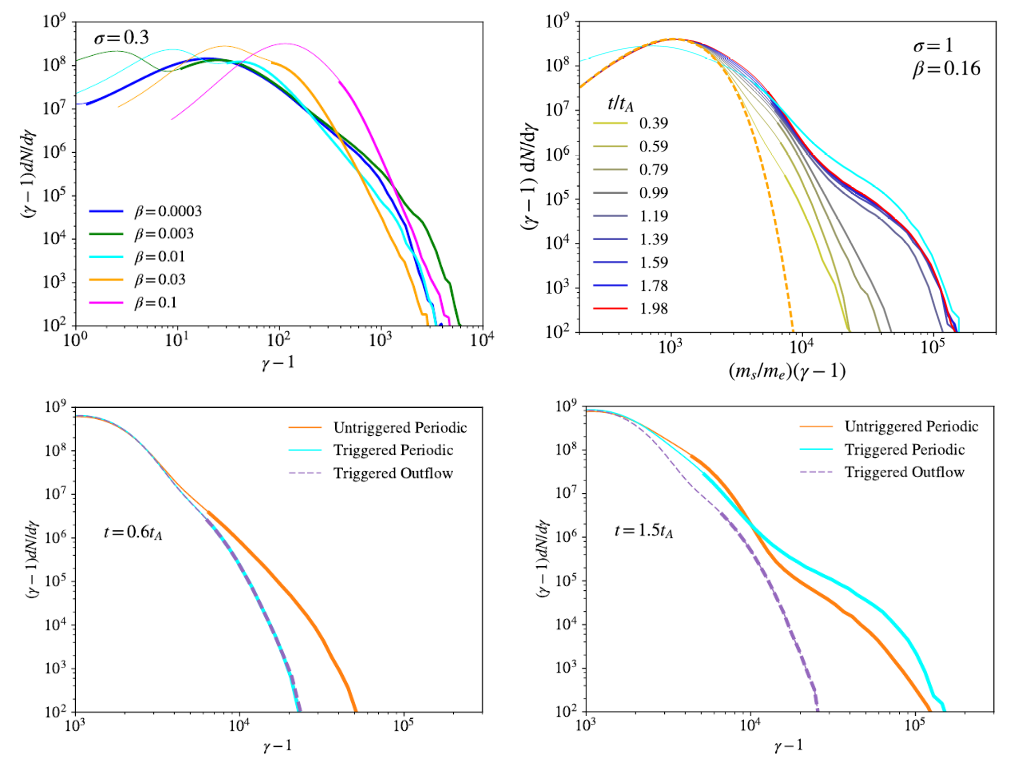}
}
\par\end{centering}
\caption{Adapted from~\citet{2018ApJ...862...80B}.  Upper left
  panel: The $\beta_{\rm p,  i}$-dependence of the final ($t/t_{\rm A} = 2$)
  electron spectra at low magnetization. Upper right panel:
  Time-evolution of the electron spectrum for $\sigma_{\rm i}=1$ and
  $\beta_{\rm p,  i}=0.16$. The spectrum starts to develop an additional
  component after about one $t_{\rm A}$. Lower panels: electron spectra for
  different initial and boundary conditions. For details, see text.}
  \label{Fig:Reconnection_Results_Ball}
\end{figure}
\citet{2018ApJ...862...80B} present another study of collisionless
relativistic REC of an electron-ion plasma, adding new facets
to the picture (see Fig.~\ref{Fig:Reconnection_Results_Ball}).
Firstly, the dependence on the $\beta_{\rm p,  i}$ parameter is systematically
explored, in addition to the dependence on $\sigma_{\rm i}$. They explore
$\sigma_{\rm i} = 0.1,\, 0.3,\, 1$ and $3$, while $\beta_{\rm p,  i}$ varies from
$\beta_{\rm p,  i} = \beta_{\rm p,  e} = 10^{-4}$ up to the maximum possible value of
$\beta_{\max} \approx 1/4\sigma_{\rm i}$. A realistic mass ratio is
used. The computational domains are larger by at least a
factor~5 than those of previous studies. The initial current sheet
width is $\delta = 80 c/\omega_{\rm p,e}$. Periodic boundaries are used
normal to the sheet. A moving injector and a dynamically-enlarging box
is used. A description of the implementation of this boundary type is
given in \citet{2009ApJ...698.1523S}. In this way, the magnetic flux
is not limited to the one present at initial times. Finally, they
trigger the start of REC by removing the pressure at one
point within the sheet's center. Two exhausts with powerful
depolarization fronts develop. As the boundaries are periodic normal
to the sheet, the fronts meet to form a large plasmoid. When the
fronts first meet, a secondary current sheet is created normal to the
first sheet. In addition, they perform a study with different initial
and boundary conditions, see below and
Fig.~\ref{Fig:Reconnection_Results_Ball}, lower panels.  All runs of
this study are integrated till $2~t_{\rm A}$, independently of the box size
and value of other relevant parameters, with $t_{\rm A}$ is given by the
ratio between the length of the sheet and the Alfv\'en speed of the
inflow, $t_{\rm A} = L_{\rm x}/U_{\rm A}$.

The initial current sheet fragments to a certain degree into secondary
plasmoids and secondary X-points. For fixed $\sigma_{\rm i}$, the
fragmentation is less pronounced for higher $\beta_{\rm p,  i}$ and, for fixed
$\beta_{\rm p,  i}$, is more pronounced for higher
$\sigma_{\rm i}$. \citet{2016MNRAS.462...48S} found the same dependences for
the ultra-relativistic case.

\citet{2018ApJ...862...80B} find that the electron spectrum in the
reconnection region is non-thermal and can be modeled as a power-law
with slope $p$ which depends on $\sigma_{\rm i}$ and $\beta_{\rm p,  i}$, as
\begin{eqnarray}
  & & p (\sigma_{\rm i}, \beta_{\rm p,  i}) = A_{\rm p} + B_{\rm p} \tanh(C_{\rm p} \beta_i) \nonumber \\
  & & A_{\rm p} = 1.8 + 0.7/\sqrt{\sigma};  \hspace{0.5cm} B_{\rm p} = 3.7 \sigma_{\rm i}^{-0.19};
  \hspace{0.5cm} C_{\rm p} = 23.4 \sigma_{\rm i}^{0.26}.
  \label{Eq:RelativisticReconnection_PowerlawShape}
\end{eqnarray}
Thus, at low $\beta_{\rm p,  i}$, the slope is (nearly) independent of $\beta_{\rm p,  i}$
and hardens with increasing $\sigma_{\rm i}$, having a (nearly) equal form
as found by~\citet{2018MNRAS.473.4840W}. At higher values of
$\beta_{\rm p,  i}$, the electron power law steepens and the electron spectrum
eventually approaches a Maxwellian distribution for all values of
$\sigma$, see Fig.~\ref{Fig:Reconnection_Results_Ball}, upper left
panel. At values of $\beta_{\rm p,  i}$ near $\beta_{\rm i, max} \approx 1/4\sigma_{\rm i}$, when
both electrons and protons are relativistically hot prior to
REC, the spectra of both species display an additional
component at high energies, containing a few percent of particles, see
Fig.~\ref{Fig:Reconnection_Results_Ball}, upper right panel.

Importantly, when using the same $\sigma_{\rm i}$ and $\beta_{\rm p,  i}$ as
\citet{2018MNRAS.473.4840W}, the non-thermal particle spectra found by
\citet{2018ApJ...862...80B} are systematically softer than the one by
\citet{2018MNRAS.473.4840W}. It is shown in
\citet{2018ApJ...862...80B} that this discrepancy may be caused by two
reasons: firstly, the authors find numerically that a larger box
size makes the spectra generally softer - though there are weak
indications for a certain saturation of the slopes at the two biggest
box-lengths they have used ($L_{\rm x} c/\omega_{\rm p,e} = 5'440$ and $
10'880$). Secondly, also found numerically, runs with a (single)
triggered initial X-point show a softer spectrum than runs in which
the tearing instability produces spontaneously many X-points (as used
by~\citet{2014A&A...570A.112M} and
\citet{2018MNRAS.473.4840W}). Indeed, \citet{2018ApJ...862...80B}
reproduce exactly the same slopes as \citet{2018MNRAS.473.4840W} when
using exactly the same setup and boundary conditions. The lower row of Fig.~\ref{Fig:Reconnection_Results_Ball} summarizes the influence of
the initial and boundary condition for the electron spectrum. Another finding is that protons have up to a magnitude larger mean
energy than electrons, though protons show a steeper slope in their
spectrum than electrons. 

An important result is that for all low $\beta_{\rm p,  i}$, the time-evolution
of non-thermal acceleration is different for electrons and
protons. While electrons immediately develop a non-thermal tail in
their spectrum, protons develop a non-thermal tail only after $t
\approx 0.8 t_{\rm A}$, corresponding approximately to the time when the two
reconnection fronts interact across the periodic boundaries. This may
indicate another acceleration mechanism for the two species.

\paragraph{Result summary:} The last few years have brought
progress in our understanding of the population of the non-thermal
high-energy particles accelerated by relativistic
REC. However, we emphasize again that all these results have
been achieved on the basis of only one particular setting, 2D Harris
or force free sheets. One should always keep in mind that this setting
is a very particular one out of many other, probably more realistic
settings. Under these conditions, the reconnection rate is given by
Eq.~(\ref{Eq:Relativistic_ReconnectionRate}). This rate is about 0.2 and
thus higher than in the non-relativistic case. Two parameters are
responsible for the formation of the power-law slope of the
distribution of non-thermal particles accelerated in the reconnection
event. The 'cold' magnetization $\sigma_{\rm i}$ and the magnetization
$\beta_{\rm p,  i}$, which includes thermal aspects of the plasma in which the
REC takes place. A pretty good expression for this power-law
slope is given by Eq.~(\ref{Eq:RelativisticReconnection_PowerlawShape}).
Note that for cold plasma (low $\beta_{\rm p,  i}$), this expression becomes
fairly independent of $\beta_{\rm p,  i}$ and matches the simpler expression
given by Eq.~(\ref{Eq:Reconnection_WERNER_PowerLawIndex}). Without a
background field -- unless in the highly relativistic limit -- the
ions gain more energy, up to 75\,\% of the released magnetic energy in
the semi-relativistic regime
(see Eq.~(\ref{Eq:Reconnection_WERNER_EnergyRepartionIonElectrons})).

There are, however, some aspects which disturb the picture and point
to the need of additional work. Firstly, the exact shape of the
power-law seems to depend, to a certain degree on the initial and
boundary conditions (Fig.~\ref{Fig:Reconnection_Results_Ball}, lower
row) and on the length of the Harris sheet. Secondly, if electrons and
ions are hot before reconnection, there seems to exist an additional
power-law slope in the electron spectra, at least at late times
(Fig.~\ref{Fig:Reconnection_Results_Ball}, upper right panel). This
may indicate that at least two different acceleration mechanisms are
dominantly at work.

While the questions just addressed can be judged as minor, there
remain two more fundamental largely open questions. One is the role of a
background field (and such a background field is always present in a
real environment). \citet{2014A&A...570A.112M} found indications that
a background field (at least up to $B_{\rm G} = B_0$) does not affect the
power-law slope of the electron distribution, but the spectrum evolves
more slowly than without a background field
(Fig.~\ref{Fig:Reconnection_Results_melzani}, right column panels). In
addition, this study found that in the presence of a background field,
electrons may gain more energy than protons, reversing the ratio found
in simulations without a background field. However, {\it one study is
  no study}, the more as the low mass ratio used for these simulations
may have spoiled the result. The second question is the exact shape
of the proton population distribution. At low $\sigma_{\rm i}$, these
spectra seem to consists of a very flat part at intermediate energies,
before a clear power-law is established at higher energies
(Fig.~\ref{Eq:Reconnection_Results_WERNER}, upper left panel). This
may again point to two different acceleration processes at work. The
slope of the high-energy power-law for protons may be close to the
one of electrons \citep{2014A&A...570A.112M, 2018MNRAS.473.4840W} or
may be steeper \citep{2018ApJ...862...80B}:

\subsubsection{Selected physical processes which accelerate particles}
\label{Sec:Kinetic_Reconnection_AccelerationProcesses}

So far there is no comprehensive picture on which physical process is
responsible for the acceleration of non-thermal particles in magnetic
reconnection -- though several reasonable ideas exist and some
processes could be identified to be active, but possibly together with
other processes. Therefore, at this place, we present those
acceleration channels most discussed in the literature and the
arguments of the authors who advocate them. We attempt a provisionally
ranking in the next section.

\paragraph{Thermal exhausts:} As discussed in Sect.~\ref{S:REC}, REC
at single X-points or on larger current-sheets produces exhausts where
the particles leave the reconnection region (consult
Fig.~\ref{fig:Collisionless_Reconnection}). Strictly speaking, the
flow in the exhaust is not completely thermal. For instance,
temperature is non-isotropic and electrons and ions do not have the
same temperature, see e.g., Fig.~10 of
\citet{2014A&A...570A.111M}. However, the magnitude of the speed in
the exhausts, $u_{out} \sim \sqrt{2}\, u_{\rm A, in}$
(Eq.~\ref{Eq:Reconnection_Speed_Exhaust}), and the temperature in the
exhausts can be understood on the basis of flow conservation laws
between the inflow and the outflow.

Note that in a highly magnetized environment such as relativistic
REC (but not exclusively), the Lorentz-factor of the exhausts
may already be of order of a few, because the Alfv\'en speed of the
inflow may already be very close to the speed of light.

Below, the dynamics of chains of X-points and current sheets are
discussed. There, exhausts of the different sites collide and form
plasmoids (Fig.~\ref{Fig:Reconnection_Acceleration_Mechanism}). But in
nature there are also reconnection sites found with just one
X-point. There, the exhausts can freely expand to form jets. This
mechanism is used to explain jets on all scales, from the solar
atmosphere and corona \citep{2011SSRv..159..357Z}, but also from
compact objects \citep{2005A&A...441..845D}. The more, small
reconnection events and associated exhaust within large scale jets may
be at the origin of fast TeV variability in blazars
\citep{2009MNRAS.395L..29G, 2015MNRAS.449...34K}.

\paragraph{Dynamics of plasmoids and chains of plasmoids:}
We now look at a longer current sheet that break apart and,
consequently, allows a variety of potential particle acceleration
mechanisms -- to be further detailed below -- to act on the plasma.
If the sheet is long enough, many X-points will develop, with
associated exhausts. Exhausts of neighboring X-points collide and form
plasmoids, plasma regions which are bound by a strong circularly
closed magnetic field (consult
Fig.~\ref{Fig:Reconnection_Acceleration_Mechanism}). The formation of
plasmoids ($\rightarrow$ Colliding plasmoids) has the potential to
accelerate particles as the border of two associated exhausts, called
dipolarization fronts, form approaching magnetic mirrors, see
e.g.~\citet{2015JPlPh..81a3209L}.

The field within the plasmoid tends to zero. Thermal plasma within the
plasmoids cannot escape as the strong encircling field deflects
particles immediately back to the interior. As REC goes on
and more of the inflowing plasma is being processed, islands
potentially grow in size and their encircling fields grow in
strength. Plasmoids will also merge and grow in size. After merging,
plasmoids will contract and give rise to an important acceleration
mechanism ($\rightarrow$ Contracting plasmoid). The process of
breaking a current sheet will also generate turbulence, another
important source of particle acceleration ($\rightarrow$ Turbulence).

At this place, the stability of current sheets, the formation and
dynamics of plasmoids cannot be reviewed in detail. This process sets
the stage for the acceleration of ultra-fast particles, but not
directly causes it. For a deeper understanding of the process, the
reader may consult the vast literature on the subject, e.g.
\citet{2007PhPl...14j0703L, 2008PhRvL.100w5001L, 2009PhRvL.103j5004S,
  2012NPGeo..19..251L, 2012NPGeo..19..297K, 2013PhPl...20h2105M,
  2013ApJ...774...41K, 2016PPCF...58a4021L, 2016MNRAS.462...48S,
  2017ApJ...838...91K}.

\paragraph{Contracting plasmoids:}

PIC simulations show that contracting plasmoids (see structure B in
Fig.~\ref{Fig:Reconnection_Acceleration_Mechanism}) can efficiently
accelerate electrons \citep{2006Natur.443..553D} and ions
\citep{2010ApJ...709..963D} to super-Alfv\'enic speeds by a Fermi-like
process. Such particles are injected into magnetic islands from the
exhausts of X-points of a reconnecting current sheet. Particles
approaching the strong magnetic fields encircling magnetic islands
will be turned around by these fields and are captured in this way
within the island, bouncing in the island backwards and forwards or
move along the magnetic field lines. When islands are freshly formed,
the encircling field also lets contract the island, up to a point
where a quasi-stationary equilibrium is reached between the pressure
of the enclosed particles and the tension force of the magnetic
structure. In this way, the particle will gain energy at each
bounce. When more energized, the particles will start to diffuse out
of the island and either escape or, in a more complex situation, may
be kept by another island held together with a stronger field.

Counting on this mechanism, \citet{2010ApJ...709..963D} can explain
the anomalous cosmic ray (ACR) energy spectrum observed by both
voyager missions in the region between the solar wind termination
shock and the heliopause (note that the same idea was also developed
in \citet{2009ApJ...703....8L}).  \citet{2010ApJ...709..963D} firstly
perform large scale MHD simulations of the solar wind which show
stripes of inversed field directions due to the non-alignment of the
magnetic dipole of the Sun with its rotation axis. Current sheets
develop in the region where the field polarity changes.  (This is very
similar to the striped neutron star winds \citep{2004PhRvL..92r1101K,
  2011ApJ...737L..40U, 2012ApJ...746..148C, 2014PhPl...21e6501C,
  2014RPPh...77f6901B}.) The MHD simulations show that stripes and
current sheets are compressed in the passage of the solar wind through
the solar wind termination shock, setting the stage for REC
in the now unstable sheets. The post-shock situation can be well
approximated by multi-layered Harris-sheets.

Subsequent PIC simulations using the plasma-parameters found in the
MHD simulations demonstrate that particles can be accelerated to
several tens Alfv\'enic speeds, forming a power law with a spectral
index of about 1.5. The multi-layered Harris-sheets break off, forming
a network of plasmoids which collide and collapse.  By a thorough
analysis of the data, \citet{2010ApJ...709..963D} show the outstanding
role contracting islands play in the acceleration process.

The model assumes that interstellar ions are pickuped by the solar
wind and advected subsequently through the solar wind termination
shock. Thus the model can also explain similarities in the spectra of
different ion species observed by the voyager missions. The result
achieved demonstrates the power of combining large-scale MHD
simulations with detailed kinetic simulations.

Subsequently, \citet{2011ApJ...735..102K} performed MHD simulations in
two and three space dimensions of the same multi-layered Harris
sheets. A dynamic network of magnetic islands develops, similar to the
one observed in the PIC simulations of
\citet{2010ApJ...709..963D}. Test-particles are injected into this
configuration and integrated with the sixth-order implicit
Runge--Kutta--Gauss method \citep{1994_NumHamProb...S} which conserves
particle energy and momentum. The accelerated test-particles show also
a super-Alfv\'enic distribution though there are differences in the
distribution as compared to \citet{2010ApJ...709..963D}. Contracting
island again play a dominant role in the acceleration process. But the
authors report also on drift acceleration along the magnetic field
discontinuity and the Fermi-process between the converging inflows
(see the corresponding paragraphs within this section). The reasons
behind these differences remain open. The paper extends the
multi-layered Harris sheet configuration to three space dimensions and
repeats the analysis. They find substantial differences in the
acceleration process and stress the need for more comparative studies
between 2D and 3D settings.

\citet{2017PhPl...24f2906M} develop a general framework which includes
compressibility and non-uniform fields to analyze the process of
contracting plasmoids. They derive an expression for the power-law
scaling of the distribution function and for the rate at which the
power-law develops in time. In analogy to the case of acceleration by
the converging inflows, the spectrum gets harder if the compression
increases, which is generally true for Fermi-like processes. The
authors also find that a guide field of order unity suppresses the
development of power-law distributions.

\paragraph{Colliding plasmoids:}

As described, plasmoids eventually collide. Subsequently, they first
grow and then contract, thereby accelerating particles.

There are two other interesting points in colliding
plasmoids. Firstly, at the contact interface between the coalescing
plasmoids secondary current-sheets with secondary REC
develop, mostly normal to the primary sheet. This secondary
REC may support or suppress particle acceleration, its effect
seems, however, to be small, see the discussion in
\citet{2012ApJ...750..129B}. Secondly, even before collision, in the phase they approach each
other, a Fermi-like acceleration process may work as particles are
reflected on the plasmoids and travel back and forth between
neighboring approaching plasmoids \citep[for
  instance]{2015JPlPh..81a3209L}.

\citet{2018MNRAS.473.4840W} use this process to explain the expression
for the power-law index they found
(Eq.~\ref{Eq:Reconnection_WERNER_EnergyRepartionIonElectrons}). The
acceleration by approaching plasmoids is a second order Fermi process
as the movement of the plasmoids can be regarded as stochastic.  The
energy gain per bounce is thus $\Delta \epsilon \simeq (u_{\rm pl}/c)^2
\epsilon$, where $\epsilon$ is the particle energy without the
rest-mass energy and $u_{\rm pl}$ the speed of the plasmoid.  The typical
bounce time can be approximated by $t_{\rm b} \sim \lambda_{\rm pl} /c$. Thus,
the acceleration time-scale is $t_{\rm acc} \equiv \epsilon \Delta t_{\rm b} /
\Delta \epsilon = const \cdot c \lambda_{\rm pl} / u_{\rm pl}^2$. The process
ends when the two plasmoids collide and the particle escapes, at
$t_{esc} \simeq \lambda_{pl} /u_{pl}$, where $\lambda_{pl}$ denotes a
typical distance between plasmoids. By this, the Fermi~II power law
index, $p$, can be written as $p = 1 + t_{\rm acc}/t_{\rm esc} = 1 + const\,
c/u_{\rm pl}$. \citet{2018MNRAS.473.4840W} assume that $u_{\rm pl} \simeq
U_A$. In the regime where $ \sigma_{\rm i} << 1$ (but still $\sigma_{\rm e} >>1$),
$U_{\rm A} \approx c \sigma_{\rm i}^{1/2}$, and thus $p = 1 + C \sigma_{\rm i}^{-1/2}$,
which has the form of
Eq.~(\ref{Eq:Reconnection_WERNER_EnergyRepartionIonElectrons}). We note
that one can repeat the exercise for any other Fermi~II process
taking place within the sheet, in particular also for acceleration due
to turbulence. \citet{2018ApJ...862...80B} account for a variant of this process to
explain the additional component in the power-law at high energies
when the inflow is hot ($\beta_{\rm p,  i}$ close to $\beta_{\rm i, \max}$).

\paragraph{Turbulence:}

As discussed in Sect.~\ref{S:REC}, the break apart of current sheets
by tearing mode instabilities (or a combination of tearing and kink
mode in 3 spatial dimensions) generates turbulence in the region where
REC takes place. This turbulence, in combination with
turbulence in the inflows, is responsible that collisional
REC becomes as fast as observed \citep{2009ApJ...700...63K,
  2012NPGeo..19..251L}. In the frame of test particles, it was shown
that MHD fast and slow compressible modes accelerate energetic
particles through the second order Fermi acceleration. Density
fluctuations in converging flows can enable first order Fermi
acceleration of particles \citep{2012NPGeo..19..297K}.
\citet{2017ApJ...838...91K} provide a thorough analysis of the
statistics of the reconnection-driven MHD
turbulence. \cite{2012SSRv..173..557L} and \citet{2016ASSL..427..409L}
review the development and the character of (self-generated)
turbulence related to magnetic reconnection, in the classic as well as
in the relativistic regime.

\begin{figure}[htbp]
\centerline{\includegraphics[width=0.99\linewidth]{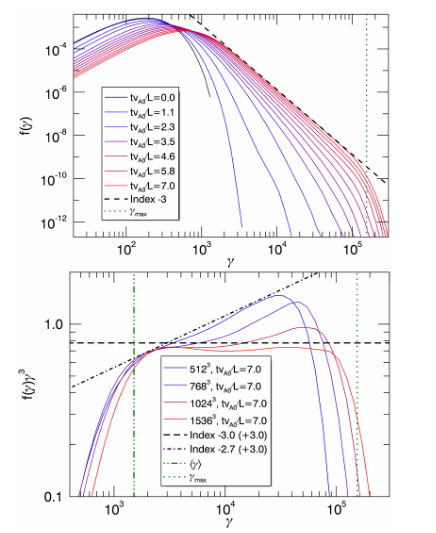}}
\caption{Top panel:
  Evolution of the particle Lorentz factor (energy) distribution $f (\gamma)$ for the
  $1536^3$ simulation of forced collisionless turbulence. The dotted
  line indicates the slope expected for a second order
  Fermi-process. Bottom panel: Compensated distribution $f (\gamma)
  \gamma^3$, at fixed time $tU_{A_0} /L = 7.0$, for varying system
  sizes. Power laws with pre-compensated index $-3.0$ (black dashed) and
  $-2.7$ (black dash-dotted) are also shown, along with the mean energy
  $<\gamma>$ (green dash-dotted) and system-size cutoff $\gamma_{\max}$
  (green dotted) for the $1536^3$ case.
  Images adapted from \citet{2018ApJ...867L..18Z}.}
  \label{Fig:Reconnection_Acceleration_Turb1}
\end{figure}

Corresponding studies for the collisionless case only recently start
to appear. \citet{2018ApJ...867L..18Z} study forced turbulence using
PIC simulations. They show that particles can indeed be accelerated to
very high energy also in the case of collisionless turbulence.
Figure~\ref{Fig:Reconnection_Acceleration_Turb1}, top panel, shows the
time-evolution of the spectrum of accelerated particles, converging to
a power law of the form $\gamma^{-3}$. The lower panel shows the
dependence of the domain resolution, from $512^3$ to $1536^3$ cubed.

In a preprint, \cite{2018arXiv180901966Z} present corresponding
results for an ion-electron plasma and come to the conclusion that
non-thermal acceleration is efficient for both species in the fully
relativistic regime. In the semi-relativistic regime, while still
efficient for cosmic ray production, the mechanism is less efficient
for electrons. For the production of hard non-thermal radiative
signatures a (very) low $\beta_{\rm p,  i}$ is necessary, or nearly relativistic
ion-temperatures. The authors emphasize that the result is still
somehow preliminary as it is not yet established how this result
scales to a large system size. Related to these simulations, the
authors have statistically analyzed collisionless turbulence
\citep{2017PhRvL.118e5103Z} and make a connection to the Fokker--Planck
framework \citep{2019arXiv190103439W}.

A word of caution, equally raised by the authors, must be added. The
forcing of the turbulence implies a steady energy input into the
system. As no energy sinks are present, energy may pile up to give
artificial heating and non-thermal particle acceleration. More studies
are clearly needed. Also, non-global compression was present which
would rise the value for the power-law index
\citep{1978MNRAS.182..147B,2012MNRAS.422.2474D}.

Nevertheless, it can be firmly stated that turbulence is present in
reconnection sites, both, collisional and non-collisional, and that
this turbulence can contribute to particle acceleration. We note these
results are also very promising to advance our understanding of
particle acceleration in shocks.

\paragraph{The reconnecting electric field:} This field is an important source of particle energization in the relativistic case.

As particles are de-magnetized in the current sheet, this field can
directly accelerate particles without perturbation by other
structures. However, the width of the region where the electric field
is strong is not too large, i.e., it is given by $E > cB$, or with a
guide field, by $\mathbf{E} \cdot \mathbf{B} \ne 0$.

In non-relativistic REC, this process cannot be too effective
as the electric field is too weak as compared to the width of the
acceleration zone and the particles escape soon
\citep{2010ApJ...709..963D, 2011ApJ...735..102K, 2012MNRAS.422.2474D}.

In the relativistic case, however, the process becomes important and
the accelerated particles will form a power law. Particles enter the
region with a large electric field at all distances from the central
X-points. The particles close to the X-point will only slowly be
turned, by $B_X$, towards the exhausts, in contrast to those particles
which enter the sheet at some distance from the X-point. The longer
the particle is held in the central region, the longer it can be
accelerated by the reconnection electric field. As emphasized by
\citet{2001ApJ...562L..63Z} the gyro-radius of relativistic particles
grows with their Lorentz-factors. Already accelerated particles thus
stay even longer within the region of a large electric field and hard
tails of the distribution can grow. These authors predict, for a pair
plasma, a power-law distribution of non-thermal particles created by
the reconnection electric field of $p \propto c B_x / E_{Rec}$
\citet{2012ApJ...750..129B} refined the analysis in two aspects, also
for a pair plasma. Firstly, they looked closer at the creation of the
power law distribution but then included also the process when the
particles finally have to leave the region immediately around the
X-point. The latter produces an exponential cut-off such that the
particle distribution function can be written in the form
\begin{equation}
  f(\gamma) \sim \gamma^{-1/4} \exp{-a \sqrt{\gamma}},
  \label{Eq:KineticReconnection_Bessho}
\end{equation}
with $a$ a constant of order $c B_x /E_{\rm Rec}$. Note that this is a very
hard distribution. The authors emphasize that particles accelerated by
the reconnection electric field may be further accelerated by other
processes, in the field when they swirl around contracting islands or
when they encounter other X-points. We add that any other of the
processes described in this section could be tapped. PIC simulations
of relativistic collisionless REC confirm that this process
significantly contributes to the high-energy spectrum, for both cases,
of a electron/positron pair plasma \citep{ 2012ApJ...750..129B,
  2014ApJ...783L..21S} and an ion-electron \citep{2014A&A...570A.112M,
  2018ApJ...862...80B}.

\paragraph{Fermi process within the inflow region:}

Recall that the inverse field components float, due to E\,$\times$\,B-drift,
towards the region where they will reconnect
(Fig.~\ref{fig:ReconnectionTopology_Sweet-Parker}, right panel).

\citet{2005A&A...441..845D} point out that this may, in principle,
accelerate particles by a first order Fermi-process (see
Sect.~\ref{S:DSA-Fermi}), if particles are allowed to cross the
reconnection region and to be ping-ponged between the lower to the
upper inflow into the sheet. Repeating the argumentation given by
\citet{1978MNRAS.182..443B} for the shock case,
\citet{2005A&A...441..845D} derive an emerging spectrum of accelerated
particles of $N(E) \propto E^{-5/2}$ for this acceleration mechanism.

An important point is raised by \citet{2010MNRAS.408L..46G} who states
that this mechanism does not depend on the not well known reconnection
mechanism but works solely on the converging inflows to the
reconnection region. Be $\delta$ the width of the dissipation region
and $R_{\rm G}$ the gyro-radius associated to the undisturbed magnetic field
and $\theta$ the angle under which the particle enters the
reconnection layer. Be this angle moderately small (as compared to the
length of the sheet, $L$) and consider the case where the particle can
freely pass the reconnection region. Then, the particle will gyrate
around the field drifting into the reconnection site and will cross
back the sheet to its other side, where it gyrates back and forth. Of
course, turbulent field fluctuations or collisions may scatter the
particle as well, but their presence is not really necessary. In
addition, the particle is gaining energy (though less) at each passage
in the reconnection electric field. \citet{2010MNRAS.408L..46G} refers to this process 
as the betatron effect.

The energy amplification factor, A, for one cycle can then be derived
to:
\begin{equation}
  A(\theta) = \Gamma_{\rm r}^2 (1 + \beta_{\rm r} \cos \theta)^2,
\end{equation}
where $\theta$ denotes the angle at which the particle enters the
reconnection layer and $\Gamma_{\rm r}, \beta_{\rm r}$ the Lorentz-factor and
normalized speed of the inflow. We note that with $\theta=0$ and
$\beta_{\rm r} <<1$, the formula of \citet{2005A&A...441..845D} is
recovered. The time-scale of the acceleration process is
\begin{equation}
  t_{Acc} = \frac{2 \pi \gamma m c^2}{(1-1/A) e B c}
\end{equation}
and is thus of the order of the gyration period of the largest
$\gamma_{\rm e}$ factor. \citet{2010MNRAS.408L..46G} shows that for an
isotropic particle distribution (and thus $\theta$), the amplification
amplitude, $A$, is only about one quarter smaller as for the case
$\theta=0$.

The process is limited by two factors, the size of the current sheet
and by radiative losses. These are dominated by synchrotron losses
while photo-pion production turn out to be much less important
\citep{1995PhRvL..75..386W}. Assuming jets carrying field-reversed
components of length of the jet diameter, \citet{2010MNRAS.408L..46G}
estimates that protons can reach energies up to $10^{20}$~eV in GRBs
or luminous AGN jets. For iron, an important species in very high
energy cosmic rays, the same limit applies. For electrons, however,
the acceleration process stops much earlier due to synchrotron losses. The author emphasizes that an important pre-requisite
for the process to work is that it can only be initiated if the
particles are sufficiently pre-accelerated, because, otherwise, the
gyro-radius of the particles is not large enough to let the particle
stream over the reconnection region.

\citet{2012MNRAS.422.2474D} points out that the spectrum derived by
\citet{2005A&A...441..845D} is probably much too weak (i.e., the
power-law slope should be much smaller than $5/2$). This, because the
compression of the plasma within the sheet is not considered. The
paper provides arguments that, in the case of REC, the
compression is probably larger than in the shock case, where mostly a
compression ratio of $4$ is assumed, which corresponds to an adiabatic
shock. For REC, spectra like $f(p) \propto p^{-2}$ or $N(E)
\propto E^{-1}$ have to be expected. This was the first indication
that spectra of particles accelerated in magnetic reconnection events
may indeed be very hard, much harder than for particles accelerated by
the Fermi process in shocks.

\citet{2012A&A...542A.125B}, considering compressible effects,
realistic cross B-field diffusion coefficients, and accounting for
synchrotron cooling, shows that for protons a maximum energy of
\begin{equation}
  E_{\max} \approx 60 \left(\chi/k\right)^{1/2} \left(v/c\right)^{1/2} \frac{1}{\sqrt B_0} \ \mathrm{TeV}.
\end{equation}
can be reached. k and $\chi$ depend on the concrete microphysics, with
good arguments for $\chi/k \approx 1/10$. The spectrum of the
accelerated particles is harder than standard Fermi~I.

\citet{2016ApJ...825...55P} analyze possible compression rates of
REC in the solar corona. They look at configurations such as
a Harris current sheet, a force-free current sheet, and two merging
flux ropes. Plasma parameters are taken to be characteristic of the
solar corona. It is found that plasma compression is expected to be
strongest in low-beta plasma $\beta \sim 0.01$--$0.07$ at reconnection
magnetic nulls and can be as high as a factor 10.
\cite{2018ApJ...860..138P} derive magnetic wave solutions in
compressible reconnection sites and claim that such waves may act as
scattering centers for the Drury-mechanism.

In conclusion one may say that this is a promising process whenever
there are pre-accelerated particles present which can initiate the
back- and forth-bouncing.

\paragraph{Drift acceleration:} If the particles enter the diffusion region
or if they leave the current sheet the magnetic field changes
substantially. In this situation, the particles encounter magnetic
drifts and magnetic drift acceleration (see Sect.~\ref{S:SDASSA} for
the description of the equivalent mechanism in
shocks). \citet{2009ApJ...700...63K} and \citet{2016ApJ...818L...9G}
report that this mechanism contributes to the acceleration of
particles, both in the collisional \citep{2009ApJ...700...63K} and
non-collisional \citep{2016ApJ...818L...9G} case.

\paragraph{Associated shocks:}

\cite{2005A&A...441..845D} point to another interesting possibility:
analogous to the sun, eruptive reconnection may push matter away and
produce a shock-wave in which particles can be accelerated by the
'ordinary' Fermi-mechanism related to shocks, leading to a power-law
slope of 2.

\paragraph{Dimensionality:} The role of dimensionality in the particle acceleration
process is largely unexplored. 

2D spatial setups can easily be extended to 3D spatial setups by just
expanding the 2D configuration to the third dimension in a planar
way. Such configurations will still be subject of the tearing mode,
but, in addition, the 3D extension will be subject to the kink
instability. \citet{2014ApJ...782..104C}, using PIC simulations with
radiative feedback, compare the cases where either the kink or the
tearing instability grows faster and discuss application to the Crab
nebula. In their setup they found the kink mode to dominate, leading
to a disruption of the current sheet and associated turbulence unless
the kink-mode is stabilized by a background magnetic field. 
The same result is also found by \citet{2015ApJ...806L..12O} in 3D MHD simulations.
Other authors, using different configurations, found that rather the kink
mode develops slower than the tearing mode even without a guide
field. Here, REC develops essentially similarly as in two
spatial dimension. Oblique modes, a combination of tearing and kink
modes, are possible and may even grow fastest \citep{Daughton2011}.

The results of different 3D kinetic studies of current-sheet
reconnection do not yet converge to a unique picture. Some authors
claim that there are small differences in the reconnection rate
between 2D and 3D, e.g.,  \citep[for non-relativistic
  REC]{2012PhPl...19b2110L, 2014PhPl...21e2307D} and
\citep[for the relativistic case]{2014PhRvL.113o5005G}. On the other
hand \citet{2014ApJ...783L..21S} found a four times lower rate for 3D
REC of a relativistic pair plasma as compared to 2D REC.

In a series of papers it was advocated that the drift kink instability
can modify the electric and magnetic field structures in an
anti-parallel reconnection layer and prohibit  non-thermal acceleration
\citep{2005ApJ...618L.111Z, 2005PhRvL..95i5001Z,
  2007ApJ...670..702Z, 2008ApJ...677..530Z}. But other authors found
that the kink mode cannot suppress the acceleration of particles
\citep{2011PhPl...18e2105L,2014ApJ...783L..21S}

There are very few generic 3D configurations which include 3D nulls.
Based on observations, \citet{2012ApJ...759L...9B} reconstruct a field
configuration within the solar corona and use it as initial condition
for 3D PIC simulations. \citet{PhysRevLett.111.045002} simulate
REC, starting from a cluster of eight null points. Much more
work will be necessary in future to get a more comprehensive picture
of 3D reconnection events.

\subsubsection{A critical discussion and outlook}

The first point to state is that there has been a tremendous progress
in our understanding of magnetic reconnection in the last few
years. Given a correct environment, both collisional and
non-collisional magnetic reconnection proof to be fast. For the
collisional case, both, self-generated and/or external turbulence is
the key-ingredient to make the process fast. The collisionless case
turns out to be always fast. With this progress, we now can understand
qualitatively, and even to a good degree quantitatively the 'thermal
aspects' of REC.

This review has for subject the non-thermal, ultra-energetic
particles. Mostly PIC-simulations have shown that such particles can
originate from reconnection sites. Many questions remain, however. We
want to address three of them, A--C, in the following paragraphs.

\paragraph{A) Open questions concerning small scale kinetic simulations:}

As summarized at the end of
Sect.~\ref{Sec:Kinetic_Reconnection_Results}, we have, on the basis of
2D current sheet simulations some definitive results. However, and as
emphasized also there, many details remain un-answered also for this
case. There remains, in our view, three large questions which
need to be answered before we can state that our understanding even
for this simple case is sufficiently secured.

\begin{enumerate}
\item{There is only one study which includes a guide field (at least
  for ion/electrons). This study has brought interesting results, the
  independence of the spectral slope on the presence and the strength
  of a guide field and that the energy partition between electrons and
  ions may critically depend on the presence of a guide field. Guide
  fields are necessarily present in a current sheet like reconnection
  event in space. Corresponding large parameter studies, which derive
  the relations for ion-electron energy partition, the dependence of
  the power-law slope on the magnetization and the thermal state of
  the inflow when a guide field is present are urgently needed.}
\item{The situation in three spatial dimensions remains unclear, even
  for the most simple case, the extension of a 2D current sheet
  towards the third dimension and a comprehensive study is lacking.

  Beyond this simple case, there are much more complicated
  reconnection topologies present in 3D than the simple extension of a 2D
  current sheet to 3D. An overview of such topologies can be found in
  \citet[Chapter~2]{2007rmfm.book.....B} and
  \citet{2011AdSpR..47.1508P}. None of these topologies have been
  systematically addressed by kinetic simulations.}
\item{Finally, REC is always accompanied with radiative
  emission. In a magnetized environment, synchrotron radiation is
  always present and cool the particles which emit. In addition, these
  photons undergo inverse Compton scattering with the hot particles,
  cooling them again. Many reconnection sites are embedded in a strong
  external cold radiative source, e.g., companion stars in X-ray
  binaries, leading to additional cooling. Radiative effects introduce
  new parameters into REC not discussed here so far. For
  instance, REC in a micro-quasar corona close to the hole
  and in the $\gamma$-ray emitting region of an extragalactic jet
  takes place with the same magnetizations (see
  Table~\ref{Tab:Parameters_RelElIonRecon}). Without considering
  radiative effects, REC in both environments features the
  same reconnection rate, particle spectra, or energy distribution.
  However, the field strength differs by six orders of magnitude and
  thus the effect of synchrotron radiation is largely different.

  There are first attempts to account for radiative emission in
  PIC-simulations. \citet{2018MNRAS.473.4840W} perform a relativistic
  2D Harris sheet study of a pair plasma and describe effects of
  external inverse Compton cooling on the basic dynamics, the
  non-thermal particle acceleration, and radiative signatures. They
  find the reconnection rate and the overall dynamics basically
  unchanged. Important differences are found for the particle
  spectra. They still show a hard power law (index $\ge -2$) as in
  nonradiative REC, but transition to a steeper power law
  that extends to a cooling-dependent cutoff. The steep power-law
  index fluctuates in time between roughly $-3$ and $-5$. Some other
  studies which include radiative losses address mostly REC
  in pulsar winds \citep{2014ApJ...782..104C, 2014PhPl...21e6501C,
    2016MNRAS.457.2401C}. Also the community of the laser-plasma
  facilities starts to study radiative effects with PIC codes as
  newer, more powerful lasers establishes a regime where radiative
  losses essentially co-determine the dynamics, see for instance
  \citet{2015PhPl...22c3117W, 2018PhPl...25i3109G,
    2018PPCF...60e4009B}.}
\end{enumerate}

\paragraph{B) How are particles accelerated?:}

So far, it is not possible to rank the relative importance of the
different acceleration processes discussed in
Sect.~\ref{Sec:Kinetic_Reconnection_AccelerationProcesses}. It seems
likely that different mechanisms are at work even for a simple
setting, but the more if one also considers the prevailing large scale
physical conditions, e.g., hot non-relativistic pair plasma or cold
ultra-relativistic electron proton plasma (see also part~C below).
Most papers mentioned above come to this conclusion. In the same
direction points the finding that the distribution functions of the
non-thermal particles, electrons and protons, often show broken
power-law slopes, with up to three different slopes and an exponential
cut-off. Even if some studies suggest that the shape of the
distribution functions for protons and electrons are close to
identical, other studies shows differences between the shapes of
electron and proton distribution functions. In relativistic
REC, the direct acceleration by the reconnection electric
field is unambiguously identified as one important ingredient -- in
contrast to the non-relativistic case. However, all studies showed
that it is accompanied by some stochastic Fermi-process. Whether this
process operates between magnetic islands or other magnetic structures
or is just a consequence of kinetic turbulence in the diffusion region
and in the current sheet, is not yet clear.

First order Fermi processes can also be present. Such a process
certainly works in contracting islands and seems to accelerate the
population of particles which is originally located in the current
sheet. If particles can cross island-boundaries, this process may
contribute to the acceleration of other particles as well. Whether, or
under what physical conditions, the first order Fermi process between
the converging inflows can work is an open question. In none of the
relativistic simulations it has been observed. The reason may be that
a Harris sheet is a too symmetric constellation where it is very hard
to initiate particle motions normal to the direction of the sheets
which are sufficiently fast to move the particle out of the diffusion
region. However, if one would be able to find a mechanism to
pre-accelerate particles, this mechanism could become
operational. Corrugated sheets, spine-fan or other topologies may
support the initialization of this mechanism. This question remains
open and is part of the next point of discussion.

\paragraph{C) Magnetic reconnection in a large scale environment and in real objects:}

Understanding microphysical processes is decisively important but, on
the other hand, only one part of the game to understand REC
and to what degree REC contributes to the emission of real
objects via both, the leptonic and hadronic channel. And it does not
answer the question whether REC contributes to the cosmic ray
flux. To get the answers to such questions, we definitely need to
combine microphysics with large scale flow.

It is important to understand what triggers a reconnection event and
the nature of the event. Are, in real astrophysical objects,
quasi-stationary configurations present, like Harris-sheets or other
magnetic nulls which eventually get unstable? Or are reconnection
events driven by large-scale flows more important. Related, does
REC start at one point or will a network of reconnection
points establish? All this is largely unexplored.

We therefore advocate simulations of large scale MHD flows, entire
accretion disks and large portions of jets in high-energy
objects. Such MHD simulations must control magnetic diffusion, to be
able to identify reconnection sites and topologies. Ideally, such
large scale simulations are also resistive. Only on the basis of such
MHD simulations we will obtain a good estimate of the spatial and
temporal distribution of reconnection events in such
objects. Subsequent microphysical studies which adapt the
configurations found by the MHD solutions will then allow to model
photon and possible neutrino emission. MHD simulations of magnetic reconnection
will be reviewed in a future version of the review. 



\newpage
\section{Macro scale numerical particle acceleration studies}\label{S:MAC}
At scales comparable with the system size dynamics are often treated in the magnetohydrodynamics (MHD) approximation. In this approximation different methods have been developed to handle the acceleration and propagation of energetic supra-thermal particles, which are listed and described below. Before treating this aspect we present in Sects.~\ref{S:MHD} and \ref{S:MHD-num} the main MHD solvers used in most of modern codes. A discussion concerning relativistic MHD is also included. Detailed monographs and reviews on the subject can be found, e.g., in \citet{1998cmaf.conf....1L, 2015LRCA....1....3M} and references therein. The next sections treat the way CRs can be coupled with MHD. Section \ref{S:MHDM} discusses the multi-fluid approach where energetic particles are treated as a fluid. Section~\ref{S:MHD-KIN} describes the procedure to combine kinetic and HD/MHD methods, in particular the way to treat energetic particles back-reaction over fluid solutions (see Sect.~\ref{S:KIN}). Section~\ref{S:PICMHD} presents the P(MHD)IC method in some details, this method combines simulation techniques exposed in Sect.~\ref{S:PICCodes} to investigate microscale physics but uses the electromagnetic field derived from the MHD equations. Section~\ref{S:SAN} discusses semi-analytical calculations of the problem of DSA, introduced in Sect.~\ref{S:DSA}. 

\subsection{The equations of magnetohydrodynamics}\label{S:MHD}

The use of a magnetohydrodynamic (MHD) approach is crucial for the description of the large scale astrophysical phenomena involving collisional plasma, i.e. if the timescale associated to collision is shorter than the system dynamical time. This circumstance can occur for instance in supernova remnants, stellar bubbles and accretion discs. However, most of astrophysical shocks (including SNR shocks) are non-collisional as previously explained and their structure can not be dynamically described using a MHD model. However, if collective interactions are sufficiently frequent to keep the system isotropic, if electroneutrality can be assumed (which is the case for scales larger than the Debye length), in the cold plasma approximation the MHD equations reduce to the one-fluid system of equation derived in Eqs.~\ref{Eq_sec7_mhd1} below. This system describes long wavelength and low frequency perturbations. 

\subsubsection{Classical magnetohydrodynamics}\label{S:MHD-eq}

 MHD equations couple fluid mechanics equations and Maxwell's equations. They are obtained by averaging the moments of the Boltzmann equation over the velocity space. Particle density conservation is deduced by taking the zeroth order moment of the Boltzmann equation, the momentum conservation equation is obtained by taking the first order moment, and the energy conservation equation is obtained by taking the second order moment. Each of these moment equations introduces a new unknown function: the continuity equation introduces the velocity, the moment conservation equation introduces the pressure and the energy conservation equation introduces the internal energy. Additional assumptions are then required to close the system. To that aim, an equation of state is usually introduced which expresses the internal energy as function of density and pressure, but other assumptions can be used concerning energy fluxes.

The full MHD equations for a single fluid are given in a conservative form as follows:
 \begin{eqnarray}\label{Eq_sec7_mhd1}
 \partial_{t}\rho+\nabla.\left(\rho\,\vec{u}\right) & = & 0, \\
 \partial_{t}\left(\rho\,\vec{u}\right)+\nabla.\left(\rho\vec{u}\vec{u}- \vec{B}\vec{B}+\left(p+\frac{B^2}{2}\right)\vec{I}\right) & = & \vec{0}, \nonumber\\
 \partial_{t}\left(e\right)+\nabla.\left(\left(e+p+\frac{B^2}{2}\right)\,\vec{u}-\left(\vec{B} \cdot \vec{u}\right)\vec{B} \right) & = & 0, \nonumber \\
\partial_{t}\left(\vec{B}\right)-\nabla \times (\vec{u}\times\vec{B}) &=& \vec{0}, \nonumber
 \end{eqnarray}
 where $\rho$ is proper rest mass density, $p$ is the thermal pressure, $\vec{u}$ is the Eulerian fluid velocity vector, $\vec{B}$ is the magnetic field, and $e=\frac{p}{\gamma_{\rm ad}-1}+\frac{\rho u^2}{2}+\frac{B^2}{2}$ is the total energy density, where $\gamma_{\rm ad}$ is the gas adiabatic index.
 The ideal Ohm's law (perfect conductivity) is retained: $\vec{E}=-\vec{u}/c \times \vec{B}$.

These equations are justified for a plasma where the relevant time scales are long in comparison with microscopic particle motion time scale and spatial scales are large in comparison with the thermal ion gyroradius and Debye length.

\subsubsection{Equation of state (EOS)}\label{Sec:mhd_EOS}

The set of partial differential Eqs.~(\ref{Eq_sec7_mhd1}) governing compressible fluid dynamics is incomplete. There are more unknowns than equations and then an additional closure equation is required. Usually, this equation involves the internal energy, the thermal pressure, and the density, and in some cases the temperature as well. This closure equation describes the thermodynamical processes in the fluid. In a real plasma the different particle populations can have different thermodynamical behaviour. For simplicity, the most used closure equation is the polytropic equation of state,
\begin{equation}\label{Eq:Sec7_EOS_1}
\frac{\rm d}{\rm dt}\left(\frac{p}{\rho^{\gamma_{\rm ad}}}\right)=0,
\end{equation}
in an adiabatic gas with one degree of freedom $\gamma_{\rm ad}=5/3$ and in an isothermal gas $\gamma_{\rm ad}=1$.

\subsection{Numerical solutions}\label{S:MHD-num}
The finite volume method (FVM) is widely applied to solve problems described by hyperbolic partial differential equations (PDEs) as is the case for MHD equations. In this method the set of Eqs.~(\ref{Eq_sec7_mhd1}) is written in a shortened conservative form:
\begin{equation}\label{Eq_sec7_short_mhd}
\partial_{t} \vec{U}+\partial_{\rm i}  \vec{F}^{\rm i}(\vec{U}) = \vec{0}\,,
\end{equation}
where $i=(x,y,z)$ are the space variables, the conserved variables are given by
\begin{equation}
\vec{U}=
\begin{bmatrix}
\rho\\
\rho\vec{u}\\
e\\
\vec{B}\\
\end{bmatrix},
\end{equation}
and the associate fluxes are
\begin{equation}
\vec{F}=
\begin{bmatrix}
\rho\vec{u}\\
\rho\vec{u}\vec{u} - \vec{B}\vec{B}+\left(p+\frac{B^2}{2}\right)\vec{I}\\
\left(e+p+\frac{B^2}{2}\right)\,\vec{u}-\left(\vec{B} \cdot \vec{u}\right)\vec{B}\\
\vec{u}\vec{B} - \vec{B}\vec{u}\\
\end{bmatrix}
\end{equation}

\begin{figure}[htb]
\begin{centering}
\includegraphics[width=1\linewidth]{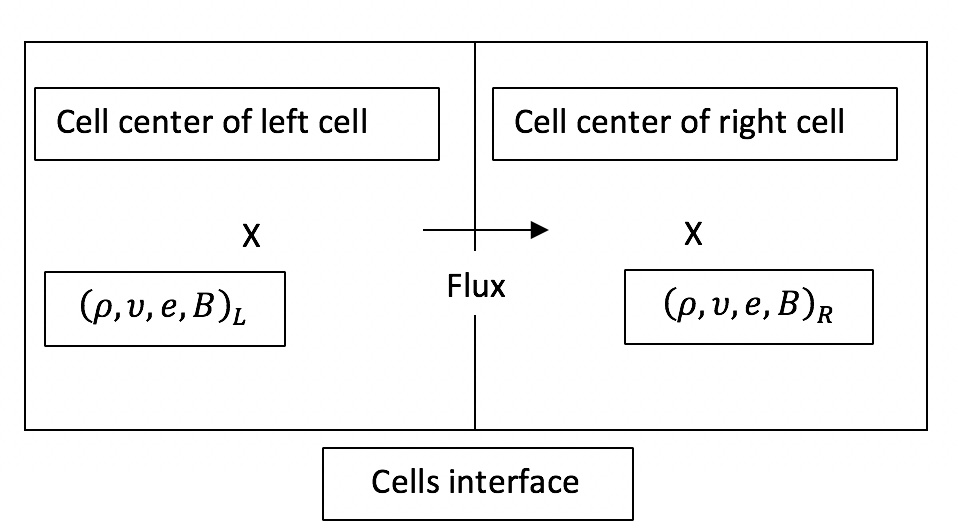}
\end{centering}
\caption{Finite volume scheme with variables set at cell center and flux computed at face center
\label{Fig:S5:Macro_flux}}
\end{figure}

Numerically Eqs.~(\ref{Eq_sec7_short_mhd}) are discretized in space and time as 
\begin{equation}\label{Eq_sec7_mhddiscretization}
\vec{U}^{\rm n+1} = \vec{U}^{\rm n} + \Delta t^{\rm n}\left[\sum_{\rm idim=1}^{\rm idim=ndim}\frac{\vec{F}^{\rm i+\frac{1}{2}}_{\rm idim}-\vec{F}^{\rm i-\frac{1}{2}}_{\rm idim}}{\Delta x_{\rm idim}}\right],
\end{equation}
where $U^{\rm n}$ and $U^{\rm n+1}$ are the conserved variables respectively at time $t^{\rm n}$ and $t^{\rm n+1}=t^{\rm n}+\Delta t^{\rm n}$ where $\Delta t^{\rm n}$ is the time step. The time-averaged fluxes $\vec{F}^{\rm i+\frac{1}{2}}_{\rm idim}$ in the time interval $\left[t^{\rm n},t^{\rm n+1}\right]$ at the interface between a cell with indices~$\rm i$ and its neighbour~$\rm i+1$ in the direction idim are calculated from the solution of Riemann problems\footnote{A Riemann problem is an initial value problem of a conservative equation in fluid dynamics which involves a discontinuous distribution of the conserved variables.  It leads to the derivation of characteristics or the eigenvalues of the problem corresponding to different wave solutions.}. This is illustrated in Fig.~\ref{Fig:S5:Macro_flux}.

In MHD, the Riemann problem is described by a 7-waves pattern. These seven eigenvalues correspond to the left and right going Alfv\`en waves and four magnetosonic waves (two fast and two slow), and between these two propagating waves there is the entropy wave. These waves are defined in a direction $\rm idim$ as
\begin{eqnarray}\label{Eq:S:Marco_Lambda}
\lambda^{\rm idim}_{2,6} &=& u^{\rm idim} \mp U^{\rm idim}_{\rm{a}}\;\nonumber\\
\lambda^{\rm idim}_{1,7} &=& u^{\rm idim} \mp U^{\rm idim}_{\rm{f}}\;\nonumber\\
\lambda^{\rm idim}_{3,5} &=& u^{\rm idim} \mp U^{\rm idim}_{\rm{s}}\;\nonumber\\
\lambda^{\rm idim}_{4}   &=& u^{\rm idim}
\end{eqnarray}
where the Alfv\`en speed $U^{\rm idim}_{\rm{a}}$, the fast $U^{\rm idim}_{\rm{f}}$ and the slow $U^{\rm idim}_{\rm{s}}$ speed in direction ${\rm idim}$ in ideal MHD case are defined as \citep{1995ApJ...452..785R}
\begin{eqnarray}
U^{\rm idim}_{\rm{a}} &=& \frac{ \vert B^{\rm idim}\vert}{\sqrt{\rho}},\nonumber\\
\left(U^{\rm idim}_{\rm{f,s}}\right)^2 &=& \frac{c^2_{\rm s}+U^2_{\rm a}\pm\sqrt{\left(c^2_{\rm s}+U^2_{\rm a}\right)^2-4\,c^2_{\rm s}\,{U^{\rm idim}_{\rm a}}^2}}{2}.
\end{eqnarray}
Here $c_{\rm s}=\sqrt{\left(\frac{\partial p}{\partial \rho}\right)}$ is the local sound speed and $U_{\rm{a}}=\frac{\vert B\vert}{\sqrt{\rho}}$ is the local Alfv\`en speed. We can note here that the characteristic MHD waves are direction dependent. These waves are depicted in Fig.~\ref{Fig:S5:Macro_RiemannFan}. The seven eigenvalues satisfy the inequalities
\begin{equation}
\lambda_{1}\leq \lambda_{2}\leq \lambda_{3}\leq \lambda_{4}\leq \lambda_{5}\leq \lambda_{6}\leq  \lambda_{7}.
\end{equation}
However, some eigenvalues can coincide depending on the direction and strength of the magnetic field. Therefore, the MHD equations form a non-strictly hyperbolic system \citep{1988JCoPh..75..400B}.\\

\subsubsection{The Courant--Friedrichs--Lewy condition}\label{S:CFL}

Courant, Friedrichs and Lewy (\citeyear{1928MatAn.100...32C}) showed that the stability of numerical schemes requires the use of all the information contained in the initial state that will influence the solution in a given spatial cell. To satisfy this condition, the ratio between the spatial discretization $\Delta x$ and the time step $\Delta t $ should be smaller than the largest velocity of the signal solution of the PDEs; i.e. $\max(\mid\lambda_1\mid,\mid\lambda_7\mid)$ (maximum speed propagating to the left and to the right). This inequality is called the CFL condition:
\begin{equation}\label{Eq_Macro_CFL}
\Delta t \leq \frac{\Delta x}{ \max(\mid\lambda_1\mid,\mid\lambda_7\mid)}.
\end{equation}
Satisfying this condition is necessary for the convergence of explicit difference schemes.

\subsubsection{Riemann solvers}\label{S:MHD-RIE}

\begin{figure}[htb]
\begin{centering}
\includegraphics[width=1\linewidth]{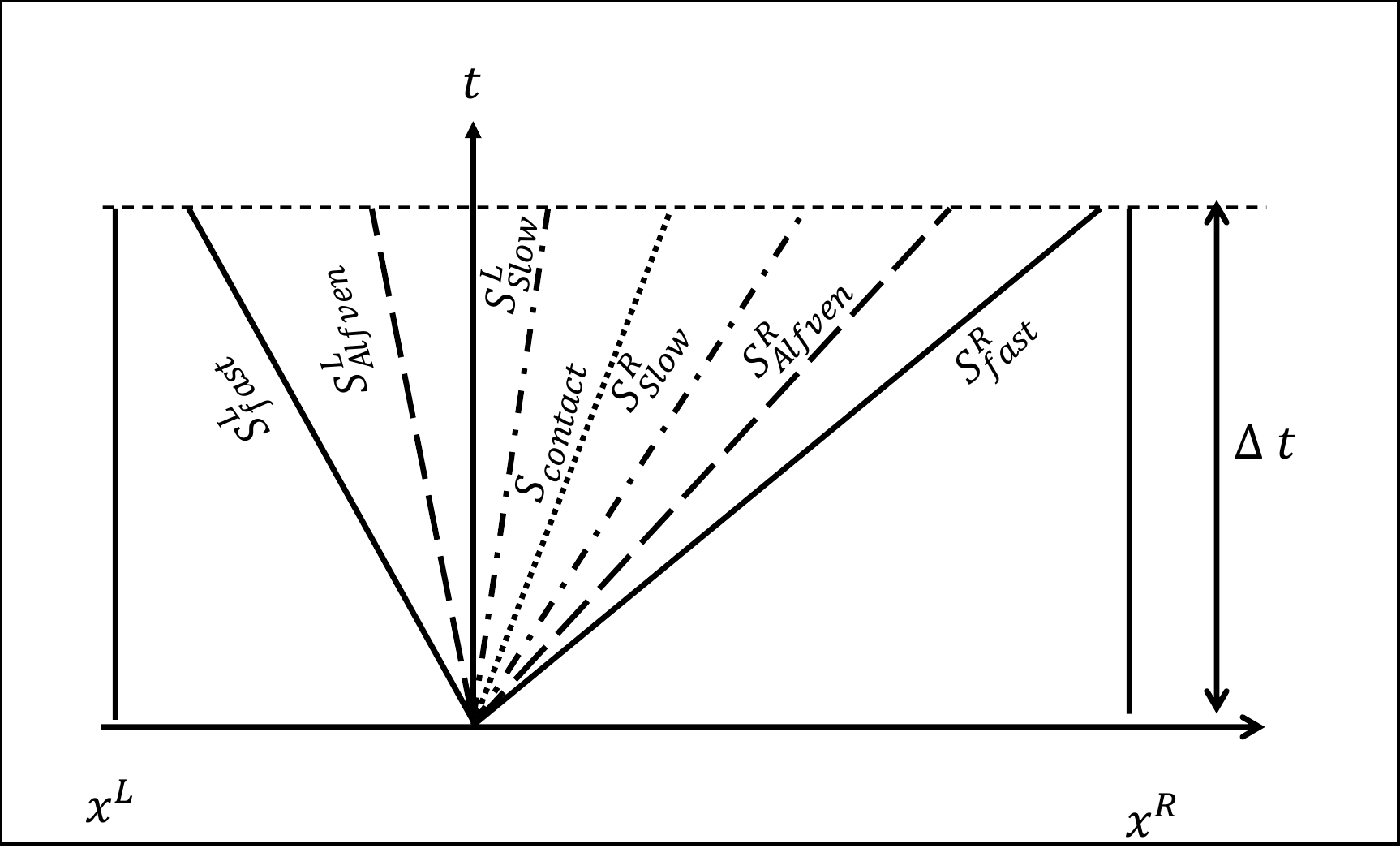}
\end{centering}
\caption{The Riemann fan. See text for the nomenclature of the waves.
\label{Fig:S5:Macro_RiemannFan}}
\end{figure}

The Riemann solvers are a fundamental tool in the development of FVM.  They are based on a simplification of the hyperbolic equations. Moreover, at each step they simplify the  physical state to constant piecewise values with jump discontinuities at some or all eigenvectors (and associated eigenvalues) that characterize the hyperbolic equations. In the case of MHD equations, the solution  of the Riemann problem is controlled by seven waves, either discontinuities or rarefaction fans. Each wave is associated with one eigenvalue (characteristic velocities (Eq.~\ref{Eq:S:Marco_Lambda})), where $\lambda_{1,3,5,7}$ are associated with shock or rarefaction waves, $\lambda_{2,6}$ are associated with rotational discontinuity and $\lambda_{4}$ is associated with a contact discontinuity (see Fig.~\ref{Fig:S5:Macro_RiemannFan}).

There are different types of Riemann problem solvers (see \citealt{1998cmaf.conf....1L, 2015LRCA....1....3M}): 
\begin{itemize}
\item An exact Riemann solver requires an iterative method and thus is impractical for a MHD code (e.g., \citealt{TAKAHASHI&YAMADA_2014JPlPh..80..255T,Torrilhon_2003JPlPh..69..253T}). However, the results of the exact Riemann problem are used as reference solution to check the numerical  precision and  performance of approximate Riemann solvers.
\item A linearized Riemann solver such as the Roe solver \citep{1981JCoPh..43..357R} requires a decomposition of the left and right eigenvectors into characteristics. This solver is complex and time-consuming in MHD and HD cases. Moreover, the Roe solver can lead to negative solutions  \citep{Einfeldt_1991JCoPh..92..273E}. 
\item A guess-based Riemann solver such as the HLL solver (Harten--Lax--van Leer) introduces an estimate of the wave speed and the solution is averaged over the Riemann fan.
\end{itemize}
In this review, we focus on the most used Riemann solvers for MHD, all in the guess-based category. There are three main such solvers:
\begin{itemize}
\item The TVDLF (Total Variation Diminishing Lax--Friedrich Rusanov) Riemann solver 
(e.g., \citealt{BouchutBook2004}) is constructed by assuming a mean state which is given by the fastest wave. Indeed, the flux at cells interface $\rm i+1/2$ is given by,
\begin{equation}
F^{\rm i+1/2}=\frac{1}{2}\left[F\left(U_{\rm L, i+1/2}\right)+F\left(U_{\rm R, i+1/2}\right)- S_{\max}\left(U_{\rm R, i+1/2}-U_{\rm L, i+1/2}\right)\right],
\end{equation}
where the fastest propagating speed at cell interface is:
\[
S_{\max}={\max}\left(S_{\rm L, \max},\,S_{\rm R, \max}\right),
\] with $S_{\rm L, \max},\,S_{\rm R, \max}$ are the eigenvalues associated with the fastest wave propagating respectively at the left and at the right of cell interface $i+1/2$. They are defined as,
\begin{equation}
S_{\rm L \max} =\underset{(1\le k\le 7)}{\max}\left(\mid\lambda_{k, \rm L}\mid\right),\;\;S_{\rm R \max} =\underset{(1\le k\le 7)}{\max}\left(\mid\lambda_{k, \rm R}\mid\right).
\end{equation}
\item The HLL solver proposed by \citet{Hartenetal83} is constructed by assuming an average intermediate state between the fastest and slowest waves.
\begin{equation}
F^{\rm i+1/2}=\begin{cases}
F\left(U_{\rm L, i+1/2}\right)& S_{\rm L, i+1/2}>0\\
F\left(\rm HLL, i+1/2\right)& S_{\rm L, i+1/2}\le 0 \le S_{\rm R, i+1/2}\\
F\left(U_{\rm R, i+1/2}\right)& S_{\rm R, i+1/2}<0
\end{cases}
\end{equation}
where
\begin{eqnarray}
F\left(\rm HLL, i+1/2\right)&=& \nonumber \\ \frac{S_{\rm R, i+1/2}\,U_{\rm R, i+1/2}-S_{\rm L, i+1/2}\,U_{\rm L, i+1/2}+F\left(U_{\rm L, i+1/2}\right)-F\left(U_{\rm R, i+1/2}\right)}{S_{\rm R, i+1/2}-S_{\rm L, i+1/2}} &&
\end{eqnarray}

The outermost wave speed $S_{\rm L, i+1/2}$ and $S_{\rm R, i+1/2}$ are estimated using the left and right states,
\begin{eqnarray}
S_{\rm L} &=& \underset{(1\le k\le 7)}{\min}\left(\lambda_{k, \rm L},\lambda_{k, \rm R}\right)\nonumber \\
S_{\rm R} &=&\underset{(1\le k\le 7)}{\max}\left(\lambda_{k, \rm L},\lambda_{k, \rm R}\right)
\end{eqnarray}
\item The HLLC solver proposed by \cite{1994ShWav...4...25T} is a two-state HLL Riemann solver. It introduces sub-structures associated with a contact discontinuity into the sub-slow state of the HLL Riemann solver.

\item The HLLD solver \citep{2005JCoPh.208..315M} is a four-state HLL Riemann solver. The HLLD Riemann solver introduces sub-structures associated with the two rotational discontinuities $\lambda_{\rm a, L}, \,\lambda_{\rm a, R}$ separated by the contact discontinuity.
\end{itemize}

In the Riemann solvers presented above, the numerical estimation of the fluxes at interface $F^{\rm i+1/2}$ requires the values of the conserved variables to the left $U_{\rm L, i+1/2}$ and to the right $U_{\rm R,i+1/2}$ of the cell interface located at $x_{\rm i+1/2}$. These are reconstructed from cell centred values $U$. These spatial reconstructions can be performed  by using a slope limited scheme to keep the reconstruction monotonic. There are various slope limited schemes such as minmod, super-bee and monotonized central difference limiter (MCD) \citep{toro2013riemann}, ppm \citep{Colella:1984JCoPh..54..174C}, KOREN \citep{Koren:1993LNP...414..110K}, van Leer \citep{vanleer:1979JCoPh..32..101V}. The minmod limiter is the most stable in the presence of strong discontinuities and is very efficient in decreasing numerical instabilities. MCD and Super-bee limiters are more efficient in the vicinity of smooth flows because they permit to retrieve a centered slope. PPM and KOREN are higher order limiters and thus with lower numerical dissipation, they can be used on large classes of problems with smooth flow, intermediate discontinuities and some strong shocks.

Numerical resolution of the partial differential Eqs.~(\ref{Eq_sec7_mhd1}) also requires high order temporal accuracy. This is realized by using a second order predictor-corrector scheme or a higher order scheme such as the strong stability preserving Runge-Kutta scheme.

\subsubsection{Semi-implicit and implicit schemes}\label{S:MHD-sch}
The various complex physical processes  associated with astrophysical plasma phenomena act on widely different time scales. The precise treatment of these phenomena with an explicit MHD approach relies on the accuracy to which one can capture the dynamics as imposed by the CFL condition. In some cases, the resulting short time-resolution can make the MHD simulation computationally intractable with the explicit method. To make these simulations computationally feasible, it is necessary to integrate the MHD equations with larger time steps. This is possible when we are not interested in tracking all fast waves in the system and we can step over some unimportant propagating waves. This can be realized by the use of implicit schemes, which are preferable for complex astrophysical MHD simulations. However, parallel computation with implicit schemes becomes less efficient than with explicit schemes since implicit schemes use iterative algorithms that request more communication between the different processes. 

In the implicit method the original system of differential Eqs.~(\ref{Eq_sec7_short_mhd}) is rewriten as follows:
\begin{equation}\label{Eq:Sec5.2_mhd_implicit0}
\partial_{t} \vec{U}=R(\vec{U}) \,,
\end{equation}
where $R(U)=-\partial_{\rm i}  \vec{F}^{\rm i}(\vec{U}) +S(U)$ and $S(U)$ represents source terms.  $R(U)$ is a non-linear function of $U$.
Eq.~(\ref{Eq:Sec5.2_mhd_implicit0}) is usually discretized in time using a third-level Backward Differentiation Formula method 
\begin{equation}\label{Eq:Sec5.2_mhd_implicit}
U^{n+1}=U^{n}+\Delta {t}\left[\beta {\cal R}(U^{n+1})+\left(1-\beta\right)\frac{U^{n}-U^{n+1}}{\Delta {t^{n-1}}}\right]\,,
\end{equation}
where $\beta=\left(\Delta {t^{n}}+\Delta {t^{n-1}}\right)/\left(2\,\Delta {t^{n}}+\Delta {t^{n-1}}\right)$. In the case of constant time steps $\beta=2/3$ and Eq.~(\ref{Eq:Sec5.2_mhd_implicit}) is simplified to a second-order Backward Differentiation formula; with $\beta=1$ Eq.~(\ref{Eq:Sec5.2_mhd_implicit}) corresponds to a backward-Euler scheme. In the case $\beta=0.5$ the scheme is fully implicit.

In the implicit method, at each time step, the multidimensional nonlinear system
\begin{equation}\label{Eq:Sec5.2_mhd_implicit1}
G(U^{n+1})=U^{n+1}-\left\{U^{n}\Delta {t}\left[\beta {\cal R}(U^{n+1})+\left(1-\beta\right)\frac{U^{n}-U^{n+1}}{\Delta {t^{n-1}}}\right]\right\}=0,
\end{equation}
must be solved to determine the time-updated solution for the state $U^{n+1}$.

In some specific simple cases, such as 2D isothermal hydrodynamics, Eq.~(\ref{Eq:Sec5.2_mhd_implicit1}) can be preconditioned analytically to obtain a diagonal matrix, leading to a system of linear equations that can be solved by relaxation techniques \citep{1968JCoPh...3...80H}. 
The linearization (analytical preconditioning) of Eq.~(\ref{Eq:Sec5.2_mhd_implicit1}) can be done either about the initial equilibrium $G(U^{n})$ or about the current state in order to construct the implicit operator $G(U^{n+1})=0$ needed to advance to the next time step. These classes of implicit schemes are used for specific problems and geometry (e.g., \citealt{1986JCoPh..65...57H}). Another class of implicit schemes uses Newton--Krylov techniques for the resolution of nonlinear systems like Eq.~(\ref{Eq:Sec5.2_mhd_implicit1}). Newton's method consists of the local linearization of Eq.~(\ref{Eq:Sec5.2_mhd_implicit1}) for each state $U^{\rm n\,(m)}$ at iteration $m$ according to
\begin{equation}\label{Eq:Sec5.2_mhd_implicit_newton}
G(U^{\rm n\,(m)})=(U^{\rm n\,(m-1)})+\frac{\partial G(U)}{\partial\,U}\left(U^{\rm n\,(m)}-U^{\rm n\,(m-1)}\right)+{\cal O}(\Delta t^2),
\end{equation}
which can be substituted into Eq.~(\ref{Eq:Sec5.2_mhd_implicit1}) giving the equation to be solved,
\begin{equation}\label{Eq:Sec5.2_mhd_implicit_newton2}
\left(I-\Delta t \,\beta \,\frac{\partial G(U)}{\partial\,U}\right)\delta U=\Delta t \left(\beta U^{\rm n\,(m-1)}+\left(1-\beta\right)\frac{U^{\rm n\,(m-1)}-U^{\rm n\,(m-2)}}{\Delta t^{\rm n\,(m-2)}}\right),
\end{equation}
and after the values of U is updated as
\begin{equation}\label{Eq:Sec5.2_mhd_implicit_newton3}
U^{n+1}=U^{n}+\delta U\ .
\end{equation}
However, this nonlinear resolution method benefits tremendously from accurate initial guesses, $U^{n(0)}$. In many cases an explicit predictor is used to provide an initial guess to the implicit scheme (e.g., \citealt{ReynoldsSW06}). The iterative resolution of the linear Eq.~(\ref{Eq:Sec5.2_mhd_implicit_newton2}) is performed in general by using preconditioned Krylov (sub)space solvers (e.g., \citealt{Toth:2006:PET:1217520.1217543}) with the non-restarted generalized minimum residual method (GMRES) iterative solver \citep{Brown:1989:RSM:2613951.2613987}. 
These linear solvers are very efficient for large-scale problems since they do not require storage of the matrix.

\subsubsection{Magnetic divergence-free algorithms} \label{S:MHD-div}

The  resolution of Euler and Maxwell equations using the standard Godunov schemes does not work by default in maintaining the divergence-free property of the magnetic field $\nabla\cdot \vec{B}=0$. The resulting error accumulated during the simulation may grow to the point that it produces unphysical forces \citep{2000JCoPh.161..605T}. 
Several strategies have been undertaken to handle the  magnetic field evolution in numerical MHD. They are classified into two main categories. 

We first find the divergence-cleaning schemes, where the evolution of magnetic field components is treated as any other MHD variables and only in a second step a divergence-cleaning procedure is applied. In these schemes the magnetic field components are defined at cells center as others variables. Then, the MHD equations are solved by adding source terms function of $\nabla\cdot\vec{B}$. There are various methods for divergence-cleaning schemes, such as the Generalized Lagrange Multiplier (GLM) \citep{2002JCoPh.175..645D}. In the GLM method, a new transport variables $\Psi$ and its governing equation is introduced into the MHD equations system, which plays the role of advection and dissipation of the local divergence error. The  divergence-cleaning can be treated as well by the eight-wave formulation approach \citet{1999JCoPh.154..284P}. There is also the projection method \citep{1980JCoPh..35..426B}. In this scheme at each iteration the Poisson equation $\nabla^2\Phi=-\nabla\cdot \vec{B}$ is solved and in the second step the magnetic divergence part is removed form the magnetic field $\vec{B}=\vec{B}-\nabla{\Phi}$. Let us also mention the vector divergence-cleaning scheme \citep{1998ApJS..116..133B} which is an extension of the projection method. Finally, another divergence-cleaning scheme based on the use of an artificial diffusivity added at each time step, following the completion of the TVD Lax--Friedrich scheme \citep{2007JCoPh.226..925V},  where a term $\eta \nabla\cdot\nabla\cdot\vec{B}$ is added to the magnetic field to diffuse the magnetic divergence.

A second category of schemes is based on constrained transport methods, originally introduced by \cite{1988ApJ...332..659E}. Such schemes use a staggered mesh formulation which is  inherently divergence-free. In this method the magnetic field is defined at face centers and the remaining fluid variables are defined at cell centers. In this approach, the electric field is set along the cell edges. This method sustains a specified discretization of the magnetic field divergence around machine round off error.

 The  constrained transport method is attractive from a physical point of view, however, it requires specific treatment  for magnetic field variables different from others variables, which is inconvenient for implementation specifically in the AMR. The diffusive method reduces the numerical error of $\nabla \cdot \vec{B}$ by adding a source term in the induction equation and the energy equation. The projection method involves an additional Poisson equation which significantly increases the computational cost. The GLM method is based on the use of central cell magnetic field  and thus it can easily be applied on general grids.
 
\subsubsection{Adaptive mesh refinement techniques} \label{S:MHD-AMR}

The numerical resolution of the PDEs (Eq.~\ref{Eq_sec7_mhddiscretization}) 
 uses a discrete domain. Therefore its precision depends on the mesh resolution (spacing), determined according to the scales of the phenomenon under study. In fluid mechanics a broad variety of spatial perturbations exist and can interact with each other. The complexity of these interactions requires the resolution of the problem at all scales. With uniform meshes, if high resolution is required throughout the computational domain the simulation can become computationally extremely costly.
 
 Adaptive Mesh Refinement (AMR) addresses the problem of resolving this wide range of scales by increasing the spatial resolution around small scale structures. It is achieved by increasing locally the mesh resolution and then adjusting the computational effort locally to maintain a uniform level of accuracy throughout the computational domain. This type of AMR approach is called the $h$-type refinement. It consists in the splitting of existing elements into smaller ones. The development and use of the AMR starts with \citet{1984JCoPh..53..484B}, the transition from serial to parallel computing occurred after \citet{Griebel:1999:PMA:329112.329118}. Various AMR approaches exist, depending on the cells shape  and the logical grouping of the cells on the mesh: 
 gathering the cells according to their size $h$, inducing a  particularly strict ansatz in the hierarchy, $h=n^{-\rm L}$, corresponding to some refinement level $L$ (L is an integer). The cells volume at level $l$ are $h^{-{\rm ND}}$ (ND = number of dimensions) smaller than the cells at coarse level $l=1$. The most used hierarchy is $n=2$ where the coarse cell is divided by $2^{\rm ND}$ when it is refined.
 
 In the block structured methods the cells are arranged in blocks according to their levels only. This method does not have any constraints on the size or shape of these blocks. Tree-base methods impose constraints on the block size and shape. With this method, the blocks are organized hierarchically as a quadtree in 2D (octree in 3D). In this distribution, the blocks in use at level~$l$ represent the tree leaves since they have no children and they have an associated parent (ascendant) blocks. The parent blocks that are at the lowest refinement level~$l=1$ represent the tree roots. In some simulations, there is a need to use multiple tree roots arranged in an unstructured AMR, giving rise to a forest of trees.
 
\subsubsection{Errors estimator}

The AMR consists in the use of a coarse grid over the entire computational domain and refined grids only in some specific regions where the local truncation errors are judged to be too large to maintain a given numerical accuracy. These errors can be computed using the  Richardson-estimator which compares the evolution of the variables at two successive levels. This estimator is accurate however it requires a lot of memory and is time consuming. Another estimator is the L\"{o}hner-estimator \citep{Lohner:1987CMAME..61..323L}, a modified  central second derivative normalized by the sum of first-order forward and backward gradients. It has the advantage of using mostly local calculations of any variables of the simulation and their combinations.

\subsubsection{Load balance}

 In the simulation box, the distribution of blocks across the processors requires the use of a space filling curve. The most used space filling curves (SFCs) are Hilbert, Peano, and Sierpinski curves \citep{Bader:2012:SCI:2412036}. The Hilbert and Peano curves use recursive algorithms. The Morton order is also widely used due to its ease of implementation in the space filling. However the curves it generates are  not continuous and thus do not fit into the family of finite SFCs.
 
 With this curve the blocks are organized over a forest of trees and thus they are distributed over all the processors. There are two approaches for this forest partition. In the first approach, each tree and its leaves are associated to one owner process. In the second approach, the tree leaves can belong to multiple processes. The first approach is simple to implement however it does not provide a right load balance since the number of blocks per process may differ. The second approach, even if it presents issues with shared tree between processes, provides a perfect load balance, the difference in block distribution over processes is at most one. This last approach provides the best scalability \citep{Keppens:2012JCoPh.231..718K}. 

\subsubsection{Standard numerical tests}\label{S:TESTMHD}

 The development of numerical tools requests extended tests for all implemented  physics and algorithms. Many comprehensive and well documented sets of tests have been presented in the literature. Tests are set for all physics, all dimensions, and implemented geometries and AMR schemes. For many 1D tests an exact solution exists and it is possible to compute the deviation of the numerical result from it using some error norm (e.g. norm error $L_{1} = \Sigma_{i}\|q_{i}-q_{i}^0\|/N$, where $N$ is number of point, $q_{i}^0$ is the exact solution at cell $i$, $q_{i}^0$ is the numerical result at cell $i$).
 
\begin{figure}[htb]
\begin{centering}
\includegraphics[width=1\linewidth]{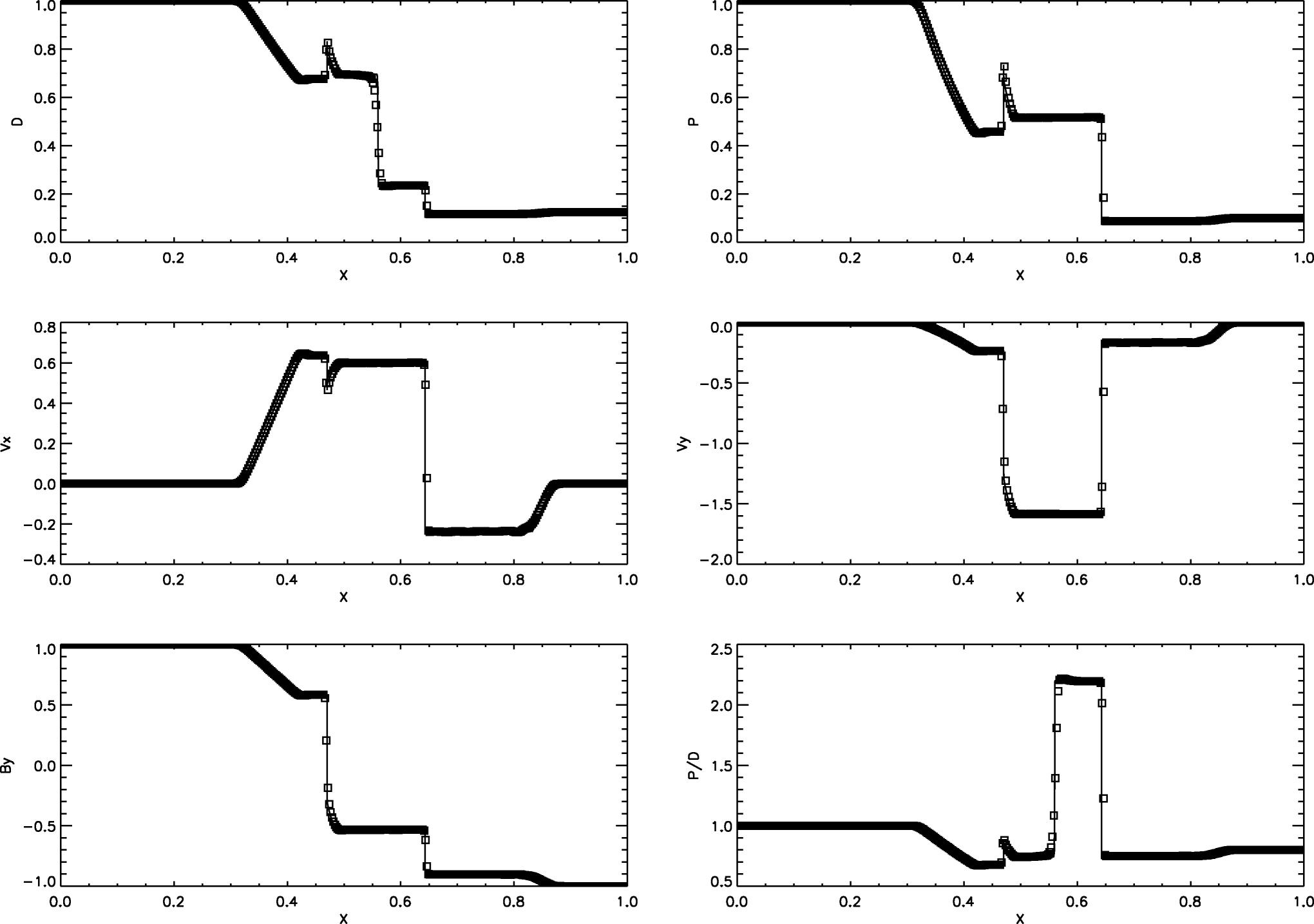}
\end{centering}
\caption{Sod shock tube test in MHD. The density, pressure, velocity components, transverse component of the magnetic field, and specific internal energy (scaled by $\left(\gamma_{\rm ad} -1\right)$) for the Brio \& Wu (\citeyear{1988JCoPh..75..400B}) shock tube problem are plotted at t=0.08, computed with 400 grid points, second-order spatial reconstruction, and Roe fluxes. The solid line is a reference solution computed with 104 grid points \citep{Stone:2008ApJS..178..137S}.
\label{Fig:SOD}}
\end{figure}

One of the standard one-dimensional tests is the Sod MHD shock tube \citep{1988JCoPh..75..400B}. The shock tube test consists of two constant states, one on the left and the second on the right, separated by a discontinuity. This test allows to check the ability of a numerical code to treat correctly the Riemann problem in 1D and resolve the Riemann fan evolution, see Fig.~\ref{Fig:SOD}.
 
\begin{figure}[htb]
\begin{centering}
\includegraphics[width=1\linewidth]{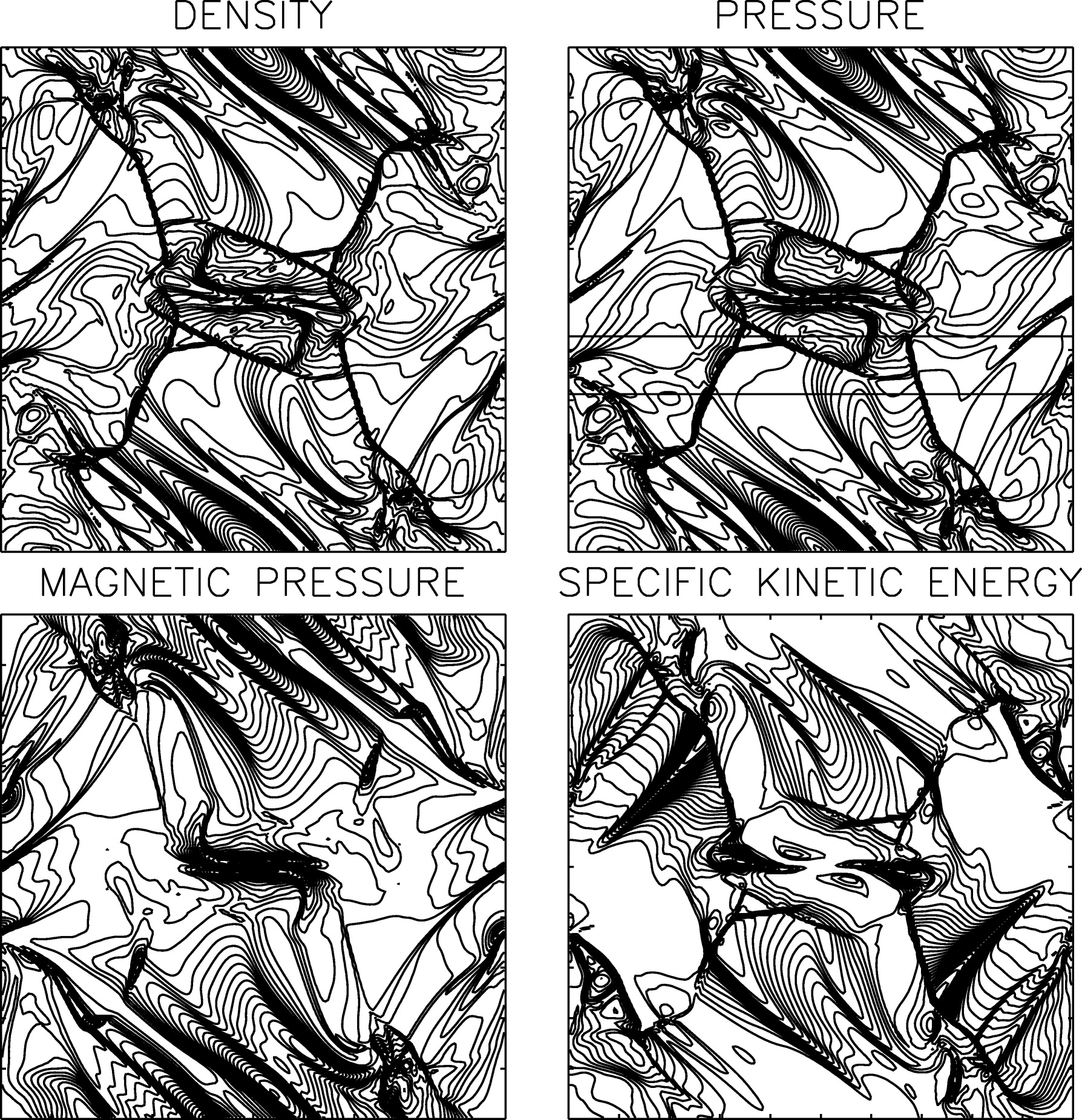}
\end{centering}
\caption{Orszag-Tang vortex test. Contours of selected variables at $t_{f}=1/2$, computed using a grid of 192$\times$ 192 cells, third-order reconstruction, and Roe fluxes. Thirty equally spaced contours between the minimum and maximum are used for each plot. From \citet{Stone:2008ApJS..178..137S}.
\label{Fig:S5:OrszagTang}}
\end{figure}

For two-dimensional MHD tests, one of the standard tests in the vortex of Orszag-Tang (\citeyear{OrszagTang1979JFM....90..129O}), see Fig.~\ref{Fig:S5:OrszagTang}. This test consists in a doubly periodic fluid configuration leading to 2D supersonic MHD turbulence. The density and pressure are set to constant values, while the velocity and magnetic field are set as
\begin{eqnarray}\label{Eq5:OrszagTang}
 \vec{u} &=& \left(-\sin{y},\sin(   x),0\right), \nonumber\\
 \vec{B} &=& \left(-\sin{y},\sin(2\,x),0\right). 
\end{eqnarray}
This test does not have an analytical solution and the results have to be compared between different codes.

\subsubsection{Relativistic magnetohydrodynamics} \label{S:MHD-rel}

The MHD description introduced in the previous paragraphs is relevant for non-relativistic plasma velocities and energies. It is adequate for plasmas in the interstellar medium and in the vicinity of stars. However, some astrophysical phenomena involve relativistic plasma flows with energies of the order of the mass energy. In order to model these plasmas in the framework of the fluid model, the MHD formulation has to be revised using the special relativity (SR) framework. In flat space-time, the plasma is described by the following SR-MHD equations (see \citet{2015LRCA....1....3M})
\begin{eqnarray}
\label{Eq_sec7_srmhd1}
\partial_{\rm t} \left(\gamma \rho\right)+\vec{\nabla}.\left(\rho \Gamma \vec{u}\right) &=& 0, \nonumber\\
\partial_{\rm t}\left(\gamma^2
 h\vec{u}+\vec{E}\times\vec{B}\right)+\vec{\nabla}.\left(\gamma^2  h \vec{u}\vec{u}-\vec{E}\vec{E}-\vec{B}\vec{B}+ p{\vec{I}} \right) &=& \vec{0}, \nonumber \\
 \partial_{t}\left(h\gamma^2-p-\gamma\rho+\frac{E^2+B^2}{2}\right)+\nabla.\left(\left(h\gamma^2-p-\gamma\rho\right)\vec{u}+\vec{E}\times\vec{B}\right) &=& 0, \nonumber \\
\partial_{\rm t}\vec{B}+ \vec{\nabla}.\left(\vec{u} \vec{B}-\vec{B}\vec{u}\right) &=& \vec{0},
\end{eqnarray}
where the closure of this system of equations is provided by the equation of state expressed with the enthalpy $h=h(\rho,p)$. The total pressure is $p_{\rm t} = p+\frac{E^2+B^2}{2}$ where the electric field is given by Ohm's law. In the ideal case $\vec{E}=-\vec{u}/c\times\vec{B}$, finally $\Gamma=(1-(u/c)^2)^{-1/2}$ is the Lorentz factor of the flow.\\
The numerical resolution of  SR-MHD equations exploits the same type of algorithms as presented before, by solving Eqs.~(\ref{Eq_sec7_srmhd1}) in conservative form
 \begin{equation}\label{Eq_sec7_short_srmhd}
 \partial_{\rm t}U+\partial_{\rm i} F^{\rm i}(U)=0,
 \end{equation} 
 where the conserved variables are
 \begin{equation}\label{Eq_sec7_U}
 U=
 \begin{pmatrix}
 \gamma\rho\\
 h\gamma^2 \vec{u}+\vec{E}\times\vec{B}\\
 h\gamma^2-p-\gamma\rho+\frac{E^2+B^2}{2}\\
 \vec{B}\\
 \end{pmatrix},
 \end{equation}
 and the corresponding flux tensor
 \begin{equation}\label{Eq_sec7_Flux}
 F(U)=
 \begin{pmatrix}
 \gamma\rho\vec{u}\\
 h\gamma^2 \vec{u}\vec{u}-\vec{E}\vec{E}-\vec{B}\vec{B}+p \vec{I}\\
 \left(h\gamma^2-p-\gamma\rho\right)\vec{u}+\vec{E}\times\vec{B}\\
\vec{u} \vec{B}-\vec{B}\vec{u}\\
 \end{pmatrix}.
 \end{equation}
Most of the complications in relativistic MHD comes from the non-linear relation between the primitive variables $\left(\rho,\vec{u},p\right)$ and conserved variables $U$:
\begin{eqnarray}\label{Eq:Sec_5.2_CONS_PRIM}
\xi-p-\tau-D+B^2-\frac{1}{2}\left(\left(\frac{B}{\gamma}\right)^2+\left(\frac{\vec{S}\cdot \vec{B}}{\xi}\right)^2\right)=0,\nonumber\\
\vec{u}\cdot\vec{B}=\frac{\vec{S}\cdot\vec{B}}{\xi},\nonumber\\
\frac{1}{\gamma^2}=1-\frac{\left(\vec{S}+\left(\vec{u}\cdot\vec{B}\right)\vec{B}\right)^2}{\left(\xi+B^2\right)^2},\nonumber\\
\vec{u}=\frac{\vec{S}+\left(\vec{u}\cdot\vec{B}\right)\vec{B}}{\xi+B^2},
\end{eqnarray}
where $D=\gamma\rho$ is laboratory frame density, $\xi=\gamma^2\rho\,h$, $\vec{S}=h\gamma^2 \vec{u}+\vec{E}\times\vec{B}$ the momentum and $\tau=h\gamma^2-p-\gamma\rho+\frac{E^2+B^2}{2}$.
Eqs.~(\ref{Eq:Sec_5.2_CONS_PRIM}) can be handled only numerically, the most used scheme is the Newton-Raphson method \citep{2006ApJ...641..626N}. The iteration is performed on the pressure, on the enthalpy or the velocity. In SR hydrodynamics the iterative method can be avoided and instead a quartic equation can be solved \citep{1993JCoPh.105...92S}.

As in classical MHD, the SRMHD schemes exploits characteristic speed of plasma normal modes, but these are limited by the light speed, i.e.
\begin{equation}
-1\leq\lambda_{1}\leq \lambda_{2}\leq \lambda_{3}\leq \lambda_{4}\leq \lambda_{5}\leq \lambda_{6}\leq  \lambda_{7}\leq1.
\end{equation}
For a specific direction $\rm idim$, the entropy wave as in classical MHD travels with speed $\lambda_{4}=v^{\rm idim}$, the Alfv\'en wave has speed 
\begin{equation}
\lambda_{2,6}=u^{\rm idim}\mp\frac{B^{\rm idim}}{\sqrt{(\rho h+B^2)\mp\left(v\cdot B\right)}},
\end{equation}
and the  magneto-acoustic speeds are found from the quartic equation

\begin{eqnarray}\label{Eq:Sec5.2_Lambda_SRMHD}
& &\rho h \left(1-c_{\rm s}\right)\gamma^4\left(\lambda-u^{\rm idim}\right)^4 -\left(1-\lambda^2\right) \times \nonumber\\
&&\left(\gamma^2\left(\rho\,h\,c_{s}^2-B^2\right)\left(\lambda-u^{\rm idim}\right)^2-c_{s}^2\left(\Gamma\left(u\cdot B\right)\left(\lambda-u^{\rm idim}\right)-\frac{B^{\rm idim}}{\gamma}\right)^2\right)=0
\end{eqnarray}
where $c_{\rm s}$ is the sound speed.
In the hydrodynamics case, Eq.~(\ref{Eq:Sec5.2_Lambda_SRMHD}) becomes a single quadratic expression, and then the use of exact analytic formulae for the root evaluation is straightforward. In SRMHD Eq.~(\ref{Eq:Sec5.2_Lambda_SRMHD}) is quartic, although an analytical solution exists \citep{DelZanna2003A&A...400..397D}, it is more easily obtained by numerical iteration, for which a Laguerre method can be used.

\begin{figure}[htbp]
\begin{centering}
\includegraphics[width=\textwidth]{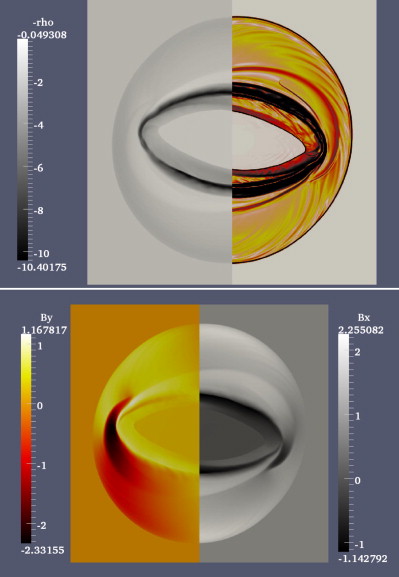}
\end{centering}
\caption{The relativistic rotor problem. The top panel shows the density structure (both in linear scale as well as using a Schlieren plot), the bottom panel the magnetic field components, at time $t = 0.4$. 
From \citet{Keppens:2012JCoPh.231..718K}.
\label{Fig:S5:RotortestSRMHD}}
\end{figure}

\paragraph{Standard numerical tests}
In special relativistic hydrodynamics and magnetohydrodynamics several numerical experiments are used as benchmark tests. Like in classical HD and MHD there is the Sod shock tube and the equivalent of the Orzag Tang vortex (see above). Here we show the rotor test \citep{DelZanna2003A&A...400..397D}, in Fig.~\ref{Fig:S5:RotortestSRMHD}. It consists in a disk of radius 0.1 with higher density, $\rho=10$, positioned at the center of the computational domain $ \left[0, 1\right] \times \left[ 0, 1\right] $, rotating at high relativistic speed, $\Omega=9.95$, thus the Lorentz factor at the disk edge $\Gamma_{\max}\simeq 10.0$, the rotor is embedded in a static background with $\rho=1$, $p=1$, and uniform magnetic field $B_x=1$ with polytropic index $\gamma=5/3$.

\subsection{Multi-fluid methods}\label{S:MHDM}

The previous section presented the case of a single fluid. However, in astrophysics the thermal plasma pressure is usually in competition with the pressure in other components: cosmic rays, radiation fields, dust, neutral species, or requires to include an electron fluid. In bi- or multi-fluid models it is in principle necessary to add as many fluid equations as the number of species to be included in the simulations, and to account for the friction forces induced by collisions among the species in the momentum conservation equation (see, e.g., \citealt{2007MNRAS.376.1648O}). 

As in this review we are mostly interested in non-thermal particle acceleration and transport we restrict our discussion to the case of the interaction of a thermal fluid and a non-thermal (or cosmic ray) component \citep{1981ApJ...248..344D, 2008ApJ...685..105R}. 

\subsubsection{Model equations}
CRs impact fluid dynamics through the effect of the gradient of their pressure $p_{\rm CR}$ (Eq.~\ref{eq:CR-pressure}). The momentum equation is modified as
\begin{equation}
  \partial_{t}\left(\rho\,\vec{u}\right)+\vec{\nabla}.\left(\rho\vec{u}\vec{u}- \vec{B}\vec{B}+\left(p+p_{\rm CR}+\frac{B^2}{2}\right)\vec{I}\right)=0 \ ,
\end{equation}
and the energy equation now includes the work of this force plus another component produced by the CR diffusion
\begin{equation}\label{Eq:CRMHDE}
   \partial_{t}\left(e\right)+\vec{\nabla}.\left(\left(e+p+p_{\rm cr}+\frac{B^2}{2}\right)\,\vec{u}-\left(\vec{B} \cdot \vec{u}\right)\vec{B}+ \vec{F}_{\rm CR} \right)=0 \ ,
\end{equation}
where the CR flux is $\vec{F}_{\rm CR}=-\ull{D}.\vec{\nabla} e_{\rm CR}$ and $e_{\rm CR}$ is the CR energy density. The CR diffusion coefficient can be decomposed into parallel and perpendicular components with respect to the background magnetic field, namely 
\begin{equation}
    D_{ij}= D_{\perp} \delta_{ij}+(D_\parallel-D_\perp) b_ib_j
\end{equation}
where $\vec{b}=\vec{B}/B$ \citep{2003A&A...412..331H}. An energy equation for the CR energy density $e_{\rm CR}$ is now required. It reads
\begin{equation}\label{Eq:CRMHDECR}
    \partial_{t}\left(e_{\rm CR}\right)+\nabla.\left(e_{\rm CR} \vec{u}+ \vec{F}_{\rm CR} \right)=-p_{\rm CR} \vec{\nabla}.\vec{u} \ . 
\end{equation}
CR pressure and energy density are linked by $p_{\rm CR}=(\gamma_{\rm CR}-1)e_{\rm CR}$, where a CR gas adiabatic index $\gamma_{\rm CR}$ is introduced.
The above energy equation does not have a flux-conservative form. \citet{2016MNRAS.462.4517K} (see also \citet{2006MNRAS.367..113P}) propose an alternative approach leading to a full set of flux-conservative equations for the CR-MHD system. To proceed the authors introduce the CR mass density $\rho_{\rm CR} = p_{\rm CR}^{1/\gamma_{\rm CR}}$, thus approximating CRs as a polytropic gas. With this assumption the CR energy equation can be recast into a continuity equation for the CR gas
\begin{equation}
     \partial_{t}\left(\rho_{\rm CR}\right)+\nabla.\left(\rho_{\rm CR} \vec{u}\right) = 0 \ .
\end{equation}
The previous equation implies that that $\rho_{\rm CR}/\rho$ is conserved along a streamline.

\subsubsection{Specific numerical schemes of CR-fluid systems}

The new set of equations can be solved using FVM as for standard MHD equations. The solver now has to account for the CR pressure which modifies the local sound speed $c_{\rm s} = \sqrt{\gamma_{\rm ad} P/\rho+ \gamma_{\rm CR} P_{\rm CR}/\rho}$. The CR pressure is dominated by the relativistic part of the CR distribution hence $\gamma_{\rm CR} \simeq 4/3$. The standard solvers detailed in Sect.~\ref{S:MHD-RIE} can be used to treat the above CR-MHD system.
Working in the framework of the fluid-kinetic approach (Sect.~\ref{S:MHD-KIN}), \cite{2007JCoPh.227..776M} develop a modified Glimm--Godunov solver where the CR mediation is included in the Riemann problem.
\citet{2016MNRAS.462.4517K} propose a CR+MHD solver, second order accurate in space and time for a bi-fluid system based on a Roe solver.

\paragraph{Semi-implicit and implicit methods} An important difficulty of explicit schemes for CR-HD or CR-MHD systems comes from the CFL stability criterion for the CR diffusion, which imposes a timestep $\Delta t \le \rm{X}_{\rm CFL}~\Delta x^2/(2 D)$, where $\rm{X}_{\rm CFL}$ is the CFL number. This criterion, because the time step scales non-linearly with the grid resolution, imposes severe slowing down limitations, especially in multi-scale problems where AMR is active. Several semi-implicit or implicit approaches have been proposed to cure this issue, usually in the context of thermal conduction studies \citep{2008MNRAS.386..627B}. The CFL constraint from CR diffusion is alleviated by calculating the diffusion operator using an implicit method (see Sect.~\ref{S:MHD-sch} and next). Another way to reduce the computation time is the so-called super-time stepping technique \citep{2007MNRAS.376.1648O, 2008MNRAS.386..627B} where the CFL condition is imposed over a large time interval composed of multiple elementary substeps over which the stability condition can be relaxed. To increment the guessed solutions at every substep from the previous guess a Runge--Kutta method is applied, using a polynomial recursion relation (using either Chebyshev or Legendre polynomials). 

Here we discuss more specifically the implicit scheme proposed by \citet{2016A&A...585A.138D} which is well-adapted to simulations with AMR. The diffusion operator in the energy Eq~(\ref{Eq:CRMHDE}) can be discretized as (we write it here in 2D for a Cartesian grid but it can easily be generalized to 3D)
\begin{equation}
    e^{\rm n+1}_{\rm i,j}= e^{\rm n}_{\rm i,j}-\frac{\Delta t}{\Delta x}\left(F^{\rm n+1}_{\rm i+1/2,j}+F^{\rm n+1}_{\rm i,j+1/2}-F^{\rm n+1}_{\rm i-1/2,j}-F^{\rm n+1}_{\rm i,j-1/2}\right) \ 
\end{equation}
where the energy density is calculated at the cell center (i,j) and the fluxes are obtained at cell interfaces. The quantities are expressed at the final time step $t^{\rm n+1}$, forming a linear system that can be solved by matrix inversion. In this scheme the anisotropic part of the fluxes at the cell interfaces are calculated from the fluxes at the cell corners, i.e., $F^{\rm n+1}_{\rm i+1/2,j}= 0.5(F^{\rm n+1}_{\rm i+1/2,j+1/2}+F^{\rm n+1}_{\rm i+1/2,j-1/2})$. The anisotropic part of the flux corresponds to the diffusion along the background magnetic field $D_\parallel \vec{b}\vec{b}.\vec{\nabla} e$. 
 
\paragraph{Cosmic-Ray streaming regularization} 
CRs are not exactly advected with the background fluid but with the scattering centers (MHD waves) carried by the background fluid \citep{1975MNRAS.172..557S}. Considering slab-type waves that CRs can self-generate as they stream along the background magnetic field, we find a streaming velocity
\begin{equation}\label{Eq:STS}
    \vec{u}_{\rm st}= \vec{u}+\left\langle \frac{3}{2} (1-\mu^2) \frac{\nu_+-\nu_-}{\nu_++\nu_-}\right \rangle \vec{u}_{\rm A}\ ,
\end{equation}
where $\vec{u}$ and $\vec{u}_{\rm A}$ are resp. the fluid and waves speed, $\mu$ is the CR pitch-angle cosine, and $\nu_{\pm}$ are the angular scattering frequency produced by the forward/backward (+/-) propagating waves along the background magnetic field. This calculation assumes that the quasi-linear theory of CR transport applies \citep{2002cra..book.....S}. A detailed calculation of this velocity requires to know the scattering frequencies, which is only possible by adding two other energy equations for each type of propagating wave. This is the purpose of the next paragraph. However, self-generated waves are preferentially produced when the local CR pressure is in excess with respect to the gas and magnetic pressure (as is likely the case close to CR sources), along a CR gradient. In that case, waves are preferentially triggered in one direction and one can write $\vec{u}_{\rm st}=\vec{u}-\vec{u}_{\rm a} \vec{B}.\vec{\nabla} p_{\rm CR}/|\vec{B}.\vec{\nabla} p_{\rm CR}|$ \citep{2017MNRAS.465.4500P}. The CR energy Eq.~(\ref{Eq:CRMHDECR}) needs to be modified according to 
\begin{equation}\label{Eq:CRMHDECR2}
    \partial_{t}\left(e_{\rm CR}\right)+\nabla.\left(e_{\rm CR} \vec{u}_{\rm st}+ \vec{F}_{\rm CR} \right)=-p_{\rm CR} \vec{\nabla}.\vec{u}_{\rm st} \ . 
\end{equation}
This equation is non-linear as $\vec{u}_{\rm st}$ depends on $\vec{\nabla} e_{\rm cr}$. The LHS of Eq.~(\ref{Eq:CRMHDECR2}) resembles an advection equation but with a speed dependent on the sign of the gradient of the CR energy density, which introduces some spurious oscillations at extrema where the gradient changes its sign \citep{Sharma:2010:NIS:2078574.2078591}. These authors propose a regularization of the energy equation by replacing the sign of $\vec{B}.\vec{\nabla}p_{\rm CR}$ in Eq.~(\ref{Eq:CRMHDECR2}) with a smooth function. The drift speed is rewritten as
\begin{equation}
    u_{\rm drift}= u_{\rm A} \tanh\left(\frac{X}{\epsilon}\right) \ ,
\end{equation}
where $X=\vec{B}.\vec{\nabla} p_{\rm CR}/|\vec{B}.\vec{\nabla} p_{\rm CR}|$, and $\epsilon$ is a small parameter to be adjusted. The energy equation becomes diffusive at streaming speed extrema with a diffusion coefficient dependent on~$\epsilon$. \citet{Sharma:2010:NIS:2078574.2078591} show that an implicit non-linear integration scheme (see their Eq.~3.9) produces a rapid convergence, but requires a sufficiently small time step to be adjusted with the value of~$\epsilon$. \citet{2018ApJ...854....5J} propose an alternative approach where Eq.~(\ref{Eq:CRMHDECR2}) is replaced by a system of two equations
\begin{eqnarray}\label{Eq:JIANG}
 \partial_{t}\left(e_{\rm CR}\right)+\nabla.\left(\vec{\Phi}_{\rm CR} \right)&=&-p_{\rm CR} \vec{\nabla}.\vec{u}_{\rm st} \nonumber \\ 
 \frac{1}{u_{\rm m}^2} \partial_t \left(\vec{\Phi}_{\rm CR}\right) + \vec{\nabla} P_{\rm CR} &=& -\frac{1}{D} \vec{\Phi}_{\rm CR} \ ,
\end{eqnarray}
where $\vec{\Phi}_{\rm CR} = \vec{F}_{\rm CR}+ e_{\rm CR} \vec{u}_{\rm st}$ is the total CR flux and $U_{\rm m}$ is a speed in practice taken larger than the maximum natural mode speed of the plasma. In this approach the problematic term $\rm{sgn}(X)$ is replaced by $\vec{\nabla}. \vec{\Phi}_{\rm CR}$. The system is closed by choosing the diffusion coefficient as $D= D_0- \vec{u}_{\rm st} (e_{\rm CR}+ p_{\rm CR})/(\vec{b}.\vec{\nabla}p_{\rm CR})$, where $D_0$ is a background coefficient produced by large-scale injected turbulence. \citet{2018arXiv180511092T} criticize the two previous methods. They present CR transport mediated by self-generated waves including an accurate description of CR pitch-angle scattering up to the second order in accuracy in $(u_{\rm a}/c)$. The latter is possible because of the addition of two supplementary energy equations respectively for forward and backward propagating waves (see the next paragraph). The authors employed a system of equations similar to Eq.~(\ref{Eq:JIANG}) but with supplementary terms derived from the effect of CR scattering off self-generated waves.

\paragraph{A four-fluid approach}
A more complete description of the CR-fluid system includes the description of the forward and backward propagating CR self-generated waves \citep{1975MNRAS.172..557S}. It includes two supplementary fluid energy equations \citep{1992A&A...259..377K}. Now the total pressure includes the contribution of forward (backward) waves $P_{\rm{w},+}$ ($P_{\rm{w},-}$), and the total energy density includes wave terms as well $e_{\rm{w},+}$ ($e_{\rm w,-}$), see \citet{1992A&A...259..377K, 2018arXiv180511092T} for a complete derivation of the new system of equations. This system is restricted to the quasi-linear theory framework. Hence, coupled with a MHD code the amplitude of self-generated waves has to be small with respect to the amplitude of the background magnetic field. The four-fluid system has been numerically solved by \citet{2018arXiv180511092T} using an explicit scheme separating the CR and wave fluid equations.

The interest of this approach resides in the more accurate calculation of the CR streaming speed given by Eq.~(\ref{Eq:STS}) as well as the CR diffusion coefficient in space and energy, both dependent on the amplitude of scattering frequencies $\nu_{\pm}$ off forward and backward waves. In particular, it is possible to account for a fluid description of a stochastic acceleration term in the CR fluid energy density equation. This term has the form \citep{2018arXiv180511092T}
\begin{equation}
4 \frac{\bar{\nu}_+ \bar{\nu}_-}{\bar{\nu}_+ + \bar{\nu}_-} \frac{v_{\rm a}^2}{c^2} \left(e_{\rm CR} + P_{\rm CR} \right) \ ,
\end{equation}
where $\bar{\nu}$ is the momentum-averaged scattering frequency. 

\paragraph{Other numerical strategies} \citet{2006MNRAS.367..113P} and \citet{2007A&A...473...41E} develop a method based on a smooth particle hydrodynamics (SPH) approach. In SPH fluid dynamics is treated using Lagrangian particles. To each particle is attached relevant fluid properties (e.g. density, pressure ...) calculated using a SPH kernel dependent on a smoothing length \citep{1992ARA&A..30..543M}. The CR distribution follows a power-law distribution whose normalization and power-law index vary under the effect of adiabatic gas variation (compression/expansion) as a function of the gas density (itself calculated in the SPH code).  

\subsubsection{Numerical tests}

Aside the standard numerical tests for HD and MHD codes described in Sect.~\ref{S:TESTMHD}, we detail here some specific setups aiming at testing the transport (either passive or active) of CRs. 

\begin{figure}[htb]
\begin{centering}
\includegraphics[width=1\linewidth]{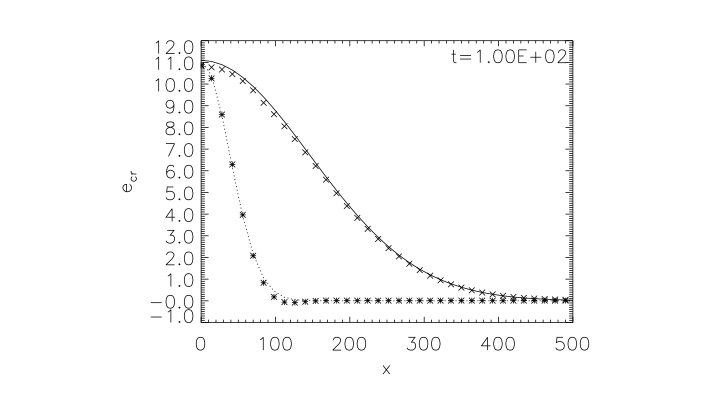}
\end{centering}
\caption{CR propagation along an inclined magnetic field. The two curves show cuts of the ellipsoid solution at a given time along the
major axis (crosses) and along the minor axis (asterisks). The solid and
dotted lines represent the analytical solutions corresponding cuts of the fitted 2D Gaussian profile. 
\label{F:CR1D} 
CR diffusion is treated using an explicit scheme with $D_\parallel=100$ (in units of of pc$^2$Myr$^{-1}$ and $D_\perp$=0. (from \citet{2003A&A...412..331H})}
\end{figure}

\begin{figure}[htb]
\begin{centering}
\includegraphics[width=1\linewidth]{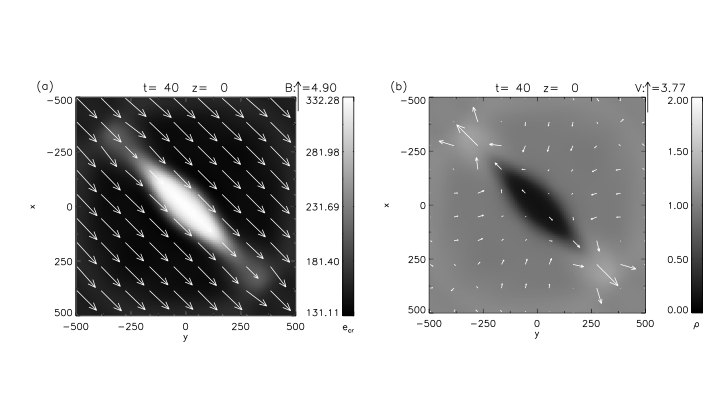}
\end{centering}
\caption{CR propagation along an inclined magnetic field in the case CR backreaction is active. Equipartition between gas, magnetic field and background CR pressures is assumed. An over pressure of a factor 100 in CR is injected at the center at the start of the simulation. The diffusion coefficients are: $D_\parallel=100$ and $D_\perp$=4 (in units of of pc$^2$Myr$^{-1}$). Left: CR pressure map and magnetic field lines. Right: gas density map and velocity vectors. (from \citet{2003A&A...412..331H}).\label{F:CR1Da}}
\end{figure}

\citet{2003A&A...412..331H} propose a series of tests of the CR flux term in Eq.~(\ref{Eq:CRMHDE}) using the \textsc{Piernik} code \citep{2010EAS....42..275H}. A~first, straightforward test is 1D diffusion of CRs along the background magnetic field directed in one direction of the Cartesian grid, setting the perpendicular diffusion to zero, and no CR backreaction (turning off the CR pressure gradient), in a static medium. 
Diffusion can also be tested along an inclined magnetic field with contributions of the different directions to the CR flux.
Figure~\ref{F:CR1D} shows the profile of the CR energy density along and perpendicular to the ellipsoid solution of the propagation of an initial spheroidal distribution in 3D (see Fig.~2 in \citet{2003A&A...412..331H}). 
The authors propose also the same test but now turning on the effect of CR pressure gradient. Figure~\ref{F:CR1Da} shows two plots: CR energy density and magnetic field lines (left), gas density and velocity field lines (right). It can be seen that the gas accelerates up to a few km/s under the effect of a strong CR gradient preferentially along the background magnetic field line. This gas motion leads to a drift of CRs along the magnetic field. Perpendicular diffusion imposes a broadening of the CR profile perpendicular to the magnetic field lines. Similar tests are proposed in \citet{2006MNRAS.373..643S} (see their Fig.~5).

\begin{figure}[htb]
\begin{centering}
\includegraphics[width=1.1\linewidth]{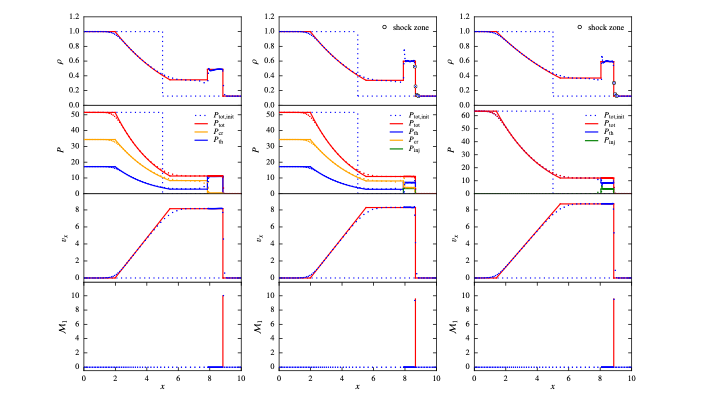}
\end{centering}
\caption{1D shock-tube test problems including CRs. From top to bottom are shown: the gas mass density, gas pressure, gas velocity and shock Mach number. Left: Shock propagation in a composite gas including CRs but without the effect of CR acceleration. Middle: Same as the left case but now including CR acceleration. Right: Same as the middle case but without background CRs. Solid lines: analytical solutions, dotted lines: numerical solutions. The different pressures are: gas (blue), CRs (orange), total (red), the pressure in the CR injected at the shock front is in green. In the left and middle cases the initial CR pressure is two times the gas pressure on the left part and equal to it on the right part. (From \citet{2017MNRAS.465.4500P})\label{F:ShocktubeCR}}
\end{figure}

A~second type of tests used in the context of CR acceleration at SNR blast waves involves Sod shock-tube simulations (restricted to HD). \citet{2017MNRAS.465.4500P} (see also \citet{2014MNRAS.437.3312S}) derive an analytical solution of the shock-tube problem including a CR gas. Figure~\ref{F:ShocktubeCR} shows a solution of a 1D Riemann shock tube problem. Three cases are shown: on the left the solutions for a shock propagating in a composite gas of thermal plasma and CRs but without any CR acceleration, in the middle the same case but now including CR acceleration, on the right a shock propagating in the thermal gas only but with CR acceleration. \citet{2018ApJ...868..108B} propose a 1D Brio-Wu shock-tube test (hence in MHD). 

\subsubsection{Cosmic Ray physics using a bi-fluid approach} 

We illustrate here the ability and the limits of the bi-fluid approach in treating three important aspects of CR physics: CR acceleration at shocks, CR-driven winds and CR-induced magnetic production in the Galaxy.

\paragraph{Cosmic Ray acceleration at shocks:}
 After the derivation of the test-particle solution produced by DSA (see Sect.~\ref{S:DSA}), it appeared that the pressure carried by the high energy CRs is large enough to modify the shock structure and hence the Fermi acceleration process itself. \citet{1981ApJ...248..344D, 1977ICRC...11..132A} derived self-consistent fluid solutions including the CR pressure in the Rankine-Hugoniot conditions using a bi-fluid model. They find that the non-linear model has up to three stationary shock solutions when the adiabatic indices of the gas and CR are $\gamma_{\rm ad}=5/3$ and $\gamma_{\rm c}=4/3$ respectively. The selection among these solution depends on the injection efficiency \citep{1997ApJ...491..584M}. \citet{2013ApJ...775..130S}\footnote{The section 2 of \citet{2013ApJ...775..130S} provides a short updated review of the main assumptions and issues of the bi-fluid model.} discuss the time-dependent stability of these solutions and confirm that the high- and low-efficiency solutions are stable against large-amplitude perturbations, while the intermediate one is not. At high-injection efficiency and high-shock Mach numbers the shock may be completely smoothed in the two-fluid approach. \citet{2001RPPh...64..429M, 1997ApJ...485..638M} show that a subshock must always exist in a kinetic model and hence that these smooth solutions are an artefact of the model because the maximum CR energy is always finite. 

 Acknowledging these issues, the main interest of the bi-fluid approach to investigate CR back-reaction resides in its simplicity and also in the fact that this formalism can be easily combined with a MHD model \citep{1986A&A...160..335W, 1994ApJ...429..748J, 1995ApJ...441..629F} and used to follow CR-MHD fluid non-linear dynamics with respect to the shock obliquity. The bi-fluid model also allows to combine effects of CRs and radiation \citep{2007A&A...463..195W}. It is particularly useful to diagnose the profile of H$_\alpha$ lines in Balmer-dominated shocks occurring while a SNR propagate in a partially ionized medium \citep{2009ApJ...690.1412W}. 
 However, only a kinetic calculation permits to follow the energy dependence of the particle distribution and the instabilities particles may generate. Non-linear kinetic models have since superseded the bi-fluid approach in this context. Numerical non-linear kinetic models are presented in sections \ref{S:MHD-KIN} and \ref{S:PICMHD} and semi-analytic models in Sect.~\ref{S:SAN}.
 
\paragraph{CR driven winds} CRs by the interplay of the generation of waves for instance by the streaming instability can convert a part of the CR bulk momentum into fluid momentum and hence drive winds \citep{1975ApJ...196..107I}. \citet{1991A&A...245...79B, 1993A&A...269...54B} investigate both analytically and numerically the launch of CR-driven winds in the framework of the bi-fluid model. Actually as their model includes also self-generated waves it proposes a three-fluid description. The authors use the flux tube approximation where stationary MHD equations are solved along the outwardly-directed magnetic field lines. The flux tube has a variable area cross section A(z), z marking the height above the galactic disk. Solutions, especially the sonic point of the MHD system, are searched for as in the case of the solar wind. The numerical solutions start from the calculation of the fluid speed gradient at the critical point and then are propagated out- and downward to match the inner and outer boundaries. Stationary CR-driven outflows are obtained if the inter galactic gas pressure is low enough. \citet{1996A&A...311..113Z} extend this work by including galactic rotation. They show that the wind forces the gas in the halo to corotate with the galactic disk up to a distance of a few kpc. The waves generated by the CRs also contribute to heat the halo. \citet{2008ApJ...674..258E} adopted a similar modelling as in previous works although with different boundary condition assumptions. 
They find that thermal and CR pressures are equally important to drive the wind. Their solutions are found to be consistent with an injection of CRs by SNe at a standard rate although with a bit high efficiency. \citet{2012MNRAS.423.2374U} carry SPH simulations including a treatment of adiabatic gain/loss of the CR gas and the physics of streaming. \citet{2016ApJ...824L..30P, 2017ApJ...834..208R} add the effect of anisotropic diffusion coefficients in the disk in a bi-fluid formalism. All these works show that CRs have a systematic negative feedback impact over the star formation rate in our Galaxy while they can launch powerful winds from the galactic disk. 

\paragraph{Galactic magnetic dynamo}
Dynamo is likely at the origin of large scale magnetic fields generated in the Galaxy [see \citet{1999ARA&A..37...37K} and section~6 in \citet{2018JPlPh..84d7304B}]. Bi-fluid models have been widely used to investigate the Parker or buoyancy instability thought to participate fast dynamo processes in spiral galaxies \citep{1992ApJ...401..137P, 1993A&A...278..561H}. CRs produce an inflation of magnetic loops anchored in the galactic disk that induces magnetic reconnection and disconnects disk and halo magnetic fields. The closed loops in the halo and the anchored loops in the disk can be subject to cyclonic rotation at the origin of an $\alpha\Omega$ dynamo process, the $\alpha$ effect resulting from the Parker instability. Notice that CRs can provoke $\alpha$ dynamo because of their current, strong enough to trigger the non-resonant streaming instability \citep{2014ApJ...788..107B}. 

A numerical investigation of the Parker instability using a bi-fluid model has been proposed by \citet{2000ApJ...543..235H} using a flux-tube approximation. In that approximation CR are injected from a SNR at its lifetime end in the flux tube composed of the magnetic field lines threading the SNR. CRs are injected with an over-pressure of $\sim 30$ with respect to their background pressure and then inflate the flux tube in the vertical direction (with respect to the disk), eventually leading to a flux-tube explosion in a runaway process and ultimately to the production of galactic winds.\footnote{Figure~1 in \citet{2003A&A...401..809L} lists a series of physical processes which result from the Parker instability.} The authors were able to find $\alpha$ coefficients large enough to ensure efficient Parker instability including magnetic field back-reaction. \citet{2003ApJ...589..338R} investigate the growth rate of the Parker instability as function of the CR diffusion coefficient and find that the growth rate decreases with the parallel coefficient whereas the perpendicular coefficient has no strong impact. \citet{2004ApJ...607..828K} conducte a 2D analysis of the Parker and explosion instabilities. Later \citet{2004ApJ...605L..33H} and \citet{2009ApJ...706L.155H} propose 3D simulations of the Parker instability using the \textsc{Piernik} code in a 3D box and in global Galactic geometry respectively. The global simulations show the growth of the magnetic field strength saturates at about $t = 4$ Gyr, reaching values of 3--5 $\mu$G in the disk. The magnetic field, initially randomly oriented, shows at the end of the simulation a toroidal component forming a spiral structure with reversals in the plane of the disk. 

\subsection{The fluid-kinetic framework}\label{S:MHD-KIN}


Solving the Fokker--Planck equation (FPE) is usually a very demanding task for computer simulations (see Sect.~\ref{S:KIN}). Then, coupling a FPE with a MHD code has only been scarcely attempted yet. In this section we discuss approaches developed to couple a fluid model (HD or MHD) and a kinetic model (Vlasov or Fokker--Planck or approximations of these), appropriate to describe respectively the shock and the particles in the context of CR acceleration at shocks. In Sect.~\ref{S:HYKIN} we discuss methods that couple (M)HD equations and the diffusion-convection equation. In Sect.~\ref{S:VFP} we present a different approach that relies on a more general Vlasov-Fokker--Planck equation for description of the evolution of the distribution function~$f(\vec{p})$.

\subsubsection{Coupling hydro and diffusion-convection equations}\label{S:HYKIN}

Works reviewed in this section model separately the evolution of the astrophysical flow (e.g., the blast wave of a supernova remnant, or a cosmological structure formation shock), using HD or MHD equations (see Sect.~\ref{S:MHD}), and the acceleration of energetic particles, using the diffusion-convection equation, that is the angle-averaged Fokker--Planck equation (see Sect.~\ref{S:KIN}) of the form
\begin{equation}
\frac{\partial f\ensuremath{}}{\partial t}+\frac{\partial}{\partial x}\left(uf\right)=\frac{\partial}{\partial x}\left(D\frac{\partial f}{\partial x}\right)+\frac{1}{3p^{2}}\frac{\partial p^{3}f}{\partial p}\frac{\partial u}{\partial x}\,,\label{eq:transport-particles-f}
\end{equation}
which includes advection terms in space and in momentum, and a diffusion
term in space (we are here restricting ourselves to first order acceleration,
and so have not included diffusion in momentum). 
When written in a conservative form the spatial advection term 
may be included in the hydrodynamical solver, using the operator-splitting technique. 
The right-hand side is commonly solved with Finite Differences methods (FDM), 
using the techniques presented in Sect.~\ref{S:KIN}. 
The diffusion term is the most difficult to treat. As stressed before, explicit schemes suffer from the constraint due to a CFL condition that is quadratic
in the resolution $\Delta t<\Delta x^{2}/2D$, 
and which for physically-motivated values of the diffusion coefficient imposes time-steps
that are much shorter than the hydrodynamic time-step. As in Sect.~\ref{S:MHDM},
this leads to the use of implicit or semi-implicit schemes, or possibly
accelerated explicit schemes. 
The diffusion coefficient~$D$, although it should in principle be computed from~$f$ itself, 
is commonly prescribed as a function of~$p$, usually as a power-law $D(p)\propto p^{\alpha}$,
with optionally a dependence on position~$x$ (the
Bohm coefficient reduces to such a form with $\alpha=2$ in the non-relativistic
regime and $\alpha=1$ in the relativistic regime).
Compared to the bi-fluid approach of previous Sect.~\ref{S:MHDM}, the methods presented in this section aim at describing the spectrum of particles $f(p)$, using various approximations and techniques to handle the large range of scales to be resolved.

\paragraph{The Piecewise power-law method.}\label{S:PPL}

This method was introduced by \citet{1999ApJ...512..105J,2001ApJ...557..475T} to investigate the time evolution of relativistic synchrotron-emitting electrons in radio galaxies.\footnote{An alternative approach was proposed by \citet{1999A&A...349..323M} where a set of Lagrangian test particles, a set of relativistic electrons is followed along a jet including acceleration and losses. Note that simulations combining mixed Lagrangian (for energetic particles) and Eulerian (for the magnetized fluid) coordinates have also been developed in the framework of relativistic flows by \citet{2018ApJ...865..144V}. The authors solve the diffusion-convection equation for relativistic electrons transported in a relativistic flow calculated using a relativistic version of the PLUTO code.} The basic idea is to approximate the electron distribution $f(p)$ as a piecewise power-law with N~bins in momentum, $f(p) = \sum_{\rm i} f_{0,i} (p/p_{\rm i})^{\alpha_{\rm i}}$ with $i\in[1,N]$ (on a fixed, logarithmic momentum grid), and to calculate $f_{\rm 0,i}$, $\alpha_{\rm i}$ as function of the position in the simulation space (here a jet). The number of relativistic electrons in the bin~i is $n_{\rm i} = \int_{p_{\rm i}}^{p_{\rm i+1}} f(p) 4\pi p^2 dp$, which can be normalized to the grid mass density of the gas. We note $b_{\rm i}=n_{\rm i}/\rho$. At shocks, the standard DSA theory in the test-particle limit is applied (see Sect.~\ref{S:DSA-Fermi}), the index is fixed to $\alpha = 3r/(r-1)$ where r is the shock compression ratio evaluated from the shock Mach number. At shocks $b=\sum_{\rm i} b_{\rm i}$ is calculated solving a simple equation $db/dt = Q_{\rm inj}/\rho$ where $Q_{\rm inj}=\epsilon u_{\rm sh, L}/\mu_{\rm e} m_{\rm p}$. Here $u_{\rm sh,L}$ is the Lagrangian shock speed, $\mu_{\rm e}$ the electron mean mass, $m_{\rm p}$ the proton mass. The parameter $\epsilon$ is the fraction of thermal electrons injected as non-thermal particles. Away from shocks, the parameters $b_{\rm i}$ and $\alpha_{\rm i}=3-(b_{\rm i+1}/b_{\rm i})/\Delta \ln(p)_{\rm i}$ evolve accounting for adiabatic and radiative losses [see \citet{2001CoPhC.141...17M}]. 

Adiabatic losses, which produce a shift in energy to lower energy, are of particular importance for CR-MHD systems discussed in Sect.~\ref{S:MHDM}. These systems can be extended to account for more than one CR population. This extension is reflected in the choice of the mean diffusion coefficient corresponding to a particular CR energy range of a given population. This effect can be treated with the piecewise power-law method \citep{2001CoPhC.141...17M, 2014arXiv1406.4861G}. The adiabatic loss rate is
$d\ln(p)/dt= (\gamma_{\rm CR}-1) \vec{\nabla}.\vec{u}$. Here the adiabatic CR index $\gamma_{\rm CR}$ changes depending on the CR energy, from 5/3 in the non-relativisitc case to 4/3 in the relativistic case. The CR density can be computed from the relation $n_{\rm i}=E_{\rm i}/\langle E_{\rm i} \rangle$, where $E_{\rm i}$ is the CR energy in the i$^{\rm{th}}$ bin, and the average energy per CR particle in bin i $\langle E_{\rm i} \rangle= \int_{p_i}^{p_{i+1}} E(p) p^2 dp/\int_{p_i}^{p_{i+1}} p^2 dp=e_{\rm CR,i}/n_{\rm i}$ does not depend on~$f$. The CR energy density in the bin i can then be updated accounting for the loss term
\begin{equation}
    e_{\rm CR,i}^{n+1}=e_{\rm CR,i}^n + \Delta t (\Phi_{\rm i-1/2}-\Phi_{\rm i+1/2})
\end{equation}
where the energy flux is $\Phi_{i+1/2}=4\pi/\Delta t \int_{t}^{t+\Delta t} b(E) E p^2 f(p)_{i+1/2}$. The final energy $E(t+\Delta t)$ is expressed in terms of the adiabatic loss term and $\Delta t$ and has to be in the interval $[E_{i-1/2}, E_{i+1/2}]$, otherwise a sub-cycling is required (see \citet{2014arXiv1406.4861G} for further details).

\paragraph{Including backreaction effects.}

In efficient DSA the evolution of the shock and of the particles are
non-linearly coupled (NLDSA). In the hydro-kinetic approach the connection
between the fluid and the particles is made via extra terms 
in the equations, that model the injection and the back-reaction of particles. 
First, a source term is required in the diffusion-convection equation at the shock front, that depends on the shock jump conditions, 
and a corresponding loss term must be added in the set of hydrodynamic equations to ensure energy conservation (and in principle conservation of mass, but the inertia of energetic particles is normally negligible). 
Secondly, similar to the bi-fluid approach, one has to take into account the pressure
of particles $P_{\mathrm{CR}}$ given by Eq.~(\ref{eq:CR-pressure}).
The force $-\mathbf{\nabla}P_{\mathrm{CR}}$ exerted by the particles
on the flow is added as a source term to the equation of conservation
of momentum, and the corresponding work $-\mathbf{u}.\mathbf{\nabla}P_{\mathrm{CR}}$
is added as a source term to the equation of conservation of energy.
At this level of modeling it is the gradient of particle pressure
upstream of the shock that will cause the appearance of the shock
precursor,\footnote{At the PIC/hybrid level of modelling of Sect.~\ref{S:VMAX}, this is done via the Lorentz force.} which in turn will produce concave particle spectra.

The first investigations of time-dependent NLDSA using this approach were 
performed by \cite{1987MNRAS.225..399F} and \cite{1987MNRAS.225..615B}. 
Knowing that the canonical result for~$f(p)$ is a power-law of index $s=4$, 
\cite{1987MNRAS.225..399F} work with the quantity $g(p)=p^{4}\,f(p)$, 
and since the momentum gain is proportional to the current momentum 
they replace~$p$ with the quantity $y=\ln p$. 
Equation~(\ref{eq:transport-particles-f}) is then rewritten as 
\begin{equation}
\frac{\partial g}{\partial t}+\frac{\partial\left(ug\right)}{\partial x}=\frac{\partial}{\partial x}\left(D\frac{\partial g}{\partial x}\right)+\frac{1}{3}\left(\frac{\partial g}{\partial y}-g\right)\frac{\partial u}{\partial x}\,.\label{eq:transport-particles-g}
\end{equation}
These early works assumed a small dependence of the diffusion coefficient on~$p$,
using resp. $\alpha=1/4$ and $\alpha=1/2$, because of numerical limitations.
As already noted in Sect.~\ref{S:DSA}, it is the diffusion
coefficient that sets the spatial and temporal scales of the simulation.
The spatial scales range from the microscopic scale where the particles
decouple from the fluid (of the order of a few thermal gyration lengths)
to macroscopic scales (like the radius of the supernova remnant).
The resolution of the numerical grid is dictated by the diffusion
of the lowest energy particles, whereas the size of the grid is dictated
by the diffusion of the highest energy particles. 
The ratio $D(p_{\max})/D(p_{\min})$, 
and thus the number of cells, may exceed ten orders of magnitude
if $D(p)\propto p$, which is too demanding in terms of memory requirements
and computing time. Fortunately we need very high resolution only
around the shock, since low energy particles cannot travel far away
from the shock. More generally, particles of a given momentum~$p$
require a certain spatial resolution over a certain extent around
the shock. This leads to the implementation of adaptive mesh refinement
(AMR), reviewed in Sect.~\ref{S:MHD-AMR}. In the context of DSA simulations, 
this technique was pioneered by \cite{1992A&A...262..281D},
to be able to use the true Bohm scaling $D(p)\propto pv$, and
later used by~\cite{2001ApJ...550..737K} and by~\cite{2008MNRAS.383...41F}. 
The latter authors also parallelized the scheme in the $p$-direction, 
and studied repeated acceleration by multiple shocks.
\cite{2014ApJ...792..133F} use a different diffusion coefficient, 
proposed for perpendicular shocks, while DSA simulations are usually focused on the case of parallel shocks.

The hydro-kinetic approach was mostly developed by the team of Kang
and Jones and collaborators with their code \textsc{Crash}, starting from \cite{1991MNRAS.249..439K},
with a number of publications produced until this date. 
An example of results can be seen in Fig.~\ref{fig:NLDSAcomp}, 
compared with two other methods discussed elsewhere in this review.
Some important numerical developments include the following. 
\cite{2000A&A...364..911G} introduced an injection scheme based on the thermal leakage model.
\cite{2001ApJ...550..737K} implemented a grid-based AMR scheme, as
well as sub-zone shock tracking, in order to address realistic diffusion
coefficients at a manageable computational cost. \cite{2005APh....24...75J} applied
a Coarse-Grained Finite Momentum-Volume Scheme (CGFMV), 
an extension of the piecewise power-law method introduced by \citet{1999ApJ...512..105J} 
already presented above. 
The basic idea is to lower the numerical resolution in momentum (down to as
few as two to three bins per decade), but prescribe the shape of the
spectrum in each bin so as to maintain reasonable accuracy; 
the numerical spectrum is then no longer a piecewise constant 
function but a piecewise linear function.
All the works cited so far were restricted to slab geometry, 
\cite{2006APh....25..246K} simulated shocks in spherical geometry---still effectively one-dimensional, under the assumption of spherical symmetry---in order to study CR feedback at SNR shocks. 
For this the authors employed a frame comoving with the outer shock, 
which was found to lower the convergence requirements. 
Following the formalism of cosmology, they use coordinate $\tilde{r}=r/a$
where $a(t)$ is the expansion factor, with expansion rate obtained
from the measured shock speed in the Cartesian grid. Then density and pressure (or energy density) are re-scaled as $\tilde{\rho}=\rho a^{3}$ and $\tilde{P}=Pa^{3}$
(the time variable is left unchanged). 
For quasi-parallel plane shocks, \cite{2007APh....28..232K,2009ApJ...695.1273K} found a self-similar evolution and proposed analytic forms for the solutions during that stage.
All these improvements allowed the team to investigate the time evolution of 
CR-modified shocks and particle spectra, and eventually compute
the non-thermal radiation from the accelerated particles in SNRs  \citep{2011MNRAS.414.3521E,2012ApJ...745..146K}.
The latest developments are the inclusion of prescriptions for magnetic
field amplification (MFA, see Sect.~\ref{S:DSA-NL}) and of the effects of the Alfv\'enic drift (see Eq.~(\ref{Eq:STS})) by 
\cite{2012JKAS...45..127K,2013JKAS...46...49K,2013ApJ...777...25K}, 
with a formalism similar to the approaches discussed later in Sect.~\ref{S:SAN-Blasi}.

\paragraph{Other approaches in spherical geometry.}

\begin{figure}[htb]
\begin{centering}
\includegraphics[width=0.45\linewidth]{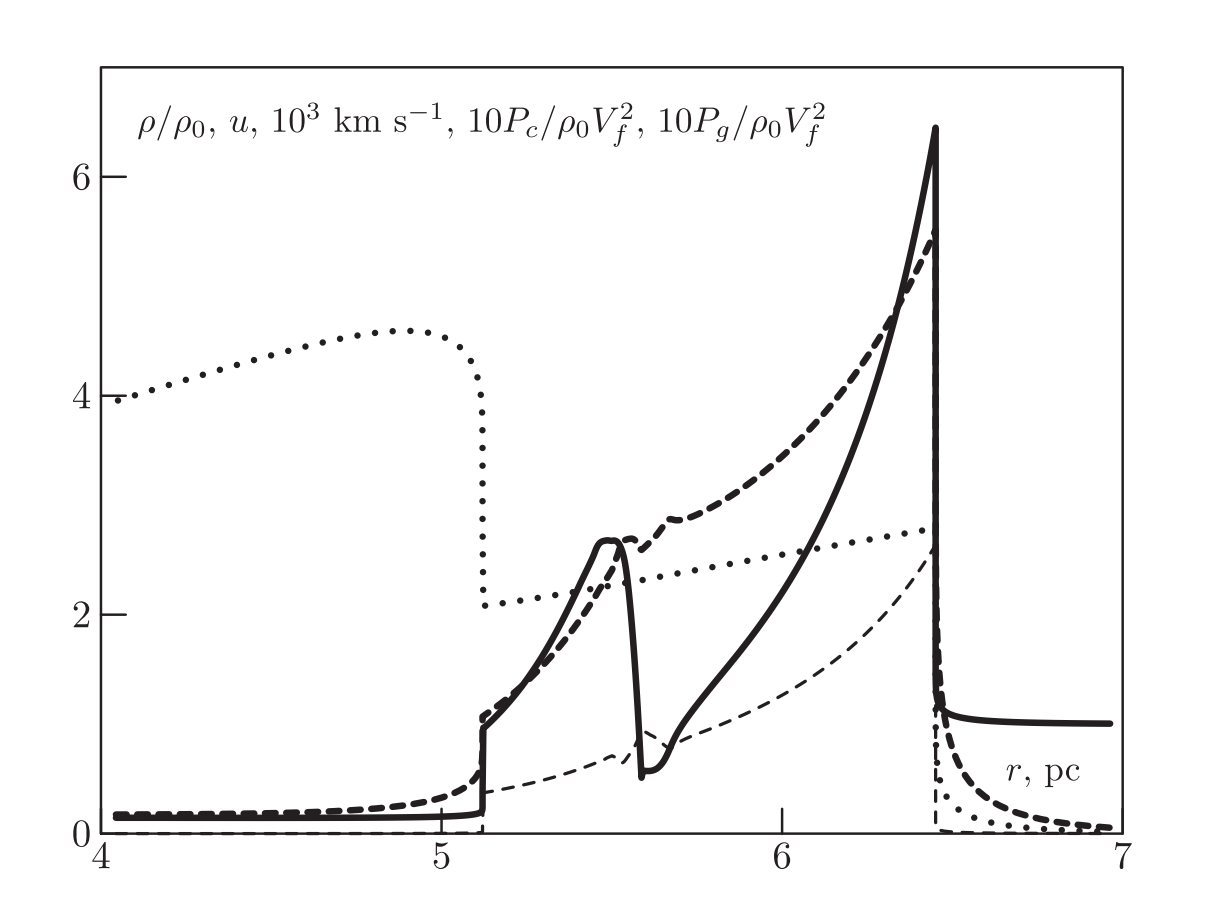}
\includegraphics[width=0.50\linewidth]{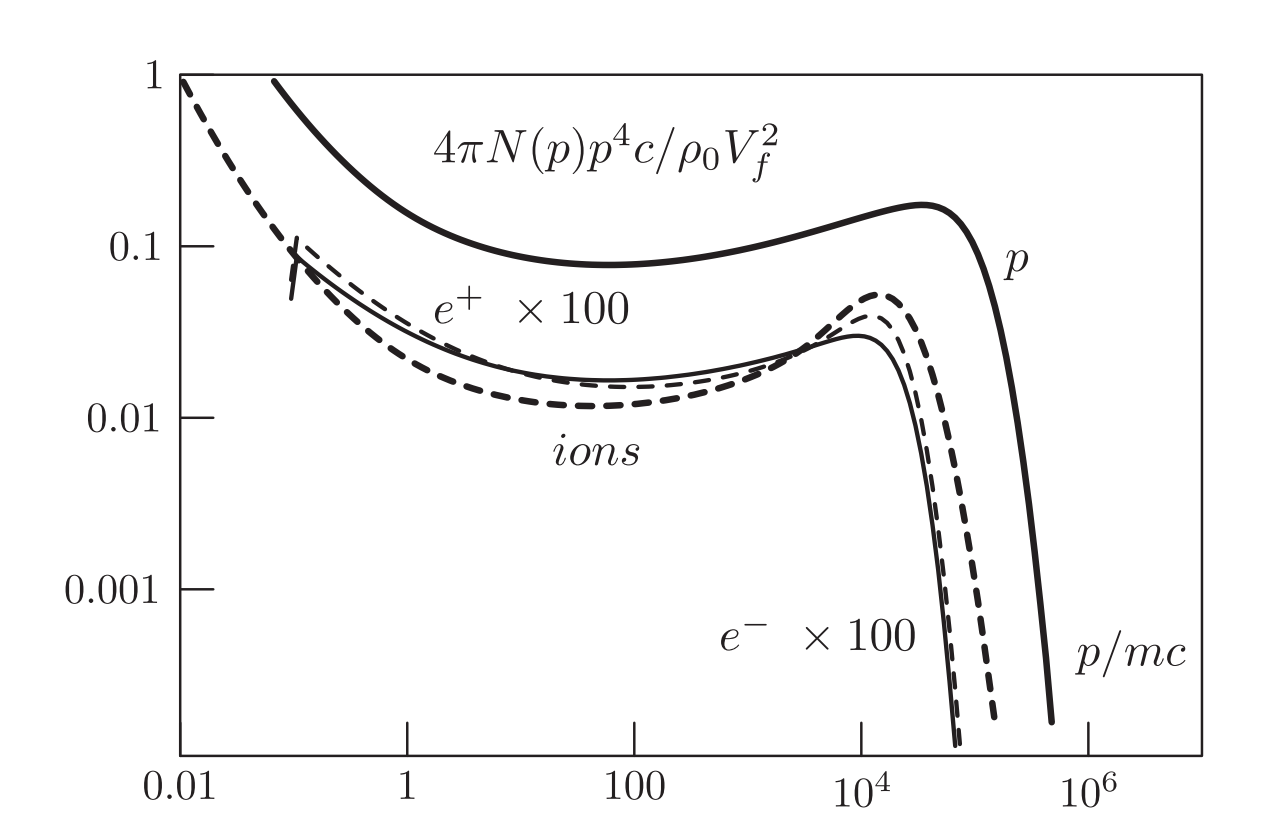}
\par\end{centering}
\caption{Representative results from a hydro-kinetic simulation of a SNR in spherical geometry, at $t=1000$~yr. 
The supernova has typical mass $1.4 M_\odot$ and energy $10^{51}$ erg, 
released in a uniform medium of density $0.1$~cm$^{-3}$, temperature $10^4$~K, and magnetic field $5~\mu G$.
Left: radial dependencies of gas density (thick solid line), gas velocity (dotted line), CR pressure (thick dashed line), and gas pressure (dashed line). The reverse and forward shocks are visible as discontinuities in the hydro profiles. An extended CR precursor is visible ahead of each shock.
Right: spectra of accelerated particles: protons at the forward shock (thick solid line), ions at the reverse shock (thick dashed line), electrons at the forward shock ($\times$100, thin solid line), and positrons at the reverse shock ($\times$100, thin dashed line) present from radioactive decay in the ejecta. Note the concavity of the spectra, plotted as $p^4f(p)$.
From \cite{2012APh....39...12Z}.
}
\label{fig:ZP2012}
\end{figure}

To account for the spatially-dependent diffusion of energetic particles, 
an alternative approach to AMR is to perform a change of variables, 
as done by \cite{2010ApJ...718...31P,2012APh....39...12Z}
and by \cite{2012A&A...541A.153T,2012APh....35..300T}. These simulations
are performed in spherically symmetric geometry, and notably include 
the reverse shock and the contact discontinuity as well as the forward shock 
(located at radius $R_{b}$, $R_{c}$, and $R_{f}$ respectively). 

In \cite{2010ApJ...718...31P,2012APh....39...12Z}
the radius~$r$ is replaced by $r/R_{b}$ for $r<R_{b}$ and $r/R_{f}$
for $r>R_{f}$ (unshocked medium), and by $(r-R_{c})/(R_{c}-R_{b})$
for $R_{b}<r<R_{c}$ and $(r-R_{c})/(R_{f}-R_{c})$ for $R_{c}<r<R_{f}$
(shocked region). Also time is replaced by a dimensionless parameter
that scales as $\ln(R_{f})$. The hydrodynamic equations are solved using FDM,
separately in the upstream regions (using an implicit scheme since
the flow is supersonic) and in the downstream regions (using an explicit
scheme), and the three discontinuities between the different regions
are moved manually in time. As in other methods, the diffusion-convection
equation for particles is recast as a set of tri-diagonal equations.
In these works the CR pressure is included in the HD equations, 
and the velocity used for particles includes the Alfv\'enic drift. 
An example of the results is shown in Fig.~\ref{fig:ZP2012}, 
for a fiducial SNR at age 1000~yr.

In \cite{2012A&A...541A.153T,2012APh....35..300T}
the radius is first normalized to the outer shock radius: $x=r/R_{f}$
(a comoving coordinate, as in \cite{2006APh....25..246K}) and then
transformed according to $(x-1)=(x_{*}-1)^{3}$, so that $dx/dx_{*}=3(x_{*}-1)^{2}$,
with a uniform binning in new coordinate~$x_{*}$. The transport
equation for the particles is solved using an implicit FDM, while
the HD equations are solved with the \textsc{VH-1} code. These simulations
were made in the test-particle regime, they do not include the back-reaction
of energetic particles on the hydrodynamics. 
\cite{2012APh....35..300T} discuss the possible contribution 
to the SNR emission of CRs accelerated at the reverse shock w.r.t.\ those
accelerated at the forward shock.

\subsubsection{Coupling MHD and Vlasov--Fokker--Planck equations}\label{S:VFP}

\cite{2011MNRAS.418.1208B,2013MNRAS.430.2873R} develop a method based on performing a spherical harmonics expansion of the distribution~$f$ in momentum~$\mathbf{p}$, using the \textsc{Kalos} code previously developed for simulations of laser--plasma interactions \citep{2006PPCF...48R..37B}. The authors consider a 1D problem with a planar shock (with the normal to the shock front oriented in the x direction with $x < 0$ the upstream medium) and keep the information about the momentum vector~$\vec{p}$. They solve the following Vlasov--Fokker--Planck (VFP) equation \citep{2011MNRAS.418.1208B}
\begin{eqnarray}\label{Eq:VFP}
\partial_t f(x,\vec{p},t) + (v_x + u) \partial_x f - \partial_x u\, p_x\, \partial_{p_x} f -\frac{u}{c} \partial_x u\, p_x\, \partial_{p_x} f && \nonumber \\
+ q \vec{v} \times \vec{B}. \partial_{\vec{p}} f = {\cal C}(f) \, ,
\end{eqnarray}
where v and u stand for the particle and the fluid speeds, and q is the particle charge. The particle distribution function is measured in the fluid rest-frame. $\vec{B}$~is the local magnetic field. The term ${\cal C}(f)$ is a collision term; in the collisionless plasmas investigated here this term is due to wave-particle interactions. It is written
\begin{equation}
{\cal C}(f) = \partial_\mu \nu_{\rm s} (1-\mu^2) \partial_\mu f(x,\vec{p},t) \, ,
\end{equation}
where $\mu =\cos(\theta)$ and $\nu_{\rm s}$ is a parameter to be scaled with respect to the particle gyro-frequency $\Omega = qB/\gamma mc$.
In this model particle angular deflection is assumed to be small. This justifies the use of the term Fokker--Planck. 

The particle distribution is then expanded into spherical harmonics as
\begin{equation}
f(x,\vec{p},t) = \sum_{l,m} f^m_l(x,p,t) P^{|m|}_l(\cos{\theta}) \exp(-im\phi) \, ,
\end{equation}
where $l$ is a positive integer, $m$ is an integer between $-l$ and $+l$, $\theta$ is the particle momentum pitch-angle with respect to the shock normal, $\phi$ is the particle momentum azimuthal angle, and $P^{m}_l(\cos{\theta})$ are the associated Legendre polynomials of order $l$. We have $f^{-m}_l = (f^m_l)^*$, the complex conjugate of $f^m_l$. 
By applying this expansion to the VFP equation, one obtains a hierarchy of equations for each component $f^m_l$, not reproduced here [see Eq.~(2) of \citet{2011MNRAS.418.1208B}]. Usually, as in shocks the particle distribution is close to isotropy due to efficient wave-particle scattering it is only necessary to retain the first few terms $f_0^0$, $f_1^0$, and $f_1^1$. $f_0^0$ represents the isotropic part of the distribution function while $f_1^0$ represents the particle flux along the shock normal. $f_1^1$ complements the construction of the CR current. In the limit of a small ratio $u/c$
the system reduces to a simple system using the Chapman--Enskog expansion $f_l^m \sim (u/c)^l f_0^0$ (with $v \simeq c$). In the upstream medium
\begin{eqnarray}
\frac{c}{3} \partial_x f_0^1 + u \partial_x f_0^0 &=& 0 ,\nonumber \\
c \partial_x f_0^0 + 2 \Omega_z Re(f_1^1) &=&-\nu_{\rm s} f_1^0 ,  \nonumber \\ 
-i\Omega_x f_1^1 -\frac{\Omega_z}{2} f_1^0 &=& -\nu_{\rm s} f_1^1 ,
\end{eqnarray}
while in the downstream medium the particle distribution is isotropic, so limited to the term $f_0^0$. The solution at the shock front can be obtained using the continuity of the particle distribution and its flux at the shock front, accounting for the frame transformation across the shock. Looking for a power-law solution $f \propto p^{-\gamma}$, at the leading order in $u/c$ one finds
\begin{equation}
\gamma= 3 + 3 \frac{u_2}{u_{\rm sh}- u_2} \frac{f_0^0(+\infty)}{f_0^0(x=0)} \, ,
\end{equation}
where $u_2$ is the fluid speed and $f_0^0(+\infty)$ the particle isotropic component far downstream. One recovers $\gamma =4$ for a strong unmodified shock.

\paragraph{Oblique shock solutions.} \citet{2011MNRAS.418.1208B} apply the above formalism to an oblique flow, where the magnetic field direction is given by the angle $\theta$ with respect to the shock normal. The procedure requires to start with a guess of the particle distribution at the shock front as a power-law. The index $\gamma$ is calculated iteratively with the full set of equations for $f_l^m$. These are solved using FDM techniques (see Sect.~\ref{S:FPE}) with $f_l^m$ calculated at the cell centers for even~$l$ and at the cell boundaries for odd~$l$. The spatial resolution is finer close to the shock front and coarser at the edge of the simulation box. Non-linear CR backreaction is not considered in this work: the background magnetic field and fluid velocity are not modified by the CR pressure. The shock velocity profile is smoothed using a hyperbolic tangent profile of width $x_{\rm s}$ of 1\% of the particle Larmor radius upstream. The shock index solutions 
show soft distributions at quasi-perpendicular fast shocks. Such configurations can occur if strong magnetic field amplification occurs after the supernova shock breakout \citep{2018MNRAS.479.4470M} or/and if the shock propagates in circumstellar medium with a spiral magnetic field \citep{2008MNRAS.385.1884B}.


\paragraph{CR driven instabilities.} \citet{2013MNRAS.430.2873R, 2013MNRAS.431..415B} couple the VFP method described above to MHD solutions in order to calculate the CR pressure back-reaction over the shock solutions. The authors adopt mixed coordinate frames \citep{1975MNRAS.172..557S} where the particle momenta are evaluated in the local fluid frame. It has the advantage of considerably simplifying the collision term on the r.h.s.\ of Eq.~(\ref{Eq:VFP}), which is parametrized by a scattering frequency $\nu_{\rm s}$. The spherical harmonic expansion is stopped at the first order and $f_0$ and $\vec{f}_1$ are used to evaluate the CR charge density and current to be re-injected into the ideal MHD equations (see Sect.~\ref{S:MHD-eq}). The vector $\vec{f}_1 = f_{\rm x} \vec{e}_{\rm x} + f_{\rm y} \vec{e}_{\rm y} +f_{\rm z} \vec{e}_{\rm z}$, where $f_{\rm x} = f_1^0$, and the perpendicular components are $f_{\rm y} = - 2 {\cal R} f_1^1$ and $f_{\rm z}= 2 {\cal I} f_1^1$. The CR charge and current densities can be defined as 
\begin{eqnarray}
\rho_{\rm CR} &=& 4\pi q\int f_0^0 p^2 dp \nonumber \\
J_{\rm CR} &=& \frac{4\pi}{3} q\int 
\begin{bmatrix}
f_1^0 \\  - 2 {\cal R} f_1^1 \\  2 {\cal I} f_1^1
\end{bmatrix}
p^2 v dp \, .
\end{eqnarray}
The current drives a Lorentz force -$\vec{J}_{\rm CR}/c \times \vec{B}$ and induces a plasma heating -$\vec{J}_{\rm CR}.\vec{E}$. (We will come back on CR-MHD coupling in Sect.~\ref{S:PICMHD}.)\\
The numerical scheme relies on the following approximations:
\begin{enumerate}
    \item The particle distribution follows a power-law.
    \item A ratio  $\nu_{\rm s}/\Omega_{\rm s} < 1$ ensures the closing of the spherical harmonics expansion.
    \item Another component $f_{\rm LS}$ is added to $f_{\rm x}$ in order to mimic the effect of a large scale CR component slowly varying over the simulation domain.
\end{enumerate}
Spatially, the simulations are 3D and use periodic boundary conditions. The CFL condition requires some sub-cycling of the MHD step to account for the CR evolution.

\begin{figure}[htb]
\begin{centering}
\includegraphics[width=1\linewidth]{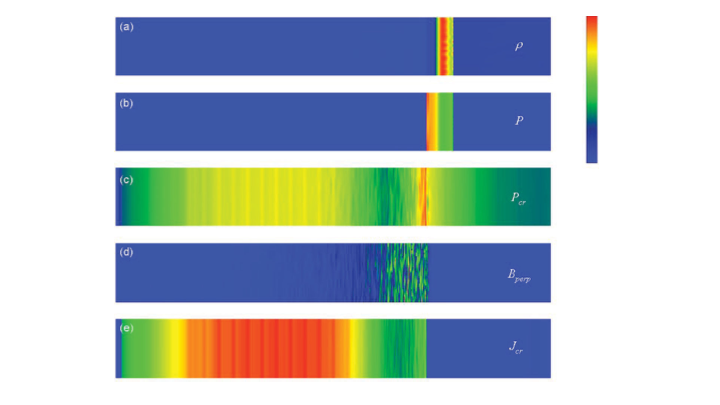}
\par\end{centering}
\caption{2D slices of 3D simulation of the VFP-MHD system. From top to bottom: gas density, gas pressure, CR pressure, magnetic field component perpendicular to the shock normal, CR current  \citep{2013MNRAS.431..415B}. All quantities are obtained at a time of 1.3 year. The scales in the z direction are compressed by a factor 24. \label{fig:Bell13}}
\end{figure}

In Fig.~\ref{fig:Bell13} we show the solution of the 3D VFP-MHD system at a particular time. The simulation has the following setup: a background gas density $n=0.1~{\rm cm}^{-3}$, a background parallel magnetic field $B=47~\mu$G, a shock speed $U_{\rm sh}= 60,000~{\rm km\ s}^{-1}$, the CR are injected at 100 TeV. The background field is oriented in the z direction. The simulation box has 5676 cells in this direction and 32 in the x and y directions. The CR current is initialized along z. Figure~\ref{fig:Bell13} clearly shows two CR populations (see panel~c). A~first one is escaping ahead of the shock and generating a current (see panel~e). The return current in the plasma then triggers non-resonant streaming modes and magnetic field fluctuations which induce a confinement of a second population of CRs at the shock front. This population is accelerated via DSA. \citet{2019ApJ...872...46I} has adapted this procedure to the case of a SNR propagating in a molecular cloud. There TeV CRs are released first and produce through their current enough magnetic perturbations to confine GeV CRs around the shock front. 

\subsection{Particle-in-cell-magnetohydrodynamics}\label{S:PICMHD}

The approach presented in this section combines a MHD approximation necessary to derive the evolution of the background thermal plasma coupled to a PIC module necessary to calculate the trajectories of supra-thermal particles imposed by the Lorentz force. Contrary to standard PIC simulations, now the Lorentz force is calculated using the electro-magnetic field derived from the system of MHD equations rather than Maxwell equations \citep{2000MNRAS.314...65L, 2012MNRAS.419.2433R}. From the distribution of macro particles obtained by the PIC module it is now possible to derive the electric charge and current distribution associated with the supra-thermal particles and to re-inject these two quantities into the MHD equations. This modifies the dynamics of the background plasma. We also need to update the Ohm's law to account for the electric field associated with the energetic particles. \citet{2015ApJ...809...55B, 2018MNRAS.473.3394V, 2018ApJ...859...13M} are three recent works which propose the PIC-MHD approach adapted to the problem of CR in astrophysical flows incluing such a modified Ohm's law. They are based on three different MHD codes: respectively \textsc{Athena}, \textsc{MIP-AMR-VAC}, and \textsc{Pluto}.

The PIC-MHD model assumes global plasma electro-neutrality, i.e., $\sum_\alpha n_\alpha q_\alpha + e(n_{\rm th, i}-n_{\rm th, e})=0$, where the index $\alpha$ runs over the different supra-thermal population species: electrons, positrons, ions and $n_{\rm th,p/e}$ stands for the thermal background electron and ion densities (here we have assumed the plasma to be composed of protons only for simplicity). Non-thermal particles carry a current density $\vec{J}_{\rm EP}= \sum_{\alpha} n_\alpha q_\alpha \vec{u}_\alpha$ to be inserted into the Amp\`ere's law. In the general Ohm's law Hall and electron pressure gradient terms can be safely neglected because they become important only at scales smaller than the ion inertial scale $c/\omega_{\rm pi}$, see \citet{2015ApJ...809...55B}.
The modified Ohm's law can the be written as 
\begin{equation}
    c \vec{E} = \left[(1-R) \vec{u} + R\, \vec{u}_{EP} \right]\times \vec{B} \, ,
\end{equation}
where 
\begin{equation}
    R= \sum_{\alpha} \frac{n_\alpha q_\alpha}{n_{\rm th,e} e}\ 
\end{equation}
is a measure of the relative density of energetic particles (the ideal Ohm's law is recovered by setting $R=0$), and 
\begin{equation}
    \vec{u}_{EP} = \frac{\sum_\alpha n_\alpha q_\alpha \vec{u}_\alpha}{\sum_\alpha n_\alpha q_\alpha} = \frac{\vec{J}_{\rm EP}}{n_{\rm EP} e}
\end{equation}
is the average velocity of the energetic particle population defined in terms of the energetic particles current $\vec{J}_{\rm EP}$ and density $n_{\rm EP}$. Energetic particles (or CRs)  then induce a specific Hall effect. Even if $n_{\rm EP}/n_{\rm th,e} \ll 1$ as it is the case for CRs, the average speed of the non-thermal particles may not be small compare to c, so it has to be retained in the modified Ohm's law.
The Lorentz force to be inserted as a source term in the Euler and energy conservation equations is now:
\begin{equation}
\vec{F}_{\rm EP}=(1-R)\left(n_{\rm EP}\, e ~\vec{u} \times \vec{B} + \frac{\vec{J}_{\rm EP}}{c} \times \vec{B} \right) \, .
\end{equation}
The Lorentz equation which controls supra-thermal particle trajectories can be written as (for particle j of type $\alpha$):
\begin{equation}
    \frac{\partial \vec{p}_{\alpha,j}}{\partial t}= \frac{q_\alpha}{c}\left(\vec{u}_{\alpha,j} -R \vec{u}_{\rm EP}-(1-R) \vec{u} \right) \times \vec{B} \, .
\end{equation}
Again setting $R=0$ gives the standard form of the Lorentz equation. Notice now that the ensemble of energetic particles produce an electric field which modifies the trajectory of each energetic particle.

Energy conservation can be obtained by imposing within each cell the condition:
\begin{equation}
    \sum_\alpha \sum_j n_\alpha \vec{u}_{\alpha,j}. \frac{\partial \vec{p}_{\alpha,j}}{\partial t} = \vec{F}_{\rm EP}.\vec{u}_{\rm EP} \, .
\end{equation}

\subsubsection{Numerical schemes}

All works cited above use a Boris pusher to integrate energetic particle tracks with time \citep{birdsall_langdon}. In order to reach second order accuracy of the PIC-MHD scheme \citet{2015ApJ...809...55B} (see also \citet{2018ApJ...859...13M}) introduce a predictor-corrector scheme to calculate the background plasma evolution under the effects of energetic particles. In the predictor (resp. corrector) part the source terms in the MHD equations are deduced from the location of individual particles to neighboring grid cell centers at the initial time (resp. at half the MHD timestep). The momentum and energy feedback are calculated at cell centers to derive new fluid solutions. This feedback calculation ensures second order accuracy. Aside the corner transport upwind (CTU) scheme \citep{Collela90},  \citet{2018ApJ...859...13M} also introduce a second order Runge-Kutta time stepping method in the predictor-corrector scheme. \citet{2018MNRAS.473.3394V} use a different strategy. Each component, i.e., the MHD fluid and the energetic particles, evolves over their own grid: the MHD grid is used as a base and the PIC grid is offset so that MHD cell centers stand as PIC cell corners.

\subsubsection{Coupling MHD and PIC time steps}

In order to properly resolve each particle trajectory the MHD timestep deduced from the CFL condition has to verify $\Delta t\, r_{\rm g} \le \zeta$, with $\zeta$ a constant to be adjusted and $r_{\rm g}$ the maximum gyroradius of the energetic particles. For instance in the case of a fixed grid, \citet{2015ApJ...809...55B} use $\zeta=0.3$ for shock problem studies. In the case of non-relativistic MHD flows, the kinetic part evolves more rapidly than the MHD part. To treat this issue it is possible to impose some sub-cycling of the PIC step. 
\citet{2015ApJ...809...55B} typically use 10 sub-cycles. \citet{2018ApJ...859...13M} improve this strategy and propose two sub-cycling schemes which recalculate the energetic particle induced Lorentz force at every sub-steps or at even sub-steps.


\subsubsection{Adaptive mesh refinement for the PIC module}

\citet{2018MNRAS.473.3394V} have adopted an adaptive mesh refinement procedure for the PIC part, conversely \citet{2015ApJ...809...55B} use a fixed grid. 
The octree system of adaptive mesh refinement of the \textsc{MPI-AMR-Vac} code is used but it involves an additional refinement condition on the energetic particles: if the number of energetic particles within a grid reaches 25\% of a pre-set maximum the grid is no longer allowed to coarsen. If the number of
particles reaches 80\% of the maximum, the grid is refined (assuming it has not yet reached the maximum allowed refinement level). A~supplementary condition is applied over the average gyroradius of the particles: if within the grid it becomes smaller than a pre-set number of times the size of the
individual grid cells the grid is allowed to be refined. Hence, the particle gyro-radius is always resolved which is a necessary condition for a correct calculation of the source terms in the MHD equations. 

A tricky aspect of mesh refinement appears when a particle is moving from one cell of a given level to another cell of a finer level. In that case it is essential to conserve charge and current. Then, the effective weight of the particle has to be increased by a factor equivalent to the reduction in effective volume. This effective weight has to be carefully calculated on the finer grid if a particle stands at the boundary of two grids at different levels. The physical extent where the weight is calculated is fixed by the coarser mesh, so the calculation of the particle weight may cover several rows (in a 2D view) of the finer grid (see Fig.~1 in \citealt{2018MNRAS.473.3394V}). 

\subsubsection{Numerical tests}

\begin{figure}[htb]
\centerline{\includegraphics[width=1\linewidth]{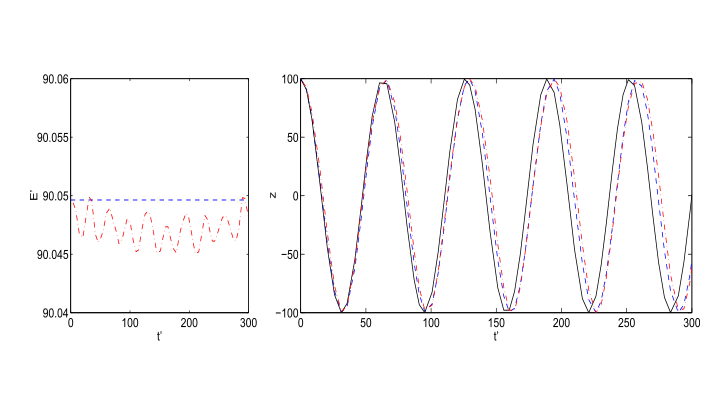}}
\caption{Gyromotion test for a relativistic particle (with a Lorentz factor $\gamma = 10$). The normalized variable time step is $\Delta t \Omega_{\rm L}= 0.5 \pm 0.1$. Dashed blue lines: numerical solutions obtained in the case of a null drift velocity. Red dashed lines: numerical solutions obtained in the case of a drift velocity with strength $u_{\rm d} = U_{\rm A}$. Left panel: time evolution of the particle energy in the co-moving frame (indicated by a prime). Right panel: particle motion in the co-moving frame. The black curve is the analytical solution \citep{2015ApJ...809...55B} . \label{fig:PICMHDT1}}
\end{figure}

\citet{2015ApJ...809...55B} propose two tests of the PIC-MHD method. The first one follows the evolution in time of a test-particle in an uniform magnetic field, so it tests particle gyromotion. Different setups have been tested: non-relativistic and relativistic particles, background plasma moving perpendicularly to the background field. Figure~\ref{fig:PICMHDT1} shows the time evolution of the particle energy in the co-moving plasma frame and the evolution of the particle position in the co-moving frame in the relativistic case. It can be seen that the Boris pusher conserves the energy perfectly in the case of no drift, and to better than 0.1\% level in the case of a small drift (a bit better than in the non-relativistic case as the energy conservation degrades with respect to the relative speed of the drift with respect to the particle speed). The particle trajectory fits with the analytical solution well for long time evolution. \citet{2018ApJ...859...13M} obtain similar results for this test. The authors propose another setup which tests the motion of a particle in a non-orthogonal electric and magnetic field. Energy conservation to 0.1\% level is found for a mildly relativistic particle with respect to the analytical solution (see their Fig.~3).

\begin{figure}[htb]
\centerline{\includegraphics[width=1\linewidth]{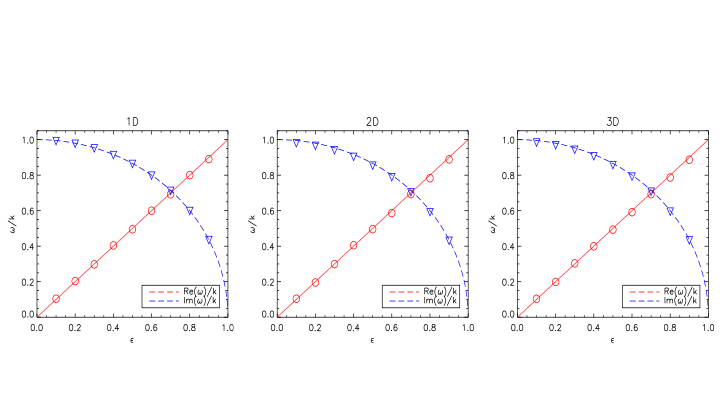}}
\caption{Real and imaginary part of the non-resonant instability relation dispersion as a function of $\epsilon= U_{\rm A}/U_{\rm CR}$. Solid lines: analytical solution. Diamond and circle: numerical solutions. \citep{2018ApJ...859...13M}. 
\label{fig:PICMHDT2}}
\end{figure}

A second setup concerns the test of the CR feedback. It captures the linear growth rate of the non-resonant (or Bell) instability \citep{2015ApJ...809...55B, 2018ApJ...859...13M}.  
In this setup the background plasma is uniform and at rest, it is pervaded by a uniform magnetic field and a CR current is propagating along the background magnetic field with a speed $U_{\rm CR}$. The results are displayed in Fig.~\ref{fig:PICMHDT2}. The plots show the real and imaginary parts of the dispersion relation as a function of the ratio of the background Alfv\'en speed and the CR drift speed. A very good agreement between analytical (solid lines) and numerical (symbols) solutions is obtained.

Finally, \citet{2018ApJ...859...13M} propose a setup to test the relative drift of the thermal gas and CRs. In this test, the two components drift in opposite direction perpendicularly to a background uniform magnetic field. This test is used to evaluate the performances of the predictor-corrector scheme and the impact of the number of CR sub-cyclings. The results show that the predictor step is mandatory in order to reproduce the analytical expected solution and to reach second order accuracy in time.  

\subsubsection{Results on cosmic ray acceleration at non-relativistic shocks}

CR acceleration and CR driven instabilities at SNR shocks can be investigated with the help of the PIC-MHD method, as, in contrast to the case of hybrid techniques, now the PIC part only has to handle supra-thermal particles. This improves a lot particle statistics, which is necessary to properly reconstruct the CR charge and current. 

\paragraph{Simulation setups} 
\citet{2012MNRAS.419.2433R} investigate the development of CR driven instabilities at shocks in the PIC-MHD framework generalizing the work of \citet{2000MNRAS.314...65L} restricted to the resonant streaming instability. In this work the background plasma is at rest and pervaded by a uniform magnetic field augmented with a turbulent spectrum composed of modes in the perpendicular plane. CR are initialized with a mono-energetic distribution, drifting with respect to the background plasma. A~total of 1024 particles per grid cell is used to ensure gyrotropy. The drift speed of CRs corresponds to a shock Alfv\'enic Mach number of initially $M_{\rm A}=10^3$. CRs are injected at an energy that allows the gyromotion to be resolved. The fraction of incoming kinetic energy imparted into CRs is $10^{-4}$.

\citet{2015ApJ...809...55B} also report on fast super-Alfv\'enic shock simulations. Their setup is as follows: a plasma flow is launched against a static, conductive wall (\citet{2018ApJ...859...13M} proceed similarly). Quasi instantaneously after the collision a shock forms which propagates in the upstream medium with a speed $u_{\rm sh} > U_{\rm A,u}$, where $U_{\rm A,u}=B_{\rm u}/\sqrt{4\pi \rho_{\rm u}}$ is the upstream Alfv\'en speed. 
The shock position is identified by evaluating the transversely-averaged fluid speed along the shock propagation direction (let say $x$) $u_{\rm x}$. This speed is further smoothed by a Gaussian kernel of typical size of 4 grid cells. Then the shock position is fixed at the location where $u_{\rm x}$ has decreased by 40\%. Supra-thermal particles are injected as CRs at a kinetic energy of $E_{\rm inj}=10 E_{\rm sh}$ at the shock position. The CR population injected is normalized to a small fraction of the incoming flux. Another important aspect is that the authors use an artificial light speed ${\cal C}$ different from c for CRs but still much larger than any fluid speed in the system. The Lorentz factor of species $\alpha$ of an individual CR is: $\gamma_{\rm \alpha j}={\cal C}/\sqrt{{\cal C}^2-u_{\rm \alpha, j}^2}$. This speed is inserted into the Lorentz equation. It has the advantage to be adapted to the investigation of different particle energy regimes (a large ${\cal C}$ is adapted to follow non- or mildly-relativistic CRs and a small one is adapted to the relativistic regime). \citet{2015ApJ...809...55B} report on the production of non-thermal particle distribution consistent with the Fermi first order process, which is a strong indication that PIC-MHD methods can handle CR scattering off self-generated waves. \citet{2018MNRAS.473.3394V} use a different shock generation procedure. The shock is set up from a configuration following the Rankine-Hugoniot conditions. Particles are then injected following the same procedure as in the Bai et al work. Table~\ref{T:PICMHD} summarizes the setups of these different works.

\begin{table*}
  \caption{Setups for shock acceleration studies. All simulations have been done in 2D space configuration. The spatial resolution for \citet{2018MNRAS.473.3394V} is the basic one, the simulations also include four levels of refinement. The maximum time $t_{\max}$ is in units of $\omega_{\rm c}^{-1}$. The last column displays the shock magnetic obliquity. The second row for \citet{2015ApJ...809...55B} and \citet{2018ApJ...859...13M} shows the setup for relativistic runs with a reduced light speed ${\cal C}$.}
 \begin{footnotesize}
\centering
\begin{tabular}{lcccc} 
\hline
\hline
Authors & \rm{Resolution} & $t_{\max}$ & $M_{\rm A}$ & $\theta_{\rm B}$ \\
        &  ($r_{\rm g}$) & ($\omega_{\rm c}^{-1}$) &  & ($^\circ$) \\
        \hline
\citet{2015ApJ...809...55B} & $1.2~10^5 \times 3000$ &  $3~10^3$  & 30  & 0 \\
                            & $3.89~10^5 \times 4800$ & $1.2~10^4$ & 30 & 0 \\
                            \hline
\citet{2018MNRAS.473.3394V}  & $240 \times 30$  & $2~10^3$ & 3-30-300 & 0-70-90 \\
\hline
\citet{2018ApJ...859...13M} &  $1.2~10^5 \times 3000$ & $3~10^3$ &  30  &   0 \\
                            &  $3.84~10^5 \times 4800$ & $1.1~10^4$ & 30 & 0 \\
\hline

\hline
\hline
\end{tabular}
\end{footnotesize}
\label{T:PICMHD}
\end{table*}

\begin{figure}[htb]
\centerline{\includegraphics[width=1.\linewidth]{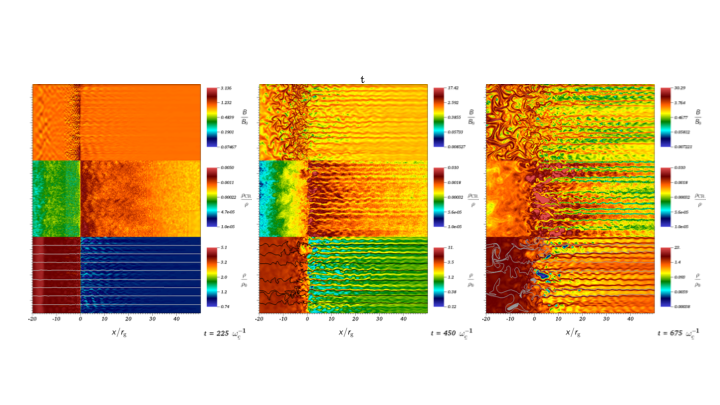}}
\caption{Super-Alfv\'enic parallel shock at time $225\omega_{\rm c}^{-1}$, $450\omega_{\rm c}^{-1}$ and $600\omega_{\rm c}^{-1}$ (from \citet{2018MNRAS.473.3394V}). The Alfv\'enic Mach number is $M_{\rm A}=300$. The upper row shows the magnetic field strength relative to the original background magnetic field. The middle row shows the non-thermal particle charge density relative to the thermal gas density. The lower row shows the  thermal gas mass density relative to the
upstream density at the start of the simulation, combined with
the magnetic field stream lines. Filamentary structures characteristic of the non-resonant streaming instability develop in the upstream medium. The shock front gets strongly corrugated with time.
\label{F:SHC}}
\end{figure}

\begin{figure}[htb]
\centerline{\includegraphics[width=1.1\linewidth]{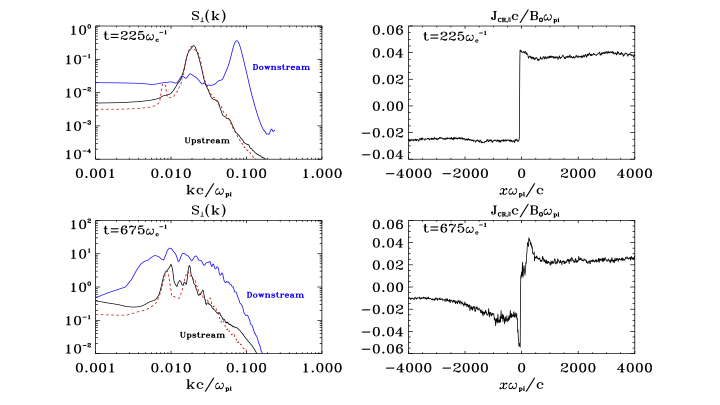}}
\caption{Fourier spectra of the turbulence in the up- and downstream media at times $225\omega_{\rm c}^{-1}$ and $600\omega_{\rm c}^{-1}$ (from \citet{2018MNRAS.473.3394V}). At early times the non-resonant Bell-like mode develops at the maximum wavelength predicted by the linear theory. The turbulent scale is smaller in the downstream medium due to to shock compression of the transverse magnetic field component. At later times non-linear effects start to inject turbulent motions at larger scales and modify the upstream CR current.
\label{F:NRS}}
\end{figure}

\paragraph{Non-resonant CR driven instability studies and shock acceleration}
The main objective of these first studies is the onset of non-resonant (NR) streaming modes in the context of strong shocks.\footnote{\citet{2012MNRAS.419.2433R} discuss a filamentation instability generated by the CRs drifting ahead the shock front. This instability, contrary to the non-resonant streaming instability, is able to generate large scale perturbations which helps to confine high energy CRs.} All these studies have considered the case of parallel super-Alfv\'enic shocks and conclude that indeed CRs trigger a streaming instability in the non-resonant regime (see Fig.~\ref{F:SHC}). The theoretical linear growth timescale is recovered by the simulations. A~Fourier analysis shows the destabilization of the main NR modes in the upstream medium at a scale fixed by the CR current (see Fig.~\ref{F:NRS}). The interest of the PIC-MHD method with respect to previous MHD simulations (e.g., \citealt{2004MNRAS.353..550B, 2008ApJ...678..939Z}) is that it accounts for the time evolution of the CR-MHD including CR back reaction. A clear transfer to smaller wave numbers (hence large scales) can be noticed on Fig.~\ref{F:NRS}. The back-reaction of CR also induces a strong corrugation of the shock front as can be seen on Fig.~\ref{F:SHC}. The corrugation is so strong that it becomes difficult to identify the position of the shock front. Longer term simulations in 3D are required to investigate the transition from a non-relativistic to relativistic regime and to capture the propagation of the CRs in the self-generated turbulence \citep{vanMarle19}. 

\citet{2018MNRAS.473.3394V} have also considered an oblique shock configuration with a magnetic obliquity angle of $70^{\circ}$ with respect to the shock normal. The authors find that particle acceleration is delayed but still present. Particle acceleration proceeds in two steps in this configuration. First SDA occurs that injects a population of charged CRs upstream, which trigger non-resonant streaming modes parallel to the background magnetic field, which in turn perturb the magnetic field downstream. This perturbation leads to a corrugation of the shock front, which changes the orientation of the magnetic field and ultimately allows particles to be injected in a parallel configuration. Here again long-term 3D evolution are necessary to investigate the dynamics of particle acceleration and transport. At perpendicular (super-luminal) shocks the electromotive electric field cannot be compensated upstream and the authors do not observe any particle acceleration. The results on the oblique shock configuration have triggered some discussions. The main argument against is \citep{2014ApJ...783...91C, 
2018JPlPh..84c7101C} that at high obliquity CR injection is not efficient because incoming ions are reflected at the shock overshoot, a structure which can not be captured by a MHD code. The main argument in favour is that hybrid simulations have a small simulation box and cannot be performed over duration much exceeding a few hundreds of $c/\omega_{\rm c,i}$ so they are not able to capture instabilities that develop at larger scales in the downstream medium once the non-resonant streaming instability is triggered. Moreover, hybrid simulations treat the entire ion population (both thermal and supra-thermal) as kinetic and hence have much lower statistics to reconstruct the CR current at the origin of the streaming instability. Both techniques however agree at early timescales for high Alfv\'enic Mach number shocks and both show the development of a supra-thermal tail associated with the SDA mechanism. 

\subsubsection{Other astrophysical applications}

We discuss here recent setups designed to investigate CR or energetic particles transport in specific astrophysical contexts. Two subjects are considered: other type of CR driven instabilities and energetic particles acceleration near X~points in relativistic magnetic reconnection. The number of scientific cases will likely rapidly increase with the availability of numerical resources. 

\paragraph{CR driven instabilities.}
In \citet{2000MNRAS.314...65L} perturbations are produced by the drifting of a mono-energetic or Gaussian distribution of CRs in a background medium at rest. The background magnetic field is composed of a uniform component of strength~B and Alfv\'en waves of amplitude $\delta B =0.1 B$. CR drift with a speed ten times the local Alfv\'en speed. The authors consider the growth of the resonant streaming instability and the energy/impulsion transfer to the background plasma using 1-, 2- and 3-D simulations. Rapid magnetic field generation is obtained. The linear growth rate can be sustained even in the non-linear regime. Magnetic field generation saturates when the level of perturbations are $\sim B$.
  
\citet{2018MNRAS.476.2779L} use a similar numerical approach as the one developed in \citet{2015ApJ...809...55B} (also using the MHD code \textsc{Athena}) to investigate the growth of the gyro-resonant instability produced by an anisotropic pressure of the CR gas with respect to the background magnetic field. If the strength of the magnetic field changes over scales larger than particle Larmor radius then the first adiabatic invariant $p_\perp^2/B$ is conserved along the particle trajectory. In this instability the driving term is given by the relative perpendicular to parallel CR pressure $A=P_\perp/P_\parallel -1$. The simulation setup involves a background uniform magnetic field on which is superimposed a flat spectrum of circularly polarized MHD waves, and an anisotropic CR energy distribution scaling as $E^{-2.8}$ consistent with the Galactic CR spectrum. The main parameters in the simulations are the ratio of CR density to gas density $n_{\rm CR}/n_{\rm g} \ll 1$, and the initial anisotropy parameter $|A| < 1$. The main findings are: the linear growth phase well reproduces the quasi-linear growth rate deduced from \citet{2013SSRv..178..201B}, the non-linear saturation of the instability occurs due to particle isotropization, a process faster at low CR energies which contain most of the CR pressure. \citet{2019ApJ...876...60B} perform PIC-MHD simulations to study the resonant streaming instability. The authors use the $\delta f$ method (see \citealt{Kunz14} and references therein) which consists in affecting a weight $w_{\rm i}$ to each particle~i as: $w_{\rm i}= 1- f(\vec{x}_{\rm i}(t), \vec{p}_{\rm i}(t))/f$. This method permits a drastic reduction of the noise inherently associated with particle simulations. In their fiducial setup the authors inject 2048 particles per cell. The results show that the quasi-linear theory of wave growth is well reproduced for both right- and left-handed mode polarizations. The technique permits to investigate the problem of $90^{\circ}$ pitch-angle scattering which involves non-linear wave particle interactions. 

\paragraph{CR acceleration near an X-point.} 
\citet{2018ApJ...859...13M} propose a study of particle acceleration near a 2D X-point in relativistic flows. The simulations are performed in the test-particle limit. The out-of plane electric $E_{\rm z}$ and guiding magnetic field $B_{\rm z}$ strength are varied. Initially particles are distributed over a Maxwellian with a thermal speed of 0.1 $U_{\rm A}$. Particles are accelerated near the null point where the electric field intensity is the highest. An energy distribution with a power-law approximately $E^{-2}$ is obtained. The particle energy shifts to high energies as the guiding magnetic field strength increases because the acceleration zone where $\vec{E}.\vec{B} \neq 0$ extends with respect to the case when $B_{\rm z}$ is smaller than the in-plane magnetic field strength. 

\subsection{A list of HD and MHD codes with CR physics}

In Table~\ref{Tab:fluid_codes} we present the MHD codes in use in various astrophysical applications which include a module (fluid or kinetic) to treat CR physics, a more complete list can be found in the appendix of \citet{2015LRCA....1....3M}.

\begin{table*}[th]
  \caption{A list of the main fluid codes used in astrophysics. The table displays: the code name, the CR treatment module CR = fluid or PIC-MHD, if or not a relativistic option exists (SR = special relativistic), if or not an AMR version does exist, the geometrical options (Ca= Cartesian, Cy= Cylindrical, Sp= Spherical, Po= Polar coordinates), the parallelization model. Code presentation can be found in \citet{Stone:2008ApJS..178..137S} for \textsc{Athena}, \citet{1999JCoPh.154..284P}  for \textsc{BAT-R-US}, \citet{2014ApJS..211...19B} for \textsc{Enzo},  \citet{Keppens:2012JCoPh.231..718K} for \textsc{MPI-Amrvac}, \citet{Hanasz10}, for \textsc{Piernik}, \citet{Mignone:2007ApJS..170..228M} for \textsc{Pluto}, \citet{2002A&A...385..337T} for \textsc{Ramses}. \label{Tab:fluid_codes}}
 \centering
\begin{tabular}{l|c|c|c|c|c}
 Code name & physics modules & AMR  & geometry & parallelization \\
\hline
 \hline
 \textsc{Athena}  & CR, PIC-MHD & y  & Ca, Cy, Sp, Po & OpenMP/MPI \\
 \hline
  \textsc{BATS-R-US}    & PIC-MHD & y & Ca & OpenMP/MPI \\
   \hline
  \textsc{Enzo} & CR & y & Ca, Sp & MPI \\
   \hline
   \textsc{MPI-Amrvac}    & CR, PIC-MHD, PIC-RMHD   & y     & Ca, Cy, Sp, Po & OpenMP/MPI   \\
   \hline 
   \textsc{Piernik} & CR & y& Cy & MPI &  \\
   \hline 
   \textsc{Pluto}  & CR, PIC-MHD & y  & Ca, Cy, Sp, Po  & OpenMP/MPI   \\
   \hline
    \textsc{Ramses}    &  CR    & y    & Ca & OpenMP/MPI \\
 \hline
  \hline
\end{tabular}
\end{table*}


\subsection{Semi-analytical approaches for cosmic ray acceleration}\label{S:SAN}

Because direct simulations of DSA are numerically expensive, simpler
approaches are desirable for the modelling of large-scale and multi-physics
problems. Most of the methods described above deal with the microphysics
of collisionless shocks and particle acceleration in some way, and are not readily
applicable to simulate a macroscopic object like, say, a supernova remnant. 
The most precise methods, PIC simulations, are restricted
to tiny scales and narrow dynamical ranges. For instance, in the most
advanced numerical simulations of DSA relevant to SNRs to date, by \cite{2014ApJ...794...46C},
lengths are normalized by $l_{p}=c/\omega_{p}$ where $\omega_{p}$
is the ion plasma frequency, which for a typical density $n_{p}=1\,\mathrm{cm^{-3}}$
evaluates to $l_{p}\simeq2\times10^{7}\,\mathrm{cm}=2\times10^{-12}\,\mathrm{pc}$.
So in their figures~6 and~7, the ``far upstream'' reaches about
$3\times10^{-7}\,\mathrm{pc}$, whereas in state of the art simulations
of the 3D evolution of a young SNR, by \cite{2010A&A...509L..10F},
the smallest length resolved is $\simeq5\,\mathrm{pc}/1024\simeq5\times10^{-3}\,\mathrm{pc}$,
nearly 5~orders of magnitude larger. Similarly in the same figures
the range of magnetic turbulence scales probed is $\sim10^{-11}\text{\textendash}10^{-8}\,\mathrm{pc}$,
whereas in one of the most recent study of the effect of turbulence
on the SNR emission, by \cite{2017ApJ...849L..22W}, the smallest resolved
scale is $4\times10^{-2}\,\mathrm{pc}$, 6 to 9 orders of magnitude
larger. It~is therefore necessary to make use of sub-grid models
in simulations of astrophysical objects. In~hydro-kinetic approaches of Sect.~\ref{S:HYKIN},
the complex interaction between the particles and the magnetic turbulence
is encapsulated in the diffusion coefficient~$D$. With such approaches
it is possible to operate on space- and time-scales that are relevant
to an object like a SNR (e.g. \citealt{2015PKAS...30..545K}), although
simulations still have a high computational cost when using a realistic
dependence of~$D$ on~$p$, and to our knowledge their use has been
restricted to 1-dimensional problems (in either slab or spherically symmetric
geometries). When the focus is on properly describing the geometry
of the SNR, and a 3-dimensional modeling is required, the treatment
of DSA needs to be simplified even further.

\subsubsection{General considerations\label{S:SAN-general}}

To fill this need, \cite{1999ApJ...526..385B} proposed a simple analytical
model of NLDSA, where the spectrum of the particles is assumed to
be a three-parts power-law, with slopes linked to the shock properties.
This model was used to make the first studies of the effect of NLDSA
on young SNRs, by coupling it to 1D self-similar solutions \citep{2000ApJ...543L..57D}
then to 1D hydrodynamic simulations \citep{2004A&A...413..189E}. A~more
physical, semi-analytical model was proposed by \cite{2002APh....16..429B},
in the framework of the hydro-kinetic treatment. The~key idea (developed in the next section), which 
allows to greatly simplify the mathematics, is that the energy-dependent
diffusion of the particles allows to establish a one-to-one correspondence
between the particle energy~$E$ (or equivalently momentum~$p$),
the position variable~$x$, and the fluid velocity~$u$. This trick
was already used in \citet{1979ApJ...229..419E} and developed by
\citet{1984ApJ...277..429E, 1984ApJ...286..691E, 1996APh.....5..367B}, but somehow its usefulness was not fully
realized until Blasi (\citeyear{2004APh....21...45B,2005MNRAS.361..907B}) published their NLDSA model. 

Being physically motivated while computationally extremely fast, Blasi's model
was quickly adopted by the community doing SNR simulations: it replaced
the Berezhko and Ellison model in the 1D hydrodynamic simulations
of \cite{2007ApJ...661..879E} and following works, and it was used
for the 3D hydrodynamical simulations of \cite{2010A&A...509L..10F}
and following works. It allowed \cite{2014MNRAS.443.1390K} to perform 
parametric studies of the efficiency of CR acceleration in young SNRs.
Note that an important limitation of the model
is that it is looking only for stationary solutions, and so needs
to be re-run at each time step in order to compute the time evolution
of the coupled shock-particles system, assuming quasi-stationarity
is reached at each step. This is justified at most energies, but will
break down close to the highest energies when the acceleration time
becomes of the same order as the age of the simulated shock. 
The model jointly solves the particle spectrum and the
fluid velocity profile as functions of the momentum of particles.
As inputs, it requires basic information on the shock (speed~$V_{\rm sh}$
and Mach number~$M_{\rm sh}$), which can be determined from a hydrodynamic simulation,
as well as an injection recipe at some $p_{\mathrm{inj}}$ and a cutoff
recipe to set $p_{\max}$. Amongst the outputs, it provides
the total shock compression ratio~$r_{\mathrm{tot}}$, that can be
used to determine an effective adiabatic index~$\gamma_{\mathrm{eff}}$
for the fluid+particles system. The back-reaction of the particles
on the flow can then be imposed by tweaking the value of~$\gamma$
in the hydro model according to the prediction of the NLDSA model. 
\cite{2004A&A...413..189E} showed good agreement between 
this pseudo-fluid approach and two-fluid calculations in 1D. 
It is worth mentioning that in the original (and most popular) version
of Blasi's model one does not deal explicitly with the diffusion
coefficient~$D$ (although often an assumption on $D(p)$ is made in order to estimate
$p_{\max}(t)$). Accordingly, hydrodynamic simulations typically
do not need to explicitly resolve the shock precursor generated by
the particles, although for a given $D(p)$ law the velocity profile
in the precursor may be reconstructed if desired (the position~$x_{p}$ where
the fluid velocity is~$u_{p}$ is given by $x_{p}=D\left(p\right)/u_{p}$).
In a subsequent model \cite{2005MNRAS.364L..76A} introduced
the explicit spatial dependence of the distribution~$f$ and the
diffusion coefficient~$D$. This generalization is more complex to
derive and significantly longer to compute, and \cite{2008MNRAS.385.1946A}
showed that the two models provide similar results. 

The simple model was gradually improved to incorporate other physical
processes. Of particular importance is the improvement of the treatment
of the magnetic turbulence \citep{2008ApJ...679L.139C,2009MNRAS.395..895C}.
A~recipe for magnetic field amplification (MFA) was included, 
and the fate of magnetic waves generated
by the particles was considered: they may either be damped in the
plasma upstream of the shock or be carried through the shock, which leads to very different magnetic fields in the downstream region, although similar overall
levels of back-reaction on the shock. These effects were included in
SNR simulations as well (by \citealt{2012ApJ...750..156L} in 1D and \citealt{2014ApJ...789...49F} in 3D). 
Another point of interest, regarding SNRs as sources of cosmic
rays, is the role of particles escaping the accelerator. In the 
base model only the spectrum at the shock is computed and escape is
treated implicitly, but if desired the escape of particles can be
treated explicitly, in two ways \citep{2009MNRAS.396.2065C,2010APh....33..307C}.
Other recent developments include the role of ionization \citep{2011MNRAS.412.2333M}
and the role of neutrals \citep{2013A&A...558A..25M} in the DSA process. 

\begin{figure}[htbp]
\begin{centering}
\includegraphics[width=\textwidth]{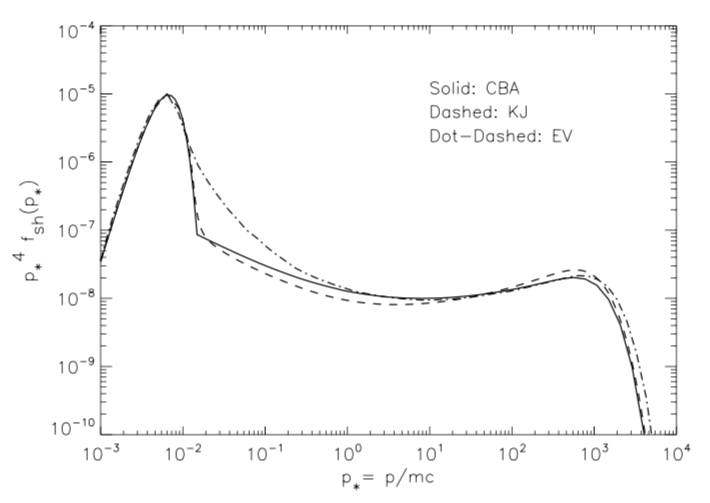}
\includegraphics[width=\textwidth]{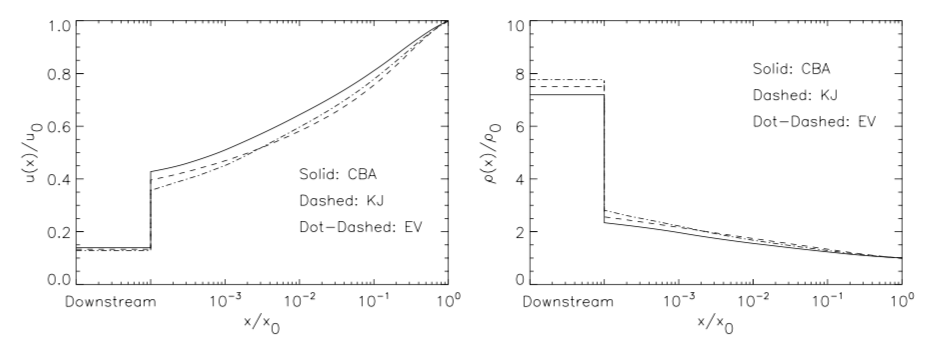}
\par\end{centering}
\caption{Comparison of different methods for non-linear diffusive shock acceleration (taken from \citealt{2010MNRAS.407.1773C}). The plot at the top shows the distribution function $f(p)$ of particles, the Maxwellian component and a non-thermal tail are apparent. The two plots at the bottom show the hydro profiles of the shock: velocity $u(x)$ and density $\rho(x)$. The models assume efficient acceleration and include backreaction effects: the reduced sub-shock and extended precursor are apparent, and accordingly the non-thermal spectrum is concave. The three different sets of curves show the results of three different methods to solve the coupled fluid and particles system: a semi-analytical model (solid lines, labelled CBA for Caprioli--Blasi--Amato, described in Sect.~\ref{S:SAN-Blasi}), a numerical model coupling hydro equations and a diffusion-convection equation (dashed lines, labelled KJ for Kang \& Jones, described in Sect.~\ref{S:HYKIN}), and another numerical model relying on a Monte Carlo approach (dot-dashed lines, labelled EV for Ellison \& Vladimirov, described in Sect.~\ref{S:MCSHOCK}).
Copyright MNRAS.}
\label{fig:NLDSAcomp}
\end{figure}

A~comparison between Blasi's semi-analytical model and two popular numerical approaches, 
hydro-kinetic simulations {\`a} la Kang \& Jones (Sect.~\ref{S:HYKIN}) 
and Monte Carlo simulations {\`a} la Ellison \& Eichler (Sect.~\ref{S:MCSHOCK}), 
can be found in \cite{2010MNRAS.407.1773C}. 
Typical results are shown in Fig.~\ref{fig:NLDSAcomp}.

In the following last section, we outline the inner workings of Blasi's NLDSA model. 

\subsubsection{Construction of a NLDSA model\label{S:SAN-Blasi}}

We restrict ourselves to 1D slab geometry along direction~$x$, with
a velocity discontinuity (sub-shock) located at $x=0$, and a velocity
ramp (precursor) extending for $x<0$ over a distance $x_{\max}$.
Subscripts distinguish between three distinct media, with usual notations:
0~denotes the far upstream (unperturbed) medium ($x<-x_{\max}$),
1~denotes the region immediately upstream of the sub-shock ($x=0^{-}$),
and 2~the region downstream of the sub-shock ($x=0^{+}$).

\paragraph{Distribution of accelerated particles at the shock\label{S:SAN-Blasi-f}}

The evolution of the distribution function~$f$ is described by a
convection-diffusion equation which, under the assumption of stationarity 
${\partial f}/{\partial t}=0$ reads 
\begin{equation}
-u\frac{\partial f}{\partial x}+\frac{\partial}{\partial x}\left(D\frac{\partial f}{\partial x}\right)+\frac{1}{3}\frac{\mathrm{d}u}{\mathrm{d}x}p\frac{\partial f}{\partial p}+Q=0\label{eq:transport_f}
\end{equation}
where~$D$ is the diffusion coefficient, and~$Q$ represents injection
of particles, assumed to occur only at the shock front: 
$Q(x,p)=Q_{1}(p)\,\delta(x)$.
By integrating Eq.~(\ref{eq:transport_f}) across the shock (from
$x=0^{-}$ to $x=0^{+}$), using the continuity of the distribution
function $f_{2}=f_{1}$, then integrating it from far upstream
($x<-x_{\max}$) to just ahead of the sub-shock ($x=0^{-}$), 
and assuming homogeneity downstream and of the shock: 
$\left({\partial f}/{\partial x}\right)_{2}=0$, 
the equation can be recast as 
\begin{equation}
\frac{1}{3}\left(u_{p}-u_{2}\right)p\frac{\mathrm{d}f_{1}}{\mathrm{d}p}-\left(u_{p}+\frac{1}{3}p\frac{\mathrm{d}u_{p}}{\mathrm{d}p}\right)f_{1}+\left(u_{0}f_{0}+Q_{1}\right)=0\:,\label{eq:df[0,1]}
\end{equation}
where we have introduced the key function 
\begin{equation}
u_{p}=u_{1}-\frac{1}{f_{1}}\int_{-x_{\max}}^{0^{-}}\mathrm{d}x\:\frac{\mathrm{d}u}{\mathrm{d}x}\:f\left(x,p\right)\:,\label{eq:u(p)}
\end{equation}
which is the average fluid velocity experienced by particles with
momentum~$p$ while diffusing upstream of the shock front. Assuming
that~$D$ is a growing function of~$p$, particles of a given momentum~$p$
explore a region of a certain extent~$x_{p}$ upstream of the shock,
and thus sample only a part of the precursor in the velocity profile.
Hence $u_{p}$ can be thought of as being the typical velocity of the fluid
at the point~$x_{p}$ that particles of momentum~$p$ can reach. 

The solution of Eq.~(\ref{eq:df[0,1]}) can be written
in implicit form
\begin{equation}
f_{1}\left(p\right)=\frac{3}{U_{p}-R_{\mathrm{tot}}^{-1}}\int_{p_{\min}}^{p}\frac{\mathrm{d}p'}{p'}\:\left(f_{0}\left(p'\right)+\frac{Q_{1}\left(p'\right)}{u_{0}}\right)\:\exp\left(-\int_{p'}^{p}\frac{\mathrm{d}p''}{p''}\:\frac{3U_{p''}}{U_{p''}-R_{\mathrm{tot}}^{-1}}\right)\label{eq:f(U)}
\end{equation}
where we have noted $p_{\mathrm{min}}$ the minimum momentum of particles
and we have introduced the total compression of the shock
$R_{\mathrm{tot}}={u_{0}}/{u_{2}}$
and, in a similar way, the normalized velocity everywhere in the precursor:
$U_{p}={u_{p}}/{u_{0}}$.

If we further assume that there are no pre-existing particles: $f_{0}=0$,
and that a fraction~$\eta$ of the particles crossing the sub-shock
are ``injected'' in the acceleration process at a single momentum~$p_{\mathrm{inj}}$:
\begin{equation}
Q_{1}\left(p\right)=\frac{\eta n_{1}u_{1}}{4\pi p_{\mathrm{inj}}^{2}}\,\delta\left(p-p_{\mathrm{inj}}\right)\:,\label{eq:Q_inj}
\end{equation}
then Eq.~(\ref{eq:f(U)}) simplifies to
\begin{equation}
f_{1}\left(p\right)=\frac{3}{U_{p}-R_{\mathrm{tot}}^{-1}}\:\frac{\eta n_{0}}{4\pi p_{\mathrm{inj}}^{3}}\:\exp\left(-\int_{p_{\mathrm{inj}}}^{p}\frac{\mathrm{d}p'}{p'}\:\frac{3U_{p'}}{U_{p'}-R_{\mathrm{tot}}^{-1}}\right)\:.\label{eq:f0(p,U(p))_int}
\end{equation}
The injection momentum~$p_{\mathrm{inj}}$ can be parametrized as
\begin{equation}
p_{\mathrm{inj}}=\xi\:p_{\mathrm{th},2}\:.\label{eq:p_inj}
\end{equation}
where $p_{\mathrm{th},2}=\sqrt{2m_{p}kT_{2}}$ is the mean downstream
thermal momentum. Continuity of the thermal and non-thermal distributions
at $p_{\mathrm{inj}}$ imposes that
\begin{equation}
\eta=\frac{4}{3\sqrt{\pi}}\left(R_{\mathrm{sub}}-1\right)\xi^{3}\exp\left(-\xi^{2}\right)\label{eq:eta(xi)}
\end{equation}
where we have introduced the compression of the sub-shock
$R_{\mathrm{sub}}={u_{1}}/{u_{2}}$.
The factor $R_{\mathrm{sub}}-1$ acts as a regulator: injection is
switched off when the sub-shock gets smoothed. We have a single parameter~$\xi$
to describe injection, but note that the value of~$\eta$ is extremely
sensitive to the value of~$\xi$.

\paragraph{Velocity profile of the fluid in the precursor\label{S:SAN-Blasi-u}}

In the previous paragraph we have expressed the distribution of accelerated
particles~$f$ as a function of the velocity profile of the thermal
fluid~$U$. As particles back-react on the shock dynamics, $U$~is
itself a function of~$f$. To find this second relation, we make
use of conservation of momentum, which involves 4~terms: dynamical
pressure~$\rho u^{2}$, thermal pressure~$P_{\mathrm{th}}$, non-thermal
pressure~$P_{\mathrm{CR}}$, and waves pressure~$P_{\mathrm{w}}$.
We write it from a point far upstream ($x<-x_{\max}$), where
the fluid velocity is~$u_{0}$, to the point~$x_{p}$ (reached by
particles of momentum~$p$), where the fluid velocity is~$u_{p}$:
\begin{equation}
\rho_{p}u_{p}^{2}+P_{\mathrm{th},p}+P_{\mathrm{cr},p}+P_{\mathrm{w},p}=\rho_{0}u_{0}^{2}+P_{\mathrm{th},0}+P_{\mathrm{cr},0}+P_{\mathrm{w},0}\:.\label{eq:cons_rho.u=00005B0,p=00005D}
\end{equation}

The upstream fluid pressure~$P_{\mathrm{th},0}$ can be expressed as
\begin{equation}
{P_{\mathrm{th},0}}{}=\frac{\rho_{0}u_{0}^{2}}{\gamma_{\mathrm{th}}M_{S,0}^{2}}\:.\label{eq:P_th,0}
\end{equation}
Assuming adiabatic compression, the fluid pressure~$P_{\mathrm{th},p}$ 
at any point~$x_{p}$ in the precursor is given by 
\begin{equation}
\frac{P_{\mathrm{th},p}}{P_{\mathrm{th},0}}=\left(\frac{\rho_{p}}{\rho_{0}}\right)^{\gamma_{\mathrm{th}}}=U_{p}^{-\gamma_{\mathrm{th}}}\label{eq:P_th_p}
\end{equation}
where in the second equality we have made use of the conservation
of mass. 
Various processes may lead to non-adiabatic compression in the precursor. 
One of the most discussed is heating through the damping of Alfv\'en waves, 
for which \cite{1999ApJ...526..385B} propose the following recipe (obtained for large $M_{A,0}$):
\begin{equation}
\frac{P_{\mathrm{th},p}}{P_{\mathrm{th},0}}=U_{p}^{-\gamma_{\mathrm{th}}}\left(1+\zeta\left(\gamma_{\mathrm{th}}-1\right)\frac{M_{S,0}^{2}}{M_{A,0}}\left(1-U_{p}^{\gamma_{\mathrm{th}}}\right)\right)\label{eq:P_th_p_Alfven}
\end{equation}
where $M_{A,0}$ is the upstream Alfv\'enic Mach number, and $\zeta\in[0,1]$ is a free parameter added by \cite{2009MNRAS.395..895C} and discussed more below.

The particle pressure 
$P_{\mathrm{cr},p} = P_{\mathrm{cr},p,0}+P_{\mathrm{cr},p,1}$ 
at point~$x_{p}$ is the sum of two terms. 
The first term is the pressure contributed by the adiabatic compression of an upstream population~$f_{0}$:
\begin{equation}
P_{\mathrm{cr},p,0}=U_{p}^{-\gamma_{\mathrm{cr}}}\times P_{\mathrm{cr},0}\label{eq:P_CR,ext_p}
\end{equation}
where $\gamma_{\mathrm{cr}}\simeq4/3$ is the adiabatic index of the particles ``fluid''
and
\begin{equation}
P_{\mathrm{cr},0}=\int_{p_{\min,0}}^{p_{\max,0}}\frac{p'v\left(p'\right)}{3}\,f_{0}\left(p'\right)\,\mathrm{4\pi p'^{2}d}p'=\frac{4\pi}{3}m_{p}c^{2}\int_{p_{\min,0}}^{p_{\max,0}}\frac{p'^{4}\,f_{0}\left(p'\right)}{\sqrt{1+p'^{2}}}\,\mathrm{d}p'\label{eq:P_CR,ext_0}
\end{equation}
(with momenta expressed in $m_{p}c^{2}$ units in the right expression).
The second term is is the pressure of the particles accelerated at the shock 
(with distribution~$f_{1}\left(p\right)$ extending up to $p_{\max,1}$) 
and able to reach the position~$x_{p}$ (that is  those of momenta $\geq p$):
\begin{equation}
P_{\mathrm{cr},p,1}=\int_{p}^{p_{\max,1}}\frac{p'v\left(p'\right)}{3}\,f_{1}\left(p'\right)\,\mathrm{4\pi p'^{2}d}p'=\frac{4\pi}{3}m_{p}c^{2}\int_{p}^{p_{\max,1}}\frac{p'^{4}\,f_{1}\left(p'\right)}{\sqrt{1+p'^{2}}}\,\mathrm{d}p'\:.\label{eq:P_CR,int_p}
\end{equation}

Finally we turn to the pressure in magnetic waves. 
Far upstream, we assume that the magnetic field is not turbulent:
$P_{\mathrm{w},0}=0\label{eq:Pw,0}$. 
In the precursor, particles are believed to generate themselves the turbulence required for their scattering, hence as a first approach we may parametrize the pressure of waves as being some fraction~$\alpha<1$ of the pressure of particles :
\begin{equation}
P_{\mathrm{w},p}=\alpha\:P_{\mathrm{cr},p}\:.\label{eq:Pw,p(Pcr,p)}
\end{equation}
According to quasi-linear theory, $\alpha\sim v_{A}/u_{0}$ for the
resonant streaming instability, and $\alpha\sim u_{0}/c$ for the
non-resonant modes. \cite{2009MNRAS.395..895C} propose the following
recipe for the resonant instability (obtained for large $M_{S,0}$
and $M_{A,0}$):
\begin{equation}
\frac{P_{\mathrm{w},p}}{\rho_{0}u_{0}^{2}}=\frac{1-\zeta}{4M_{A,0}}\,U_{p}^{-3/2}\left(1-U_{p}^{2}\right)\:.\label{eq:Pw,p_Alfven}
\end{equation}
The term $U_{p}^{-3/2}$ represents adiabatic compression. The factor~$1-\zeta$
is introduced to balance the factor~$\zeta$ in relation~(\ref{eq:P_th_p_Alfven}): the amount of wave damping has to remain reasonably small for the magnetic field to be substantially amplified. 
Different parametrizations of~$\delta B$
in the precursor are possible, see \cite{2013ApJ...777...25K} for
a comparison of four different models.

Using the above relations for~$P_{\mathrm{th}}$, $P_{\mathrm{cr}}$,
$P_{\mathrm{w}}$, together with mass conservation,
Eq.~(\ref{eq:cons_rho.u=00005B0,p=00005D}) can be written as
\begin{eqnarray}
U_{p}+\left(\begin{array}{l}
{\displaystyle \frac{U_{p}^{-\gamma_{\mathrm{th}}}}{\gamma_{\mathrm{th}}M_{S,0}^{2}}\left(1+\zeta\left(\gamma_{\mathrm{th}}-1\right)\frac{M_{S,0}^{2}}{M_{A,0}}\left(1-U_{p}^{\gamma_{\mathrm{th}}}\right)\right)}\\
{\displaystyle +\frac{U_{p}^{-\gamma_{\mathrm{cr},0}}}{\gamma_{\mathrm{th}}M_{S,0}^{2}}\frac{P_{\mathrm{cr},0}}{P_{\mathrm{th},0}}+\frac{4\pi}{3}\frac{m_{p}c^{2}}{\rho_{0}u_{0}^{2}}\int_{p}^{p_{\max,1}}\frac{p'^{4}\,f_{1}\left(p'\right)}{\sqrt{1+p'^{2}}}\,\mathrm{d}p'}\\
{\displaystyle +\frac{1-\zeta}{4M_{A,0}}\,\frac{1-U_{p}^{2}}{U_{p}^{3/2}}}
\end{array}\right)&& \nonumber \\
=1+\frac{1} {\gamma_{\mathrm{th}}M_{S,0}^{2}}\left(1+\frac{P_{\mathrm{cr},0}}{P_{\mathrm{th},0}}\right).&&
\label{eq:cons_rho.u=00005B0,p=00005D_bis}
\end{eqnarray}
Deriving relation~(\ref{eq:cons_rho.u=00005B0,p=00005D_bis}) with
respect to~$p$, we finally obtain

\begin{eqnarray}
\left(1-\frac{U_{p}^{-\left(\gamma_{\mathrm{th}}+1\right)}}{M_{S,0}^{2}}\left(1+\zeta\left(\gamma_{\mathrm{th}}-1\right)\frac{M_{S,0}^{2}}{M_{A,0}}+\frac{\gamma_{\mathrm{cr},0}}{\gamma_{\mathrm{th}}}\frac{P_{\mathrm{cr},0}}{P_{\mathrm{th},0}}\frac{U_{p}^{-\gamma_{\mathrm{cr},0}}}{U_{p}^{-\gamma_{\mathrm{th}}}}\right)-\frac{1-\zeta}{8M_{A,0}}\frac{U_{p}^{2}+3}{U_{p}^{5/2}}\right)\nonumber \\
\times \frac{\mathrm{d}U_{p}}{\mathrm{d}p} =\frac{4\pi}{3}\frac{m_{p}c^{2}}{\rho_{0}u_{0}^{2}}\frac{p^{4}f_{1}\left(p\right)}{\sqrt{1+p^{2}}}\:.\:\:\label{eq:U(f)}
\end{eqnarray}
The distribution of particles~$f_{1}$ at the shock being known,
the velocity profile~$U_{p}$ of the fluid can be computed by integrating
Eq.~(\ref{eq:U(f)}) from one of these two boundary conditions to the other:
\begin{equation}
U_{p}\left(p=0\right)=U_{p}\left(x=0^{-}\right)=\frac{1}{R_{\mathrm{prec}}}\:,\label{eq:BC_U_min}
\end{equation}
\begin{equation}
U_{p}\left(p=p_{\max}\right)=U_{p}\left(x=-x_{\max}\right)=1\:,\label{eq:BC_U_max}
\end{equation}
where we have introduced the compression factor of the whole precursor
$R_{\mathrm{prec}}={u_{0}}/{u_{1}}$. 
In practice, we will be looking for a $R_{\mathrm{prec}}$ such that, starting
from condition~(\ref{eq:BC_U_min}), condition~(\ref{eq:BC_U_max}) is matched after integration of Eq.~(\ref{eq:U(f)}).

\paragraph{Compression at the sub-shock\label{S:SAN-Blasi-r}}

So far we have expressed~$f_{1}$ as a function of~$U_{p}$, $R_{\mathrm{tot}}$
and~$R_{\mathrm{sub}}$ (Eq.~(\ref{eq:f(U)}) with injection
recipe~(\ref{eq:Q_inj})-(\ref{eq:eta(xi)})), and~$U_{p}$ as a
function of~$f_{1}$ and~$R_{\mathrm{prec}}$ (Eq.~(\ref{eq:U(f)})
with boundary conditions~(\ref{eq:BC_U_min})-(\ref{eq:BC_U_max})),
To solve the coupled system $f_{1}-U_{p}$, we need another independent relation between any two of the three compression ratios $R_{\mathrm{prec}}$,
$R_{\mathrm{sub}}$ and~$R_{\mathrm{tot}}$ (the third one being deduced from $R_{\mathrm{tot}}=R_{\mathrm{prec}}\times R_{\mathrm{sub}}$).
To obtain this relation, we once again use conservation of momentum,
this time across the sub-shock (from $x=0^{-}$ to $x=0^{+}$):
\begin{equation}
\rho_{2}u_{2}^{2}+P_{\mathrm{th},2}+P_{\mathrm{cr},2}+P_{\mathrm{w},2}=\rho_{1}u_{1}^{2}+P_{\mathrm{th},1}+P_{\mathrm{cr},1}+P_{\mathrm{w},1}\:.\label{eq:cons_rho.u[1,2]}
\end{equation}

The pressure of accelerated particles is always continuous across
the shock:
$P_{\mathrm{cr},2}=P_{\mathrm{cr},1}$.
For the magnetic waves, using a simplified treatment of the parallel shock and in the limit of large Alfv\'enic numbers \cite{2008ApJ...679L.139C,2009MNRAS.395..895C} estimate the jump in pressure $P_{\mathrm{w}}$ and in energy flux $F_{\mathrm{w}}$ to be
\begin{eqnarray}
\left[P_{\mathrm{w}}\right]_{1}^{2} & = & R_{\mathrm{sub}}^{2}-1\:, \label{eq:Pw,2/Pw,1(Rsub)}\\
\left[F_{\mathrm{w}}\right]_{1}^{2} & = & 2\left(R_{\mathrm{sub}}-1\right)\,P_{\mathrm{w},1}\,u_{1}\:.
\label{eq:[Fw]12(Rsub)}
\end{eqnarray}
Then the jump in fluid pressure at the shock is
\begin{equation}
\frac{P_{\mathrm{th},2}}{P_{\mathrm{th},1}}=\frac{\left(\gamma_{\mathrm{th}}+1\right)R_{\mathrm{sub}}-\left(\gamma_{\mathrm{th}}-1\right)\left(1-\left(R_{\mathrm{sub}}-1\right)^{3}\frac{P_{\mathrm{w},1}}{P_{\mathrm{th},1}}\right)}{\left(\gamma_{\mathrm{th}}+1\right)-\left(\gamma_{\mathrm{th}}-1\right)R_{\mathrm{sub}}}\:.\label{eq:Pth,2/Pth,1_Pw,1}
\end{equation}
Substituting relations~(\ref{eq:Pth,2/Pth,1_Pw,1}) and~(\ref{eq:Pw,2/Pw,1(Rsub)}) in Eq.~(\ref{eq:cons_rho.u[1,2]}), and using mass conservation, we obtain
\begin{equation}
M_{S,1}^{2}=\frac{2R_{\mathrm{sub}}}{\left(\gamma_{\mathrm{th}}+1\right)-\left(\gamma_{\mathrm{th}}-1\right)R_{\mathrm{sub}}-2R_{\mathrm{sub}}\left(\gamma_{\mathrm{th}}-\left(\gamma_{\mathrm{th}}-2\right)R_{\mathrm{sub}}\right)P_{\mathrm{w},1}^{\star}}\label{eq:Ms,1(Rsub)_waves}
\end{equation}
where we have introduced the sonic Mach number~$M_{S,1}$ of the
sub-shock, 
and where we have noted
\begin{equation}
P_{\mathrm{w},1}^{\star}=\frac{P_{\mathrm{w},1}}{\rho_{1}u_{1}^{2}}=R_{\mathrm{prec}}\frac{P_{\mathrm{w},1}}{\rho_{0}u_{0}^{2}}=\frac{1-\zeta}{4M_{A,0}}\,R_{\mathrm{prec}}^{5/2}\left(1-R_{\mathrm{prec}}^{-2}\right)\label{eq:Pw,1/Pd,1}
\end{equation}
where we used recipe~(\ref{eq:Pw,p_Alfven}) for the last equality.
In the case where the pressure of magnetic waves is negligible ($P_{\mathrm{w},1}\simeq0$,
that is $\zeta\simeq1$ with recipe~(\ref{eq:Pw,p_Alfven})), Eq.~(\ref{eq:Ms,1(Rsub)_waves})
reduces to the well-known hydrodynamics relation
\begin{equation}
M_{S,1}^{2}=\frac{2R_{\mathrm{sub}}}{\left(\gamma_{\mathrm{th}}+1\right)-\left(\gamma_{\mathrm{th}}-1\right)R_{\mathrm{sub}}}\Longleftrightarrow R_{\mathrm{sub}}=\frac{\left(\gamma_{\mathrm{th}}+1\right)M_{S,1}^{2}}{\left(\gamma_{\mathrm{th}}-1\right)M_{S,1}^{2}+2}\:.\label{eq:Ms,1(Rsub)_hydro}
\end{equation}
In the case $P_{\mathrm{w},1}>0$, Eq.~(\ref{eq:Ms,1(Rsub)_waves}) is a quadratic relation for $R_{\mathrm{sub}}$ 
as a function of~$M_{s,1}$ and~$P_{\mathrm{w},1}^{\star}$, and
thus of~$R_{\mathrm{prec}}$. We can thus solve the sub-shock.

\paragraph{Alfv\'enic drift\label{S:SAN-Blasi-drift}}

At this point we should make the distinction between the velocity
of the flow~$u$, and the velocity of the scattering centers~$\tilde{u}$.
In the MFA picture Alfv\'en waves are generated by particles counter-streaming
the flow, so that in the precursor 
$\tilde{u}_{p}=u_{p}-v_{A,p}$
where $v_{A,p}$ is the Alfv\'en speed at the location reached by particles
of momentum~$p$, while in the downstream region
$\tilde{u}_{2}=u_{2}+v_{A,2}$.
It is this velocity~$\tilde{u}$ that should be used in the transport
equation for the particles. 
The effective velocity jumps experienced by the particles
are smaller than $R_{\mathrm{sub}}$ and $R_{\mathrm{tot}}$, which leads to steeper spectra. 
In the preceding derivation we have assumed $u\simeq\tilde{u}$ for
simplicity, but if MFA is efficient this may not be true. 
Now a difficulty is that, when the magnetic field is strongly turbulent, 
it is not clear how the waves speed should be calculated.
The common approach (used e.g. by \cite{2012ApJ...750..156L,2012JKAS...45..127K,2014ApJ...789...49F})
is to parametrize the Alfv\'enic drift in the form
\begin{equation}
v_{A,p}=\frac{B_{0}+f_{A}\times\left(B_{p}-B_{0}\right)}{\sqrt{4\pi\rho}}
\end{equation}
where 
$B_{p}=\sqrt{B_{0}^{2}+\delta B_{p}^{2}}$
is the total magnetic field at point~$x_{p}$ and we have introduced
the free parameter $f_{A}\in\left[0,1\right]$. In this model,
MFA is thus described by two free parameters~$\zeta$ (equations~(\ref{eq:P_th_p_Alfven})
and~(\ref{eq:Pw,p_Alfven})) and~$f_{A}$. 
The Alfv\'enic drift is an important correction when $\zeta$ is close to~$0$ and $f_{A}$ is close to~$1$.

\paragraph{Escaping flux\label{S:SAN-Blasi-escape}}

A steady-state solution to the problem can only exist if particles
can escape above some maximum momentum~$p_{\max}$, or upstream
of some maximum diffusion length~$x_{\max}$. In the model,
the two approaches are equivalent, the two quantities being related
through the relation $x_{\max}=D\left(p_{\max}\right)/u_{0}$.
However, the two approaches do not provide the same information on the escape of particles: imposing $f\left(x_{\max}\right)=0$
allows to compute the energy spectrum~$\phi_{0}\left(p\right)$ of
particles leaving the shock around~$p_{\max}$, whereas imposing
$f\left(p_{\max}\right)=0$ only allows to compute the net
(integrated) energy flux~$F_{\mathrm{esc},0}$ at the boundary~$x_{\max}$.

If one integrates Eq.~(\ref{eq:transport_f}) from the point
$x_{\max}$ where particles are supposed to leave the system,
defined so that
$f\left(x_{\max}\right)=0$, 
then a new term~$\phi_{0}$ appears on the l.h.s. of Eq.~(\ref{eq:df[0,1]})~:
\begin{equation}
\phi_{0}=-D\left(\frac{\partial f}{\partial x}\right)_{0}\:,\label{eq:phi_0}
\end{equation}
which is the flux of particles leaving the system through the boundary~$x_{\max}$.
Assuming as before that no seed particles are present upstream and
that injection at the shock front is mono-energetic, the solution~(\ref{eq:f0(p,U(p))_int})
becomes
\begin{equation}
f_{1}\left(p\right)=\frac{3}{U_{p}-R_{\mathrm{tot}}^{-1}}\:\frac{\eta n_{0}}{4\pi p_{\mathrm{inj}}^{3}}\:\exp\left(-\int_{p_{\mathrm{inj}}}^{p}\frac{\mathrm{d}p'}{p'}\:\frac{3\left(U_{p'}+\Phi_{0}\left(p'\right)\right)}{U_{p'}-R_{\mathrm{tot}}^{-1}}\right)\label{eq:f0(p,U(p))_int_esc}
\end{equation}
where we have noted the normalized escape flux
$\Phi_{0}\left(p\right) = {\phi_{0}\left(p\right)}/{\left(u_{0}\,f_{1}\left(p\right)\right)}$.
According to~\cite{2010APh....33..307C}, to a very good approximation
we have 
\begin{equation}
\frac{1}{\Phi_{0}\left(p\right)}=\int_{0}^{x_{\max}}\mathrm{d}x'\,\frac{u_{0}}{D\left(x',p\right)}\exp\left(-\int_{0}^{x^{'}}\mathrm{d}x"\frac{u\left(x"\right)}{D\left(x',p\right)}\right)\:.\label{eq:W(p)}
\end{equation}
The net flux of energy through the upstream boundary is
\begin{eqnarray}
F_{\mathrm{esc},0}&=&\int_{p_{\mathrm{inj}}}^{p_{\max}}K(p')\,\phi_{0}\left(p'\right)\,4\pi p'^{2}\mathrm{d}p'\nonumber \\
&=&4\pi\,m_{p}c^{2}\int_{p_{\mathrm{min}}}^{p_{\max}}\left(\sqrt{1+p'^{2}}-1\right)p'^{2}\,\phi_{0}\left(p'\right)\,\mathrm{d}p'\label{eq:Fesc(phi_esc)}
\end{eqnarray}
where $K(p)$ is the kinetic energy of a particle of momentum~$p$
(expressed in $m_{p}c^{2}$ units in the right expression).

If one integrates Eq.~(\ref{eq:transport_f}) from sufficiently
far upstream, one can assume that the upstream gradient of particles
vanishes, so that $\phi_{0}=0$. One can still compute the flux of
escaping energy, by requesting that the particle distribution vanishes
at the maximum momentum~$p_{\max}$: 
$f\left(p_{\max}\right)=0$.
From~\cite{2009MNRAS.396.2065C}, this condition imposes that 
\begin{equation}
F_{\mathrm{esc},0}=\frac{4\pi}{3}\,\left(u_{2}-u_{0}\right)\,p_{\max}^{3}\,K\left(p_{\max}\right)\,f\left(p_{\max}\right)
\label{eq:Fesc(pmax)}
\end{equation}
where $K(p)$ is the kinetic energy of a particle of momentum~$p$.

Finally we note that, to obtain all the required relations between
hydrodynamic and kinetic quantities, we have only made use of the conservation
of mass and of the conservation of momentum. Once the particle distribution,
shock velocity profile, and escape flux have been obtained, the third
conservation law, namely conservation of energy, can be checked, which
provides a way to assess the precision of the model. We write it between
upstream ($x=-x_{\max}$) and downstream of the shock ($x=0^{+}$):
\begin{equation}
\begin{array}{ll}
 & {\displaystyle \frac{1}{2}\rho_{2}u_{2}^{3}+\frac{\gamma_{\mathrm{th}}}{\gamma_{\mathrm{th}}-1}P_{\mathrm{th},2}u_{2}+\frac{\gamma_{\mathrm{cr}}}{\gamma_{\mathrm{cr}}-1}P_{\mathrm{cr},2}u_{2}+F_{\mathrm{w},2}}\\
= & {\displaystyle \frac{1}{2}\rho_{0}u_{0}^{3}+\frac{\gamma_{\mathrm{th}}}{\gamma_{\mathrm{th}}-1}P_{\mathrm{th},0}u_{0}+\frac{\gamma_{\mathrm{cr}}}{\gamma_{\mathrm{cr}}-1}P_{\mathrm{cr},0}u_{0}+F_{\mathrm{w},0}-F_{\mathrm{esc},0}}
\end{array}\label{eq:cons-E}
\end{equation}
where the different terms account for kinetic energy, thermal energy, CR pressure, magnetic waves pressure, and CR escape.

\paragraph{Procedure for solving the coupled problem\label{S:SAN-Blasi-algo}}

As a summary, we outline the practical way for the numerical resolution
of the system. For a given compression~$R_{\mathrm{prec}}$ in the
precursor, the non-linearly coupled system~$\left(f,U\right)$ can
be solved iteratively as follows:\\
\\
\texttt{
compute~quantities~upstream~of~the~sub-shock\\
compute~$R_{\mathrm{sub}}$~and~$R_{\mathrm{tot}}=R_{\mathrm{prec}}\times R_{\mathrm{sub}}$\\
compute~quantities~downstream~of~the~sub-shock~\\
set~injection:~$p_{\mathrm{inj}}$~and~$\eta$, and~$p_{\max}$~\\
set~$f_{1}=0$\\
set~$U_{p}=1/R_{\mathrm{prec}}$\\
(set $\phi_{0}=0$)\\
repeat~until~convergence~of~$\left(f_{1},U_{p}\right)$:\\
\indent compute~$f_{1}$~from~$U_{p}$~(and~$\phi_{0}$)\\
\indent set~~$U_{p}\left(0\right)=1/R_{\mathrm{prec}}$\\
\indent compute~$U_{p}$~from~$f_{1}$\\
\indent (compute~$\phi_{0}$~from~$f_{1}$)\\
}
~\\
We have neglected the Alfv\'enic drift for simplicity, and the lines in parentheses apply only when computing particle escape by imposing a spatial boundary condition.
Possible values for~$R_{\mathrm{prec}}$ range from~$1$ (no precursor,
the shock is not modified, $M_{S,1}=M_{S,0}$ and $R_{\mathrm{sub}}=R_{\mathrm{tot}}$)
to some value~$R_{\mathrm{prec,max}}$ obtained by requesting that~$R_{\mathrm{sub}}=1$
(limit case of a totally smoothed shock, all the compression is done
in the precursor). For a pure hydrodynamical shock,
$R_{\mathrm{sub}}\rightarrow1$ is equivalent to~$M_{S,1}\rightarrow1$
(see Eq.~(\ref{eq:Ms,1(Rsub)_hydro})), although this is no
longer true for a magnetized shock, when taking into account the pressure~$P_{\mathrm{w},1}$
of waves (see Eq.~(\ref{eq:Ms,1(Rsub)_waves})), that reduces
the compressibility of the medium. 
Note that when $P_{\mathrm{w},1}^{\star}$ reaches 1/2, a~shock can no longer form. 

For each value of $R_{\mathrm{prec}}$, the couple $\left(f_{1},U_{p}\right)$
will be accepted as a solution of the model if and only if $U_{p}\left(p_{\max}\right)=1$. 
Note that a solution may be found for more than one value of $R_{\mathrm{prec}}$.
For most of the parameter space, a single solution is found, but sometimes
three solutions are found. One corresponds to a weakly modified shock,
while the other two correspond to significantly modified shocks 
(see \citealt{2005MNRAS.361..907B} and \citealt{2008MNRAS.385.1946A}). 
Multiple solutions for CR-modified shocks had already been observed before 
(using completely different methods, see Sect.~\ref{S:MHDM}) 
but their physical meaning is unclear. In reality only a single solution
will be realized, it is commonly assumed that the others will be suppressed
because they are not stable. At this point we should keep in mind
that Blasi's NLDSA model is not time-dependent, and thus cannot describe
how the modified shock structure progressively takes shape.
This requires numerical simulations of the kind presented in the previous Sects.~\ref{S:MHD-KIN} and \ref{S:PICMHD}.

\newpage
\section{Summary and conclusions}\label{S:CON}

This review addresses the numerical techniques developed in the community of high-energy astrophysics and high-energy lasers to investigate non-thermal particle acceleration and transport in magnetized turbulent flows. 
We first review the main theoretical frameworks developed for the study of particle acceleration in astrophysical flows: diffusive shock acceleration, shock drift and shock surfing processes, stochastic acceleration, and also provide a short survey of recent developments in the field of laser plasmas. We do not cover the process of shear acceleration, the reader can advantageously consult the work of \citet{2006ApJ...652.1044R} for further reference. 
We then detail the technical numerical techniques necessary to investigate problems which appear in the kinetic treatment of particle acceleration. We only give at this stage a short introduction on hybrid methods, the reader is referred to \citet{2002hmst.book.....L} for further details. Kinetic problems in cosmic ray physics can also be treated in the framework of Fokker-Planck or diffusion-convection equations. The Fokker-Planck model finds many astrophysical applications, from the study of cosmic ray transport in the Galaxy to the study of hot plasmas around compact objects. We then focus on recent developments in the theory of particle acceleration at collisionless non-relativistic or relativistic shocks and in reconnection sites based on particle-in-cell and/or hybrid simulations. 
The final part of the review addresses large-scale particle transport and acceleration studies mostly in the magnetohydrodynamic approximation. We review the rapid developments of numerical techniques coupling MHD with the kinetic description of non-thermal components. We end with the developments made to find semi-analytical solutions of the diffusion-convection equation in the context of CR acceleration at shocks.
For completeness, we recommend interested readers to consult some recent and complementary reports and monographs on particle-in-cell methods and Vlasov methods \citep{2018LRCA....4....1P}, hybrid methods \citep{2002hmst.book.....L}, and magnetohydrodynamics \citep{2015LRCA....1....3M, 1998cmaf.conf....1L}.\\

The subject of energetic particle acceleration and transport in turbulent flows is rapidly growing thanks to the increase in computational power. This applies to standard techniques for catching the propagation of energetic particles like particle-in-cell simulations. Beside this, we have seen that the real challenge is to handle the dynamics over the space, time and energy scales of the high-energy phenomena in Astrophysics. This is mandatory because as the highest energetic particles are accelerated they trigger magnetic perturbations necessary to the acceleration of lower energetic particles. This back-reaction requires numerical tools able to treat the inter-connections between large and micro scales. The recent effort in developing PIC modules in MHD codes goes in this direction. This is also true for magnetic reconnection. The microphysics of current sheets depends on the way the magnetic field lines are forced to reconnect by large scale motions. Here it seems important to have simulations which combine MHD and kinetic simulations. One major difficulty remains however to control the numerical noise inherently related to PIC simulations (either due to Cherenkov radiation, or as in the case of PIC-MHD method due to the perturbations generated by the energetic particles themselves). An alternative resides in using a Vlasov approach \citep{2018FrP.....6..113G}, but this possibility remains limited by numerical resources to investigate multi-dimensional problems properly. This aspect is crucial to a proper description of particle acceleration and turbulence around shocks and in reconnection zones. One way to make progress beyond the increase of computational power is to combine different numerical techniques to investigate different regions, as it is the case with implicit PIC simulations coupled with MHD solvers \citep{2017CoPhC.221...81M, 2015JCoPh.283..436R}. Another challenge is to adapt the simulations developed for Newtonian flows to special and now general relativistic cases. This aspect is of particular importance since the era of multi-messenger Astrophysics is now a reality, in a near future we will obtain an unprecedentedly accurate description of high-energy particle sources with the advent of high precision/high sensitivity gravitational wave, neutrino and gamma-ray detectors. 

A positive aspect we can see is that the different numerical tools discussed in this review are and will be further routinely used by very different communities, i.e. high-energy astrophysics, high-energy laboratory plasmas and space plasmas to study the energetic events from the Sun. These converging interests will undoubtedly contribute to the emergence of new fruitful interdisciplinary research subjects. 

\begin{acknowledgements}
The authors thank Fabien Casse and Laurent Gremillet for several fruitful discussions. RW thanks Mickaël Melzani who contributed a lot to his knowledge of magnetic reconnection and Doris Folini for profound discussions and proof-reading. AM and MG thank Anna Grassi for her important contributions to Secs. 2.6 and 3.1-3.4. MG also thanks Simon Bolanos for his input on magnetic reconnection experiments (Sec. 2.6.4), Francesco Califano, Daniele Del Sarto and Alain Ghizzo for various discussions on Vlasov codes (Sec. 3.4) and the SMILEI dev-team for various discussions on the PIC method. This work is supported by the ANR-14-CE33-0019 MACH project. We acknowledge support by the Programme National de Haute Énergie (PNHE).
\end{acknowledgements}

\newpage
\bibliographystyle{spbasic}
\bibliography{review}

\end{document}